\documentclass[12pt]{book}

\usepackage{amsmath}
\usepackage{amssymb}
\usepackage{amsfonts}
\usepackage{graphicx}
\usepackage{psfrag}
\usepackage{paralist}
\usepackage[numbers, sort&compress]{natbib}
\usepackage{amsthm}
\usepackage{subfigure}
\usepackage{ifthen}
\usepackage[toc]{glossaries}
\usepackage{t1enc}
\usepackage[latin2]{inputenc}
\usepackage[magyar,english]{babel}

\makeglossaries

\newcommand{\tcpip}{TCP/IP}
\newcommand{\tcp}{TCP}
\newcommand{\ip}{IP}
\newcommand{\udp}{UDP}
\newcommand{\ecn}{ECN}
\newcommand{\ftp}{FTP}
\newcommand{\tftp}{TFTP}
\newcommand{\dns}{DNS}
\newcommand{\smtp}{SMTP}
\newcommand{\nntp}{NNTP}
\newcommand{\http}{HTTP}
\newcommand{\www}{WWW}
\newcommand{\ssh}{SSH}
\newcommand{\telnet}{TELNET}
\newcommand{\rtt}{RTT}
\newcommand{\rto}{RTO}
\newcommand{\ack}{ACK}

\newcommand{\nstwo}{\textsf{ns-2}}

\newcommand{\integ}{\int\limits}

\newcommand{\cwnd}{congestion window}
\newcommand{\FRFR}{fast recovery/fast retransmit}
\newcommand{\ph}{\lambda}

\newcommand{\avg}[1]{\left\langle #1\right\rangle}
\newcommand{\wbl}{{W_{\textrm{b.l.}}}}
\newcommand{\wal}{{W_{\textrm{a.l.}}}}

\newcommand{\R}{\mathbb{R}}
\newcommand{\N}{\mathbb{N}}

\newcommand{\A}{A\bigl(\ph\tilde{B}^{m+1}/\alpha\bigr)}

\newlength{\figwidth}
\newcommand{\NN}{\mathbb{N}}

\newcommand{\PP}{\mathbb{P}}

\newcommand{\Att}{A}
\newcommand{\probi}[2]{\PP_{#1}(#2\mid \tau_e)}
\newcommand{\prob}[2]{\PP_{#1}(#2)}
\newcommand{\ccdf}[2]{\bar{F}_{#1}(#2)}
\newcommand{\transprob}[1]{W_{#1}}
\newcommand{\transprobq}[1]{W'_{#1}}

\newcommand{\Ordo}[1]{\mathcal{O}\left(#1\right)}

\newcommand{\mean}[2]{\mathbb{E}_{#1}\left[#2\right]}
\newcommand{\meanq}[2]{\mathbb{E}_{#1}\left[#2\mid q\right]}
\newcommand{\meann}[2]{\mathbb{E}_{#1}\left[#2\mid n\right]}
\newcommand{\poch}[2]{\left(#1\right)_{#2}}
\newcommand{\sumterm}[1]{\Phi_\alpha(#1)}
\newcommand{\Perf}{Q}

\newacronym{ack}{\ack}{acknowledgment packet}
\newacronym{aimd}{AIMD}{additive increase, multiplicative decrease}
\newacronym{as}{AS}{Autonomous System}
\newacronym[description=Barab\'asi--Albert model]{ba}{BA}{Barab\'asi--Albert}
\newacronym{ca}{CA}{congestion avoidance}
\newacronym{ccdf}{CCDF}{complementary cumulative distribution function}
\newacronym{cdf}{CDF}{cumulative distribution function}
\newacronym{cwnd}{cwnd}{congestion window}
\newacronym[description=Erd\H{o}s--R\'enyi model]{er}{ER}{Erd\H{o}s--R\'enyi}
\newacronym{erd}{ERD}{Early Random Drop}
\newacronym{frfr}{FR/FR}{fast recovery, fast retransmission}
\newacronym{iid}{IID}{independent and identically-distributed}
\newacronym{isp}{ISP}{Internet Service Provider}
\newacronym{ip}{IP}{Internet Protocol}
\newacronym{lan}{LAN}{Local Area Network}
\newacronym[description=Network Simulator version 2]{ns}{\nstwo}{Network Simulator}
\newacronym{red}{RED}{Random Early Detection}
\newacronym{rto}{RTO}{retransmission timeout}
\newacronym{rtt}{RTT}{round-trip time}
\newacronym{rwnd}{rwnd}{receiver's advertised window}
\newacronym{ssthresh}{ssthresh}{slow start threshold}
\newacronym{tcp}{\tcp}{Transmission Control Protocol}
\newacronym{udp}{\udp}{User Datagram Protocol}
\newacronym{wan}{WAN}{Wide Area Network}
\newacronym{www}{WWW}{World Wide Web}

\DeclareMathOperator*{\Res}{Res}

\begin{document}
\frontmatter
\begin{titlepage}
\begin{center}
{\large
\textsc{Dissertation}\\
for the degree of\\
\textsc{Doctor of Philosophy}\\
in\\
\textsc{Physics}}\\[1em]

{\huge \bf Traffic Dynamics of\\[0.2em]
Computer Networks}\\[2em]

{\LARGE \sc Attila Fekete}\\[2em]

{\Large Supervisor: Prof.~G\'abor Vattay, D.Sc.}\\[2em]

{\large E\"otv\"os Lor\'and University, Faculty of Science\\
Graduate School in Physics}\\[0.5em]
{\large Head: Prof.~Zal\'an Horv\'ath, MHAS}\\[2em]

{\large Statistical Physics, Biological Physics and\\
Physics of Quantum Systems Program}\\[0.5em]
{\large Head: Prof.~Jen\H{o} K\"urti, D.Sc.}\\[1em]
\vfill

\resizebox{3cm}{!}{\includegraphics{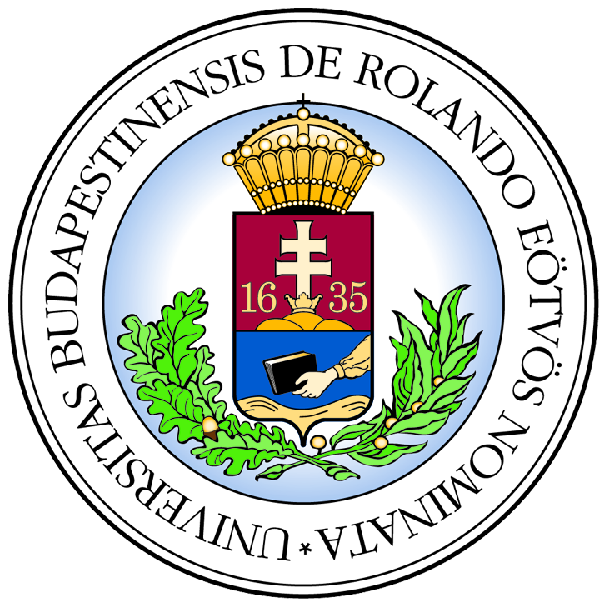}}\\[1em]
{\large Department of Physics of Complex Systems\\
E\"otv\"os Lor\'and University\\
Budapest, 2008}
\end{center}
\end{titlepage}

\cleardoublepage
\vspace*{\fill}

\begin{flushright}
\begin{tabular}{r}
\textsc{\large For little Bori}\\
\hline
\end{tabular}
\end{flushright}
\vfill
\vfill

\chapter*{Acknowledgments}

First and foremost I would like to thank my supervisor Prof. G\'abor Vattay for
his help in guiding me through my research.  I also thank him for his patience
for waiting until I finally finished this thesis.  I would also like to thank
Prof.  Ljupco Kocarev for his kind invitation to the University of California,
which was an invaluable experience.  I am also grateful to the members of the
Department of Physics of Complex Systems for their courtesy.  I am deeply
indebted to M\'at\'e Mar\'odi for many fruitful discussions and his comments on
my thesis.  I would also like to express my sincerest thanks to the staff of
Collegium Budapest for the peaceful atmosphere and unflagging support. 

Without the comfort, help and encouragement of my family I would not have been
able to accomplish my study.  I thank my wonderful wife for her love, her
wholehearted support, for proofreading the manuscript, and for lending me her
favorite desk.  I thank my little daughter Bori for the joy of being with her,
for providing me with extra energy and for proving that I do not need that
much sleep at all.  I am also thankful to my parents and to my brother for
encouraging me at all times.  I am also grateful to my in-laws for their
selfless assistance, especially for the continual baby-sitting.  Special thanks
go to Thomas Cooper for professional proofreading.

Last but not least I would like to thank the permanent members of the ``Tarokk
Department'', and my other friends for tolerating my prolonged absence from
their social life.


\tableofcontents
\mainmatter

\chapter{Introduction}
\label{cha:Introduction}

The objects, laws, and phenomena of Nature have been the subject of physics for
hundreds of years \cite{Simonyi78}.  In the second half of the 20th century,
however, new interdisciplinary and applied branches of physics were developed
that merged a wide range of scientific disciplines with physics including
economy, biology, chemistry and geology.  Most of the new branches of physics
could not have evolved as they did without a specific new technological
invention, namely computer technology, which developed independent from and
parallel to physics.  With the help of computers new research methods became
available, e.g., time-series analysis, computer simulation, and data mining. 

As the use of computers was spreading across the globe, computers themselves
not only became increasingly useful tools for the research community, but their
evolving network attracted growing academic interest.  As a research tool the
computer become the subject of research itself. In a pioneering work by Csabai
in 1994 \cite{Csabai94} the traffic fluctuations of the then Internet as it
existed at the time was investigated.  The author found that the power spectrum
of the traffic delays is $1/f$-like, similarly to other collective phenomena,
e.g., highway traffic.  Nowadays, a new interdisciplinary science is forming to
explore and model complex networks \cite{Barabasi02}, in particular the
Internet. 

The Internet is an exceptional example of complex networks in a number of
aspects.  Firstly, the structure of complex networks is often the subject of
research.  The Internet's infrastructure makes it possible to carry out
measurements on the network cheaply and easily on an incomparable scale.
Secondly, data traffic runs in the network, which adds another level of
complexity to the system. Thirdly, the Internet is a human engineered physical
network, which matches the complexity of some biological systems.  

A useful mathematical abstraction of a network is a graph, because the number
of elements of real networks is finite. However, the number of elements of
complex networks is too large for the individual consideration of each network
element.  Moreover, the exact principles governing connections between
different network components are usually unknown.  Therefore, one should rely
on statistical methods, specifically the tools of statistical physics, in order
to describe the structure of complex networks.

The optimization of traffic performance has great practical importance.
The data flows can be regarded as interacting dynamical systems superposed onto
the network infrastructure.  The theory of dynamical systems can therefore
prove to be a useful tool for studying network traffic.

All in all the Internet is an interesting new area of academic research and
several well established tools of physics can be quite useful for studying it.
Since the Internet has many layers, a number of different components and,
moreover, it is in constant development, it would be an impossible task to
cover all aspects of its operation.  Instead, I will concentrate on the
dynamical modeling of the \gls{tcp}, the most important traffic regulatory
algorithm of the current Internet.  After the introductory chapter where the
most important concepts of the Internet are introduced I begin my survey with
the investigation of \gls{tcp} operating on an elementary network configuration: a
single buffer serving a link.  This scenario comprises the building blocks of
Internet traffic.  I proceed further with the refinement of the first model,
and I consider the finite storage capacity of routers in the next chapter.
After the analytic and simulation study of the previous elementary single
buffer models a more complex model of the Internet follows.  Specifically, in
the last chapter I examine the problem of efficient capacity distribution in a
growing tree-like network.

\section{The Internet}
\label{sec:Internet}

\subsection{The short history of the Internet}
\label{subsec:InternetHistory}

The price and sheer size of the first computers restricted their applicability
in the military and academic sphere. Motivated by the military needs of the
United States in the cold war era a novel concept, the theory of
packet-switching, was proposed by the Advanced Research Projects Agency (ARPA)
to connect distant computers in a decentralized manner.  The concept of
packet-switching means that, contrary to connection-based circuit-switching,
resources are not reserved for communication between host and destination, but
data is split into small datagrams which are transmitted through the network
individually.  The first physical network was constructed in 1969 between four
US Universities: the University of California Los Angeles, Stanford Research
Institute, University of Utah and University of California Santa Barbara.  This
small network, called \textsc{ArpaNet}, is commonly perceived as the origin of
the current Internet.  Over the course of the following years the network grew
gradually and connected more and more universities.  By 1981 the number of
hosts had grown to more than 200.

Based on ARPA's research, and that of its successor DARPA the International
Telecommunication Union (ITU) started developing the packet-switched network
standards.  In 1976 the ITU standard was approved as \textsf{X.25}, and
provided the basis of the international and public penetration of packet
switched network technology.  Using the \textsf{X.25} and related standards, a
number of industrial companies created their own networks.  The most notable
was the first international packet-switched network, referred to as the
International Packet Switched Service (IPSS).  In 1978 IPSS was launched in
Europe and the US with the collaboration of the British Post Office, Western
Union International and Tymnet.  By 1981 it covered Europe, North America,
Australia and Hong Kong.  The \textsf{X.25} standard also allowed the
commercial use of the network, as opposed to ArpaNet, which being a government
founded project restricted its use to military and academic purposes.

In the first packet switched networks the network infrastructure itself assured
reliable packet transfer between hosts.  This approach made it impossible to
connect different networks with different network protocols.  In order to
overcome this difficulty a novel concept of internetwork protocol, the
\gls{tcp}, was developed.  With \gls{tcp} the differences between different
network protocols were hidden and the hosts became responsible for the
reliability of the data transfer.  The first specifications of \gls{tcp} were
given in 1974.  After several years of development and testing the \gls{tcp}
standards were published in 1981.  This paved the way for the current Internet.
Since then every subnet of the Internet has adopted \gls{tcp}.  A detailed
introduction to the protocol will be presented in the next section. 

The pure network infrastructure would have been useless without user
applications.  The basis of many early Internet applications was Unix to
Unix Copy (UUCP), developed in 1979.  The most notable services using UUCP
were electronic mail, Bulletin Board Systems (BBS) and Usenet News.  At the
dawn of the Internet era the most important service was, without doubt, email.
Most of the early Internet traffic was generated by emails, but even in recent
years email constitutes a significant share of Internet traffic.  BBS and
Usenet services were popular among home users with slow modem connections who
did not have direct Internet connections.  Messages, news, articles, programs
or data could be uploaded and/or downloaded after the user dialed into a
server.  BBS and Usenet servers then periodically exchanged data via UUCP.

By the beginning of the 1990's BBSs and Usenet had declined in importance,
mainly due to the new information medium, the \gls[format=textbf]{www}.  The
\www\ was born of the merging of the Internet and the paradigm of hypertext in
the European particle physics laboratory, the CERN.  The \www\ started
conquering the Internet after the debut of the Mosaic web browser in 1993.

The Mosaic browser was such an enormous success that it even affected the
development of the Internet itself.  The Internet crossed the borders of the
academic and industrial research domain and opened up to the wider public.
The process was accelerated by rapid technological advances in computer
technology that made personal computers a part of people's everyday lives.
The combined effect of the above led to the Internet boom in the 1990's, when a
whole new industry formed around the Internet.  

By now the Internet has expanded even further than computers.  Internet
telephony (Voice over IP), mobile Internet (GPRS, UMTS), web cameras, wireless
networks, personal digital assistants (PDAs), and sensor networks are a few
examples of the current trends.  The new technologies make both the structure
and the traffic of the Internet more and more complex.  I review these issues
in the following.

\subsection{The structure of the Internet}
\label{subsec:InternetStructure}

Since the development of the Internet was not regularized by any central
authority and it has been influenced by a number of random effects the
structure of the network is highly irregular.  Nevertheless, the Internet can
be divided into smaller segments, called \glspl[format=textbf]{as}.  Each
AS is administered by a separate organization, e.g. a university, an
\gls{isp}, or a government, and is usually organized in hierarchical, tree-like
structure.  ASs are connected to one another via the \emph{Internet
backbone}.  The Internet backbone is built from high capacity links, currently
up to a couple of $10 \mathrm{Gbps}$.  On the other end of the hierarchy end
users connect to the network.  The available bandwidth for end users can be in
the range of $56 \mathrm{Kbps}$ modems to $20 \mathrm{Mbps}$ business ADSL.  If
we consider AS as the unit of the network, and interconnections between
them as links, then we speak of AS level topology.

Internet also can be viewed on a much smaller scale consisting of two basic
components: nodes and links.  Nodes are devices, e.g. computers, cell phones,
PDAs, routers, switches or hubs, and links are connections between them, e.g.
cable (Ethernet, optical fiber), radio (WiFi, Bluetooth), infrared (IrDA), or
even satellite connections.  Those nodes which have multiple connections must
decide in which direction they forward the through traffic.  These nodes are
usually referred to as \emph{routers}.  This detailed view is called the router
level topology.

Internet topology has been studied both on AS
\cite{PastorVazquezVespignani01,SiganosFaloutsosFaloutsosFaloutsosFaloutsos03}
and router level
\cite{FaloutsosFaloutsosFaloutsos99,CaldarelliMarchettiPietronero00,
YookJeongBarabasi02,LiAldersonWillingerDoyle04}. On both level the Internet can
be modeled as a \emph{graph} from graph theory.  One of the most fundamental
quantities used for describing the structure of a graph is the degree sequence,
which is to say the number of the neighbors of nodes.  It has been found that
the distribution of the degree sequence follows a power law $P(k)\sim
k^{-\delta}$ on both level.  The appearance of a power law indicates the
scale-free nature of a particular object, so a graph the degree distribution of
which follows power law is called \emph{scale-free graph}.  Note that in recent
years the statistical properties of other scale-free networks have been
investigated by the physics community as well
\cite{Strogatz01,AlbertBarabasi02,DorogovtsevMendes02,Newman03}.  Examples of
such networks vary from social interconnections and scientific collaborations
\cite{Newman01} to the \gls{www} \cite{BarabasiAlbertJeong00}.

\begin{figure}[tb]
  \begin{center}
    \includegraphics{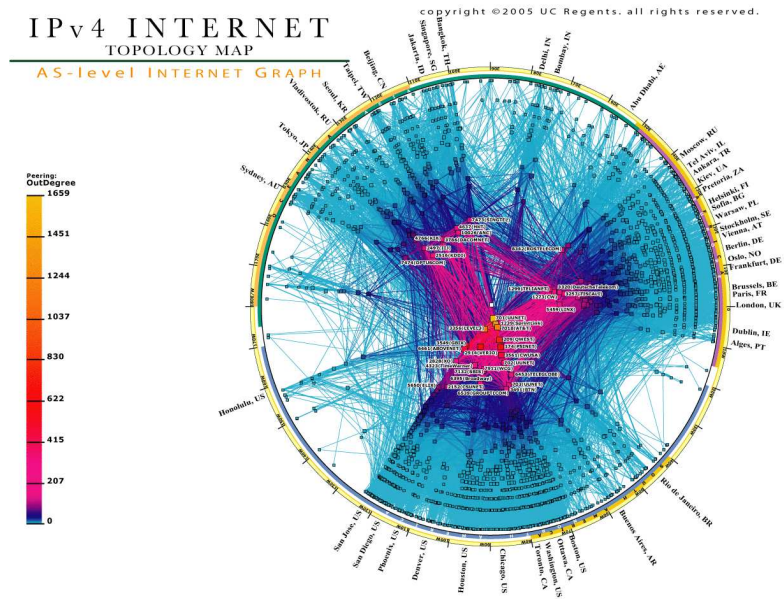}
  \end{center}
  \caption{Internet AS level topology collected between 4--17 April 2005 
  by CAIDA~\cite{CAIDA_map}.  The angular position of nodes corresponds 
  to the geographical longitude of the AS headquarters.  The radial 
  position is calculated from the out-degree of ASs.}
  \label{fig:CAIDA_map}
\end{figure}
Several projects have been launched over the past decade in order to map the
Internet topology.  For example, the Macroscopic Topology Measurements project
of CAIDA, a research group located at the University of California San Diego,
surveys the Internet continuously with probe packets from a couple of dozen
monitoring hosts.  The visualization of the \gls{as} level map produced by
CAIDA is shown in Fig.~\ref{fig:CAIDA_map}.  Rocketfuel is a Internet mapping
engine, developed at the University of Washington, which aims at discovering
\gls{isp} router level topologies \cite{SpringMahajanWatherall02}.  The engine
makes use of routing tables to focus measurements to certain \glspl{isp},
exploits the properties of \gls{ip} routing to eliminate redundancy, and uses data
from nameservers in order to classify routers.

It should be noted that the known Internet topology is only a sample of the
real one.  The surveyed topology is obtained from measurements, mostly via a
program called \texttt{traceroute}.  The program can discover routes between
the \texttt{traceroute} source and given destination hosts.  Since the number
of sources is limited only a section of the real network can be visible in one
experiment.  It is therefore questionable whether the observed topology
resembles the actual Internet topology.  Recently it has been shown that a
\texttt{traceroute}-based experiment can produce strong bias towards scale-free
topology \cite{LakhinaByersCrovellaXie03}, especially when the number of
sources is one or two.  Moreover, it has been shown that a badly designed
measurement can show scale-free topology even if the original network is
regular \cite{AchlioptasClausetKempeMoore05}. 

\subsection{Traffic on the Internet}
\label{subsec:InternetTraffic}

The properties of the Internet traffic are as important as the structure of the
network itself.  Since the time-scale of the evolution of the network
infrastructure is much larger than the time-scale of the traffic flow the
network infrastructure can be considered as a static background behind the
traffic dynamics.  In comparison with the changes in the network traffic the
dynamic changes in the network structure can be neglected.

The Internet traffic is governed by communication protocols, which can be
classified into separate abstract layers according to their functionality.
Each layer takes care of one or more separate tasks of data transfer and
handles data towards a lower or an upper layer.  User applications usually
communicate with the topmost layer, whilst the lowest layer deals with the
physical interaction of the hardware.

The most important classification regarding the Internet is the \tcpip\
protocol suite \cite{Comer88,RFC1180}, which includes five or four layers.  A
more general and detailed model is the OSI model, which includes seven layers.
The concept of layers is quite important, since it provides transparency for
user applications in a very heterogeneous environment.  In order to overview 
the mechanisms behind the Internet traffic let us introduce the four-layer 
model of the \tcpip\ suite:
\begin{itemize} 
  \item The topmost, fourth layer of \tcpip\ suite is called \emph{Application
    layer}.  It provides well-known services such as \telnet, \http, \ftp, 
    \ssh, \dns, and \smtp.  User programs should provide data to an application
    layer protocol in a suitable format.
  \item The next layer is the \emph{Transport layer}, which is responsible
    among other things for flow control, error detection, re-transmission, and
    connection handling.  The two most important protocols in this layer are
    \gls{tcp} and \gls{udp}, which will be discussed in more detail in
    Section~\ref{sec:TransportMechanisms}.  They represent two conceptually
    quite different transport mechanisms:  \gls{tcp} provides reliable,
    connection-based data transfer, while \gls{udp} serves as an unreliable,
    connectionless, best effort transport mechanism.  Other protocols at this
    layer are SCTP developed for Internet telephony, and RTP
    designed for real-time video and audio streaming.
  \item The following layer is referred to as the \emph{Internet layer}.  This
    layer solves the problem of addressing and routing of packets.  The 
    \gls[format=textbf]{ip} and the obsolete \textsc{x.25} protocols reside in 
    this layer.  \ip\ hides the details of the network infrastructure, and allows 
    the interconnection of different network architectures.
  \item Finally, the lowest layer of the \tcpip\ suite is the \emph{Network
    access layer}, which handles physical hardware and devices.  Notable
    examples on this layer are the ethernet, WiFi, and modems.
\end{itemize}

In order to understand the workings of the Internet, let us take the example of
a typical Internet application:  let us suppose that Alice wants to download a
file from Bob.  Since Alice wants to get an exact copy of the file, she starts
an \ftp\ session.  First, the \ftp\ protocol builds a connection between the
two computers.  Then the file is split into small datagrams, which are passed
on to the \gls{tcp} protocol on Bob's computer.  The \gls{tcp} protocol adds a
header to the datagrams, including a sequence number, a timestamp, and some
other information which ensures reliability.  Then \gls{tcp} passes the
datagrams on to the \gls{ip} protocol, which adds its own header.  The \gls{ip} header
contains addressing information.  The resulting \gls{ip} packet is put into the
outgoing queue of Bob's computer.  If the queue is empty, then the packet is
sent to the Network Interface Card (NIC), otherwise it has to wait until the
preceding packets have been served.  The NIC card disassembles the packet into
ethernet frames and puts them onto the physical cable.  The frames travel to
the default router in Bob's network and the router's NIC assembles them back
into an \gls{ip} packet.  Based on the destination address in the \gls{ip} header, the
router decides in which direction the packet should be forwarded and the packet
is put into the outgoing queue of the corresponding direction.  The packet is
then disassembled and transferred again over the next cable. The procedure is
repeated until the packet arrives at its final destination.  The actual method
of data transfer on the \emph{Network access layer} can differ from the above
mentioned ethernet method.  If Alice uses a dial-up connection, for instance,
the last step of the packet's path is over a telephone line via a modem.  At
Alice's computer the \gls{ip} protocol takes the packet and passes it on to the
\gls{tcp} protocol.  The \gls{tcp} acknowledges the packet and inserts it into
the missing part of the file.  Finally, when Alice's computer has received all
the pieces of the file, the \ftp\ protocol saves the whole file to its
destination on her computer.

Although both packet-switched and connection-based data transfer are present in
the above example, the Internet is called a packet-switched network because the
\emph{Internet layer}, which is the fundamental core of the Internet, utilizes
solely packet-switched technology.  Other layers can be either packet or
circuit-switched.  Ethernet traffic is packet-switched, for example, but modem
traffic is carried through circuit-switched telephone lines.  Higher level
protocols (e.g. \ftp, \telnet, \ssh) are usually connection oriented, too.

Let us study the \emph{Internet layer} in more detail.  First of all, packets
are injected into the \emph{Internet layer} randomly by higher level protocols
at certain source nodes.  Then packets are served sequentially and forwarded to
neighboring nodes by routers or, if they have arrived to their destination,
removed from the network. If a router is busy serving a packet then any
incoming packet is placed into a buffer and has to wait for serving.  If the
queue in the buffer has reached the buffer's maximum capacity then all incoming
packets are dropped until the next packet in the queue is served and an empty
space becomes available in the buffer.  The event when a buffer becomes full is
called \emph{congestion}.  The above described router policy, called
\emph{drop-tail}, is the most wide-spread nowadays.  Other router policies are
also in use.  The \gls[format=textbf]{erd} and \gls[format=textbf]{red}
polices, for instance, drop incoming packets randomly before the buffer becomes
fully occupied in order to forecast possible congestion to upper level
protocols.  The difference between the two policies is that the drop
probability depends on the instantaneous queue length in the former case and
the average queue length in the latter. It is possible to give priority to
certain packets in order to provide Quality of Service (QoS) for certain
applications, but routers usually serve packets in First In, First Out (FIFO)
order.  The serving rate of packets depends on the actual packet size and the
bandwidth of the link after the buffer.  Packets obviously suffer propagation
delay during their delivery, is a consequence of two factors: link and from
queuing delay.  The former is constant for a given route, but the later varies
randomly with queue lengths along the packet's path.

The product of the link delay and link capacity, in short the
\emph{bandwidth-delay product}, equals the number of packets that a link can
transfer simultaneously.  If this quantity is large compared to the buffer size
then the constant link delay is the dominant constituent of the propagation
delay.  \gls{wan} links are typical examples of this.  On the other hand, if
the bandwidth delay product is small compared to the buffer size then the
varying buffering delay is the dominant component.  Such links can be found in
\gls{lan}.  We will see later that the two scenarios induce different \gls{tcp}
dynamics.

It is evident that queuing theory plays an important role in the modeling of
packet-switched networks in general and the Internet in particular.  However,
queuing theory has been developed much earlier than the advent of
packet-switching technology.  The first motivation and important application of
queuing theory was actually a circuit-switched network, the classical telephone
system.  

The properties of two quantities, namely the inter-arrival and the service
times of customers, affect the behavior of queuing systems most fundamentally.
Other quantities, e.g. the size of the customer population, the number of
operators, the system capacity etc., also have an impact on the behavior of the
system, but they do not affect the essential properties of the queuing system.
Both the inter-arrival and the service time series can be modeled by discrete
time stochastic processes.  It is usually assumed that both the inter-arrival
and the service times are \gls{iid} random
variables.  Furthermore, in the most simple case, both inter-arrival and
service times are memoryless processes, that is they are exponentially
distributed random variables.  This model is called Poisson queue, since both
the number of arrivals and the number of departures in a finite time interval
follow Poisson distribution.  Poisson queues have been studied extensively and
they proved to be excellent models of telephone call centers and telephone
exchange centers.  Most of the arising questions regarding Poisson queues have
been answered analytically \cite{Cooper81}.  

Internet traffic has been analyzed on various layers of the above \tcpip\
suite.  In a pioneering work by \citet{LelandTaqquWillingerWilson94} the
authors collected and studied several hours of ethernet traffic with
$20$--$100\mu s$ resolution.  They found that autocorrelations in the captured
traffic decayed slower than exponential, that is the system has long-range
memory.  This result indicated problems with Poisson queuing models for
packet-switched networks, since in a Poisson queuing system autocorrelations
would fall exponentially \cite{Morse55}.  Furthermore, it has been shown that
the time series of the aggregated Ethernet traffic is statistically
self-similar, and has fractal properties.  \citet{PaxsonFloyd95} studied the
usability of Poisson models for application layer protocols and the
corresponding \gls{ip} traffic.  They found that, though the traffic followed a
24-hour periodic pattern, Poisson processes with fixed arrival rates are
acceptable models for user initiated sessions (\ftp, \telnet) for intervals of
one hour or less.  For machine initiated sessions (\smtp, \nntp), however, the
Poisson model failed even for short time-scales.  Furthermore, packet level
traffic deviated considerably from Poisson arrivals as well.  Similar evidence
has been found in \gls{www} traffic~\cite{CrovellaBestavros97}.  Furthermore,
it has been shown  that the distribution of the packet inter-arrival times
follows power law.  \citet{FeldmannGilbertWillingerKurtz98} have presented the
wavelet analysis of \gls{wan} traffic samples captured around the birth of the
World Wide Web between '90 and '97.  It has been found that as \gls{www}
traffic started dominating the network traffic gradually different scaling
behavior appeared in short- and long-time scales.  The authors concluded that
\gls{tcp} dynamics might be responsible for short-time scaling and application
layer traffic characteristics for long-time scaling.

All the above properties are in strong contrast with the properties of the
Poisson queuing systems, e.g. telephone networks, where both the correlations
and the inter-arrival time distribution decay exponentially.  It implies that
well developed classical models, which provide excellent descriptions of
circuit-switched traffic, are essentially useless for the description of the
Internet.  New traffic models, which provide realistic synthetic traffic, were
required.  A few important traffic models of the Internet will be presented in
Section~\ref{sec:Traffic_modeling_review}.

There are several theories which explain the origins of the observed long-range
dependent traffic.  One explanation can be that the observed traffic is the
superposition of individual effects which happen on separate network layers and
on very different time-scales; from several minutes of user interaction through
a couple of seconds of application response until the microsecond-scale of
network protocol operation.  Further assumptions are that heavy-tailed file
size distribution
\cite{CrovellaBestavros97,WillingerTaqquShermanWilson97,WillingerPaxsonTaqqu98},
or heavy-tailed processor time distribution is behind the phenomena.  There has
also been some debate on whether the \gls{tcp} protocol in itself is able to
generate long-range dependent traffic \cite{VeresBoda00} or not
\cite{FigueiredoLiuFeldmannMisraTowsleyWillinger05}.  The \gls{tcp}'s
exponential backoff mechanism is also a possible source of heavy-tailed
inter-arrival times \cite{GuoCrovellaMatta00}.


\section{Data transport mechanisms}
\label{sec:TransportMechanisms}

The Internet is an enormous data highway between computers, where data packets
play the role of vehicles and links serve as the road system.  As on normal
highways, congestions can form at bottlenecks if the capacity of a junction is
exceeded by the traffic demand. 

The dynamics of the Internet traffic is governed by protocols of the
\emph{Transport layer}.  Protocols on this layer control directly the injection
rate of \gls{ip} packets into the network. Almost all the Internet traffic is
governed by two protocols, namely the \gls{tcp} and the \gls{udp}.   Therefore,
understanding the operation of these protocols is very important from the point
of view of traffic modeling.  For example, fundamental questions are how
distant hosts utilize the network infrastructure and whether they can cause
persistent traffic congestion or not.

The performance of the network can be severely degraded as a result of
persistent congestion.  Congestion should therefore be avoided.  Just such a
congestion collapse did indeed occur in 1986 in the early Internet, when the
useful throughput of NFSnet backbone dropped three orders of magnitude.  The
cause of this collapse was the faulty design of the early \tcp.  Instead
of decreasing the sending rate of packets after detecting congestion, the early
\tcp\ actually started retransmitting lost packets, which led to an
increasing sending rate and positive feedback.

\subsection{The User Datagram Protocol}
\label{subsec:UDP} 

\Gls[format=(textbf]{udp} is a very simple protocol, which provides a procedure for
applications to send messages to other applications with a minimum of protocol
mechanism \cite{RFC768}.  Neither delivery nor duplicate protection is
guaranteed by \udp.  Furthermore, no congestion control is implemented in
it either.  \udp realizes an open-loop control design, that is no feedback
about a possible congestion is processed.

The principal uses of \udp are the Domain Name System (\dns), streaming
audio and video applications (e.g. VoIP, IPTV), file sharing applications, the
Trivial File Transfer Protocol (\tftp), and on-line multiplayer games, to name
a few.

Since \udp lacks any congestion avoidance and control algorithm,
application level programs or network-based mechanisms are required to handle
congestion.  In streaming applications, for example, users are often asked for
the bandwidth of their access link, and \udp packets are sent with the
corresponding fixed rate.  Since \udp does not have any feedback mechanism
congestion collapse of the network due to \gls[format=)textbf]{udp} network overload is
unlikely.  However, aggressive network utilization should be avoided, because
it can block other protocols, mainly \tcp.

\subsection{The Transmission Control Protocol}
\label{subsec:TCP}

\glsadd[format=(textbf]{tcp}
The \tcp\ protocol is complementary to the \udp protocol in many
sense.  Contrary to \udp, \tcp\ is connection oriented, it guarantees
in-order delivery and duplicate protection, congestion control and avoidance.
In addition, \tcp\ is a closed-loop design which can process feedback from
packet delivery.  Accordingly, \tcp\ is a much more complex design than
\udp.  In this section we present an overview of \tcp.

Among the applications using \tcp\ are the \gls{www}, email, Telnet, File
Transfer Protocol (\ftp), Secure Shell (ssh), to name a few.  Since these
applications are responsible for most of the current Internet traffic \tcp\
is the most dominant transport protocol at the moment.  Accordingly,
understanding the workings of the \tcp\ protocol has great importance in
traffic modeling.

Since \tcp\ is connection oriented, it does not start sending data
immediately, like \udp.  Rather it uses a three-way handshake for
connection establishment.  If the connection establishment phase is successful
the data transfer phase follows.  Finally, when all the data has been sent, the
connection is terminated in the final phase.  The connection establishment and
termination phases are usually short and involve only negligible amount of data
compared to the data transfer phase.  I will therefore focus solely on the main
phase, neglecting the other two phases.

In the data transfer phase the \tcp\ receiver acknowledges every arrived
packet by an \gls[format=textbf]{ack}.  The \ack\ contains the sequence number of the last
data packet arrived in order.  If a data packet arrives out of order, then the
receiver sends a duplicate \ack, that is an \ack\ with the same
sequence number as the previous one.  Duplicate \ack s directly notify the
\tcp\ sender about an out-of-order packet.

If all the packets are lost beyond a certain sequence number, then duplicate
\ack\ cannot notify the sender about packet losses.  In order to recover
from such a situation, the \tcp\ sender manages a retransmission timer.  The
delay of the timer, the \gls[format=textbf]{rto}, is updated after each
arriving \ack.  The \tcp\ sender measures the \gls[format=textbf]{rtt},
the elapsed time between the departure of a packet and the arrival of the
corresponding \ack.  The updated value of the \rto\ is calculated from
the smoothed \rtt, and the \rtt\ variation as defined in
\cite{RFC2988}. 

Packets are acknowledged after \rtt\ time period from packet departure if
the transmission is successful.  The data transfer would be very inefficient if
the \tcp\ sender waited for the \ack\ of the last packet before it sent
the next packet.  On the other hand, sending packets all at once would cause
congestion.  In order to reach optimum performance without causing congestion,
\tcp\ manages two sliding windows with the associated variables.  On the
sender side the \gls{cwnd} limits the allowed number of unacknowledged packets.
This way a \gls{cwnd} number of packets is transmitted on average during a
round-trip time period.  Since \gls{cwnd} is used directly for congestion
control it is changed dynamically.  

The other variable, the \gls{rwnd}, is managed on the receiver side.
\Gls{rwnd} is the size of a receiver buffer which can store out-of-order
packets temporally.  The value of \gls{rwnd} is included in every \ack,
though it usually does not change.  Although the limit of the unacknowledged
packets is the minimum of \gls{cwnd} and \gls{rwnd}, the later is usually large
enough not to affect data transfer in practice.  \Gls{rwnd} therefore plays a
much less important role than \gls{cwnd}.  For the sake of simplicity I will
assume that \gls{rwnd} equals infinity.  Accordingly $\min(\textit{cwnd},
\textit{rwnd})$ will be replaced with \gls{cwnd} in all the equations below
where applicable.  Let us keep in mind, however, that this is an approximation.

Internet's packet-switched technology implies that there are no reserved
resources for \tcp.  This approach is also called \emph{best effort}
delivery.  Moreover, the Internet lacks any central management authority.
Accordingly, \tcp\ does not have precise information about its fair share
of the network bandwidth in the ever-changing network conditions.  In the
previous section we have seen that buffers are able to store excess traffic
temporarily, but pockets are dropped when a buffer becomes full.  Flow control,
the alteration of rate at which packets are sent in order to get a fair share
of the network bandwidth without causing severe congestion, is one of the most
important tasks of the \tcp.  This goal is achieved by the continuous
adjustment of the congestion window and eventually the rate at which packets
are sent.

Several \tcp\ variants have been developed in recent years in order to
enhance its performance in different environments
\cite{BarakatAltmanDabbous00}.  These variants differ mainly in the congestion
avoidance algorithm.  The core concept, however, is the same in all \tcp\
variants and has not changed significantly since its first specification in
1974.  The classical \tcp\ variants (e.g.  Tahoe, Reno) try to find the
fair bandwidth share by the following method: for every successfully
transmitted and lost packet they increase and decrease their sending rate,
respectively.  This method is based on the observation that a packet loss is
most likely the result of a congestion event.  Note that these \tcp\
variants obviously cause temporary congestions in the network in the long run.
More recent variants often try to detect upcoming congestions beforehand via
explicit congestion notifications (\ecn) from routers or by detecting
increasing queuing delays from \gls{rtt} fluctuations (e.g. Fast \tcp).

\begin{figure}[tb]
  \begin{center}
    \psfrag{ 0}[c][c][0.8]{$0$}
    \psfrag{ 5}[c][c][0.8]{$5$}
    \psfrag{ 10}[c][c][0.8]{$10$}
    \psfrag{ 15}[c][c][0.8]{$15$}
    \psfrag{ 20}[c][c][0.8]{$20$}
    \psfrag{ 25}[c][c][0.8]{$25$}
    \psfrag{ 30}[c][c][0.8]{$30$}
    \psfrag{ 35}[c][c][0.8]{$35$}
    \psfrag{ 40}[c][c][0.8]{$40$}
    \psfrag{0}[c][c][0.8]{$0$}
    \psfrag{5}[c][c][0.8]{$5$}
    \psfrag{10}[c][c][0.8]{$10$}
    \psfrag{15}[c][c][0.8]{$15$}
    \psfrag{20}[c][c][0.8]{$20$}
    \psfrag{25}[c][c][0.8]{$25$}
    \psfrag{30}[c][c][0.8]{$30$}
    \psfrag{35}[c][c][0.8]{$35$}
    \psfrag{40}[c][c][0.8]{$40$}
    \psfrag{45}[c][c][0.8]{$45$}
    \psfrag{cwnd}[c][c][1]{\emph{cwnd}}
    \psfrag{rtt}[c][c][1]{\rtt}
    \psfrag{Slow start}[r][r][0.8]{Slow start}
    \psfrag{RTO expire}[r][r][0.8]{\rto\ expire}
    \psfrag{Congestion avoidance}[r][r][0.8]{Congestion avoidance}
    \psfrag{max packets}[r][r][0.8]{Maximum packets}
    \psfrag{ssthresh}[c][c][0.8]{\emph{ssthresh}}
    \psfrag{Fast retransmit / Fast recovery}[r][r][0.8]{Fast retransmit / Fast recovery}
    \resizebox{\figwidth}{!}{\includegraphics{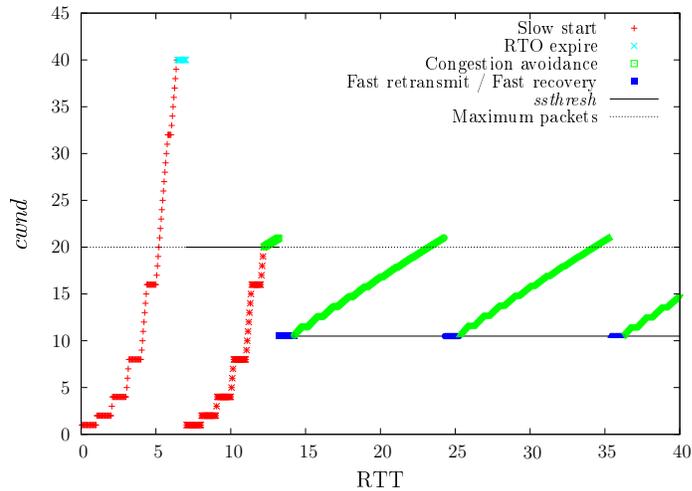}}
  \end{center}
  \caption{Schematic plot of the development of the congestion window.  The
  hypothetical network can handle 20 packets simultaneously, denoted
  by the dotted line.  Cwnd might overrun this limit, because
  congestion is detected only after \rtt\ latency.  Note the small plateaus
  both in the slow start and congestion avoidance phase, which are due to the
  bursty arrival of packets.}
  \label{fig:cwnd-devel}
\end{figure}

I discuss the Reno \tcp\ variant in more detail below, since currently this is
the most widespread variant in use.  Its congestion control mechanism includes
the following algorithms: \emph{slow start}, \emph{congestion avoidance},
\emph{fast recovery}, and \emph{fast retransmission} \cite{RFC2581}.  In
figure~\ref{fig:cwnd-devel} the schematic development of \gls{cwnd} due the
above congestion control algorithms is shown.  There are two slow start periods
at the beginning of the plot.  This is possible due to the wrong initial
estimate of the \gls{ssthresh}.  After the value of \gls{ssthresh} has been set
to approximately half of the maximum window the \gls{frfr} algorithms are able
to take care of the upcoming packet losses.  Note the small steps both in the
slow start and the congestion avoidance phase.  The steps are due to the bursty
departure of packets.

\subsubsection{Slow start and congestion avoidance}

\glsadd[format=(textbf]{ca}
The core of the \tcp\ congestion control mechanism is the slow start and
the congestion avoidance algorithms.  A state variable, the \gls{ssthresh}, is
used to determine whether the slow start or the congestion avoidance algorithm
is used to control data transmission.  When \gls{cwnd} exceeds \gls{ssthresh}
the slow start ends, and \tcp\ enters congestion avoidance.  \Gls{ssthresh}
is recalculated when congestion is detected by the following formula:
\begin{equation}
  \label{eq:ssthresh}
  \textit{ssthresh}=\max(\textit{cwnd}/2, 2).
\end{equation}

The slow start algorithm is used at the beginning of data transfer to probe the
network and determine the available capacity.  Slow start is used after
repairing losses detected by the retransmission timer as well.  In slow start
phase \tcp\ begins sending at most two packets, which is a ``slow start''
indeed.  Despite what the name might suggest, however, the growth of the packet
sending rate in this phase is quite fast actually:  the \gls{cwnd} is increased
by one for every \gls{ack}.  This way the sending rate is doubled in every
\gls{rtt}, which means exponential growth in time.

In congestion avoidance phase \gls{cwnd} is increased by one every
\gls{rtt} period.  This implies linear growth in time, which is a much more 
moderate development than the exponential growth in slow start.  One
common approximating formula for updating \gls{cwnd} after every 
non-duplicate \gls{ack} is:
\begin{equation}
  \label{eq:cwnd}
  \textit{cwnd}\to\textit{cwnd}+\frac{1}{\textit{cwnd}}.
\end{equation}
This formula is not precisely linear in time, but the advantage of this 
formula is that no auxiliary state variable is required for its application.
\glsadd[format=)textbf]{ca}

\subsubsection{Fast retransmit and fast recovery}

\glsadd[format=(textbf]{frfr}
The packet sending rate is reduced drastically at the beginning of each slow
start phase.  Although the slow start algorithm restores \gls{cwnd} to
\gls{ssthresh} at an exponential rate, its application might cause unnecessary
performance deterioration.  In order to circumvent slow start algorithm when
possible, fast retransmit and fast recovery algorithms were introduced to the
Reno version of \tcp\ in 1990 \cite{Jacobson90}.

The fast retransmit algorithm uses the arrival of three duplicate \glspl{ack}
as an indication that a packet has been lost.  After the arrival of the third
duplicate \gls{ack} the sender retransmits the missing segment without waiting
for the retransmission timer to expire.  \tcp\ does not enter slow start
after fast retransmission, but instead starts the fast recovery algorithm.
Skipping slow start is possible because each duplicate \gls{ack} indicates that
a packet has been removed from the network.  Therefore, newly sent packets do
not stress the network further.

After fast retransmission \gls{ssthresh} is set according to
Eq.~(\ref{eq:ssthresh}).  In addition, \gls{cwnd} is halved, 
\begin{equation}
  \label{eq:cwnd_halve}
  \textit{cwnd}\to\frac{\textit{cwnd}}{2}
\end{equation}
and for each duplicate \gls{ack} a new segment is sent if possible.  After the
first non-duplicate \gls{ack} \gls{cwnd} is set to \gls{ssthresh} again, and
\tcp\ returns to congestion avoidance.  Note that slow start might be
forced when \gls{cwnd} is small and duplicate \glspl{ack} are not accessible.
Furthermore, if more than one packet is lost within one \gls{rtt} time period,
then the FR/FR algorithms may not recover from the loss either, and
\tcp\ can enter slow start algorithm instead.  However, if the packet loss
rate is low and \gls{cwnd} is large enough, then the slow start algorithm is
used only at the beginning of the \tcp\ session, and \gls{cwnd} is governed
in an \gls{aimd} manner by the congestion avoidance and FR/FR algorithms,
respectively.  
\glsadd[format=)textbf]{frfr}

The idea behind the \gls{aimd} rule comes from the following simple control
theoretical arguments \cite{JainRamakrishnanChiu88,ChiuJain89}.  In general,
the control of the $\lambda$th \tcp's \gls{cwnd} can be given by
$w_{\lambda}(t_{i+1})=f(w_{\lambda}(t_i), y(t_i))$ where $f(w,y)$ is the
control function, which depends on the feedback (e.g. an \gls{ack}) from the
system $y(t_i)\in\{-,+\}$, and the last value of the window $w_{\lambda}(t_i)$.
The feedback is binary: $+$ and $-$ indicates whether to increase or decrease
traffic demand, respectively.  If we restrict our study to control functions,
which are linear in $w_{\lambda}(t_i)$, then we obtain 
\begin{equation} 
  w_{\lambda}(t_{i+1}) = a_{y(t_i)} + b_{y(t_i)} w_{\lambda}(t_i),
  \label{eq:control_func}
\end{equation}
where the coefficients $a_{\pm}$ and $b_{\pm}$ are constants.  It is obvious
that the control equation (\ref{eq:control_func}) is additive if $b_{\pm}=1$,
and multiplicative if $b_{\pm}\neq1$.  The most important special cases of the
possible control algorithms are collected in Table~\ref{tab:control_algs}.  A
feasible control algorithm must satisfy two important criteria:
\emph{convergence to efficiency} and \emph{fairness}.  Efficiency in this
context means maximum possible usage of the available resources and fairness
means equal share of the bottleneck capacity.  These criteria give constraints
on the coefficients $a_{\pm}$ and $b_{\pm}$.  It has been shown in
\cite{ChiuJain89} that the convergence to efficiency and fairness is provided
by the constraints $a_+>0$, $b_+\ge1$, and $a_-=0$, $0\le b_-<1$.  Moreover, it
has been shown that the convergence is fastest, when $b_+=1$.  Therefore,
the additive increase, multiplicative decrease control, which is implemented in
\tcp, is the optimal control algorithm.
\begin{table} \centering \begin{tabular}{|c|c|c|}
    \hline
                & $b_+=1$ & $b_+>1$ \\
                & $a_+>0$ & $a_+=0$ \\
    \hline
    $b_-=1$   & Additive increase & Multiplicative increase \\
    $a_-<0$   & Additive decrease & Additive decrease \\
    \hline
    $0<b_-<1$ & Additive increase & Multiplicative increase \\
    $a_-=0$   & Multiplicative decrease & Multiplicative decrease \\
    \hline
  \end{tabular}
  \caption{Possible control algorithms with a linear control function.}
  \label{tab:control_algs}
\end{table}

\subsubsection{The backoff mechanism}

Normally in slow start or in congestion avoidance mode, the \tcp\ estimates
the \gls{rtt} and its variance from time stamps placed in \glspl{ack}.  In some
cases the retransmission timer might underestimate \gls{rtt} at the beginning
of the data transfer, and the retransmission timer might expire before the
first \gls{ack} would arrive back to the \tcp\ sender.  In order to avoid
the persistent expiration of the retransmission timer the so-called Karn's
algorithm \cite{KarnPartridge91} is applied.  According to the algorithm, if
the retransmission timer expires before the first \gls{ack} would return, then
the value of the \gls{rto} is doubled.  If the timer expires again, then the
timer is doubled repeatedly a maximum six consecutive times.  Since there is a
definite ambiguity in estimating \gls{rtt} from a retransmitted packet the
\glspl{ack} of two consecutive sent packets should arrive back successfully in
order for the \tcp\ to estimate the \gls{rtt} again and go back to the slow
start mode. 

A similar situation might occur if the packet loss rate is high.  In that case,
consecutive packets can be lost and the \tcp\ might enter the backoff
state, even if \gls{rtt} might actually be smaller than the retransmission
timer.  Since the delay between packet departure is doubled, the effective
bandwidth is halved after each backoff step.  \tcp\ can reduce its packet
sending rate with this method below one packet per \gls{rtt}.

\glsadd[format=)textbf]{tcp}

\chapter{Traffic dynamics in infinite buffer}

In this chapter I study the \tcp\ dynamics on an idealized single buffer
network model where the probability that a packet is lost at the buffer is
negligible compared to other sources of packet loss.  The case of a
semi-bottleneck buffer when the size of the buffer is limited will be discussed
in Chapter~\ref{cha:finite_buffer}.  First, I introduce the important fluid
approximation of \tcp\ congestion window dynamics in
Section~\ref{sec:Fluid_models}.  In recent years many aspects of the \tcp\
congestion avoidance phase have been clarified.  The most important results of
the literature are reviewed in Section~\ref{sec:Traffic_modeling_review}.  I
define the network model under study in Section~\ref{sec:Buffer_model}.  My
results on the analytic study of the \tcp\ congestion window dynamics are
presented in Section~\ref{sec:Window_dynamics}.  The discussion of the model is
given in Section~\ref{sec:Discussion}.  Finally, I summarize my results in
Section~\ref{sec:Conclusions}.

\section{The fluid approximation}
\label{sec:Fluid_models}

The equations of motion (\ref{eq:ssthresh})--(\ref{eq:cwnd_halve}) are defined
at \gls{ack} arrivals.  The state variables are therefore changed in discrete steps
at discrete time intervals (Fig.~\ref{fig:cwnd-devel}), often referred to as
``in \gls{ack} time''.  Note that ''\gls{ack} time'' dynamics is an essential, inherent
property of \gls{tcp}, because it is defined in the \gls{tcp} design and does not
depend on the network environment where \gls{tcp} is used.

In practice the discrete-time equations of \tcp\ dynamics can be approximated
very well by continuous-time equations.  Between two consecutive packet
losses the congestion window is changed according to the fluid ``\gls{ack} time''
equation (\ref{eq:cwnd}), that is
\begin{equation} 
  \label{eq:cwnd_ack}
  \frac{d W}{d t_{\textrm{ACK}}}=\frac1{W}.
\end{equation} 

Since the arrival of \gls{ack} packets is not uniform in time, the \gls{ack} and real
time averages of important quantities, for instance the throughput, are usually
different.  It would be difficult and rather impractical to transform the
dynamics of the state variables from \gls{ack} to real time exactly.  A usual 
approximation is that the arrival rate of \glspl{ack} is estimated by the number of 
packets in flight, that is the congestion window $W$ divided by the round trip 
time $R$: 
\begin{equation} 
  \frac{d t_{\textrm{ACK}}}{d t}=\frac{W}{R}, 
\end{equation} 
From the above equations one can obtain 
\begin{equation}
  \frac{dW}{dt}=\frac1{R}.
  \label{eq:cwnd_fluid}
\end{equation}
which is the fluid approximation of the congestion window dynamics in real 
time.  Although this real time approximation of \tcp\ dynamics is often 
sufficient, I will point out its defects.  I will also present a roundabout
solution to the problems based on the fundamental ``\gls{ack} time'' dynamics of
\tcp.

\begin{figure}[tb]
  \begin{center}
    \psfrag{W [pkt]}[c][c][1.2]{$W [\mathrm{pkt}]$}
    \psfrag{T [RTT]}[c][c][1.2]{$T [s]$}
    \psfrag{0}[r][r][1]{$0$}
    \psfrag{W}[r][r][1]{$W_m$}
    \psfrag{W/2}[r][r][1]{$\frac{W_m}{2}$}
    \psfrag{RW}[r][r][1]{$W_m R$}
    \psfrag{RW/2}[r][r][1]{$\frac{W_m}{2}R$}
    \psfrag{2RW}[r][r][1]{$2W_mR$}
    \psfrag{3RW/2}[r][r][1]{$\frac{3W_m}{2}R$}
    \resizebox{\figwidth}{!}{\includegraphics{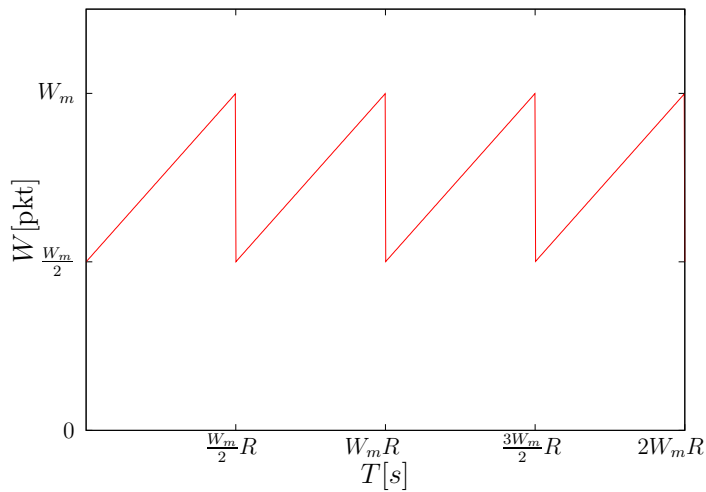}}
  \end{center}
  \caption{The \tcp\ congestion window evolution under 
deterministic packet loss and constant round-trip time.  The congestion 
window $W$ varies linearly between the maximum value $W_m$ and its half, 
$W_m/2$.}
  \label{fig:aimd_example}
\end{figure}

As a simple example of the fluid model let us calculate the average throughput,
the transmitted data per unit of time, of a single \gls{tcp} over a lossy link
\cite{MathisSemkeMahdaviOtt97}.  Let us suppose that the round-trip time is
constant and the packet loss is deterministic.  Considering these assumptions
the congestion window changes at a constant rate between consecutive packet
loss events as (\ref{eq:cwnd_fluid}).  The window is halved after each packet
loss event.  The window evolution shown in Fig.~\ref{fig:aimd_example} is
therefore a periodic sawtooth in the interval $[R W_m/2, R W_m]$ and in the
range of $[W_m/2,W_m]$.  The length of a cycle is $R W_m/2$.  The number of
transmitted packets in a cycle equals the integral of the congestion window for
one period: $N=\frac{3}{8}W_m^2$.
Since in each cycle one packet is lost, the packet loss probability can be
expressed as $p=1/N$.  Therefore, the average throughput $\bar{X}$ can be 
given by
\begin{equation}
  \bar{X}=\frac{P N}{R \frac{W_m}2}
  =\frac{P/p}{R \sqrt{\frac{2}{3p}}}
  =\frac{P}{R} \frac{c_0}{\sqrt{p}}
  \label{eq:loss_formula}
\end{equation}
where $P$ is the size of the data packets and $c_0=\sqrt{\frac{3}{2}}$ is a
constant.  The resulting formula, often referred to as the ``\emph{inverse
square-root law}'', expresses the impact of a network on \tcp\ dynamics.  The
formula establishes a connection between throughput, an important
characteristic of \gls{tcp}, and packet loss probability, an attribute of the
network on which \gls{tcp} operates.  The formula becomes inaccurate for large
$p$, because multiple packet losses, which force \gls{tcp} into the neglected
slow start phase, are more probable in this case.  The effect of multiple
losses on different \gls{tcp} variants is diverse, so the validity range of the
formula depends on the \gls{tcp} variant under consideration.

\section{Preliminary results of traffic modeling}
\label{sec:Traffic_modeling_review}

\subsection{Single session models}

The simple model given above can be extended considerably in a number of
aspects.  In a paper by \citet{AltmanAvrachenkovBarakat00} the \gls{tcp}
throughput for generic stationary congestion sequence was studied.  The model
extends the previous deterministic loss model to arbitrarily correlated loss
sequences.  The model is based on the following difference equation 
\begin{equation}
  X_{n+1}=\alpha S_n + \beta X_n,
  \label{eq:AIMD_gen_loss}
\end{equation}
where $X_n$ is the value of the throughput just prior to the arrival of loss
signal at $T_n$, $S_n=T_{n+1}-T_n$ is the time interval between consecutive
losses, and $\alpha$ and $\beta$ are the linear growth rate and multiplicative
decrease factor, respectively.  From the time average of the throughput the
following loss formula was derived
\begin{equation}
  \bar{X}=\frac{P}{R\sqrt{p}}
  \sqrt{\frac{1+\beta}{2\left(1-\beta\right)}+\frac12\hat{C}(0)
  +\sum_{k=1}^{\infty}\beta^k \hat{C}(k)},
  \label{eq:loss_formula_correlation}
\end{equation}
where 
$\hat{C}(k)=\left(\mathbb{E}\left[S_nS_{n+k}\right]
-\mathbb{E}\left[S_n\right]^2\right)/\mathbb{E}\left[S_n\right]^2$
is the normalized autocorrelation function of the loss interval process
$\left(S_n\right)_{n\in\mathbb{N}}$.  The derived formula, applied for
uncorrelated Poisson process with $C(k)=0$ and $\beta=1/2$, provides the same
result as (\ref{eq:loss_formula}).  A correlated loss interval scenario was
modeled with Markovian Arrival Process and the average \tcp\ throughput
was expressed with the infinitesimal generator of the arrival process.
Furthermore, the authors derived bounds for the throughput in case of limited
congestion window evolution and discussed the effects of timeouts.

\citet{PaydheFiroiuTowsleyKurose98} have studied the steady-state throughput of
\tcp\ Reno when packet loss is detected via both duplicate \glspl{ack} and timeouts,
and the throughput is limited by the receiver's window in more detail .  The
probability of timeout was estimated by the packet loss probability and
the congestion window.  It was shown that for small packet loss the
timeout probability can be approximated by $\min(1,3/W)$ and a very
comprehensive loss formula has been derived.

The performance of two classic \gls{tcp} versions, namely Tahoe and Reno, has
been analyzed by \citet{LakshmanMadhow97} when the bandwidth-delay product of
the bottleneck link is large compared to the buffer size.  The authors
estimated the average throughput for both slow start and congestion avoidance
phases with deterministic and independent random losses.

In a paper by \citet{OttKempermanMathis96} the stationary probability
distribution of the congestion window was calculated for constant packet
loss probability.  The authors mapped the ``\gls{ack} time'' point process to
a continuous ``subjective time'' process by the mapping
$W(t)=\sqrt{p}W_{\left\lfloor\frac{t}{p}\right\rfloor}$, where both the time
and the state space of the discrete process is rescaled in order to obtain a
well behaved process.  It was shown that for $p\to0$ the rescaled process 
$W(t)$ behaves as 
\begin{align}
  \frac{dW(t)}{dt}&=\frac{\alpha}{W(t)^m} & \text{if }t\neq\tau_k\\
  W\left(t^+\right)&=\beta W\left(t^-\right) & \text{if }t=\tau_k
\end{align}
where $\tau_k$ are the points of a Poisson process with intensity $\lambda$,
$\alpha>0$ and $0<\beta<1$ are the linear growth rate and multiplicative
decrease factor, respectively, $m\ge0$, and lastly $t^-$ and $t^+$ denote the
limit to $t$ from the left and from the right, respectively.  The parameter
values for \tcp\ congestion avoidance algorithm are $\beta=1/2$ and $m=1$.

The stationary complementary distribution function of the process $W(t)$ has 
been given in the following series expansion form:
\begin{equation}
  \bar{F}_W(w)=\sum_{k=0}^{\infty}R_k\left(c\right) 
  \exp\left(-\frac{\eta c^{-k}}{m+1} w^{m+1}\right),
  \label{eq:Ott_dist}
\end{equation}
where $\eta=\lambda/\alpha$, $c=\beta^{m+1}$ and for $|c|<1$
\begin{equation}
  \begin{split}
    R_k(c)&=\frac1{L(c)}\frac{(-1)^k c^{-\frac12k(k+1)}}
    {\left(1-c\right)\left(1-c^2\right)\dots\left(1-c^k\right)},\\
    L(c)&=\prod_{k=1}^{\infty}\left(1-c^k\right).
  \end{split}
  \label{eq:Ott_coef}
\end{equation}
``\gls{ack}'', ``subjective'' and real time averages and other moments of the
congestion window were calculated and an inverse square root loss formula
was derived.

The model has been extended for state dependent packet loss probability in
\cite{MisraOtt99}.  State dependent loss models the interaction of the
\gls{tcp} with \gls{erd} queuing policy, where the packet drop probability is a
function of the instantaneous queue length.  It is also applicable to \gls{red}
routers, where the drop probability depends on the average queue length.  An
iterative solution for the probability distribution function of the congestion
window was derived.  The authors found good agreement between the derived
distribution and computer simulations.

An in-depth analysis of \gls{red} queuing dynamics was presented in
\cite{MisraGongTowsley00}.  The time dependent congestion window development
was modeled with the stochastic differential equation
\begin{equation}
  dW_i(t)=\frac{dt}{a_i+q(t)/C}-\frac{W_i(t)}{2}dN_i(t),
\end{equation}
where $a_i$ denote the fix propagation delay of the bottleneck link, $C$ is its
capacity, $q(t)$ is the queue length at time $t$ and $N_i(t)$ is a Poisson
process with a rate that varies in time.  The above equations were transformed
to a system of delayed ordinary differential equations in order to obtain the
dynamics of the expectation of the congestion window.  The expectation value of
the queue length $\bar{q}(t)$ and the \gls{red} estimate of the average queue
length $\bar{x}(t)$ were approximated by two further differential equations.
As a result, the authors obtained $N+2$ coupled equations for the same number
of unknown variables $\left(\bar{W}_i(t), \bar{q}(t), \bar{x}(t)\right)$.  The
equations were solved numerically and were compared to computer simulations.
The authors pointed out the importance of the sampling frequency of the
smoothed queue length estimate.  A high frequency sampling might cause unwanted
oscillations in the system, while a low frequency sampling can increase the
initial overshoot of the average instantaneous queue length.   

\subsection{Multiple session models}

The above papers considered only a single \gls{tcp} connection.  In real
computer networks, however, a number of \glspl{tcp} might compete for the
network resources.  In particular the traffic of \gls{tcp} sources may flow, in
a parallel fashion, through a common link.  Web browsing represents a good
example of parallel \gls{tcp}, as up to four parallel \gls{tcp} sessions are
started at each page download. 

A possible result of \gls{tcp} interaction can be that parallel \glspl{tcp} are
synchronized.  The underlying reason for synchronization is \tcp's delayed
reaction for congestion events, which keeps drop-tail bottleneck buffers
congested for about an \gls{rtt} time period.  This temporary congestion can
induce further packet losses in competing \glspl{tcp}.  Based on this
phenomenon \citet{LakshmanMadhow97} supposed that the congestion window
development of parallel \glspl{tcp} is synchronized in the stationary
congestion avoidance regime.  The authors also took into consideration that the
bottleneck buffer of large bandwidth-delay product connections can be either
under- or over-utilized. The congestion window development was therefore split
into two phases accordingly.  The authors found a fixed point solution of both
the duration and the average congestion window of the two phases.  Finally, the
average throughput of each individual connection was estimated from the window
size divided by the round-trip time.

\Gls{tcp} synchronization is disadvantageous, since it causes performance
degradation.  However, this effect appears only in drop-tail queuing systems.
Active queue management, such as \gls{red} and \gls{erd}, alleviate the problem
of synchronization.  A paper by \citet{AltmanBarakatLaborde00} compared the
synchronization model to one in which only one of the parallel \glspl{tcp}
loses a packet at a congestion event.  The probability that a specific
connection is affected was proportional to the throughput of the particular
flow.  This drop policy models \gls{red} routers.  The stationary distribution
of the discretized congestion window at congestion instants was calculated.
The average throughput was estimated from the calculated window distribution
via a semi-Markov process.  The authors compared their results with simulations
of a \gls{red} buffer and found that their asynchronous model surpasses the
synchronous model presented in \cite{LakshmanMadhow97}.

Another typical effect in multiple \gls{tcp} scenario is the bias against
connections with long round-trip times \cite{Floyd91}.  This effect is the
fundamental consequence of \tcp\ dynamics, and it is not affected by the queue
management policy.  The phenomenon can be explained qualitatively by the
following simple arguments.  The growth rate of the congestion window is
inversely proportional to the round-trip time $R$.  The average congestion
window is therefore inversely proportional to $R$ as well.  Furthermore, the
average throughput $\bar{X}$ can be related to the average congestion window
$\bar{W}$ by Little's law from the queuing theory: $\bar{X}=\bar{W}/R$.
Therefore, the throughput is approximately proportional to $1/R^{\alpha}$, with
$\alpha=2$.  The exponent obtained from measurements has been shown to fall in
the range $1\le\alpha<2$ due to the queuing delay ignored in the above
arguments \cite{LakshmanMadhow97}.

\citet{FloydJacobson91} have shown that small changes in the round-trip time
might cause large differences in the throughput of different parallel \gls{tcp}
flows.  Specifically, packets of certain \glspl{tcp} can be dropped
tendentiously due to a phase-effect, causing an utterly unfair bandwidth
distribution.  Changing the relative phase of arriving packets at the
bottleneck link by slightly modifying the round-trip time can completely
rearrange the bandwidth share of different \gls{tcp} connections.  Random
effects, such as random fluctuations in the round-trip time or \gls{red}
queuing policy, also alleviate the phase-effect.

\subsection{The Network Simulator -- \nstwo}
\label{subsec:ns2}

\glsadd[format=(textbf]{ns}
New models, algorithms, and analytical calculations should be validated against
experiments.  Without doubt the most authentic data can be obtained from
Internet measurements, but the deployment of a measurement infrastructure can
be quite expensive and is still very limited.  Moreover, models often use
simplifications which make them difficult to compare with real Internet data.
Network simulators, on the other hand, provide ``laboratory'' environments,
where every parameter of the network and the traffic can be precisely
controlled.  Therefore, network simulators are important tools in the hands of
researchers endeavoring to carry out well controlled experiments.

One of the most widely used network simulators in the research community is the
Network Simulator---\nstwo\footnote{The next major version of the simulator,
\textsf{ns-3}, is under active development.} \cite{ns2}.  A short overview of
the simulator is given next, since several analytic and numeric results of this
thesis have been validated by \nstwo.
  
The \nstwo\ simulator mimics every component of a real network, e.g. links, 
routers, queues, protocols, applications and so on.  The network traffic is
simulated at packet level, which is to say the course of every packet is
followed from its injection into the network until its removal from it.  The
packet-level simulation of network traffic makes the simulator very realistic,
so fine details of the simulated network traffic can be observed.  The major
disadvantage of a packet-level simulation is the considerable amount of
computing power that it requires.

The \nstwo\ simulator is event-driven, that is every component might schedule
events into a virtual calendar.  The simulator's scheduler runs by selecting
the next earliest event for execution.  During processing of events further
events can be scheduled into the calendar.  This event-based mechanism can also
be observed in every part of the simulator, for example in the handling of data
packets in \nstwo.  Data packets do not actually travel between virtual nodes
in the simulator, but rather are scheduled for processing at different network
elements instead.  For example, when a packet is put onto a link for
transmission the link object in the simulator only schedules the packet for the
queue of the next node on the other end of the link.

The core of the simulator has been written in C++, but it also has an OTcl
scripting programming interface, the object oriented extension of Tcl.  The C++
core offers fast execution of the simulator.  However, average users do not
need to deal with C++ code in order to run simulations under \nstwo.  All the
network elements have been bound to objects in OTcl, so complex scenarios can
be built up simply and easily by writing short OTcl scripts.
\glsadd[format=)textbf]{ns}

\section{The infinite-buffer network model}
\label{sec:Buffer_model}

\begin{figure}
  \begin{center}
    \psfrag{TCP}[c][c]{TCP}
    \psfrag{TCP1}[c][c]{TCP$_1$}
    \psfrag{TCP2}[c][c]{TCP$_2$}
    \psfrag{TCP3}[c][c]{TCP$_3$}
    \psfrag{TCPn}[c][c]{TCP$_N$}
    \psfrag{Sink}[c][c]{Sink}
    \psfrag{Sink1}[c][c]{Sink$_1$}
    \psfrag{Sink2}[c][c]{Sink$_2$}
    \psfrag{Sink3}[c][c]{Sink$_3$}
    \psfrag{Sinkn}[c][c]{Sink$_N$}
    \psfrag{Buffer}[c][c]{Buffer}
    \psfrag{Link}[c][c]{Link}
    \psfrag{Delay d}[c][c]{Delay, $D$}
    \psfrag{Loss rate p}[c][c]{Loss rate, $p$}
    \psfrag{Bandwidth C}[c][c]{Bandwidth, $C$}
    \psfrag{1}[c][c]{$1$}
    \psfrag{2}[c][c]{$2$}
    \psfrag{3}[c][c]{$3$}
    \resizebox{\figwidth}{!}{\includegraphics{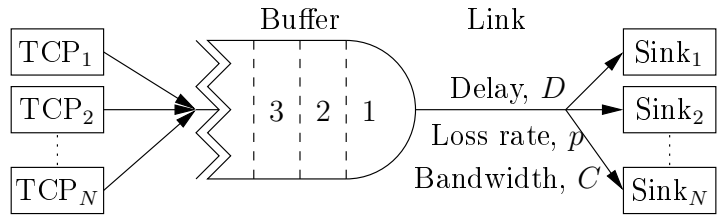}}
    \caption{Idealized network model with infinite buffer capacity.  Other 
             parts of the network are modeled with link delay $D$, bandwidth
             $C$, and packet loss probability $p$.}
    \label{fig:topology_inf}
  \end{center}
\end{figure}

A very simple model of an access router connected to a complex network consists
of a buffer and a link, shown in Fig.~\ref{fig:topology_inf}.  In this regard
the link is not a real connection between routers, but rather a virtual one.
The influence of the network on the traffic using the access router is modeled
with a few parameters of the virtual link: a fixed propagation delay $D$,
bandwidth $C$, and packet loss probability $p$.  This probability represents
the chance of link and hardware failures \cite{Bolot93}, incorrect handling of
arriving packets by routers, losses and time variations due to wireless links
in the path of the connection \cite{LakshmanMadhow97}, the likelihood of
congestion in the instantaneous bottleneck buffer, and the effect of \gls{red}
and \gls{erd} queuing policies.  In this chapter I assume that the buffer is
large enough that no packet loss occurs in it.

Two practically important limits in respect of the role of the access buffer 
are:
\begin{inparaenum}[\itshape a\upshape)]
\item \gls[format=textbf]{lan} traffic, when the bandwidth delay product of the
link is small, only a few packets can be out in the link and the buffer is
never empty; and
\item \gls[format=textbf]{wan} traffic, when the bandwidth delay product of the
link is large, packets are in the link and the buffer is mostly empty. 
\end{inparaenum}
From now on I will refer to systems with small and large bandwidth
delay products as LAN and WAN, respectively.

As I mentioned earlier in Section~\ref{subsec:TCP}, the Internet traffic is
governed mostly by \gls{tcp}.  I will therefore neglect \gls{udp} traffic in my
simple network model and will only study the behavior of \tcp\ dynamics.

In realistic networks many \gls{tcp} sources \cite{MisraOttBaras99} may share
the resources of the access network.  The difficulty of describing the parallel
\tcp\ dynamics lies in the interaction of individual \gls{tcp} flows.  It is
obvious that the number of packets in the network injected by one of the
\gls{tcp} sessions affects the networking environment of the others. In
particular, it contributes to the round trip times and packet losses felt by
the other \glspl{tcp}.  Since the congestion window controls the maximum number
of unacknowledged packets, understanding its distribution is crucial to
describe the interaction.

While an exact treatment of nonlinear interacting systems (such as this one) is
not possible in general, very efficient methods, motivated mostly by
interacting physical systems, have been developed.  One of the most established
methods is the mean field approximation.  In this approximation each subsystem
operates independently in an ``averaged'' (or mean) environment. The average
environment is calculated from the behavior of the subsystems. Finally, a fixed
point of the system has to be found where the ``mean'' environment and the
environment averaged over the independent subsystems coincide. This way we
obtain a self-consistent solution which provides an approximate but quite
accurate description of each subsystem. 

In the case of computer networks each \gls{tcp} plays the role of a subsystem,
while the environment is the round trip time.  First, the congestion window
distribution of a \gls{tcp} is calculated by assuming a given packet loss and
round trip time.  Next, the mean round trip time is calculated using the window
distribution. Finally, a fixed point value of the round trip time is
determined.  


\section{Dynamics of a single TCP}\label{sec:OneTCP}
\label{sec:Window_dynamics}

For studying the behavior of interacting \glspl{tcp} by mean field
approximation one should know the behavior of a single \gls{tcp} first.  In
this section I carry out the analysis of a single \gls{tcp} with the use of the
fluid approximation of \gls{tcp} dynamics, presented in
Section~\ref{sec:Fluid_models}.  

Recall that between two consecutive losses the congestion window is governed 
by the continuous time differential equation (\ref{eq:cwnd_fluid}):
\begin{equation}
  \label{eq:cwnd_fluid_W}
  \frac{dW}{dt}=\frac1{R(W)},
\end{equation}
where $W\in[0,\infty[$ is the congestion window, and $R(W)$ is the round-trip
time, which might depend explicitly on the value of the congestion window.

Consider, for example, a typical \gls{lan} scenario with a single \gls{tcp}
where the link delay is small and packet delay is caused mostly by buffering.
The congestion window counts the number of unacknowledged packets, and these
packets can be found on the link and in the buffer.  At each packet-shift time
unit a packet is shifted from the buffer into the link.  The round-trip time of
a freshly sent packet will be the time it should wait for the shifting of all
previously sent packets in the system, which is in turn measured by the
congestion window $R(W)=WP/C$.

\begin{figure}
  \begin{center}
    \psfrag{cwnd}[c][c][1.2]{$W(t)$}
    \psfrag{T}[ct][cB][1.2]{$t$}
    \psfrag{simulation}[rB][rb][0.8]{simulation}
    \psfrag{2}[r][r][1]{$2$}
    \psfrag{4}[r][r][1]{$4$}
    \psfrag{6}[r][r][1]{$6$}
    \psfrag{8}[r][r][1]{$8$}
    \psfrag{10}[r][r][1]{$10$}
    \psfrag{12}[r][r][1]{$12$}
    \psfrag{14}[r][r][1]{$14$}
    \psfrag{16}[r][r][1]{$16$}
    \psfrag{18}[r][r][1]{$18$}
    \psfrag{20}[r][r][1]{$20$}
    \psfrag{1000}[r][r][1]{$1000$}
    \psfrag{1050}[r][r][1]{$1050$}
    \psfrag{1100}[r][r][1]{$1100$}
    \psfrag{1150}[r][r][1]{$1150$}
    \psfrag{1200}[r][r][1]{$1200$}
    \psfrag{1250}[r][r][1]{$1250$}
    \psfrag{1300}[r][r][1]{$1300$}
    \psfrag{1350}[r][r][1]{$1350$}
    \psfrag{1400}[r][r][1]{$1400$}
    \psfrag{1450}[r][r][1]{$1450$}
    \psfrag{1500}[r][r][1]{$1500$}
    \subfigure[\nstwo\ simulation of LAN]{\resizebox{0.7\figwidth}{!}
      {\includegraphics{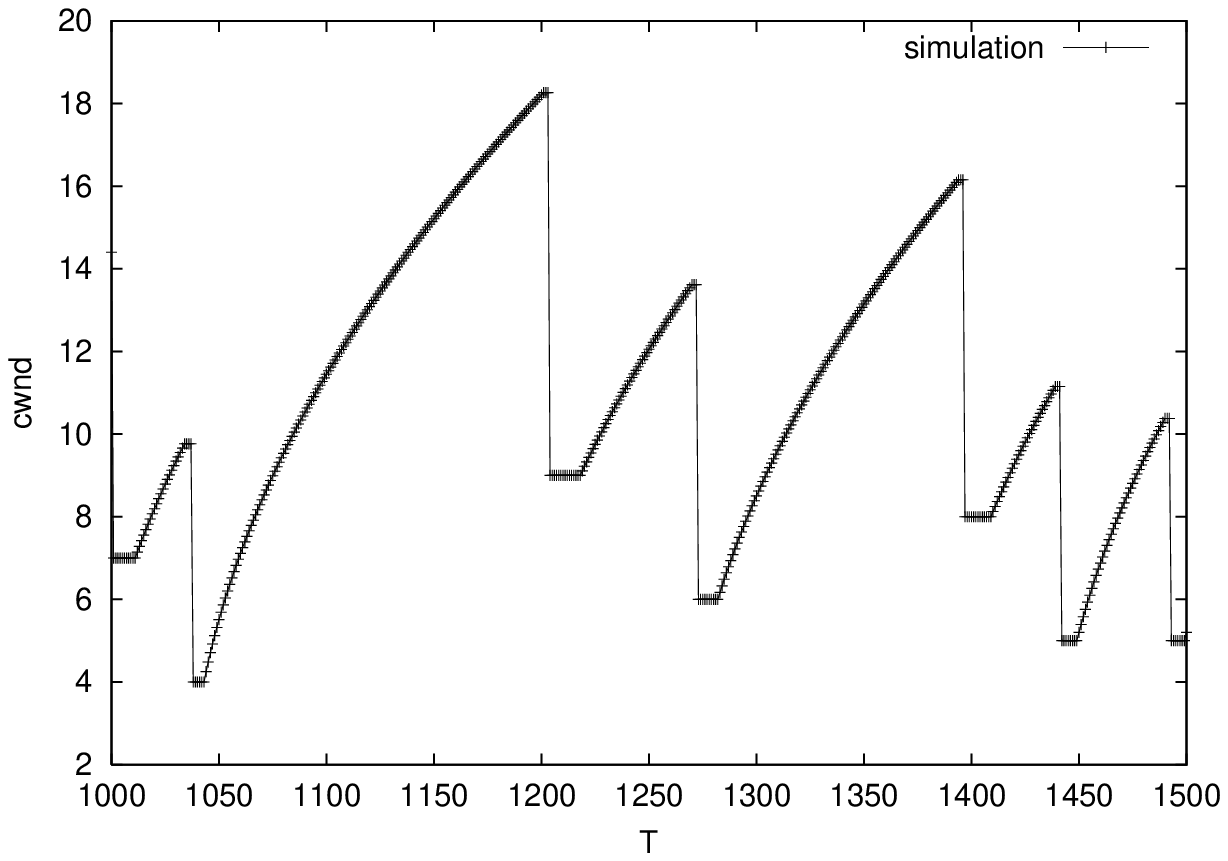}}\label{subfig:LAN_simulation}}
    \psfrag{w(t)}[cb][ct][1.2]{$W(t)$}
    \psfrag{t}[c][c][1.2]{$t$}
    \psfrag{w0}[r][r][1]{$W_0$}
    \psfrag{w1}[r][r][1]{$W_1$}
    \psfrag{d0}[c][c][1]{$\delta_0$}
    \psfrag{d1}[c][c][1]{$\delta_1$}
    \psfrag{d2}[c][c][1]{$\delta_2$}
    \psfrag{d3}[c][c][1]{$\delta_3$}
    \psfrag{d4}[c][c][1]{$\delta_4$}
    \psfrag{t0}[c][c][1]{$\tau_0$}
    \psfrag{t1}[c][c][1]{$\tau_1$}
    \psfrag{t2}[c][c][1]{$\tau_2$}
    \psfrag{t3}[c][c][1]{$\tau_3$}
    \psfrag{t4}[c][c][1]{$\tau_4$}
    \psfrag{t5}[c][c][1]{$\tau_5$}
    \subfigure[Ideal fluid model of LAN]{\resizebox{0.7\figwidth}{!}
      {\includegraphics{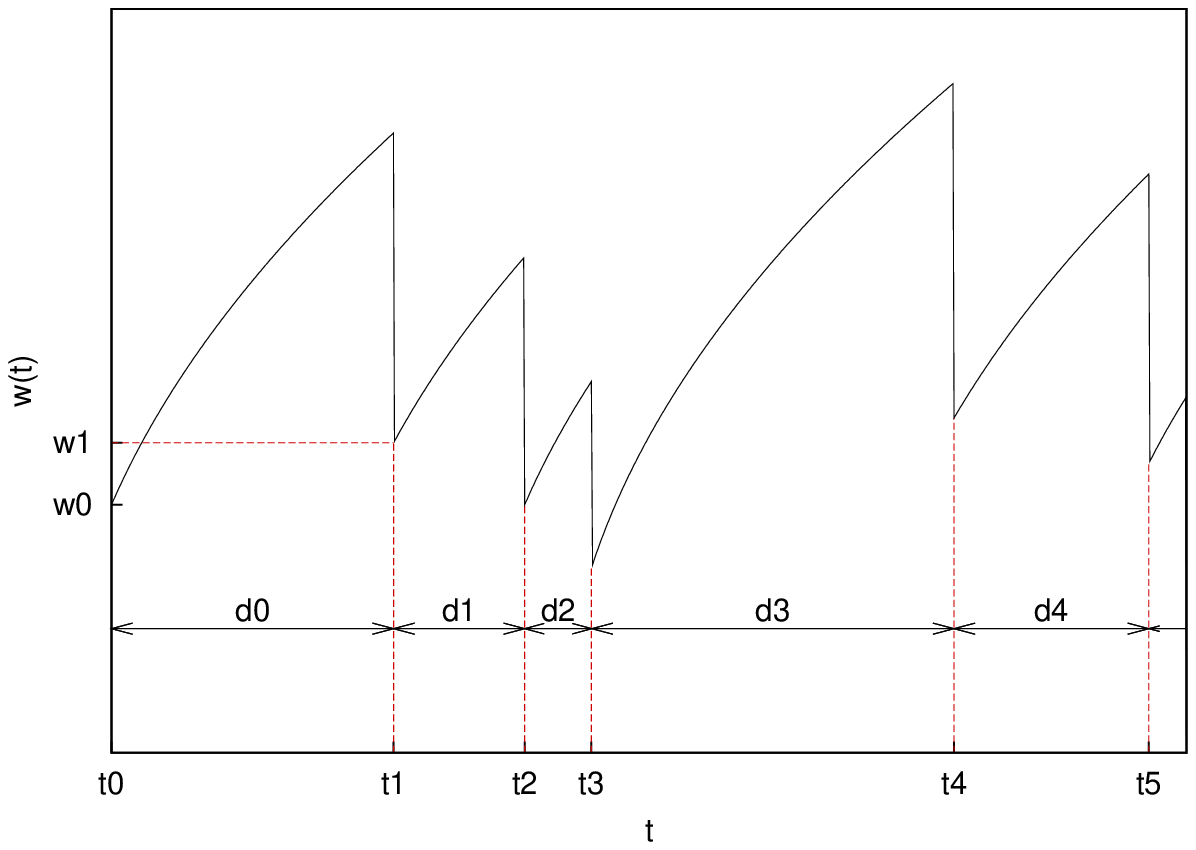}}\label{subfig:LAN_fluid}}
    \caption{Comparison of \nstwo\ simulations with the fluid 
    approximation of the congestion avoidance process of \tcp/Reno.  Note the 
    small plateaus in the simulations due to the FR/FR algorithm after each 
    period.  
    }
    \label{fig:simulation}
  \end{center}
\end{figure}

An idealized congestion window process is shown in Fig.~\ref{subfig:LAN_fluid},
while a simulated congestion window sequence can be seen on
Fig.~\ref{subfig:LAN_simulation} for comparison.  Numerical simulations were
executed by \gls{ns}, introduced in Sec.~\ref{subsec:ns2}.  Note the small
plateaus in the simulations after each cycle of the congestion window process.
These plateaus are the result of the \gls{frfr} algorithms.  First, I will
ignore the effect of the \gls{frfr} algorithms, and I will consider their
influence later.

In order to include more general---even hypothetical---\tcp\ dynamics in my 
model the round-trip time is written in the following form:
\begin{equation}
  R(W)=\alpha^{-1}W^m
  \label{eq:rtt_model}
\end{equation}
with $m\ge0$ and $\alpha>0$.  Note that this notation includes the
``\gls{ack} time'' dynamics of \tcp\ as well.

The fluid equation (\ref{eq:cwnd_fluid_W})
can be written now as $\frac{dW}{dt}=\frac{\alpha}{W^m}$, which can be
rearranged into 
\begin{equation}
  \frac{dW^{m+1}}{dt}=\alpha\left(m+1\right).
  \label{eq:cwnd_fluid_Wm}
\end{equation}
It is obvious that between two packet loss events the solution of this 
differential equation is
\begin{equation}
  W^{m+1}(t)=W^{m+1}(\tau_i)+\alpha\left(m+1\right)\left(t-\tau_i\right),
  \label{eq:cwnd_solv}
\end{equation}
where $\tau_i$ denote the instant of the $i$\textsuperscript{th} packet loss.
At $\tau_i$ the transformation
\begin{equation}
  W(\tau_i^+)=\beta W(\tau_i^-)
\end{equation}
is executed, where $W(\tau_i^-)$ and $W(\tau_i^+)$ are the congestion windows
immediately before and after the time of packet loss, and $0<\beta<1$.  The
actual value of $\beta$ is $1/2$ in most \tcp\ variants.  

Let $W_i=W(\tau_i^+)$ denote the congestion window immediately \emph{after} the
$i$\textsuperscript{th} packet loss, $\delta_i=\tau_{i+1}-\tau_i$ the length of
the time interval between two losses, and $c=\beta^{m+1}$ hereafter.  Since
\gls{tcp} is assumed to detect packet losses instantaneously in the fluid
model, $W_{i+1}^{m+1}$ can be written as
\begin{equation}
  W_{i+1}^{m+1}
  =\left[\beta W(\tau_{i+1})\right]^{m+1}
  =cW_i^{m+1}+\alpha\left(m+1\right) c\,\delta_i.
  \label{eq:cwnd_split_LAN}
\end{equation}

By repeated application of (\ref{eq:cwnd_split_LAN}) one can show that
the value of the congestion window immediately after the 
$N$\textsuperscript{th} packet loss is
\begin{equation}
  W^{m+1}_N=c^{N}W^{m+1}_0+\alpha\left(m+1\right)
  \sum_{k=0}^{N-1}c^{k+1}\delta_{N-k-1}.
  \label{eq:ws_LAN}
\end{equation}
For $N\rightarrow \infty$ the initial value $W_0$ becomes insignificant and the
sequence of congestion window values after the packet losses ($\wal$) can be
expressed as
\begin{equation}
  \wal^{m+1}=\lim_{N\to\infty}W^{m+1}_N
  =\alpha\left(m+1\right)\sum_{k=0}^{\infty} c^{k+1}\delta_k.
  \label{eq:cwnd_after_loss}
\end{equation}
Note that the indexing of the $\delta_i$ sequence is reversed.
This was allowed since, as I will show below, every $\delta_i$ had
the same statistical properties. The reversed indexing was necessary
since the infinite sum would have been meaningless without it.

In the \gls{lan} case one packet is shifted out of the buffer in each time unit.
Therefore, in the fluid approximation the times between packet losses,
$\delta_i$, are independent exponentially distributed variables.  The \gls{wan}
scenario is slightly different.  Since there is no queue in the buffer there
are periods when no packet leaves the buffer (see the small horizontal steps in
Figure~\ref{subfig:WAN_simulation}), packets cannot be lost in those intervals.
However, I assume first that times between losses are exponentially distributed
in the \gls{wan} scenario as well---
\begin{equation}
  f_{\delta_i}(x)=\ph e^{-\ph x},
  \label{eq:dist_delta}
\end{equation}
where $1/\ph$ is the average time between losses---and I will improve the model
later.  For the \gls{lan} case $\ph=pC/P$, since one packet is shifted from the
buffer in $P/C$ packet-shift time.  Combining (\ref{eq:cwnd_after_loss}) and
(\ref{eq:dist_delta}) one can obtain the distribution of $\wal^{m+1}$:
\begin{equation}
  f_{\wal^{m+1}}(w)=
  \integ_0^{\infty}\dotsi\integ_0^{\infty}\delta\left(w-\alpha\left(m+1\right)
    \sum_{k=0}^{\infty}c^{k+1}x_k\right)
  \prod_{i=0}^{\infty}f_{\delta_i}(x_i)\,dx_i,
  \label{eq:dist_ws_LAN}
\end{equation}
where $\delta()$ is the delta distribution. In reality only the distribution of
{\em generic} values of the congestion window can be measured. Therefore, their
distribution has to be derived as well.  Here I show that this can be done
analytically.

In general, between losses the congestion window is the sum of two random
variables $W^{m+1}=\wal^{m+1}+\alpha\left(m+1\right)\tau$, where $\tau$ is a
uniformly distributed random variable in the \emph{random} interval
$[0,\delta_i]$.  To obtain the probability distribution of the congestion
window at an arbitrary moment we have to derive the distribution of the random
variable $\tau$ as well.  Its distribution can be derived as follows: $\tau$ is
distributed uniformly on each interval with length $\rho$, assuming $\rho$ is
given.  This statement can be expressed mathematically with the conditional
distribution 
\begin{equation}
  f_{\tau}(t\mid\rho=x)
  =\frac1{x}\chi_{[0,x]}(t),
\end{equation}
where $\chi_H(w)$ denotes the indicator function of set $H\subset\mathbb{R}$.
Furthermore, the probability of selecting a random interval is proportional to
the length of the given interval and the distribution of the random variable
$\delta_i$.  The proportional factor can be deduced from the normalization
condition of the probability distribution:
\begin{equation}
  f_{\rho}(x)=\frac{xf_{\delta_i}(x)}{\int_0^{\infty}xf_{\delta_i}(x)\,dx}
  =\frac{x}{\mathbb{E}[\delta_i]}f_{\delta_i}(x)=\ph x f_{\delta_i}(x)
\end{equation}
Finally, the desired distribution follows from the total probability theorem:
\begin{equation}
  \begin{split}
    f_{\tau}(t)&=\int_0^{\infty}f_{\tau}(t\mid\rho=x)f_{\rho}(x)\,dx\\
    &=\ph\integ_0^{\infty}\ph e^{-\ph x}\integ_0^x\delta(t-y)\,dy\,dx
    =\ph\integ_t^{\infty}\ph e^{-\ph x}\,dx=\ph e^{-\ph t}
  \end{split}
  \label{eq:tau}
\end{equation}

In order to calculate the distribution function of a general value of the
congestion window we apply the method of the Laplace transform.  As is known,
the Laplace transform of the density function of the sum of independent random
variables is the product of the Laplace transform of their density functions.
The Laplace transform of (\ref{eq:dist_ws_LAN}) with respect to $\wal^{m+1}$
can be defined as
$\hat{f}_{\wal^{m+1}}(s)=\int_0^{\infty}e^{-sw}f_{\wal^{m+1}}(w)\,dw$.  This
can be easily evaluated and we get
\begin{equation}
  \begin{split}
    \hat{f}_{\wal^{m+1}}(s)&=\prod_{k=0}^{\infty}
    \left[\ph\integ_0^{\infty}\exp\left(-s\alpha\left(m+1\right)
        \beta^{m\left(k+1\right)}x_k-\ph x_k\right)\,dx_k\right]\\
    &=\prod_{k=1}^{\infty}\frac{\ph}{\ph+\alpha\left(m+1\right)c^{k}s}.
  \end{split}
  \label{eq:dist_ws_LAN2}
\end{equation}
The Laplace transform of the distribution of $\alpha\left(m+1\right)\tau$ 
is given by
\begin{equation}
  \hat{f}_{\alpha\left(m+1\right)\tau}(s)
  =\integ_0^{\infty}\frac{\ph}{\alpha\left(m+1\right)}
    e^{-\frac{\ph}{\alpha\left(m+1\right)} t} e^{-st}\,dt
  =\frac{\ph}{\ph+\alpha\left(m+1\right) s}.
\end{equation}
Therefore, the Laplace transform of the generic 
$W^m=\wal^m+\alpha\left(m+1\right)\tau$ distribution is the infinite product
\begin{equation}
  \hat{f}_{W^{m+1}}(s)
  =\hat{f}_{\wal^{m+1}}(s)\,\hat{f}_{\alpha\left(m+1\right)\tau}(s)
  =\prod_{k=0}^{\infty}\frac{\ph}{\ph+\alpha\left(m+1\right)c^{k}s}.
  \label{eq:w_LAN}
\end{equation}
Furthermore, we can rewrite the Laplace transform as a sum of partial fractions
\begin{equation}
  \hat{f}_{W^{m+1}}(s)=
  \frac{\ph}{\alpha\left(m+1\right)}\sum_{k=0}^{\infty}
  \frac{h_k(c)}{\frac{\ph c^{-k}}{\alpha\left(m+1\right)}+s},
  \label{eq:w_frac_LAN}
\end{equation}
where the coefficients $h_k(c)$ can be obtained from the residues of the
poles of $\hat{f}_{W^m}(s)$:
\begin{align}
    h_k(c)
    &=\Res_{-\frac{\ph c^{-k}}{\alpha\left(m+1\right)}}\frac{\alpha\left(m+1\right)}{\ph}\hat{f}_{W^{m+1}}(s)\notag\\
    &=\lim_{s\to-\frac{\ph c^{-k}}{\alpha\left(m+1\right)}}\frac{\alpha\left(m+1\right)}{\ph}
      \left(s+\frac{\ph c^{-k}}{\alpha\left(m+1\right)}\right)
      \prod_{l=0}^{\infty}\frac{\ph}{\ph+\alpha\left(m+1\right)c^{l}s}\notag\\
    &=\frac1{c^{k}}\prod_{\substack{l=0\\l\neq k}}^{\infty}\frac1{1-c^{l-k}}
    =\frac1{c^kL(c)}\prod_{l=1}^k \frac1{1-c^{-l}},
  \label{eq:h_l_LAN}
  \intertext{where}
  L(c)&=\prod_{l=1}^{\infty}\left(1-c^l\right).
  \label{eq:L_c}
\end{align}
It follows evidently from this formula that the relative strength of successive 
terms eventually decreases exponentially fast, when $k$ is large enough:
\begin{equation}
  \frac{h_{k+1}(c)}{h_k(c)}=-c^k\frac1{1-c^{k+1}}\approx-c^k\quad\text{for $k\gg-\frac1{\log c}$},
\end{equation}
therefore, only a small number of constants should be used for numerical 
purposes.

We can perform a term-by-term inverse Laplace transform on
(\ref{eq:w_frac_LAN}). The density function of $W^m$ can be given by
\begin{equation}
  f_{W^{m+1}}(w)=\frac{\ph}{\alpha\left(m+1\right)}\sum_{k=0}^{\infty}h_k(c)
  \exp\left(-\frac{\ph c^{-k}}{\alpha\left(m+1\right)}w\right).
  \label{eq:Wm_dist}
\end{equation}
The distribution of the \cwnd\ is given by a simple variable
transformation
\begin{equation}
  f_W(w)=\frac{\ph}{\alpha}w^{m}\sum_{k=0}^{\infty}h_k(c)
  \exp\left(-\frac{\ph c^{-k}}{\alpha\left(m+1\right)}w^{m+1}\right).
  \label{eq:W_dist}
\end{equation}
Finally, the complementary cumulative distribution 
$\bar{F}_W(w)=\int_w^{\infty}f_W(w')\,dw'$ can be given by
\begin{equation}
  \bar{F}_W(w)=\sum_{k=0}^{\infty}c^k h_k(c)
  \exp\left(-\frac{\ph c^{-k}}{\alpha\left(m+1\right)}w^{m+1}\right).
  \label{eq:W_cdist}
\end{equation}
Note that the above formulas do not change when $\ph$ and $\alpha$ are varied,
but the $\ph/\alpha$ ratio is kept fixed.  Furthermore, the weight of the 
$k$\textsuperscript{th} term in the probability distribution is $c^k h_k(c)$,
so the error induced by truncating terms above a threshold index can be
estimated precisely: $1-\sum_{k=0}^K c^k h_k(c)$.

Compare the results (\ref{eq:Wm_dist})\,--\,(\ref{eq:W_cdist}) with
(\ref{eq:Ott_dist})\,--\,(\ref{eq:Ott_coef}).  The calculation has reproduced
the results of \citet{OttKempermanMathis96} with a slight difference in the
notation.  However, I have not supposed in my derivation that $p\to0$, as
\citeauthor{OttKempermanMathis96} did.  Furthermore, I have shown explicitly
that $\tau$ is exponentially distributed, which is missing in the previous
derivation.

The moments of $W$ can be calculated from (\ref{eq:Wm_dist}) as
\begin{equation}
  \begin{split}
    \mathbb{E}[W^{r\left(m+1\right)}]&=\int_0^{\infty}w^r f_{W^{m+1}}(w)dw
    =\frac{\ph}{\alpha\left(m+1\right)}\sum_{k=0}^{\infty} h_k(c)
    \int_0^{\infty}w^{r}e^{-\frac{\ph c^{-k}}{\alpha\left(m+1\right)}w}dw\\
    &=\left(\frac{\alpha\left(m+1\right)}{\ph}\right)^{r}
    \sum_{k=0}^{\infty} c^{\left(1+r\right)k} h_k(c)
    \int_0^{\infty}z^re^{-z}dz\\
    &=\left(\frac{\alpha\left(m+1\right)}{\ph}\right)^{r}
    \Gamma\left(1+r\right)
    \sum_{k=0}^{\infty} c^{\left(1+r\right)k} h_k(c),
  \end{split}
  \label{eq:moment_r}
\end{equation}
where $r>0$.  The variable transformation $z=\frac{\ph
c^{-k}}{\alpha\left(m+1\right)}w$ was executed in the first integral, and the
integral definition of the Gamma function
$\Gamma(x)=\int_0^{\infty}z^{x-1}e^{-z}dz$ was applied in the second equation.
If $r=n$ is an integer the moments can be given in closed form.  For this 
end let us find the series expansion of the Laplace transform of $W^{m+1}$.
Observe---following \citeauthor{OttKempermanMathis96}---that the product form
of the Laplace transform (\ref{eq:w_LAN}) satisfies the following functional
equation: 
\begin{equation}
  \left(1+\frac{\alpha\left(m+1\right)}{\ph}s\right)\hat{f}_{W^{m+1}}(s)
  =\hat{f}_{W^{m+1}}(cs).
\end{equation}
Differentiate $n$ times both sides of this functional equation with respect 
to $s$:
\begin{equation}
  \left(1+\frac{\alpha\left(m+1\right)}{\ph}s\right)\hat{f}^{(n)}_{W^{m+1}}(s)
  +n\frac{\alpha\left(m+1\right)}{\ph}\hat{f}^{(n-1)}_{W^{m+1}}(s)
  =c^n \hat{f}^{(n)}_{W^{m+1}}(cs).
\end{equation}
Since the $n$\textsuperscript{th} derivative of the Laplace transform at
$s=0$ is related to the moments as
$\mathbb{E}[W^{n\left(m+1\right)}]=(-1)^n\hat{f}^{(n)}_{W^m}(0)$,
we find the following recurrence relation:
\begin{equation}
  \left(1-c^n\right)\mathbb{E}[W^{n\left(m+1\right)}]
  =n\frac{\alpha\left(m+1\right)}{\ph}\mathbb{E}[W^{\left(n-1\right)\left(m+1\right)}].
\end{equation}
This recursive equation with the initial condition $\mathbb{E}[W^0]=1$ 
immediately yields
\begin{equation}
  \mathbb{E}[W^{n\left(m+1\right)}]
  =n! \left(\frac{\alpha\left(m+1\right)}{\ph}\right)^n\prod_{k=1}^{n}\frac1{1-c^k}.
  \label{eq:moment_n}
\end{equation}

\section{Discussion}
\label{sec:Discussion}

\subsection{Local Area Networks}
\label{subsec:OneTCP_LAN}

I will now confirm the validity of the above results by numerical simulations.
In this section I study the \gls{lan} scenario, that is when the bandwidth-delay 
product of the link is much smaller than the size of the buffer.  As I have 
pointed out earlier the parameters of the ideal \gls{lan} scenario are $m=1$,
$c=\beta^2=1/4$, and $\alpha^{-1}\ph=p$.  The first nine numerical values of
$h_k(c)$ are shown in Table~\ref{tab:h_l_LAN}.  It can be seen that the
coefficients converge to zero so quickly that it is sufficient to keep the
first five terms in practical calculations.
\begin{table}
  \caption{$h_k\left(\beta^{m+1}\right)$ coefficients with $\beta=1/2$ and $m=1$}
  \label{tab:h_l_LAN}
  \begin{center}
    \begin{tabular}{l r@{} l l r@{} l l r@{} l}
      \hline
      $h_0$ & $ 1$&.$4523536$ & 
      $h_3$ & $-3$&.$2786819\cdot10^{-2}$ & 
      $h_6$ & $ 1$&.$9643078\cdot10^{-9}$\\
      $h_1$ & $-1$&.$9364715$ & 
      $h_4$ & $ 5$&.$1430305\cdot10^{-4}$ & 
      $h_7$ & $-4$&.$7959661\cdot10^{-13}$\\
      $h_2$ & $ 5$&.$1639241\cdot10^{-1}$ & 
      $h_5$ & $-2$&.$0109601\cdot10^{-6}$ &
      $h_8$ & $ 2$&.$9272701\cdot10^{-17}$\\
      \hline
    \end{tabular}
  \end{center}
\end{table}

The mean of the congestion window can be calculated from (\ref{eq:moment_r})
with the parameter $r=\frac1{m+1}=\frac{1}{2}$:
\begin{equation}
  \mathbb{E}[W]
  =\sqrt{\frac{2}{p}}\frac{\sqrt{\pi}}{2}\sum_{k=0}^{\infty}\frac{h_k(1/4)}{8^k}
  \approx\frac{1.5269}{\sqrt{p}},
  \label{eq:LAN_mean}
\end{equation}
which gives the well known inverse square-root formula.
The second moment can be obtained exactly from (\ref{eq:moment_n}) with $n=1$:
\begin{equation}
  \mathbb{E}[W^2]=\frac{2}{p}\frac1{1-1/4}=\frac{8}{3p},
\end{equation}
therefore the standard deviation is approximately 
\begin{equation}
  \sigma[W]=\sqrt{\mathbb{E}[W^2]-\mathbb{E}[W]^2}
  \approx\frac{0.5790}{\sqrt{p}}.
  \label{eq:LAN_stdev}
\end{equation}

\begin{figure} 
  \begin{center} 
    \psfrag{E[W]}[c][c][1.2]{$\mathbb{E}[W]$}
    \psfrag{sigma[W]}[c][c][1.2]{$\sigma[W]$}
    \psfrag{W_avg}[c][c][1.2]{$\left\langle W\right\rangle$}
    \psfrag{W_stdev}[c][c][1.2]{$\sqrt{\left\langle W^2\right\rangle-\left\langle
W\right\rangle^2}$} 
    \subfigure[Average vs. mean]{
      \resizebox{0.7\figwidth}{!}{\includegraphics{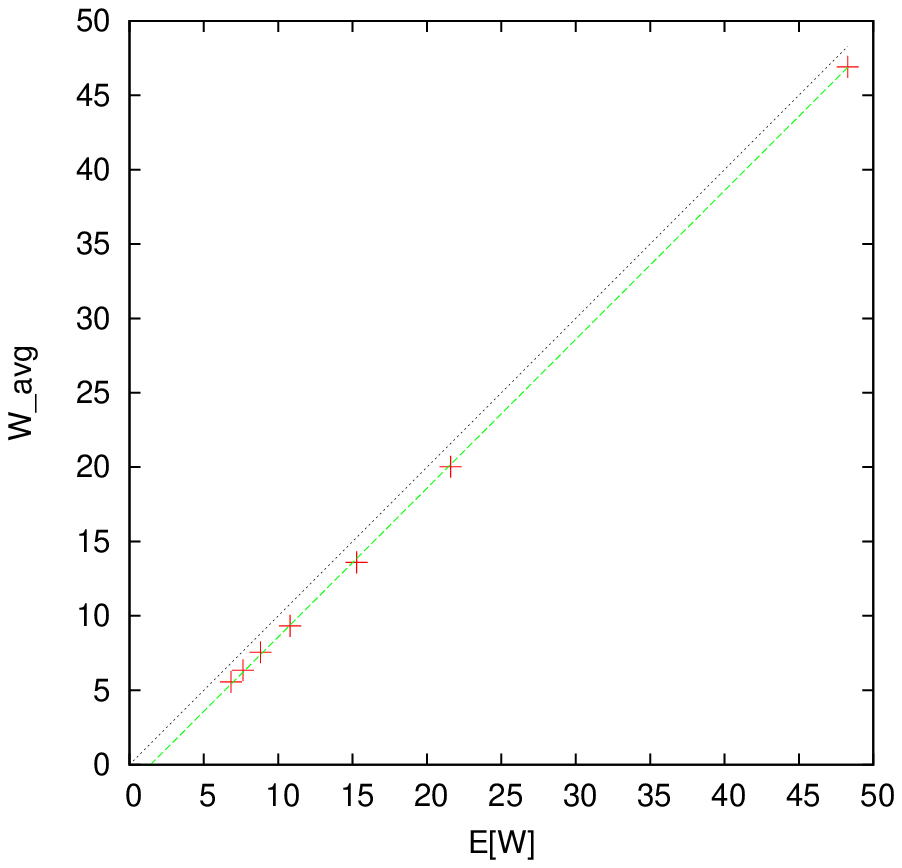}}\label{subfig:LAN_mean}}
    \subfigure[Empirical standard deviation vs. standard deviation]{
      \resizebox{0.7\figwidth}{!}{\includegraphics{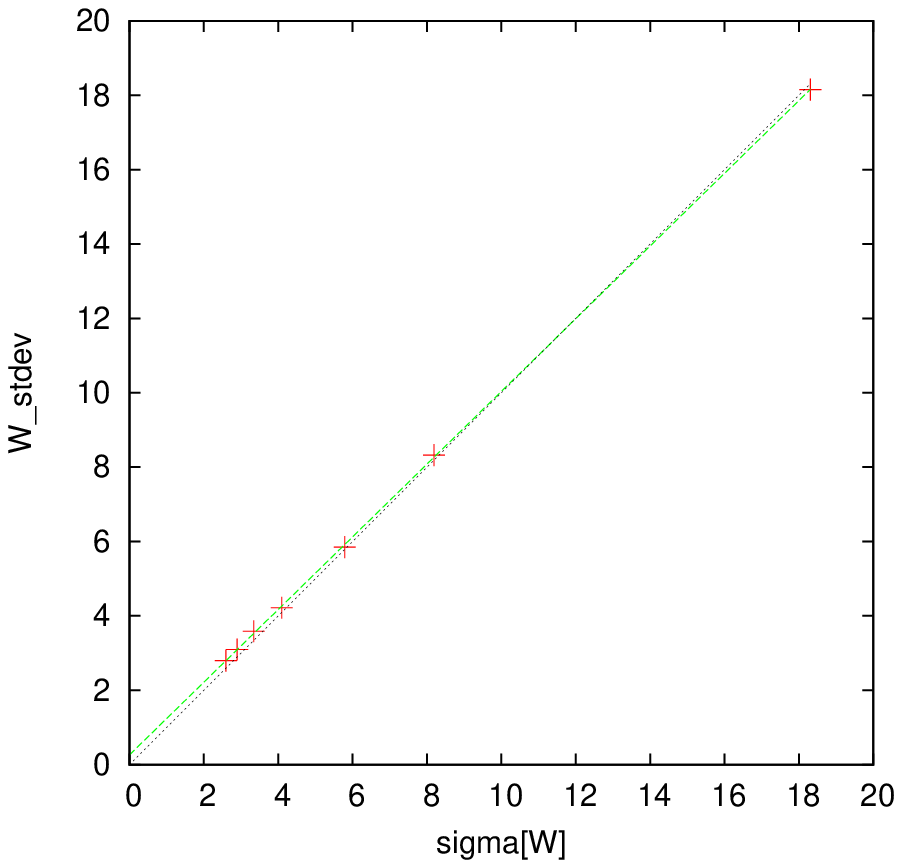}}\label{subfig:LAN_stdev}} 
  \end{center}
  \caption{Empirical mean and standard deviation of the congestion window as 
  the function of the corresponding theoretical values in the range of loss 
  probabilities $p=0.1\%$--$5\%$.} 
  \label{fig:LAN_moments} 
\end{figure} 
In Figure~\ref{fig:LAN_moments} the empirical mean and standard deviation of
the congestion window is plotted as the function of the theoretical values
(\ref{eq:LAN_mean}) and (\ref{eq:LAN_stdev}), respectively.  The model predicts
measurement points on the diagonals, shown with dotted lines.  We can see that
the empirical standard deviation agrees well with the theoretical values, but
the average congestion window is systematically smaller than predicted.  The
linear fit $f(x)=x+b$ of the data points gives an estimate for the average 
shift $b=-1.4141\pm0.0611$.

The most important source of error is that \gls{frfr} algorithms have been
neglected in my idealized model, but the simulator does use these algorithms.  
The small plateaus appearing in the congestion window after each cycle produce 
bias towards the smaller window values.  

In a refined model let us consider the \gls{frfr} algorithms as well.  Denote
$\tilde{W}$ the fluid approximation of the extended congestion window process,
which can operate in either \gls{ca} or \gls{frfr} mode.  
\renewcommand{\FRFR}{FR/FR} 
\Gls{tcp} remains in \gls{frfr} mode until the \gls{ack} of a retransmitted
packet reaches the sender, that is the round-trip time \emph{before} the
\gls{frfr} mode started: $R(\beta^{-1}W)$.  Furthermore, the probability that a
plateau forms in the window interval $[w,w+dw]$ equals the probability that the
window is reduced to the given interval after a packet loss: $f_{\wal}(w)\,dw$.
The form of the distribution $f_{\wal}(w)$ might be derived by direct
calculation, but it can also be found by a simple argument: in stationary state
of \gls{tcp}---since the packet loss process is memoryless---packet loss can
occur at every congestion window value with the same probability, supposing
that the window has reached the given value.  In other words the value of the
``after loss'' congestion window is $\wal=\beta\wbl=\beta W$.  Accordingly, the
distribution of the ``after loss'' window is
$f_{\wal}(w)\,dw=f_{W}(\beta^{-1}w)\,d\beta^{-1}w$.  On condition that
\gls{tcp} is in \gls{frfr} mode the probability distribution of the congestion
window is
\begin{equation}
  f_{\tilde{W}}(w\mid \text{\tcp\ in FR/FR mode})
  =\frac{R(\beta^{-1}w)\beta^{-1}f_W(\beta^{-1}w)}{\mathbb{E}[R(W)]}
\end{equation}
On the other hand, the window distribution in the congestion avoidance mode can
clearly be given by $f_{\tilde{W}}(w\mid\text{\tcp\ in CA mode})=f_{W}(w)$.
Now only the probabilities of the \gls{ca} and \gls{frfr} modes are required.
Since each congestion avoidance phase is followed by a \gls{frfr} mode,
probabilities of the different modes are proportional to the average length of
the corresponding mode.  The mean length of a congestion avoidance mode is
evidently the average time between two packet losses:
$\mathbb{E}[\delta_i]=1/\ph$.  The average length of a \gls{frfr} period, on the
other hand, is simply the average length of the plateaus: $\mathbb{E}[R(W)]$.
This implies that 
\begin{align}
  \label{eq:CA_prob}
  \PP(\text{\tcp\ in CA mode})
  &=\frac{1/\ph}{1/\ph+\mathbb{E}[R(W)]},
  \quad\text{and}\\
  \label{eq:FRFR_prob}
  \PP(\text{\tcp\ in FRFR mode})
  &=\frac{\mathbb{E}[R(W)]}{1/\ph+\mathbb{E}[R(W)]}. 
\end{align} 

\begin{figure}
  \begin{center}
    \psfrag{ 0}[c][c][1]{$0$}
    \psfrag{ 5}[c][c][1]{$5$}
    \psfrag{ 10}[c][c][1]{$10$}
    \psfrag{ 15}[c][c][1]{$15$}
    \psfrag{ 20}[c][c][1]{$20$}
    \psfrag{ 25}[c][c][1]{$25$}
    \psfrag{ 30}[c][c][1]{$30$}
    \psfrag{ 35}[c][c][1]{$35$}
    \psfrag{ 40}[c][c][1]{$40$}
    \psfrag{ 50}[c][c][1]{$50$}
    \psfrag{ 60}[c][c][1]{$60$}
    \psfrag{ 80}[c][c][1]{$80$}
    \psfrag{ 100}[c][c][1]{$100$}
    \psfrag{ 120}[c][c][1]{$120$}
    \psfrag{ 0.005}[r][r][1]{$0.005$}
    \psfrag{ 0.01}[r][r][1]{$0.010$}
    \psfrag{ 0.015}[r][r][1]{$0.015$}
    \psfrag{ 0.02}[r][r][1]{$0.020$}
    \psfrag{ 0.025}[r][r][1]{$0.025$}
    \psfrag{ 0.03}[r][r][1]{$0.03$}
    \psfrag{ 0.04}[r][r][1]{$0.04$}
    \psfrag{ 0.05}[r][r][1]{$0.05$}
    \psfrag{ 0.06}[r][r][1]{$0.06$}
    \psfrag{ 0.07}[r][r][1]{$0.07$}
    \psfrag{ 0.08}[r][r][1]{$0.08$}
    \psfrag{ 0.1}[r][r][1]{$0.10$}
    \psfrag{ 0.12}[r][r][1]{$0.12$}
    \psfrag{ 0.14}[r][r][1]{$0.14$}
    \psfrag{ 0.16}[r][r][1]{$0.16$}
    \psfrag{ 0.18}[r][r][1]{$0.18$}
    \psfrag{f_W(w)}[c][c][1.2]{$f_W(w)$}
    \psfrag{w}[ct][cB][1.2]{$w$}
    \psfrag{theory}[r][r][0.8]{theory CA\&FR/FR}
    \psfrag{theory0}[r][r][0.8]{theory CA only}
    \psfrag{sim}[r][r][0.8]{simulation}
    \subfigure[Loss rate $p=10^{-3}$]{
      \resizebox{0.7\figwidth}{!}{\includegraphics{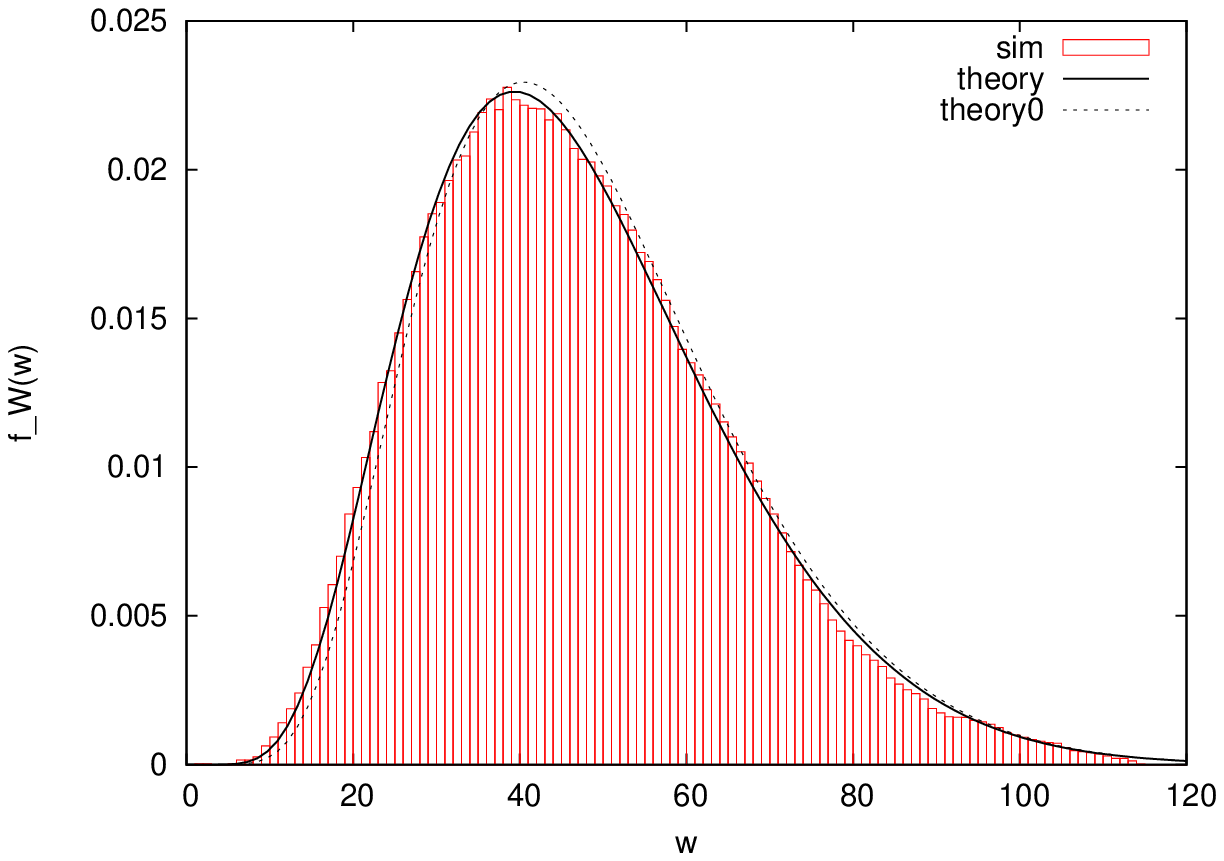}}}
    \psfrag{ 0.01}[r][r][1]{$0.01$}
    \psfrag{ 0.02}[r][r][1]{$0.02$}
    \subfigure[Loss rate $p=5\cdot10^{-3}$]{
      \resizebox{0.7\figwidth}{!}{\includegraphics{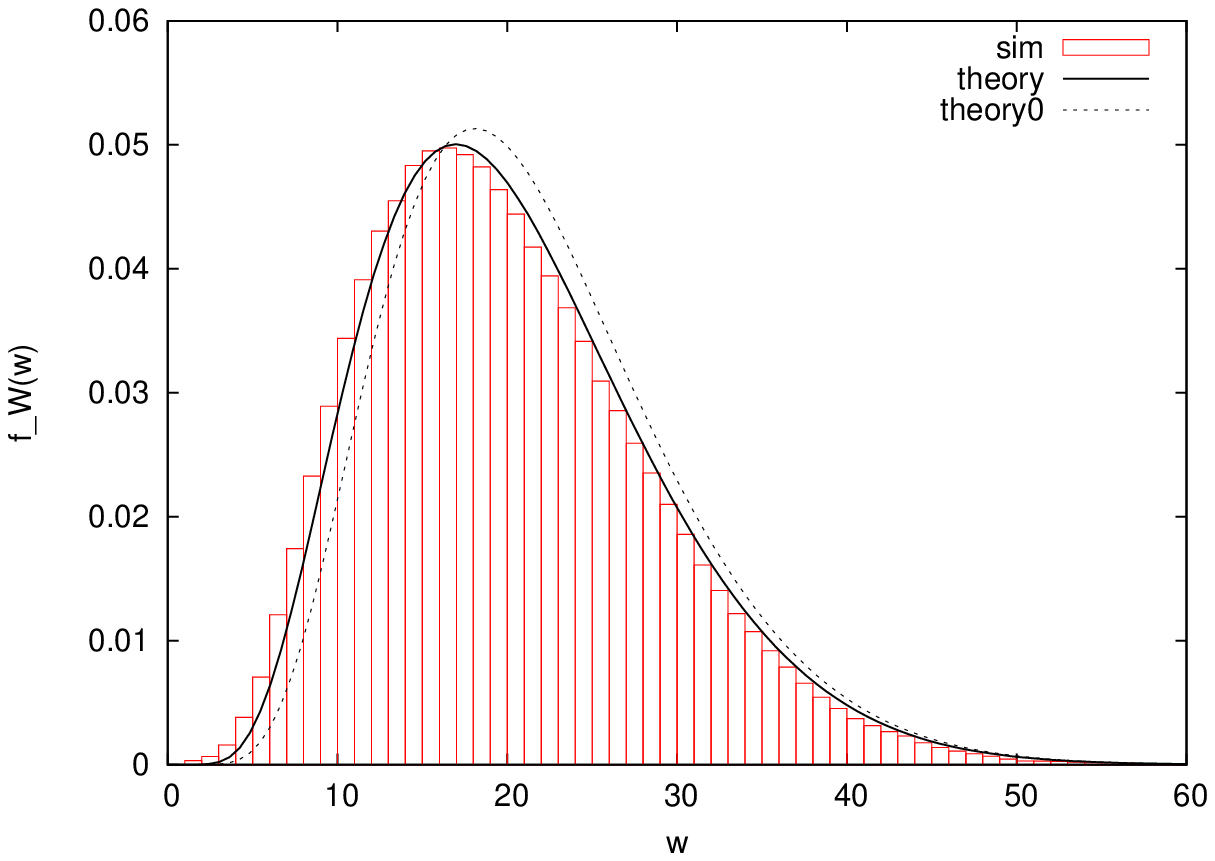}}}\\
    \subfigure[Loss rate $p=0.01$]{
      \resizebox{0.7\figwidth}{!}{\includegraphics{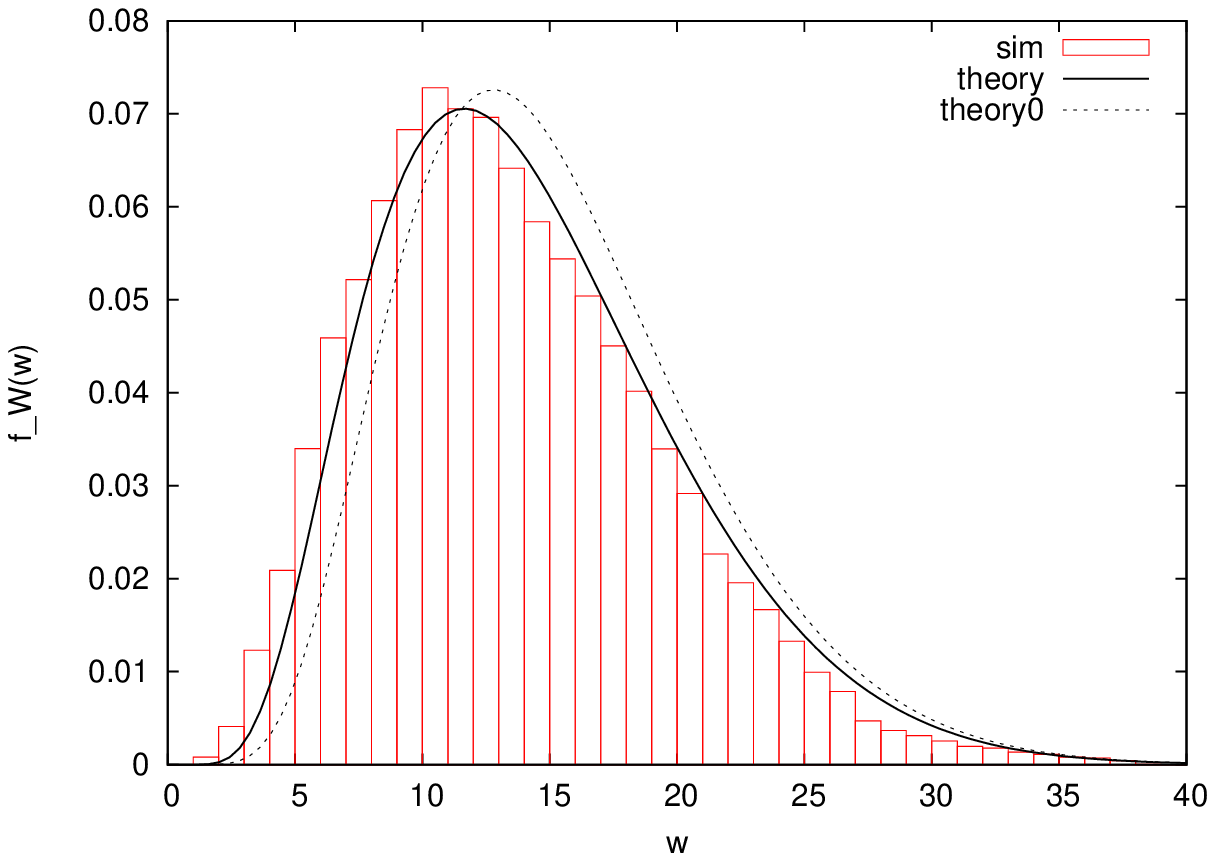}}}
    \subfigure[Loss rate $p=0.02$]{
      \resizebox{0.7\figwidth}{!}{\includegraphics{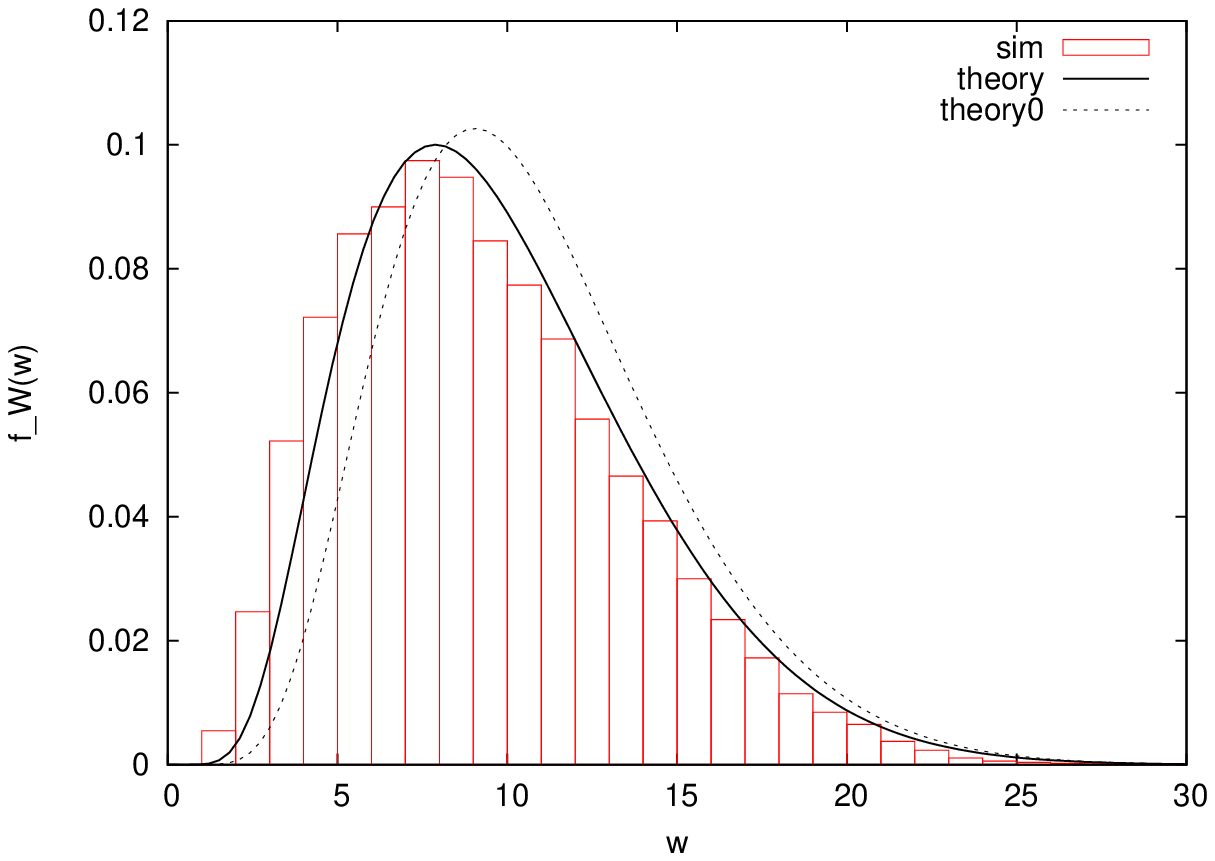}}}\\
    \subfigure[Loss rate $p=0.04$]{
      \resizebox{0.7\figwidth}{!}{\includegraphics{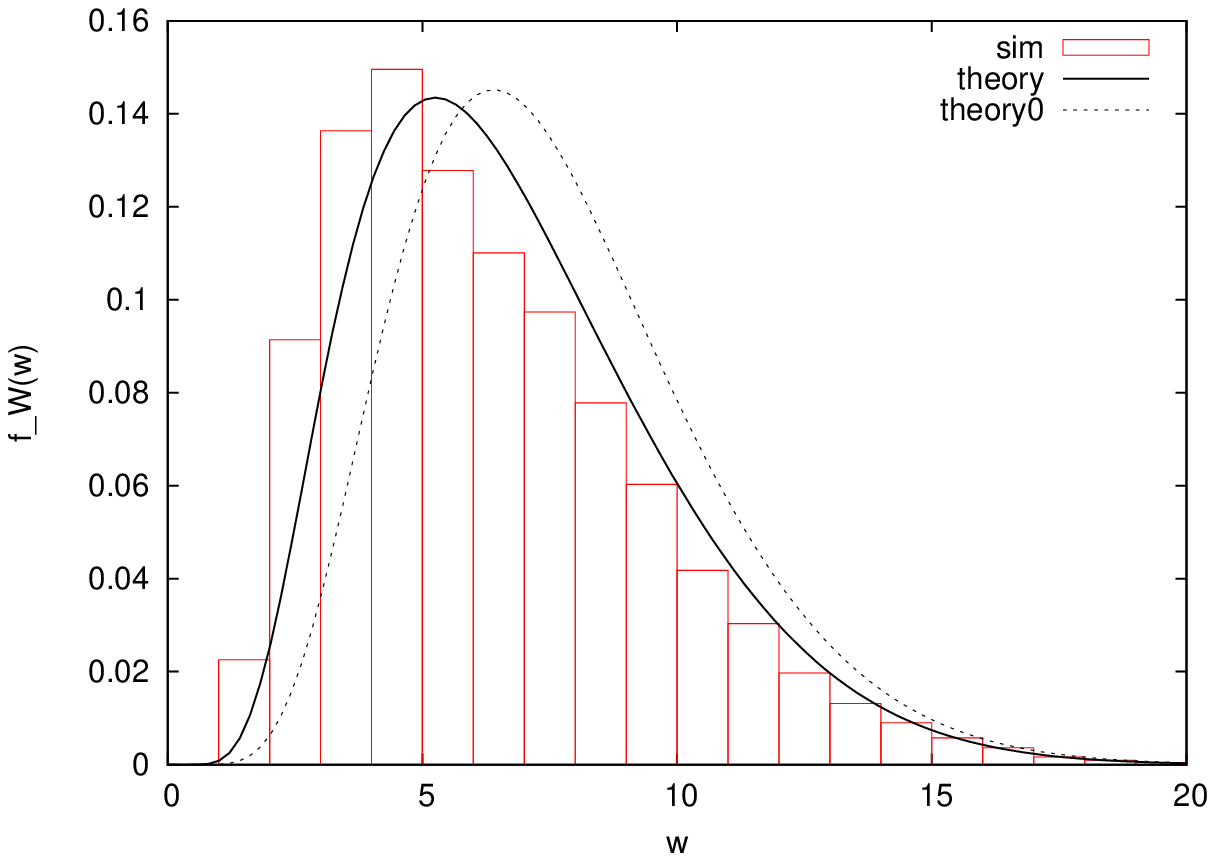}}}
    \subfigure[Loss rate $p=0.05$]{
      \resizebox{0.7\figwidth}{!}{\includegraphics{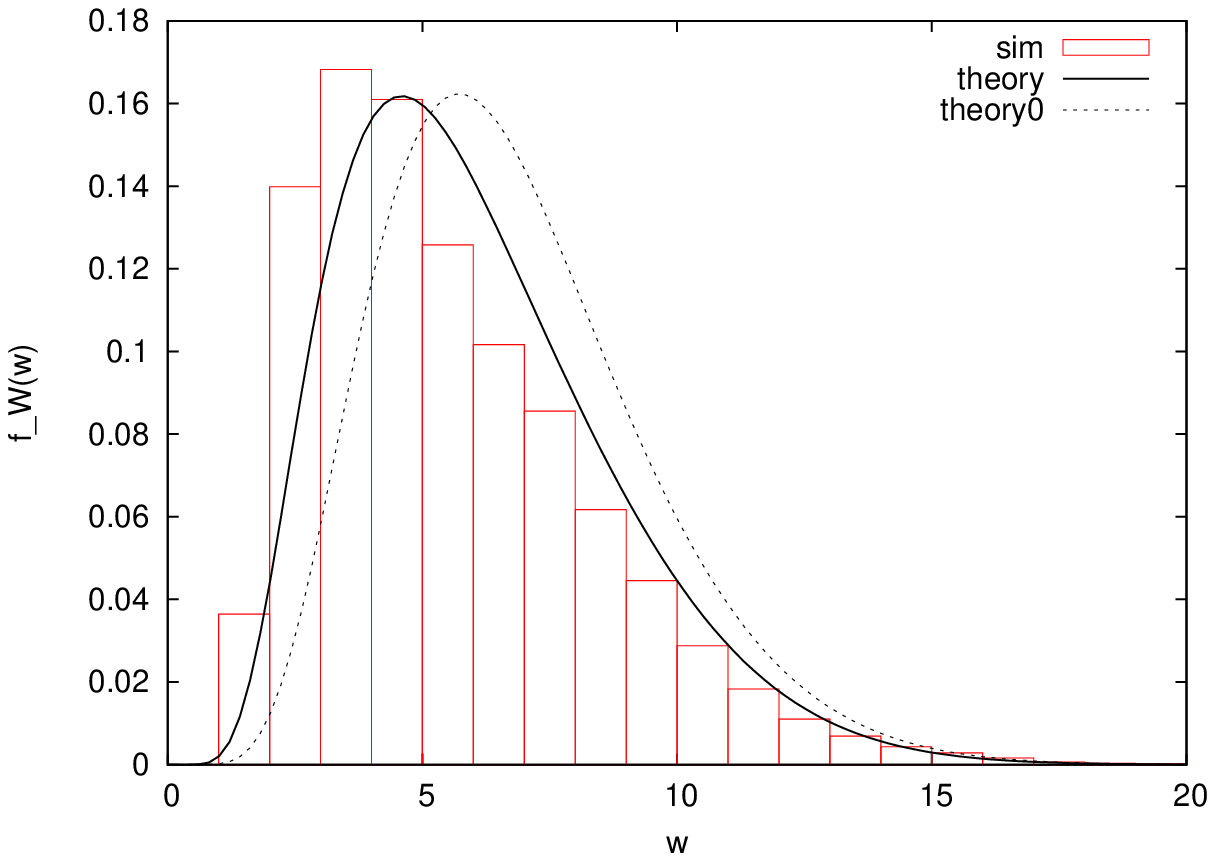}}}
  \end{center}
  \caption{Histograms and theoretical distributions of congestion windows in 
  LAN. Network parameters are $C=256\,kb/s$, $P=1500\,\mathit{byte}$, and 
  $D=0\,s$.}
  \label{fig:LAN_sim_fit}
\end{figure}

Therefore, the probability distribution of the congestion window extended by 
the \gls{frfr} algorithm is
\begin{equation}
  f_{\tilde{W}}(w)
  =\frac{f_{W}(w)+\frac{\ph}{\alpha}\beta^{-\left(m+1\right)}w^m f_W(\beta^{-1}w)}{1+\frac{\ph}{\alpha}\mathbb{E}[W^m]}
  \label{eq:cwnd_gendist}
\end{equation}
where (\ref{eq:rtt_model}), the definition of $R(W)$ has been substituted.
This formula is the main result of this section.  In
Figure~\ref{fig:LAN_sim_fit} the histogram of the congestion window simulated
with \gls{ns} and (\ref{eq:cwnd_gendist}) are compared in the range of loss 
probabilities $p=0.1\%$--$5.0\%$. We can see an almost perfect match between
theory and simulation.  In order to illustrate the improvement of the formula
(\ref{eq:cwnd_gendist}) on (\ref{eq:W_dist}), I plotted $f_W(w)$ with dotted 
lines for comparison.  I must stress here that there are no tunable 
parameters in (\ref{eq:cwnd_gendist}) and no parameter fit has been made.

Let us calculate the moments of $\tilde{W}$:
\begin{equation}
  \mathbb{E}[{\tilde{W}}^k]
  =\frac{\mathbb{E}[W^k]+\beta^k\frac{\ph}{\alpha}\mathbb{E}[W^{m+k}]}
  {1+\frac{\ph}{\alpha}\mathbb{E}[W^m]}.
\end{equation}
As an important special case we can calculate the correction of the \gls{frfr}
algorithms to the mean of the congestion window:
\begin{equation}
  \mathbb{E}[\tilde{W}-W]=
  -\frac{\ph}{\alpha}
  \frac{\mathbb{E}[W]\mathbb{E}[W^m]-\beta\mathbb{E}[W^{m+1}]}
   {1+\frac{\ph}{\alpha}\mathbb{E}[W^m]}
\end{equation}
If the formula (\ref{eq:moment_r}) is substituted into the above equation one
can obtain the dependence of the correction on $\ph/\alpha$.  Specifically, for
$m=1$:
$\mathbb{E}[\tilde{W}-W]\approx-\frac{0.9981}{1+1.5269\sqrt{\frac{\ph}{\alpha}}}$.
Interestingly, the correction tends to a constant in the small loss limit:
$\lim_{\ph/\alpha\to0}\mathbb{E}[\tilde{W}-W]\approx-0.9981$.  In the range of
loss probabilities $p=\ph/\alpha=10^{-4}$--$5\cdot10^{-2}$, investigated by
simulations, the correction to the congestion window average is between
$-0.8659$ and $0.9831$, which is less than observed in
Fig.~\ref{subfig:LAN_mean}.  The remaining discrepancy comes from the
difference between the continuous and the fluid value of $W$. In the simulation
the congestion window is not only halved its integer part is also taken.  This
discrepancy accounts for approximately $-0.5$ unit shift on average.  The slow
start mechanism, which becomes more and more dominant as the loss probability
increases, also makes the small window values more probable.  However, these
effects are beyond the scope of the applied fluid model.


\subsection{Wide Area Networks}\label{subsec:OneTCP_WAN}
\makeatletter

\begin{figure}
  \begin{center}
    \psfrag{ 0}[r][r][1]{$0$}
    \psfrag{ 4}[r][r][1]{$4$}
    \psfrag{ 5}[r][r][1]{$5$}
    \psfrag{ 6}[r][r][1]{$6$}
    \psfrag{ 7}[r][r][1]{$7$}
    \psfrag{ 8}[r][r][1]{$8$}
    \psfrag{ 9}[r][r][1]{$9$}
    \psfrag{ 10}[r][r][1]{$10$}
    \psfrag{ 15}[r][r][1]{$15$}
    \psfrag{ 20}[r][r][1]{$20$}
    \psfrag{ 25}[r][r][1]{$25$}
    \psfrag{ 200}[c][c][1]{$200$}
    \psfrag{ 250}[c][c][1]{$250$}
    \psfrag{ 300}[c][c][1]{$300$}
    \psfrag{ 350}[c][c][1]{$350$}
    \psfrag{ 374}[c][c][1]{$374$}
    \psfrag{ 376}[c][c][1]{$376$}
    \psfrag{ 378}[c][c][1]{$378$}
    \psfrag{ 380}[c][c][1]{$380$}
    \psfrag{ 382}[c][c][1]{$382$}
    \psfrag{ 384}[c][c][1]{$384$}
    \psfrag{ 400}[c][c][1]{$400$}
    \psfrag{ 450}[c][c][1]{$450$}
    \psfrag{ 500}[c][c][1]{$500$}
    \psfrag{simulation}[r][r][0.8]{simulation}
    \psfrag{cwnd}[r][r][1.2]{$W(t)$}
    \psfrag{T}[c][c][1.2]{$t$}
    \subfigure[\nstwo\ simulation of WAN]{\resizebox{0.7\figwidth}{!}
      {\includegraphics{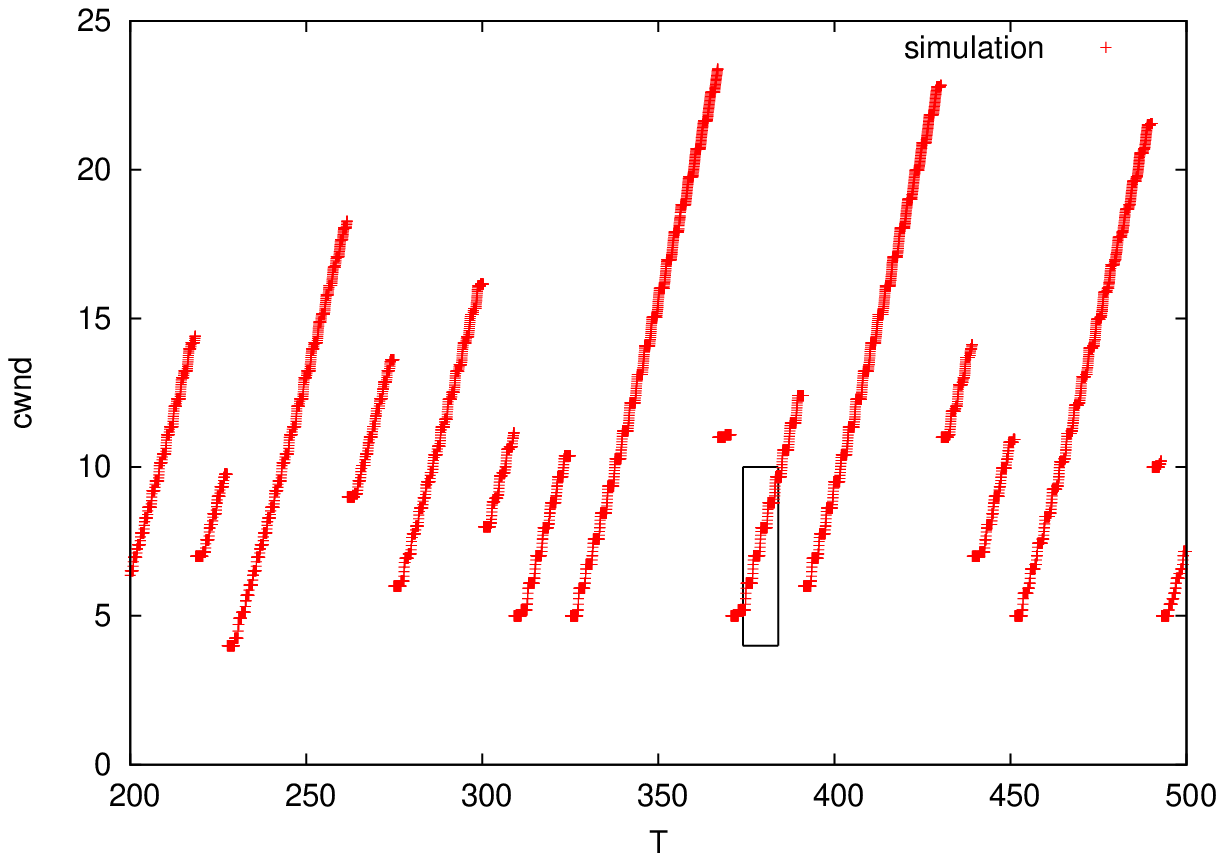}}\label{subfig:WAN_simulation}}
    \psfrag{w0}[r][r][1]{$W_0$}
    \psfrag{w1}[r][r][1]{$W_1$}
    \psfrag{d0}[c][c][1]{$\delta_0$}
    \psfrag{d1}[c][c][1]{$\delta_1$}
    \psfrag{d2}[c][c][1]{$\delta_2$}
    \psfrag{d3}[c][c][1]{$\delta_3$}
    \psfrag{d4}[c][c][1]{$\delta_4$}
    \psfrag{t0}[c][c][1]{$\tau_0$}
    \psfrag{t1}[c][c][1]{$\tau_1$}
    \psfrag{t2}[c][c][1]{$\tau_2$}
    \psfrag{t3}[c][c][1]{$\tau_3$}
    \psfrag{t4}[c][c][1]{$\tau_4$}
    \psfrag{t5}[c][c][1]{$\tau_5$}
    \psfrag{ACTIVE}[ct][cb][1]{Active}
    \psfrag{IDLE}[cb][ct][1]{Idle}
    \psfrag{JUMP1}[cb][ct][1]{Jump}
    \psfrag{JUMP2}[cb][ct][1]{Jump}
    \psfrag{JUMP3}[cb][ct][1]{Jump}
    \psfrag{JUMP4}[cb][ct][1]{Jump}
    \subfigure[Details of the cwnd in the bounded box]{\resizebox{0.7\figwidth}{!}
      {\includegraphics{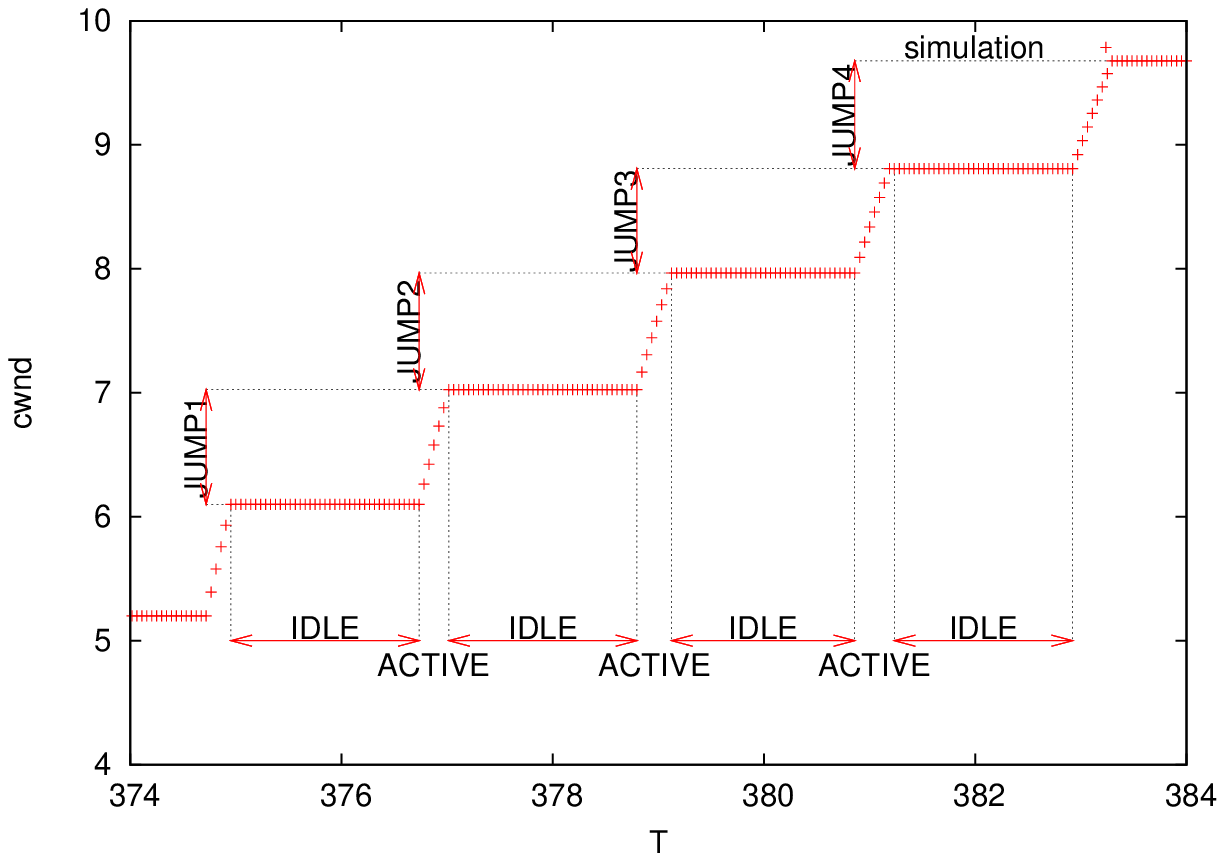}}\label{subfig:WAN_detail}}
    \caption{The congestion avoidance process of \tcp\ in WAN setup.  
             The global congestion window development is seemingly linear, but 
             the detailed plot on Figure~\ref{subfig:WAN_detail} shows a 
             different picture.  The globally linear growth is composed of 
             alternating idle and LAN-like active periods.  See the discussion
             in the text.}
    \label{fig:simulation_WAN}
  \end{center}
\end{figure}

I turn now to the \gls{wan} scenario, where buffering delay is very small
compared to the link delay.  A typical congestion window sequence is shown in
Figure~\ref{fig:simulation_WAN} with $D=1s$ and $p=0.01$.

The applicability of the developed model depends on two crucial factors: the
validity of the exponential inter-loss distribution (\ref{eq:dist_delta}) and
the validity of (\ref{eq:rtt_model}), the dependence of round-trip time on the
congestion window.  The difficulty of the \gls{wan} scenario is that---as I
mentioned earlier---there are periods when no packet leaves the buffer.  This
effect corrupts the validity of both assumptions.  Packets cannot be lost
during idle periods, so the inter-loss time distribution deviates from
exponential distribution.  Furthermore, if we assume that the round-trip time
is constant, $R(W)=2D$, then the solution of the equation of motion
(\ref{eq:cwnd_solv}) predicts linear congestion window development, which
corresponds to $m=0$ in the model.  However, a typical sequence of congestion
windows, displayed in Fig.~\ref{subfig:WAN_detail}, shows step-like growth
instead.  Another difficulty is that even if the inter-arrival times can be
approximated with an exponential distribution, the connection between the
packet loss probability $p$ and the parameter of the distribution $\ph$ is
unknown.  

Given these concerns I approach the congestion window development in a
\gls{wan} network in a manner different from simple linear growth, the model
applied exclusively in the literature.  First of all let us investigate the
window development in congestion avoidance mode in more detail.  The fine
structure of the congestion window is shown in
Fig.~\ref{subfig:WAN_detail}.  During the active period of \tcp, when
\glspl{ack} are arriving back to the sender (in ``\gls{ack} time''), the
congestion window is being increased the same way as in a \gls{lan} network.
The difference from \gls{lan} in a \gls{wan} scenario is that an idle period
follows with a constant congestion window.  Since $W$ packets are transferred
in an active period, the length of an active period is $W/\alpha$.  The
following idle period is $2D-W/\alpha$ long, because the total length of an
active and the succeeding idle period is precisely one round-trip time, $2D$.  

Let $W^*$ denote the congestion window idle periods included.  If the plateaus
corresponding to idle periods are approximated as if they were blurred evenly
on the active periods, then---analogously to the \gls{frfr} mode in
\gls{lan}---the conditional distribution of the congestion window in idle mode
of \tcp\ can be formulated as
\begin{equation}
  f_{W^*}(w\mid\text{\tcp\ in IDLE mode})
  =\frac{\frac{2D-w/\alpha}{w/\alpha} f_{W}(w)\,\Theta(2D-w/\alpha)}
  {\mathbb{E}\left[\frac{2D-W/\alpha}{W/\alpha} \Theta(2D-W/\alpha)\right]}.
\end{equation}

Only the probabilities of idle, \gls{ca} and \gls{frfr} modes are required.  The
probability of each mode is proportional to the average time \gls{tcp} spends in
the particular mode.  Considering the idle mode, the
average length of a plateau in one \gls{ack} increment is
$\mathbb{E}\left[\frac{2D-W/\alpha}{W/\alpha}\Theta(2D-W/\alpha)\right]$.
Moreover, the window is increased $\mathbb{E}[\delta_i]=1/\ph$ times in one 
loss cycle on average.  Accordingly, the probability of idle mode is
\begin{equation}
  \PP(\text{\tcp\ in IDLE mode})
  =\frac{\mathbb{E}\left[\frac{2\alpha D-W}{W}\Theta(2\alpha D-W)\right]/\ph}
  {1/\ph+\mathbb{E}[R(W)]+\mathbb{E}\left[\frac{2\alpha D-W}{W}\Theta(2\alpha D-W)\right]/\ph}.
\end{equation}
The probabilities of \gls{ca} and \gls{frfr} modes in Eqs.~(\ref{eq:CA_prob}) and 
(\ref{eq:FRFR_prob}) should be modified proportionately.  As a result we 
obtain
\begin{equation}
  f_{W^*}(w)
  =\frac{f_{W}(w)+\frac{\ph}{\alpha}\beta^{-\left(m+1\right)}w^m f_W(\beta^{-1}w)+\frac{2\alpha D-w}{w} f_{W}(w)\,\Theta(2\alpha D-w)}
  {\bar{F}_W(2\alpha D)+\frac{\ph}{\alpha}\mathbb{E}[W^m]+2\alpha D\,\mathbb{E}\left[\frac{1}{W}\Theta(2\alpha D-W)\right]}
  \label{eq:cwnd_gendist_WAN}
\end{equation}
for the congestion window distribution, where I used that
$\mathbb{E}[\Theta(2\alpha D-W)]=\int_0^{2\alpha
D}f_W(w)\,dw=1-\bar{F}_W(2\alpha D)$.  The truncated expectation of $1/W$ can 
be obtained similarly to (\ref{eq:moment_r}) with $r=-\frac1{m+1}$, but 
one should include the incomplete Gamma function
$\Gamma(z,x)=\int_x^{\infty}x^{z-1} e^{-x}\,dx$ as well:
\begin{equation}
  \begin{split}
    &\mathbb{E}\left[\frac1{W}\Theta(2\alpha D-W)\right]
    =\int_0^{2\alpha D}\frac1{w}f_W(w)\,dw\\
    =&\mathbb{E}\left[\frac1{W}\right]
    -\left(\frac{\ph}{\alpha\left(m+1\right)}\right)^{\frac1{m+1}}
    \sum_{k=0}^{\infty}
    \Gamma\left(\frac{m}{m+1},\frac{2\ph D c^{-k}}{m+1}\right)
    c^{\frac{mk}{m+1}}h_k(c).
  \end{split}
\end{equation}

The moments of $W^*$ can be given easily:
\begin{equation}
  \label{eq:cwnd_genmoments}
  \mathbb{E}\bigl[{W^*}^k\bigr]=
  \frac{\mathbb{E}[W^k]+\frac{\ph}{\alpha}\beta^k\mathbb{E}[W^{m+k}]+
  \mathbb{E}\left[\left(2\alpha D\,W^{k-1}-W^k\right)\Theta(2\alpha D-W)\right]}
  {\bar{F}_W(2\alpha D)+\frac{\ph}{\alpha}\mathbb{E}[W^m]+2\alpha D\,\mathbb{E}\left[\frac{1}{W}\Theta(2\alpha D-W)\right]}\,.
\end{equation}
 
The distribution and moments of the ideal \gls{wan} scenario can be obtained from 
(\ref{eq:cwnd_gendist_WAN}) and (\ref{eq:cwnd_genmoments}) in the  
$\alpha D\to\infty$ limit:
\begin{align}
  \lim_{\alpha D\to\infty}f_{W^*}(w)
  &=\frac1{w\mathbb{E}\left[\frac1{W}\right]}f_W(w)
&\text{and}&&
  \lim_{\alpha D\to\infty}\mathbb{E}\bigl[{W^*}^k\bigr]
  &=\frac{\mathbb{E}[W^{k-1}]}{\mathbb{E}\left[\frac1{W}\right]}.
  \label{eq:cwnd_ideal_WAN_limit}
\end{align}
Note that the formula (\ref{eq:cwnd_gendist_WAN}) at $D=0$ reduces to
(\ref{eq:cwnd_gendist}), derived earlier for an ideal \gls{lan} scenario with
\gls{ca} and \gls{frfr} modes.  Note further that (\ref{eq:cwnd_gendist_WAN})
is applicable not only for the ideal \gls{wan} or \gls{lan} scenarios, but also
for \emph{for the most generic configuration}.  Moreover, I would like to
emphasize that the parameters of the model can be obtained from the intrinsic
``\gls{ack} time'' dynamics of \tcp, and \emph{no parameter fitting is
necessary}.  Specifically, for \gls{tcp}/Reno the parameters are $m=1$,
$\ph/\alpha=p$, $\beta=1/2$ and $2\alpha D$ is the bandwidth-delay product
measured in packet units.

\begin{figure}
  \begin{center}
    \psfrag{ 0}[c][c][1]{$0$}
    \psfrag{ 5}[c][c][1]{$5$}
    \psfrag{ 10}[c][c][1]{$10$}
    \psfrag{ 15}[c][c][1]{$15$}
    \psfrag{ 20}[c][c][1]{$20$}
    \psfrag{ 25}[c][c][1]{$25$}
    \psfrag{ 30}[c][c][1]{$30$}
    \psfrag{ 35}[c][c][1]{$35$}
    \psfrag{ 40}[c][c][1]{$40$}
    \psfrag{ 50}[c][c][1]{$50$}
    \psfrag{ 60}[c][c][1]{$60$}
    \psfrag{ 80}[c][c][1]{$80$}
    \psfrag{ 100}[c][c][1]{$100$}
    \psfrag{ 120}[c][c][1]{$120$}
    \psfrag{ 140}[c][c][1]{$140$}
    \psfrag{ 150}[c][c][1]{$150$}
    \psfrag{ 160}[c][c][1]{$160$}
    \psfrag{ 180}[c][c][1]{$180$}
    \psfrag{ 200}[c][c][1]{$200$}
    \psfrag{ 250}[c][c][1]{$250$}
    \psfrag{ 300}[c][c][1]{$300$}
    \psfrag{ 350}[c][c][1]{$350$}
    \psfrag{ 400}[c][c][1]{$400$}
    \psfrag{ 0.001}[r][r][1]{$0.001$}
    \psfrag{ 0.002}[r][r][1]{$0.002$}
    \psfrag{ 0.003}[r][r][1]{$0.003$}
    \psfrag{ 0.004}[r][r][1]{$0.004$}
    \psfrag{ 0.005}[r][r][1]{$0.005$}
    \psfrag{ 0.006}[r][r][1]{$0.006$}
    \psfrag{ 0.007}[r][r][1]{$0.007$}
    \psfrag{ 0.008}[r][r][1]{$0.008$}
    \psfrag{ 0.009}[r][r][1]{$0.009$}
    \psfrag{ 0.01}[r][r][1]{$0.01$}
    \psfrag{ 0.012}[r][r][1]{$0.012$}
    \psfrag{ 0.014}[r][r][1]{$0.014$}
    \psfrag{ 0.015}[r][r][1]{$0.015$}
    \psfrag{ 0.016}[r][r][1]{$0.016$}
    \psfrag{ 0.018}[r][r][1]{$0.018$}
    \psfrag{ 0.02}[r][r][1]{$0.02$}
    \psfrag{ 0.025}[r][r][1]{$0.025$}
    \psfrag{ 0.03}[r][r][1]{$0.03$}
    \psfrag{ 0.04}[r][r][1]{$0.04$}
    \psfrag{ 0.05}[r][r][1]{$0.05$}
    \psfrag{ 0.06}[r][r][1]{$0.06$}
    \psfrag{ 0.07}[r][r][1]{$0.07$}
    \psfrag{ 0.08}[r][r][1]{$0.08$}
    \psfrag{ 0.09}[r][r][1]{$0.09$}
    \psfrag{ 0.1}[r][r][1]{$0.1$}
    \psfrag{ 0.15}[r][r][1]{$0.15$}
    \psfrag{ 0.2}[r][r][1]{$0.2$}
    \psfrag{ 0.25}[r][r][1]{$0.25$}
    \psfrag{f_W(w)}[c][c][1.2]{$f_W(w)$}
    \psfrag{w}[ct][cB][1.2]{$w$}
    \psfrag{theory}[r][r][0.8]{theory}
    \psfrag{ack-wise}[r][r][0.8]{theory---active mode}
    \psfrag{sim}[r][r][0.8]{simulation}
    \subfigure[Loss rate $p=10^{-4}$]{
      \resizebox{0.7\figwidth}{!}{\includegraphics{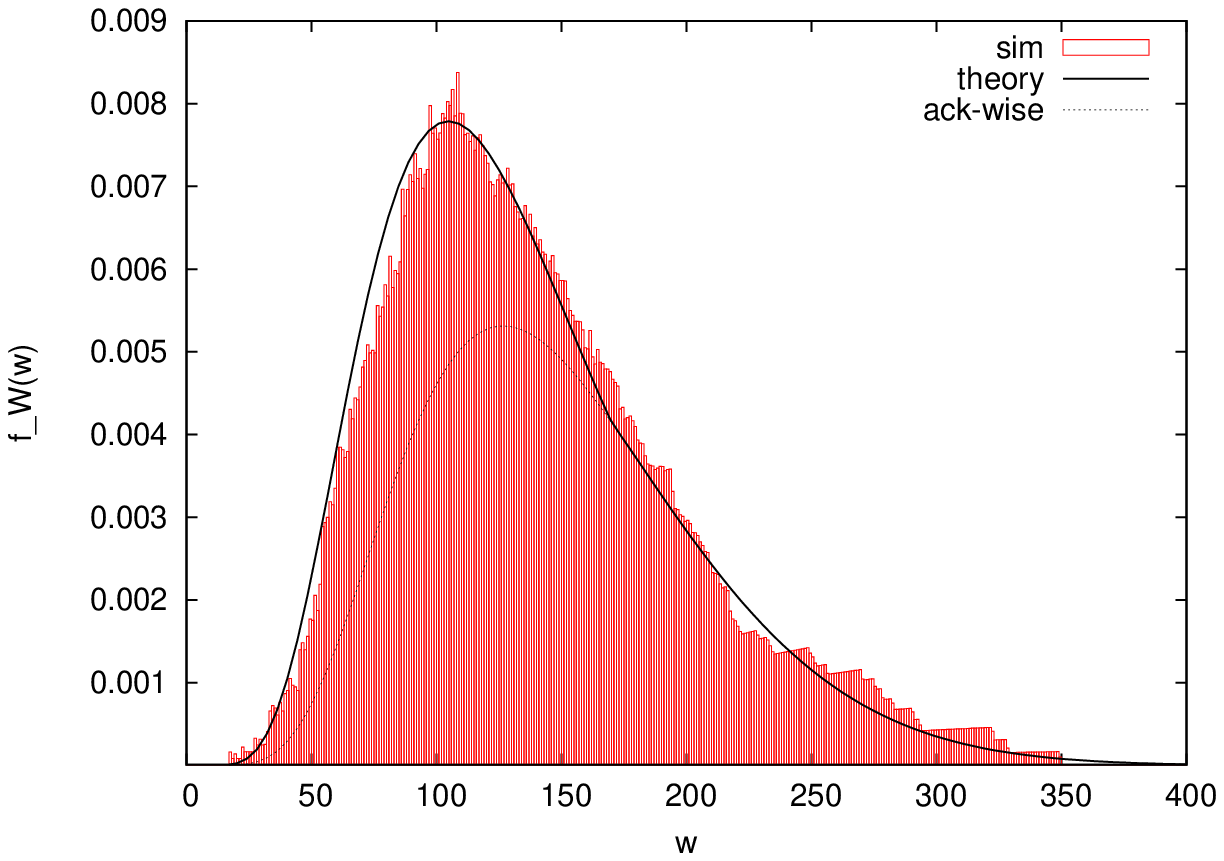}}}
    \subfigure[Loss rate $p=5\cdot10^{-4}$]{
      \resizebox{0.7\figwidth}{!}{\includegraphics{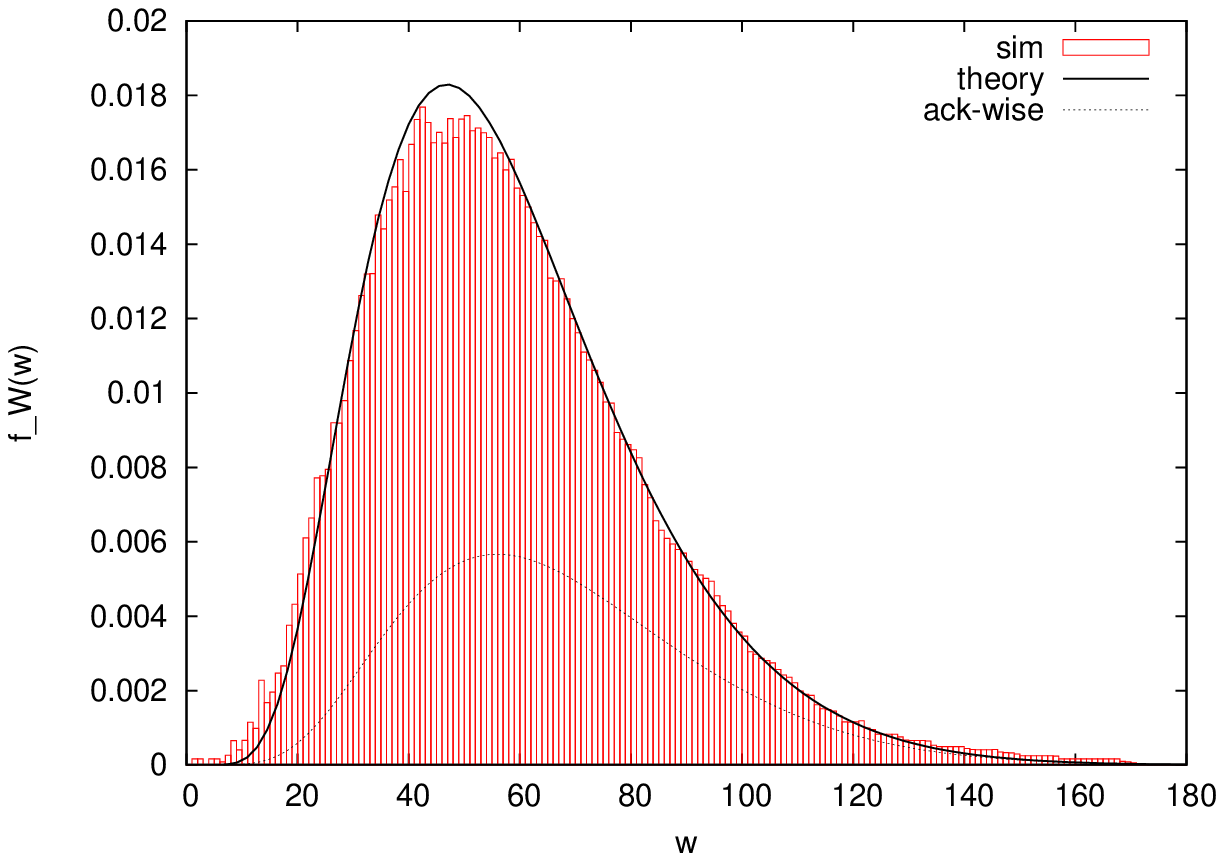}}}\\
    \subfigure[Loss rate $p=0.001$]{
      \resizebox{0.7\figwidth}{!}{\includegraphics{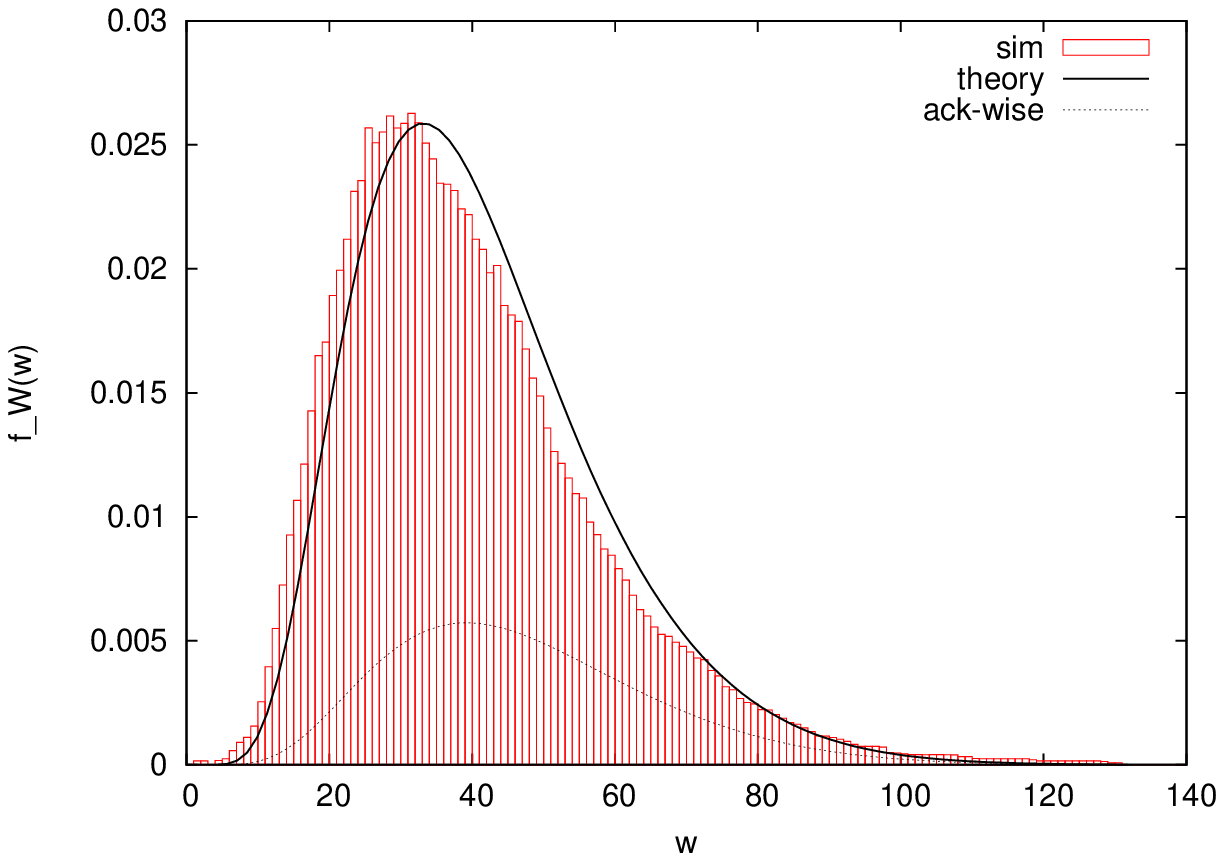}}}
    \subfigure[Loss rate $p=0.005$]{
      \resizebox{0.7\figwidth}{!}{\includegraphics{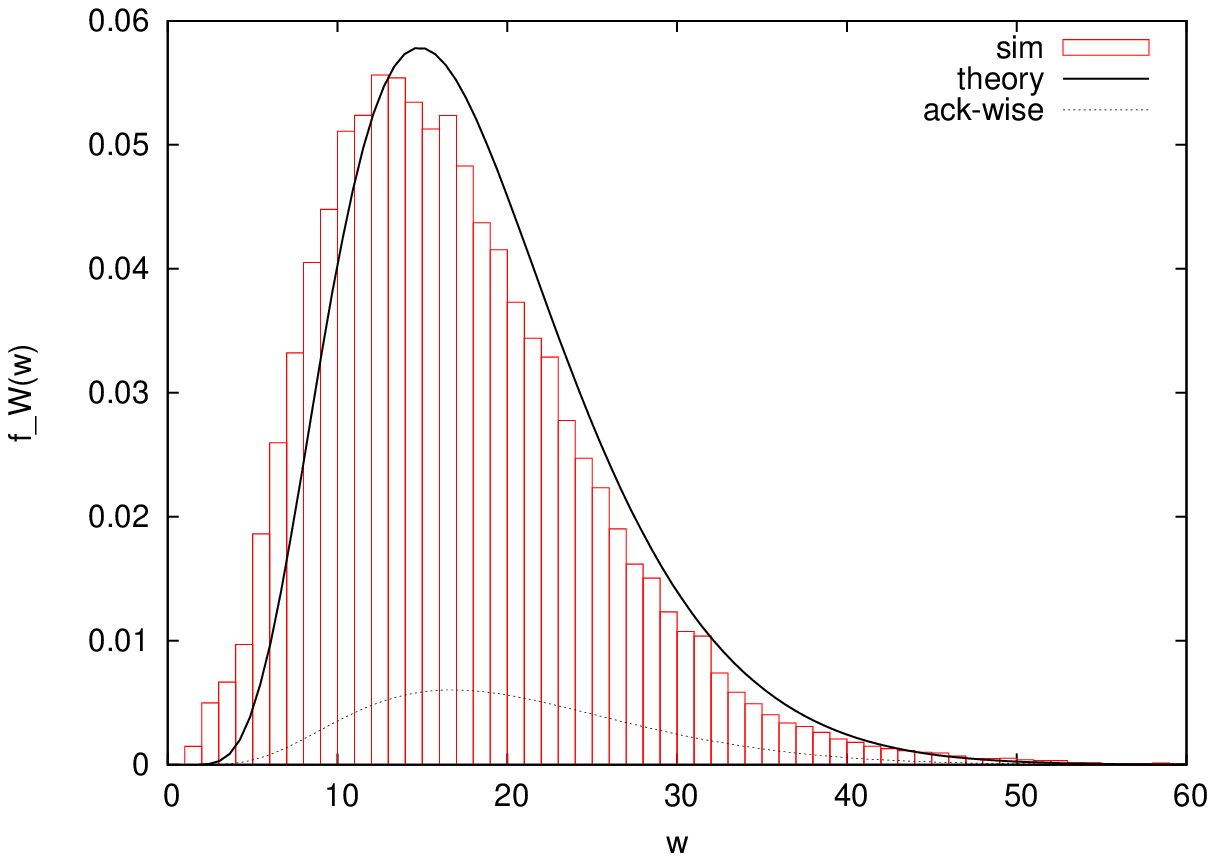}}}\\
    \subfigure[Loss rate $p=0.01$]{
      \resizebox{0.7\figwidth}{!}{\includegraphics{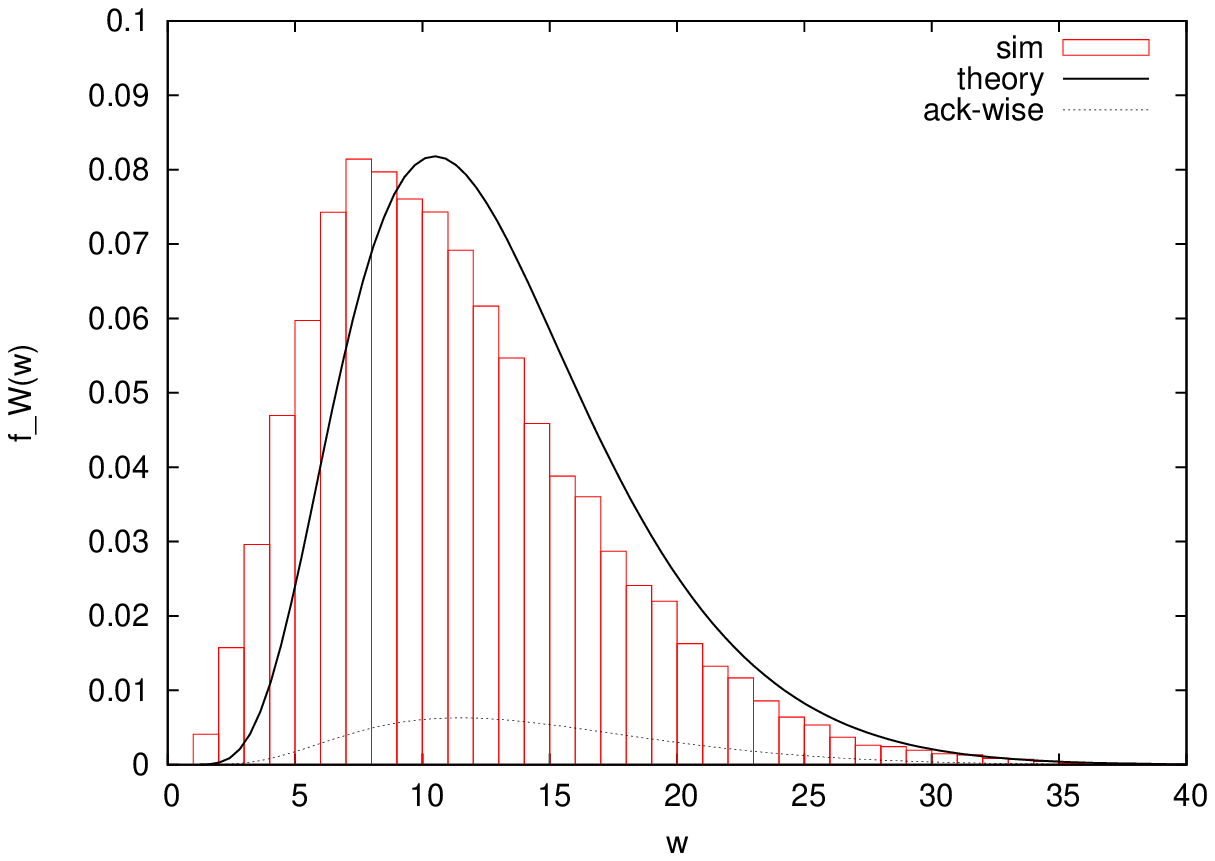}}}
    \subfigure[Loss rate $p=0.05$]{
      \resizebox{0.7\figwidth}{!}{\includegraphics{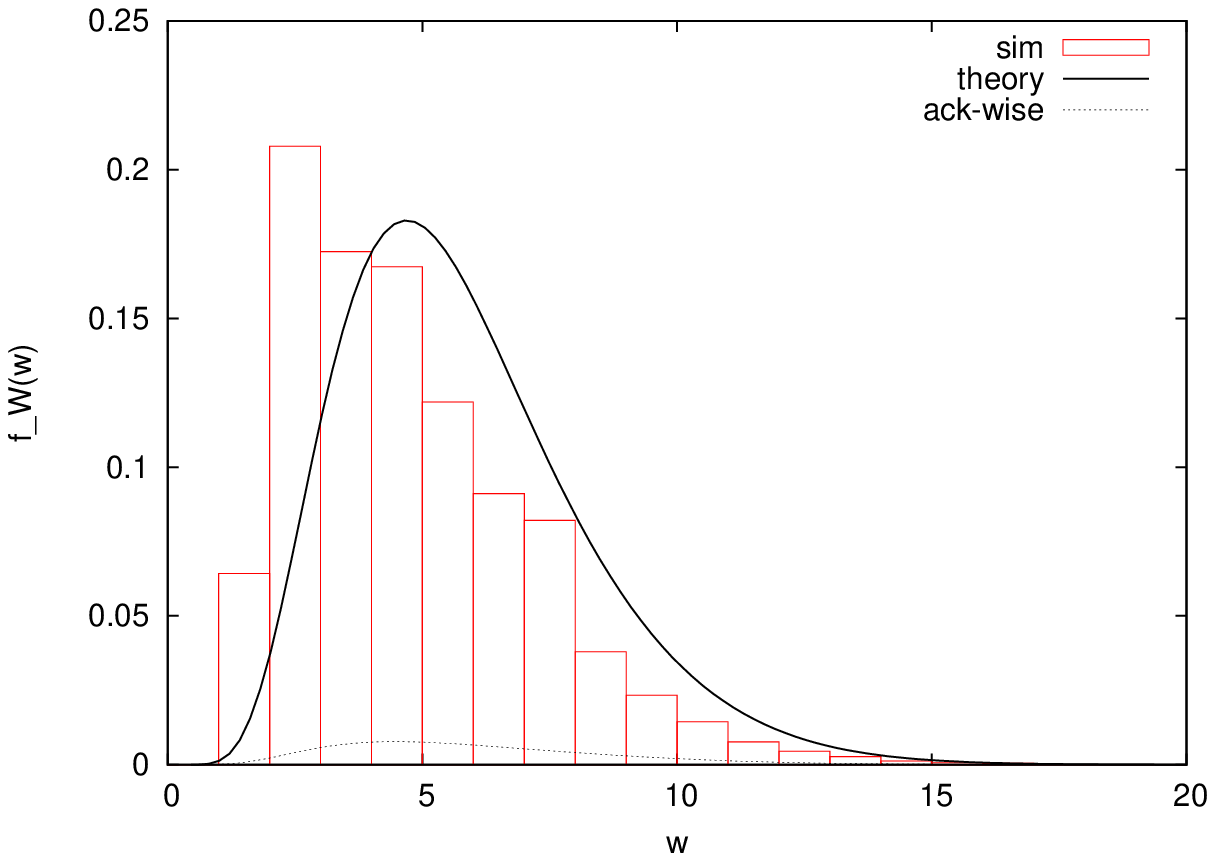}}}
  \end{center}
  \caption{Histograms and theoretical distributions of congestion windows in 
  WAN.  The bandwidth-delay product is $2DC/P=170.67$, measured in packets.  
  Note that this value falls in the bulk of the distribution at loss rate 
  $p=10^{-4}$, which means a transient between the ideal LAN and WAN.}
  \label{fig:WAN_sim_fit}
\end{figure}
In order to verify (\ref{eq:cwnd_gendist_WAN}) I carried out simulations.  The
link parameter $2\alpha D$ has been set to $170$ packets and the packet loss
has been varied in the range of $p=10^{-4}-5\cdot10^{-2}$.  Simulation results
are shown in Fig.~\ref{fig:WAN_sim_fit}.  I have plotted the contribution of
the active periods to the theoretical distribution with dotted lines for
comparison.  A transient between the ideal \gls{lan} and \gls{wan} network
configuration can be observed at $p=10^{-4}$, since the parameter $2\alpha
D=170$ falls in the bulk of the distribution.  An excellent fit can be seen at
small loss probabilities and a small discrepancy can be detected in the
mid-range $10^{-3}<p<10^{-2}$.  For probabilities $p>0.01$ the neglected slow
start mechanism becomes more and more significant.  As a result the theoretical
distribution deviates from the measured histogram more markedly.


\subsection{Dynamics of parallel TCPs}
\label{sec:ManyTCP}

Now, I am in the position to extend my results for parallel \glspl{tcp}.  Since
(\ref{eq:cwnd_gendist_WAN}) involves only the intrinsic \tcp\ dynamics, the
model parameters $m=1$, $\ph/\alpha=p$, and $\beta=1/2$ are the same for
parallel and a single \gls{tcp}.  The propagation delay parameter $2\alpha D$
might change, however, because the interaction of different \glspl{tcp} might
alter the idle periods experienced by the individual \glspl{tcp}.

\begin{figure}
  \begin{center}
    \psfrag{ 0}[c][c][1]{$0$}
    \psfrag{ 10}[c][c][1]{$10$}
    \psfrag{ 20}[c][c][1]{$20$}
    \psfrag{ 30}[c][c][1]{$30$}
    \psfrag{ 40}[c][c][1]{$40$}
    \psfrag{ 50}[c][c][1]{$50$}
    \psfrag{ 60}[c][c][1]{$60$}
    \psfrag{ 0.01}[r][r][1]{$0.01$}
    \psfrag{ 0.02}[r][r][1]{$0.02$}
    \psfrag{ 0.03}[r][r][1]{$0.03$}
    \psfrag{ 0.04}[r][r][1]{$0.04$}
    \psfrag{ 0.05}[r][r][1]{$0.05$}
    \psfrag{ 0.06}[r][r][1]{$0.06$}
    \psfrag{ 0.07}[r][r][1]{$0.07$}
    \psfrag{f_W(w)}[c][c][1.2]{$f_W(w)$}
    \psfrag{w}[ct][cB][1.2]{$w$}
    \psfrag{theory}[r][r][0.8]{mean field \& single TCP theory}
    \psfrag{mean field theory}[r][r][0.8]{mean field theory}
    \psfrag{theory LAN}[r][r][0.8]{single \tcp\ theory}
    \psfrag{sim1}[r][r][0.8]{single \tcp}
    \psfrag{sim2}[r][r][0.8]{one of two \tcp s}
    \subfigure[WAN with $D=2s$ ($2\alpha D=85.33$).]{
      \label{subfig:2TCP_WAN}
      \resizebox{0.69\figwidth}{!}{\includegraphics{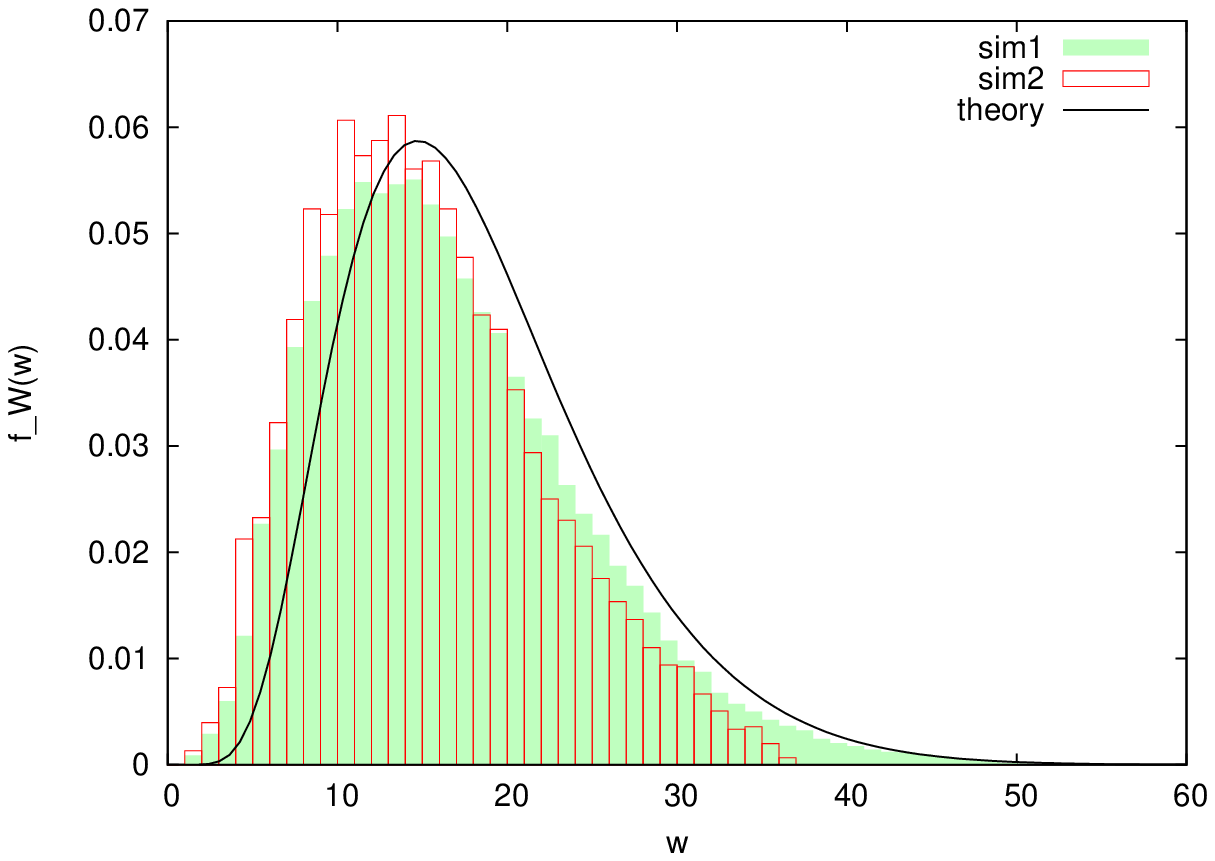}}
      }
    \subfigure[LAN with $D=0s$ ({$N\mathbb{E}[W^*]\approx36.55$}).]{
      \label{subfig:2TCP_LAN}
      \resizebox{0.69\figwidth}{!}{\includegraphics{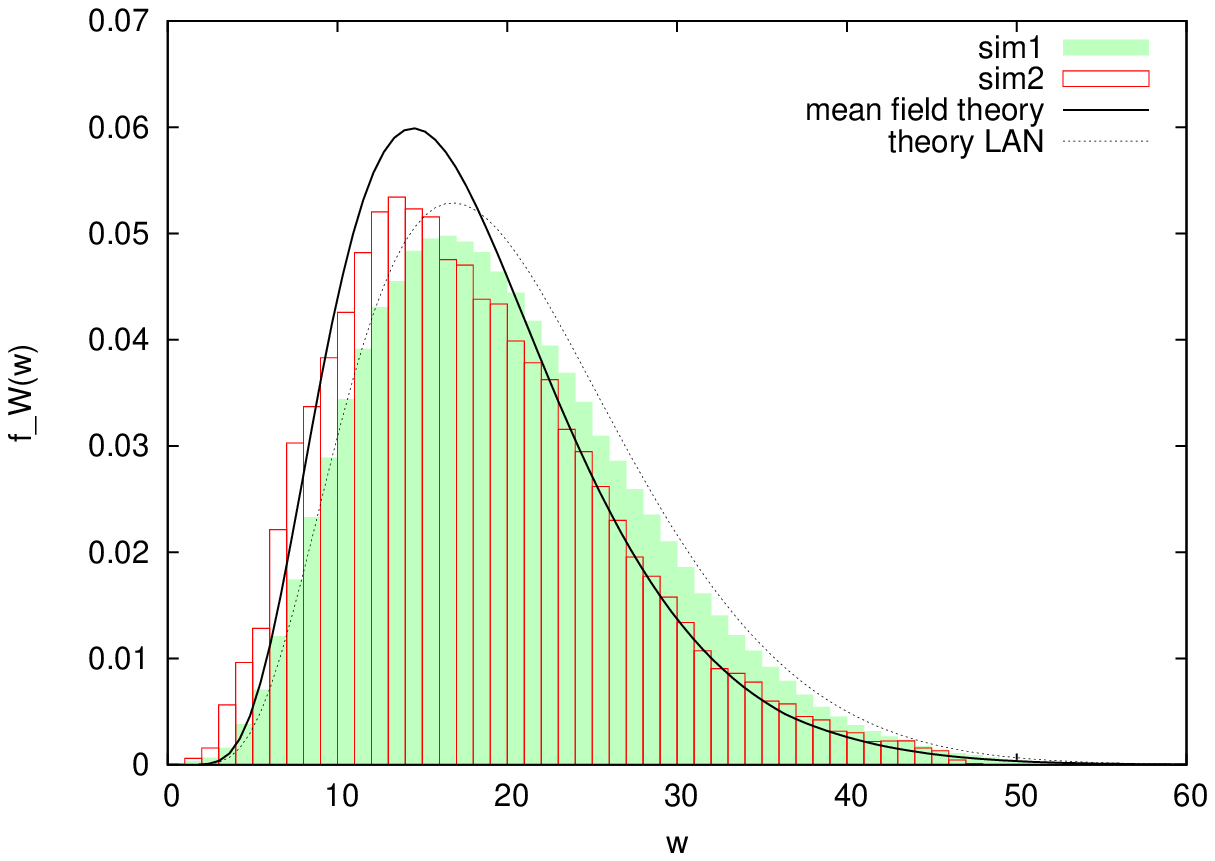}}
      }
  \end{center}
  \caption{Illustration for the mean field approximation for two parallel 
           \tcp s. The congestion window histogram of one 
           of two parallel \tcp s is shown. The histogram of a single \tcp\ is
           displayed for comparison.  Network parameters are 
           $C=256\,\textit{kb/s}$, $P=1500\,\textit{byte}$ and 
           $p=5\cdot10^{-3}$.}
  \label{fig:2TCP}
\end{figure}

The parallel \glspl{tcp} operate in \gls{wan} environment until the number of
packets in the network, that is the sum of the congestion windows
$\sum_{i=1}^NW^*_i$, is less than $2\alpha D$.  Let us consider one of the
parallel \glspl{tcp} and denote its congestion window by $W^*_n$.  The length
of the idle periods felt by the selected \gls{tcp} is $2D-W^*_n/\alpha$ in the
\gls{wan} case and the propagation delay is independent of the states of the
different \glspl{tcp}.  The congestion window distribution of each individual
\gls{tcp} can therefore be given by (\ref{eq:cwnd_gendist_WAN}).  In
Fig.~\ref{subfig:2TCP_WAN} I show the congestion window histogram of one out of
two parallel \glspl{tcp}.  The link delay is $D=2s$, large enough to leave the
buffer empty.  As a comparison I also show the histogram of a single \gls{tcp}
and the theoretical distribution function for the same network configuration.
It is apparent that the two histograms are almost identical and the discrepancy
between the theoretical distribution and the measured histogram remains almost
the same for parallel \glspl{tcp} as for a single one.

In the \gls{lan} scenario, when the sum of the congestion windows is larger
than $2\alpha D$,  a queue starts forming in the buffer and the buffering delay
becomes significant.  The selected \gls{tcp} suffers
$\sum_{i=1}^NW^*_i/\alpha-W^*_n/\alpha$ long idle periods, caused by
intermediate packets sent by the rest of the \glspl{tcp}.  Thereby, the
dynamics of the \glspl{tcp} becomes coupled and they cannot be handled as being
independent any more.  

In an attempt to solve this problem I am going to use the mean field theory,
that is I suppose that \glspl{tcp} are independent and they feel only the
average influence of other \glspl{tcp}.   For a large number of \glspl{tcp} the
sum of congestion windows will fluctuate around its average
$\mathbb{E}\left[\sum_{i=1}^{N}W^*_i\right]=N\mathbb{E}[W^*]$ and the deviation
from this average will be of order $\sim \sqrt{N}$. For sufficiently large $N$
the relative size of fluctuations will decay as $\sim 1/\sqrt{N}$. Therefore,
for large $N$ it is reasonable to replace the sum of congestion windows with
its average.  In this approximation each \gls{tcp} operates in a \gls{wan}-like
environment, since they feel a constant delay as in \gls{wan}.  So we can apply
the corresponding results of \gls{wan}.  We simply have to replace all
occurrences of $2\alpha D$ in (\ref{eq:cwnd_gendist_WAN}) with
$N\mathbb{E}[W^*]$, the mean field approximation of the sum of the congestion
windows.

The self-consistent mean field solution for $\mathbb{E}[W^*]$ can be obtained
from (\ref{eq:cwnd_genmoments}).   The occurrences of $2\alpha D$ have to be
replaced with $N\mathbb{E}[W^*]$ again, and the fixed point solution for
$N\mathbb{E}[W^*]$ should be found.  The simplest method for finding the fixed
point solution is to iterate (\ref{eq:cwnd_genmoments}): start with a good
estimate of the mean field solution, calculate the next estimate with the
equation and replace the new value to the right hand side of the equation.
This process should be repeated until the desired precision is achieved.  A
good initial value for the iteration is the mean congestion window in the 
$\alpha D\to\infty$ limit (\ref{eq:cwnd_ideal_WAN_limit}), because many
parallel \glspl{tcp} ($N\gg1$) are close to the ideal \gls{wan} scenario.

\begin{figure}
  \begin{center}
    \psfrag{ 0}[c][c][1]{$0$}
    \psfrag{ 10}[c][c][1]{$10$}
    \psfrag{ 20}[c][c][1]{$20$}
    \psfrag{ 30}[c][c][1]{$30$}
    \psfrag{ 40}[c][c][1]{$40$}
    \psfrag{ 50}[c][c][1]{$50$}
    \psfrag{ 60}[c][c][1]{$60$}
    \psfrag{ 0.01}[r][r][1]{$0.01$}
    \psfrag{ 0.02}[r][r][1]{$0.02$}
    \psfrag{ 0.03}[r][r][1]{$0.03$}
    \psfrag{ 0.04}[r][r][1]{$0.04$}
    \psfrag{ 0.05}[r][r][1]{$0.05$}
    \psfrag{ 0.06}[r][r][1]{$0.06$}
    \psfrag{ 0.07}[r][r][1]{$0.07$}
    \psfrag{f_W(w)}[c][c][1.2]{$f_W(w)$}
    \psfrag{w}[ct][cB][1.2]{$w$}
    \psfrag{theory}[r][r][0.8]{mean field theory}
    \psfrag{mean field theory}[r][r][0.8]{mean field theory}
    \psfrag{theory LAN}[r][r][0.8]{single \tcp\ theory}
    \psfrag{sim1}[r][r][0.8]{single \tcp}
    \psfrag{sim2}[r][r][0.8]{one of 20 \tcp s}
    \resizebox{0.7\figwidth}{!}{\includegraphics{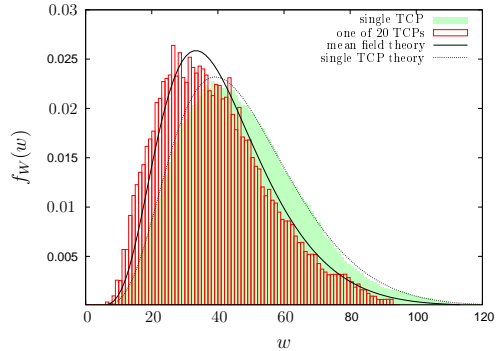}}
  \end{center}
  \caption{Congestion window histogram of one of 20 parallel \tcp s. The 
  histogram of a single \tcp\ is displayed for comparison.  Network parameters 
  are $C=256\,\textit{kb/s}$, $P=1500\,\textit{byte}$ and $p=10^{-3}$.  The 
  mean field solution of (\ref{eq:cwnd_genmoments}) is 
  {$N\mathbb{E}[W^*]\approx827.75$}, close to an ideal WAN scenario.}
  \label{fig:20TCP_LAN}
\end{figure}

\enlargethispage{1em}
In Fig.~\ref{subfig:2TCP_LAN} the congestion window histogram of one out of two
parallel \tcp s is presented in a \gls{lan} environment, when the link delay is
$D=0s$.  The mean field approximation of the distribution function shows
an excellent fit.  The histogram of a single \gls{tcp} in the same network
configuration is also plotted with the corresponding theoretical distribution.
The two histograms are rather different but both theoretical distributions are
close to the corresponding empirical values.  Fig.~\ref{fig:20TCP_LAN} shows
a similar LAN scenario with one of 20 parallel \tcp s.

\section{Conclusions}
\label{sec:Conclusions}

In this chapter I analyzed the congestion window distributions of \tcp\ in a
standalone, infinite-buffer network model.  I derived new analytical formulas
for the distribution of generic congestion window values, which take into
consideration not only the congestion avoidance mode, but also the fast
retransmit/fast recovery modes of \tcp.  My novel approach for modeling WAN
configuration made it possible to describe \tcp\ traffic with all model
parameters at hand; no parameter fitting is necessary.  Moreover, I presented
analytic calculations not only for ideal LAN and WAN scenarios, but also for
intermediate network configurations, where the queuing and link delays are
comparable.  The mean field theory has been applied for parallel \tcp\ traffic.
My analytic calculations were verified against direct simulations.  The
analytic results fit the histograms I received from the simulations when the
packet loss probability is small as well.  Discrepancies between the analytic
results and simulations become stronger when the packet loss probability
increases, however.  The differences mostly come from the neglected slow start
mode of \tcp\ and the fluid approximation of the discrete time congestion
window process.  The main virtue of my work is that it provides an analytic
description of \tcp\ traffic in more detail than previous works, without the
need to adjust parameters empirically.


\chapter{Traffic dynamics in finite buffer}
\label{cha:finite_buffer}

In the previous chapter I assumed that the common buffer under investigation
was \emph{not} a bottleneck buffer.  The model describes the dynamics of \tcp\
in the presence of external packet loss quite accurately.  However, packet loss
in current networks is generated predominantly by overloaded buffers. This is
an inherent property of \tcp\ congestion control mechanism since \tcp\
increases its packet sending rate until packet loss occurs in one of the
buffers along the route between the source and the destination.  In the
literature little or no progress has been made towards an understanding of the 
detailed mechanism of packet loss in \gls{ip} networks. 

In this chapter I give a detailed mathematical description of the packet loss
mechanism. In Section~\ref{sec:semi} the refined network model is defined.  I
investigate the dynamics of \tcp\ in the presence of a finite buffer in
Section~\ref{sec:OneFiniteTCP}.  I discuss my model in
Section~\ref{sec:Finite_Discussion}, where I will derive analytic formulas for
the packet loss and the congestion window distribution.  The new formulas and
distributions are validated by direct simulation.  Finally, I conclude this
chapter in Section~\ref{sec:Finite_Conclusions}.


\section{The finite buffer model}
\label{sec:semi}

\begin{figure}
  \begin{center}
    \psfrag{TCP}[c][c]{TCP}
    \psfrag{Sink}[c][c]{Sink}
    \psfrag{Buffer}[c][c]{Buffer}
    \psfrag{Link}[c][c]{Link delay, $D$}
    \psfrag{Delay d}[c][c]{Packet size, $P$}
    \psfrag{Loss rate p}[c][c]{Loss rate, $p$}
    \psfrag{Bandwidth C}[c][c]{Bandwidth, $C$}
    \psfrag{1}[c][c]{$1$}
    \psfrag{2}[c][c]{$2$}
    \psfrag{3}[c][c]{$3$}
    \psfrag{B}[c][c]{$B$}
    \resizebox{\figwidth}{!}{\includegraphics{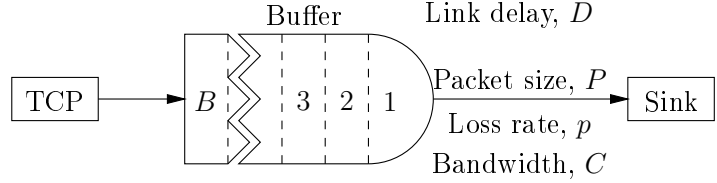}}
    \caption{The finite buffer model. In numerical
    simulations packet size $P=1500\,\textit{byte}$ and bandwidth
    $C=256\,\textit{kb}/\textit{s}$ have been fixed, and the buffer size $B$
    and the packet loss probability $p$ have been changed.} 
    \label{fig:finite_buffer_model}
  \end{center}
\end{figure}

My extended network model is very similar to the model I studied in the
preceding chapter with the decisive difference that the buffer size $B$ is 
finite now (Fig.~\ref{fig:finite_buffer_model}).  The remaining part of the 
network is---as in the previous chapter---modeled by a fixed delay $D$, 
constant bandwidth or link capacity $C$ and random loss probability of $p$ per 
packet.  In my idealized network model one \gls{tcp} injects packets into the 
buffer. 

The buffer is large enough such that \gls{tcp} can operate in
congestion avoidance mode, but it is finite, so that packet loss can
occur in it.  Furthermore, the number of packets getting
lost in the buffer is comparable with the full packet loss, including
the corrupted and lost packets in the rest of the network.

We can estimate the parameter range where the finiteness of the buffer plays an
important role in a \gls{lan} scenario, when the link delay is negligible. The
finite buffer size limits the total congestion window achievable by \tcp\ to
$w_{\max}\approx B$.  On the other hand, I have shown in the last chapter that
external packet loss in the core network would set the average congestion
window to $\avg{w}\approx c/\sqrt{p}$, where $c\approx1.5269$.  If
$c/\sqrt{p}\approx B$, that is $pB^2\approx c$ holds then the external and the
buffer loss play comparable role.  A more detailed analysis will be given in
Section \ref{subsec:A}.


\section{Dynamics of a single TCP}
\label{sec:OneFiniteTCP}

In this section I present the analysis of a single \tcp\ operating in my 
finite buffer model.  I start off by the fluid equation 
\begin{equation}
  \label{eq:cwnd_fluid_W2}
  \frac{dW}{dt}=\frac1{R(W)},
\end{equation}
similar to Eq.~(\ref{eq:cwnd_fluid_W}) of the infinite buffer model, but with
the important difference that the maximum congestion window is limited:
$w\in[0,\tilde{B}]$, where $\tilde{B}=B+2DC/P$.  The round-trip time is 
supposed to be the same as (\ref{eq:rtt_model}): 
\begin{equation}
  R(W)=\alpha^{-1}W^m,
  \label{eq:rtt_model2}
\end{equation}
where $\alpha>0$ and $m\ge0$.  The above equations can be solved the same way
as the equations of the infinite buffer scenario and we obtain the time
development of the congestion window between losses:
\begin{equation}
  W^{m+1}(\tau)=W^{m+1}(\tau_i)+\alpha\left(m+1\right)\left(\tau-\tau_i\right),
  \label{eq:cwnd_solv_finite}
\end{equation}
where $\tau_i$ denotes the instant of the $i$\textsuperscript{th} packet
loss as before.

\begin{figure}
  \begin{center}
    \psfrag{ 0}[r][r][1]{$0$}
    \psfrag{ 10}[r][r][1]{$10$}
    \psfrag{ 20}[r][r][1]{$20$}
    \psfrag{ 30}[r][r][1]{$30$}
    \psfrag{ 40}[r][r][1]{$40$}
    \psfrag{ 50}[r][r][1]{$50$}
    \psfrag{ 60}[r][r][1]{$60$}
    \psfrag{ 1000}[c][c][1]{$1000$}
    \psfrag{ 2000}[c][c][1]{$2000$}
    \psfrag{ 3000}[c][c][1]{$3000$}
    \psfrag{ 4000}[c][c][1]{$4000$}
    \psfrag{ 5000}[c][c][1]{$5000$}
    \psfrag{ 6000}[c][c][1]{$6000$}
    \psfrag{ 7000}[c][c][1]{$7000$}
    \psfrag{ 8000}[c][c][1]{$8000$}
    \psfrag{ 9000}[c][c][1]{$9000$}
    \psfrag{ 10000}[c][c][1]{$10000$}
    \psfrag{cwnd}[r][r][1.2]{$\mathit{cwnd}$}
    \psfrag{tau}[c][c][1.2]{$t$}
    \psfrag{simulation}[rB][rb][0.8]{simulation}
    \subfigure[\nstwo\ simulation of LAN]{\resizebox{0.7\figwidth}{!}
      {\includegraphics{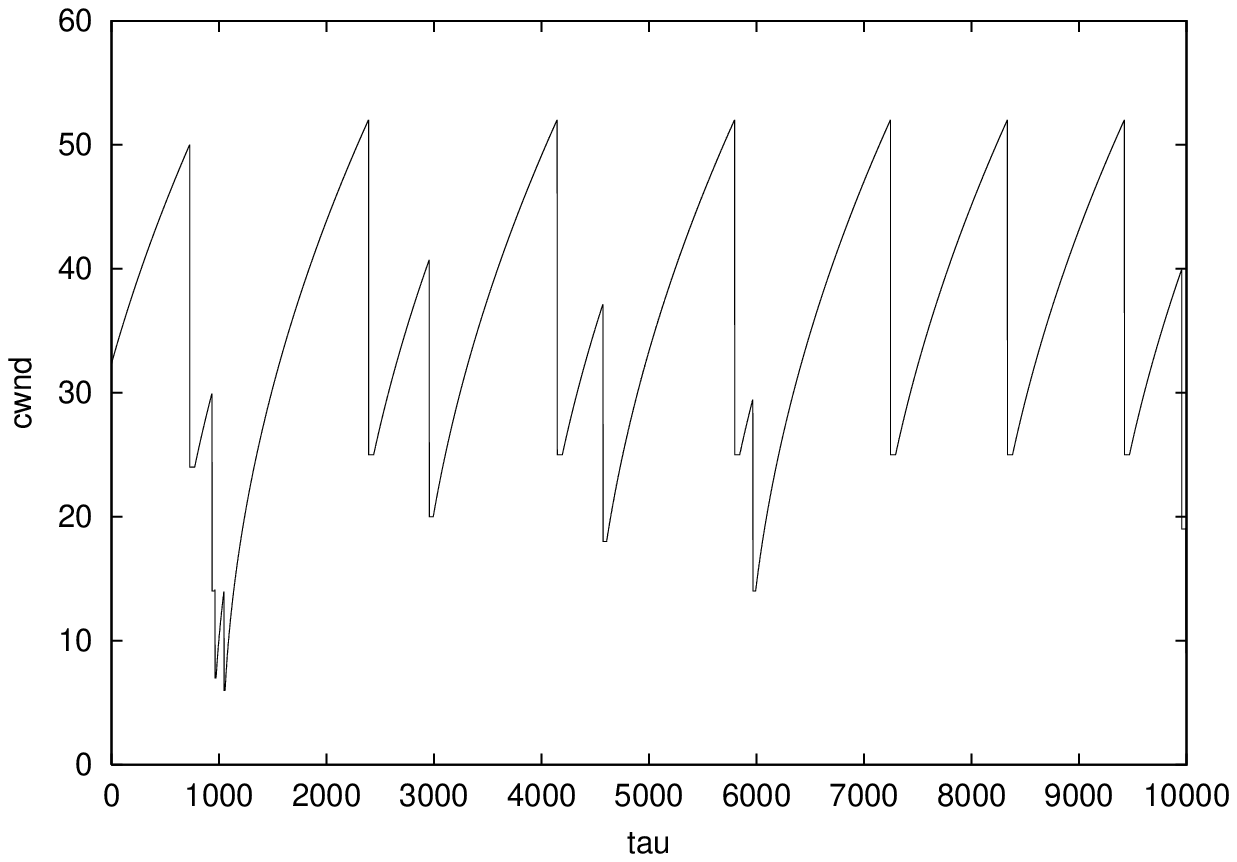}}\label{subfig:finite_LAN_simulation}}
    \psfrag{w(t)}[cb][ct][1.2]{$W(t)$}
    \psfrag{t}[c][c][1.2]{$t$}
    \psfrag{B}[r][r][1]{$B$}
    \psfrag{w0}[r][r][1]{$W_0$}
    \psfrag{w1}[r][r][1]{$W_1$}
    \psfrag{d0}[c][c][1]{$\delta_0$}
    \psfrag{d1}[c][c][1]{$\delta_1$}
    \psfrag{d2}[c][c][1]{$\delta_2$}
    \psfrag{d3}[c][c][1]{$\delta_3$}
    \psfrag{d4}[c][c][1]{$\delta_4$}
    \psfrag{t0}[c][c][1]{$\tau_0$}
    \psfrag{t1}[c][c][1]{$\tau_1$}
    \psfrag{t2}[c][c][1]{$\tau_2$}
    \psfrag{t3}[c][c][1]{$\tau_3$}
    \psfrag{t4}[c][c][1]{$\tau_4$}
    \psfrag{t5}[c][c][1]{$\tau_5$}
    \subfigure[Ideal fluid model of WAN]{\resizebox{0.7\figwidth}{!}
      {\includegraphics{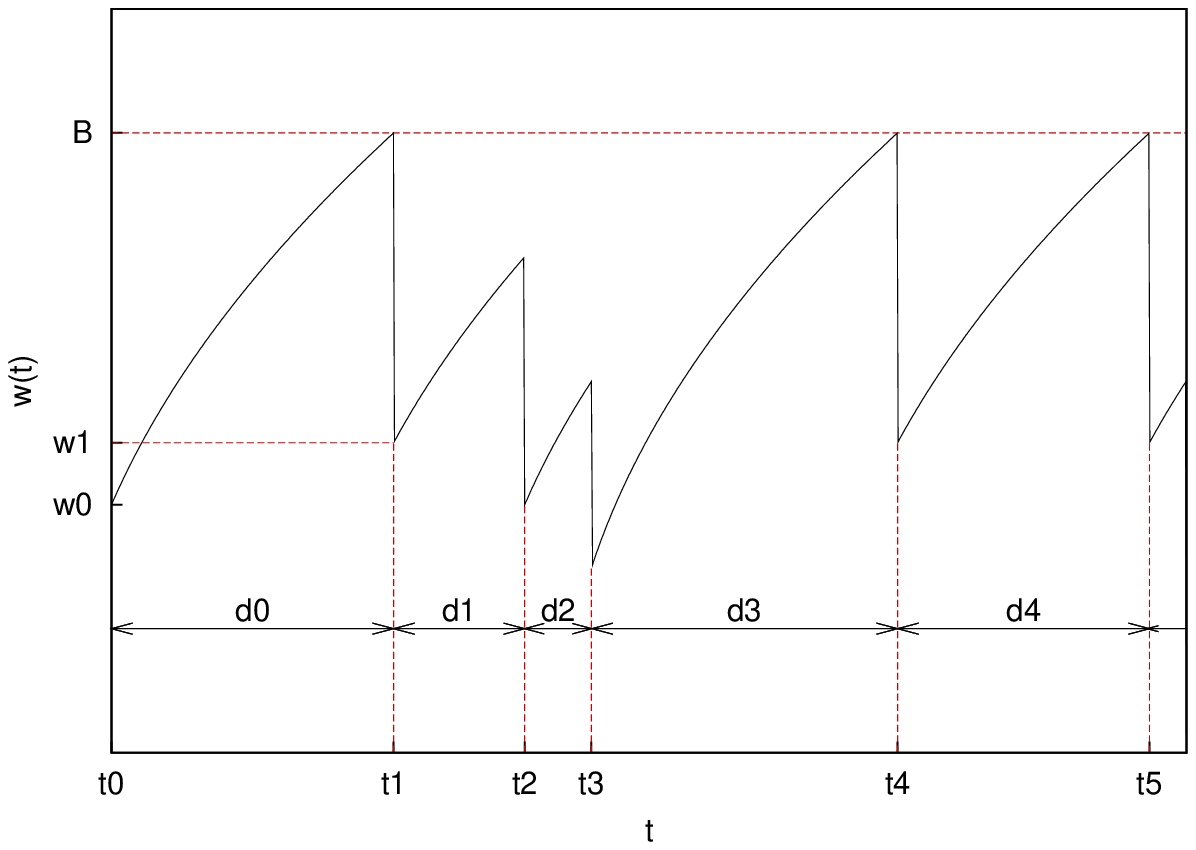}}\label{subfig:finite_LAN_fluid}}
    \caption{The congestion avoidance process of \tcp/Reno in the case of 
    finite buffer, obtained form \nstwo\ simulations .  The idealized fluid 
    approximation of the congestion window is also shown for comparison.}
    \label{fig:finite_simulation}
  \end{center}
\end{figure}

An idealized congestion window process with $m=1$ can be seen in
Figure~\ref{subfig:finite_LAN_fluid}
for a \gls{lan} network. 
In order to validate my model I implemented
it in \gls{ns}.  Simulation results of the congestion window process can
be seen in Figure~\ref{subfig:finite_LAN_simulation} 
for comparison, with $C=256\mathit{{kb}/s}$, $P=1500\mathit{byte}$, $B=50$, 
$D=0s$ and $p=0.0008$ parameter values ($pB^2=2$).
Equation (\ref{eq:cwnd_solv_finite}) gives a reasonably good description of the
window development.  
The effect of discrepancies will be discussed in Section~\ref{subsec:histo}.

Furthermore, $\delta_i$, the elapsed time between consecutive packet losses
\emph{occurring at the external link}, are supposed to be independent,
exponentially distributed random variables with mean $1/\ph$ and probability
distribution
\begin{equation}
  \label{eq:dist_inter_loss}
  f_{\delta_i}(x)=\ph\exp(-\ph x),\quad\forall i\in\N.
\end{equation}
Note the important memoryless property of the exponential distribution.  It
means that if a certain length of time has elapsed since a packet loss
then the probability distribution of the time interval remaining until the next
packet loss is still given by (\ref{eq:dist_inter_loss}) regardless of the
elapsed time.

I now derive the formula that connects consecutive window values before
losses.  Let us denote by $W_i=W(\tau^{-})$ the window value immediately
\emph{before} the $i^{\textrm{th}}$ loss event and its distribution by
$f_{W_i}(w)$.  If the value of the random variable $\delta_i$ is small enough,
the next $W_{i+1}$ can be obtained from (\ref{eq:cwnd_solv_finite}).  However,
if $\delta_i$ is so large that the window would grow above the upper limit
$\tilde{B}$ then packets will be dropped at the buffer, and the $W_{i+1}$ will
be set to $\tilde{B}$.  Accordingly, a mapping can be given that connects the
consecutive $W_i$ values:
\begin{equation}
  \label{eq:transf}
  W_{i+1}=T_{\delta_i}(W_i)=
  \begin{cases}
    \left(cW_i^{m+1}+\alpha\left(m+1\right)\delta_i\right)^{\frac1{m+1}} 
    &\text{if $\delta_i<\frac{\tilde{B}^{m+1}-cW^{m+1}_i}{\alpha\left(m+1\right)}$,}\\
    \hfil \tilde{B} 
    &\text{if $\delta_i\ge\frac{\tilde{B}^{m+1}-cW^{m+1}}{\alpha\left(m+1\right)}$.}
  \end{cases}
\end{equation}
In this manner, the time elapsed until the next packet loss might be smaller 
than $\delta_i$ if $\delta_i$ is too large.  Due to the property of the 
distribution (\ref{eq:dist_inter_loss}) noted above, at the succeeding 
application of (\ref{eq:transf}) the next $\delta_{i+1}$ time interval can be 
drawn from distribution (\ref{eq:dist_inter_loss}) again.

The next ``before loss'' window distribution $f_{W_{i+1}}(w)$ can now be
calculated by the Perron\,--\,Frobenius operator, $\mathcal{L}$, of
the mapping:
\begin{equation}
  \label{eq:FP}
  \begin{split}
    f_{W_{i+1}}(w)&=\mathcal{L}T_{\delta_i}(W_{i+1})
    =\integ_0^{\tilde{B}}\integ_0^{\infty}\delta\left(w-T_{x}(w')\right)\,
    f_{W_i}(w')\,f_{\delta_i}(x)\,dx\,dw'
  \end{split}
\end{equation}
where $\delta()$ is the Dirac-delta distribution and I averaged over
the distribution (\ref{eq:dist_inter_loss}). After substituting
(\ref{eq:transf}) into (\ref{eq:FP}) we have to consider the condition
$0\le\delta_i=\frac{W^{m+1}_{i+1}-cW^{m+1}_i}{\alpha\left(m+1\right)}$. 
This provides us with $W_i<W_{i+1}/\beta$ which should be taken into account in 
the upper boundary of the first integral. The integration in $x$ can
be carried out:
\begin{equation}
  \label{eq:rho_transf}
  \begin{split}
  f_{W_{i+1}}(w)
  &=\frac{\ph}{\alpha}w^m\,e^{-\frac{\ph w^{m+1}}{\alpha\left(m+1\right)}}
  \integ_0^{\min\left(\tilde{B},w/\beta\right)}
  f_{W_i}(w')\,e^{\frac{\ph c{w'}^{m+1}}{\alpha\left(m+1\right)}}\,dw'\\
  &+\delta(w-\tilde{B})\,e^{-\frac{\ph \tilde{B}^{m+1}}{\alpha\left(m+1\right)}}
  \integ_0^{\tilde{B}} f_{W_i}(w')\,e^{\frac{\ph c{w'}^{m+1}}{\alpha\left(m+1\right)}}\,dw'.
  \end{split}
\end{equation}

Let $\wbl=\lim_{n\to\infty}W_n$ denote the stationary limit of the ``before
loss'' window sequence.  Its stationary distribution, $f_{\wbl}(w)$, is the
fixed point solution of (\ref{eq:rho_transf}).  For finding the fixed point
solution observe that for any probability distribution $f_{W_i}(w)$ the
transformed one, $f_{W_{i+1}}(w)$, will contain a Dirac-delta term
$\delta(w-\tilde{B})$ because of the second term of (\ref{eq:rho_transf}).  I
therefore use the following ansatz for the stationary distribution 
\begin{equation}
  \label{eq:probe}
  f_{\wbl}(w)=A(\ph/\alpha,\tilde{B})\,\delta(w-\tilde{B})+\phi(w),
\end{equation}
where $\phi:[0,\tilde{B}]\to\R$ is a continuous regular function and
$A(\ph/\alpha,\tilde{B})$ is a constant. The delta function represents those
points where the packet loss occurs in the buffer and the value of the pre-loss
window is $\tilde{B}$. The constant $A(\ph/\alpha,\tilde{B})$ represents the
probability that a packet gets lost in the buffer, and it might depend on the
external loss $\ph/\alpha$ and buffer size $\tilde{B}$. I am going to 
present the detailed interpretation of $A(\ph/\alpha,\tilde{B})$ in 
Subsection~\ref{subsec:A}.

Applying the probe function (\ref{eq:probe}) in (\ref{eq:rho_transf})
and separating the regular and $\delta(w-\tilde{B})$ terms we obtain
\begin{align}
  \label{eq:delta_subst}
  A(\ph/\alpha,\tilde{B})&=
  e^{-\frac{\ph \tilde{B}^{m+1}}{\alpha\left(m+1\right)}}
  \integ_0^{\tilde{B}} \phi(w')\,e^{\frac{\ph c{w'}^{m+1}}{\alpha\left(m+1\right)}}\,dw'
  +A(\ph/\alpha,\tilde{B})\,e^{-\frac{\ph\left(1-c\right)\tilde{B}^{m+1}}{\alpha\left(m+1\right)}},\\
  \phi(w)
  &=\frac{\ph}{\alpha}w^m\,e^{-\frac{\ph w^{m+1}}{\alpha\left(m+1\right)}}
  \integ_0^{\min\left(\tilde{B},w/\beta\right)}
  \phi(w')\,e^{\frac{\ph c{w'}^{m+1}}{\alpha\left(m+1\right)}}\,dw'\notag\\
  \label{eq:regular_subst}
  &+A(\ph/\alpha,\tilde{B})\frac{\ph}{\alpha}w^m\,e^{-\frac{\ph \left(w^{m+1}
    -c\tilde{B}^{m+1}\right)}{\alpha\left(m+1\right)}}\Theta(w-\beta\tilde{B}),
\end{align}
where $\Theta(x)$ is the Heaviside step function.  Notice that for 
$w\in]\beta\tilde{B},\tilde{B}]$ the upper limit of the first integral is
independent of $w$ and the Heaviside function equals $1$.  The functional 
form of the unknown function $\phi(w)$ on this interval can therefore be
resolved.  Only the value of the definite integral---which is a
constant---should be determined.  If we look for a solution on the adjacent
interval $]\beta^2\tilde{B},\beta\tilde{B}]$ we can see that the upper bound of
the first integral falls in the range $]\beta\tilde{B},\tilde{B}]$, where the
functional form of the unknown function was previously found.  Again,
only the value of a definite integral is to be found.  Repeating these steps
recursively one can see that the solution for the integral equation
(\ref{eq:regular_subst}) would be simplified if one looked for the solution on
disjoint intervals $]\beta^{n+1}\tilde{B},\beta^n\tilde{B}]$, $n\in\N$.  
Accordingly, let us define the following functions:
\begin{gather}
  \label{eq:fn}
  \phi_n(w)=\phi(w)\,\chi_{]\beta^{n+1}\tilde{B},\beta^n\tilde{B}]}(w),\\
  \label{eq:Sn}
  S_n(s)=\integ_0^{\beta^n\tilde{B}}\phi(w)\,e^{-s\frac{w^{m+1}}{\left(m+1\right)}}\,dw,\\
  \intertext{where $\chi_{H}(w)$ denotes the indicator function of set $H\subset\mathbb{R}$, and the constant}
  \label{eq:In}
  I_n=S_n(-c\ph/\alpha)=\integ_{0}^{\beta^n\tilde{B}}\phi(w)\,e^{\frac{\lambda c{w}^{m+1}}{\alpha\left(m+1\right)}}\,dw.
\end{gather}

By applying the newly introduced definition of $I_0$ in (\ref{eq:delta_subst})
we clearly have
\begin{equation}
  \label{eq:sol_A}
  A(\ph/\alpha,\tilde{B})=\frac{e^{-\frac{\ph\tilde{B}^{m+1}}{\alpha\left(m+1\right)}}}{1-e^{-\frac{\ph\left(1-c\right)\tilde{B}^{m+1}}{\alpha\left(m+1\right)}}}I_0.\\
\end{equation}
This formula with the above mentioned properties of (\ref{eq:regular_subst}) 
in the interval $]\beta\tilde{B},\tilde{B}]$ provides us with
\begin{equation}
  \label{eq:sol_f0}
  \phi_0(w)=\frac{\ph}{\alpha}
  \frac{w^m e^{-\frac{\ph w^{m+1}}{\alpha\left(m+1\right)}}}
  {1-e^{-\frac{\ph\left(1-c\right)\tilde{B}^{m+1}}{\alpha\left(m+1\right)}}}I_0.
\end{equation}
Furthermore, for $n\in\mathbb{N}, n>0$ the recursion 
\begin{gather}
  \label{eq:sol_fn}
  \phi_n(w)=\frac{\ph}{\alpha}w^m e^{-\frac{\ph w^{m+1}}{\alpha\left(m+1\right)}}
  \biggl(I_n+\integ_{\beta^n\tilde{B}}^{w/\beta}
  \phi_{n-1}(w')\,e^{\frac{\ph c{w'}^{m+1}}{\alpha\left(m+1\right)}}\,dw'\biggr)
\end{gather}
can be derived easily, since the Heaviside function in (\ref{eq:regular_subst})
is identically zero if $w\in[0,\beta\tilde{B}[$.  In order to apply the above
recursion one should know constants $I_n$, which in turn can be obtained
from functions $S_n(s)$.  If we insert (\ref{eq:regular_subst}) into the
definition of $S_n(s)$ then we get
\begin{align}
  \label{eq:sol_S0}
  S_0(s)&=\frac1{1+\alpha s/\ph}
  \left[S_0(sc)+A(\ph/\alpha,\tilde{B})\,E\left(s\frac{\tilde{B}^{m+1}}{m+1}\right)\right]
  \intertext{where $E(x)=e^{-cx}-e^{-x}$, and for all $n\in\mathbb{N}, n>0$}
  \label{eq:sol_Sn}
  S_n(s)&=\frac1{1+\alpha s/\ph}\left(S_{n-1}(sc)
  -e^{-\left(s+\frac{\ph}{\alpha}\right)\frac{c^n\tilde{B}^{m+1}}{\left(m+1\right)}}I_{n-1}\right).
\end{align}

Using (\ref{eq:sol_A}) and (\ref{eq:sol_f0}) as initial conditions the
recursive expressions (\ref{eq:sol_fn}) and (\ref{eq:sol_Sn}) can be
solved.  In order to start the iteration the value of the initial condition
$A(\ph/\alpha,\tilde{B})$, or equivalently $I_0=S_0(-c\ph/\alpha)$ is needed.  
In the interest of finding $I_0$ I calculate the function $S_0(s)$ next.  If 
we suppose that $S_0(s)$ is continuous at $s=0$ then, using (\ref{eq:sol_S0}),
it can be proven by induction that
\begin{equation}
  \begin{split}
    S_0(s)&=S_0(0)\prod_{k=0}^{\infty}\frac{1}{1+sc^k\alpha/\ph}\\
    &+A(\ph/\alpha,\tilde{B})\sum_{k=0}^{\infty} 
    E\left(sc^k\frac{\tilde{B}^{m+1}}{\left(m+1\right)}\right)
    \prod_{l=0}^k\frac1{1+sc^l\alpha/\ph},
  \end{split}
  \label{eq:sol_S0_gen}
\end{equation}
where I have used that $\lim_{N\to\infty}S_0(sc^N)=S_0(0)$ for $c\in[0,1[$.
Furthermore, $S_0(0)=\integ_0^{\tilde{B}}\phi(w)\,dw=1-A(\ph/\alpha,\tilde{B})$,
because $f_{\wbl}(w)$ is normalized.

The function $S_0(s)$ is bounded because it is defined via the definite 
integral of the regular function $\phi(w)$.  The pole at $s=-\ph/\alpha$ on 
the right hand side of (\ref{eq:sol_S0}) must therefore be canceled 
by the subsequent factor:
\begin{equation}
  S_0(-c\ph/\alpha)+A(\ph/\alpha,\tilde{B})
  E\left(-\frac{\ph}{\alpha}\frac{\tilde{B}^{m+1}}{m+1}\right)=0.
\end{equation}
In addition, $S_0(-c\ph/\alpha)$ can be obtained from (\ref{eq:sol_S0_gen}). 
As a result,
\begin{multline}
  -A(\ph/\alpha,\tilde{B})E\left(-\frac{\ph}{\alpha}\frac{\tilde{B}^{m+1}}{m+1}\right)=
  \left(1-A(\ph/\alpha,\tilde{B})\right)\prod_{k=0}^{\infty}\frac1{1-c^{k+1}}\\
  +A(\ph/\alpha,\tilde{B})\sum_{k=0}^{\infty}
  E\left(-\frac{\ph}{\alpha}\frac{c^{k+1}\tilde{B}^{m+1}}{m+1}\right)
  \prod_{l=0}^{k}\frac1{1-c^{l+1}}
\end{multline}
is acquired.  We can express $A(\ph/\alpha,\tilde{B})$ now as
\begin{equation}
  \label{eq:A}
  A(\ph/\alpha, \tilde{B})
  =\frac{1}{1-L(c)\,G\left(\frac{\ph}{\alpha}\frac{\tilde{B}^{m+1}}{m+1}\right)},
\end{equation}
where $L(c)$ has been defined earlier in (\ref{eq:h_l_LAN}), and
\begin{equation}
  \label{eq:G}
  G(x)=\sum_{k=0}^{\infty}E(-c^kx)\prod_{l=1}^{k}\frac1{1-c^l}
\end{equation}
with the convention that the empty product equals $1$.  Note that in 
(\ref{eq:A}) the parameters appear only in the $\ph \tilde{B}^{m+1}/\alpha$
combination.  This expression is the control parameter in my model. Systems in
which external packet losses and buffer sizes differ, but the
$\ph\tilde{B}^{m+1}/\alpha$ product is the same, are similar in the sense that
they can be described with the same constant $\A$.


\section{Discussion}
\label{sec:Finite_Discussion}

\subsection{The interpretation of $A(\cdot)$ and the effective loss}
\label{subsec:A}

Now I present a brief explanation of the meaning of $\A$ and highlight its
importance.  First I calculate the average time elapsed between two packet-loss
events.  Remember that the inter-loss times \emph{on the link} $\delta_i$ are
\gls{iid} random variables with exponential distribution.  However, the buffer
can induce extra packet losses.  If the congestion window was $\wbl$ at the
previous packet loss then the maximum inter-loss time is clearly
$\frac{\tilde{B}^{m+1}-c\wbl^{m+1}} {\alpha\left(m+1\right)}$, at which time
the buffer becomes congested.  The exponential distribution of $\delta_i$ is
truncated above this upper limit, and the probability that $\delta_i$ exceeds
this limit is concentrated at the maximum inter-loss time.  Consequently, the
conditional probability distribution that a packet gets lost at either the
buffer or the link after $\delta'$ time, supposing that the value of the
congestion window was $\wbl$ at the previous packet loss, can be written as
\begin{equation}
  \label{eq:p_tau}
  \begin{split}
    f_{\delta'\mid\wbl}(x,w)
    &=\delta\left(x-\frac{\tilde{B}^{m+1}-c w^{m+1}}{\alpha\left(m+1\right)}\right) 
    e^{-\frac{\ph}{\alpha}\frac{\tilde{B}^{m+1}-c w^{m+1}}{m+1}}\\
    &+\ph e^{-\ph x}\left[1-\Theta\left(x-\frac{\tilde{B}^{m+1}-c w^{m+1}}
    {\alpha\left(m+1\right)}\right)\right],
  \end{split}
\end{equation}
With the help of the total probability theorem and (\ref{eq:p_tau}) the 
average inter-loss time can be given by
\begin{equation}
  \label{eq:avg_tau}
  \begin{split}
  \mathbb{E}[\delta']
  &=\integ_0^{\infty}\integ_0^{\infty}x\,f_{\delta'\mid\wbl}(x,w)\,
  f_{\wbl}(w)\,dx\,dw\\
  &=\integ_0^{\infty}\frac1{\ph}
  \left(1-e^{-\frac{\ph}{\alpha}\frac{\tilde{B}^{m+1}-cw^{m+1}}{m+1}}\right)
  f_{\wbl}(w)\,dw\\
  &=\frac{1-\A}{\ph},
  \end{split}
\end{equation}
where (\ref{eq:delta_subst}) has been used for replacing the last integral.

The meaning of this simple expression becomes clearer if we recognize that
$\ph'=1/\mathbb{E}[\delta']$ is the total packet loss rate---link and
buffer losses combined.  Therefore,
\begin{equation}
  \label{eq:eff_loss}
  \frac{\ph}{\ph'}=1-\A,
\end{equation}
which can be interpreted as the ratio of the number of packets that are lost
at the link and the total amount of lost packets. Similarly, $\A$ is \emph{the
ratio of the number of packets that are lost at the buffer
$N_{\mathrm{buffer}}$ and the total packet loss $N_{\mathrm{total}}$}. The
possibility that this ratio can be estimated from my model is the main result
of this section. This interpretation and the exact knowledge of the form of
$\A$ allows us to treat buffer-losses as if they were link-losses. It also
makes it possible to calculate the total loss along a multi-buffer, multi-link
route.

According to (\ref{eq:eff_loss}) the measured $1-\ph/\ph'$ expression should be
equal to $\A$ and it should not depend on $\ph$ and $\tilde{B}$ separately, but
only on the $\ph\tilde{B}^{m+1}/\alpha$ product.  In order to verify
(\ref{eq:eff_loss}) I carried out a number of simulations with different $\ph$
and $\tilde{B}$ parameter values in the $1\le \ph\tilde{B}^{m+1}/\alpha\le10$
parameter range for both \gls{lan} and \gls{wan} network configurations.

The parameter settings of the present model are the same as those of the
infinite-buffer model in the previous chapter.  In particular, for \gls{lan}
scenarios: $m=1$, $\beta=1/2$ and $\ph/\alpha=p$.  Setting the value of the new
parameter $\tilde{B}$ requires extra care, however.  When comparing simulations
and the formula (\ref{eq:eff_loss}) we have to take into account that in
reality the system can store more packets than the actual buffer size.  For
example, the receiver is processing one packet, and even if the link delay is
zero, one acknowledgment packet is traversing back to the sender during the
file transfer, increasing the maximum number of unacknowledged packets in the
system by two.  Moreover, \gls{tcp} detects packet loss one \gls{rtt} later
than it actually happens, causing overshoot of the maximum window.  The
difference between simulation and fluid approximation can also cause some
discrepancy. In other words, \gls{tcp} behaves as if the buffer would be bigger
than it really is.  The effect of this behavior can be observed in
Fig.~\ref{subfig:finite_LAN_simulation} where the congestion window
occasionally exceeds the buffer size $B$.
\begin{table}
  \caption{Fitted $b$ parameter values for different buffer sizes
    $B$. The average value is $\bar{b}=2.5354$.}
  \begin{center}
    \begin{tabular}{c c}
    \hline
    $B$ & $b$\\
    \hline
    30 & 2.5045\\
    40 & 2.5790\\
    50 & 2.6798\\
    60 & 2.4997\\
    70 & 2.4143\\
    \hline
    \end{tabular}
  \end{center}
  \label{tab:fit_c}
\end{table}

In order to treat this problem I assumed that we have to set the congestion
window limit to $\tilde{B}=B+b_L$, where $b_L$ has been fitted for different
buffer sizes $B$.  The fitted values of $b_L$ can be found in
Table~\ref{tab:fit_c}.  It can be seen that $b_L$ is constant and practically
independent of $B$.  Based on simulation results I set $b_L$ to its average
value $\bar{b}_L=2.5354$.

\begin{figure}
  \begin{center}
    \psfrag{ 0}[c][c][1]{$0$}
    \psfrag{ 1}[c][c][1]{$1$}
    \psfrag{ 2}[c][c][1]{$2$}
    \psfrag{ 3}[c][c][1]{$3$}
    \psfrag{ 4}[c][c][1]{$4$}
    \psfrag{ 5}[c][c][1]{$5$}
    \psfrag{ 6}[c][c][1]{$6$}
    \psfrag{0}[r][r][1]{$0$}
    \psfrag{0.2}[r][r][1]{$0.2$}
    \psfrag{0.4}[r][r][1]{$0.4$}
    \psfrag{0.6}[r][r][1]{$0.6$}
    \psfrag{0.8}[r][r][1]{$0.8$}
    \psfrag{1}[r][r][1]{$1$}
    \psfrag{1-p/p'}[c][c][1]{$N_{\mathrm{buffer}}/N_{\mathrm{total}}$}
    \psfrag{pB^2}[c][c][1]{$p\tilde{B}^2/2$}
    \psfrag{A(x)}[r][r][0.8]{$A(p\tilde{B}^2/2)$}
    \psfrag{B=30}[r][r][0.8]{$B=30$}
    \psfrag{B=40}[r][r][0.8]{$B=40$}
    \psfrag{B=50}[r][r][0.8]{$B=50$}
    \psfrag{B=60}[r][r][0.8]{$B=60$}
    \psfrag{B=70}[r][r][0.8]{$B=70$}
    \resizebox{\figwidth}{!}{\includegraphics{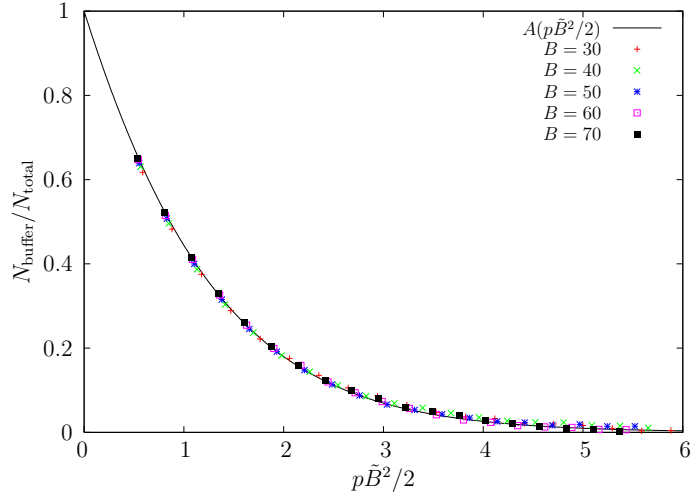}}
  \end{center}
  \caption{Comparison of the theoretical function
  $A\left(p(B+\bar{b}_L)^2\right)$ and the measured ratio 
  $N_{\mathrm{buffer}}/N_{\mathrm{total}}$ obtained from numerical simulations
  in various LAN configurations.  The control parameter $p\tilde{B}^2/2$ 
  has been varied in the range of $1$ and $6$, at various buffer sizes between 
  $B=30$ and $70$.}
  \label{fig:A-x}
\end{figure}

Simulation results are shown in Fig.~\ref{fig:A-x}, where I compare the
theoretical formula for $A\left(p\tilde{B}^2\right)$ and the ratio
$N_{\mathrm{buffer}}/N_{\mathrm{total}}$ measured by \gls{ns}. 
$N_{\mathrm{buffer}}$ and $N_{\mathrm{total}}$ are the number of packets
dropped at the buffer and the total number of dropped packets, respectively. 
In the simulated parameter range I obtained an almost perfect match.

Now I turn to the \gls{wan} scenario.  Notice that the time could have been
replaced with ``\gls{ack} time'' in the previous arguments concerning $\A$ and
the effective loss.  In addition idle periods affect neither the number of
packets dropped at the buffer nor the number of packets lost at the link.
Therefore, $\A$ is basically related to the intrinsic ``\gls{ack} time''
dynamics of \gls{tcp}.  Consequently, the ratio
$N_{\mathrm{buffer}}/N){\mathrm{total}}$ in a \gls{wan} scenario should be
equal to $\A$ with the intrinsic parameters of \gls{tcp} dynamics $m=1$,
$\beta=1/2$ and $\ph/\alpha=p$. 

In an ideal \gls{wan} network the buffer size would be zero.  However, in
reality the buffer size $B$ must be set to a positive number, otherwise packet
bursts cannot go through the buffer and \gls{tcp} shows pathological behavior.
If the size of the buffer is smaller than the maximum value of the slow start
threshold then the slow start mechanism can have a serious impact on the number
of packets lost at the buffer.  Indeed, sudden bursts of packets of the slow
start mode might cause further congestions at the buffer, which, in turn, might
induce another slow start.  This cascade of slow starts lasts until the slow
start threshold is reduced below the size of the buffer.  

\begin{figure}[tb]
  \begin{center}
    \psfrag{ 0}[c][c][1]{$0$}
    \psfrag{ 1}[c][c][1]{$1$}
    \psfrag{ 2}[c][c][1]{$2$}
    \psfrag{ 3}[c][c][1]{$3$}
    \psfrag{ 4}[c][c][1]{$4$}
    \psfrag{ 5}[c][c][1]{$5$}
    \psfrag{ 6}[c][c][1]{$6$}
    \psfrag{ 0.1}[r][r][1]{$0.1$}
    \psfrag{ 0.2}[r][r][1]{$0.2$}
    \psfrag{ 0.3}[r][r][1]{$0.3$}
    \psfrag{ 0.4}[r][r][1]{$0.4$}
    \psfrag{ 0.5}[r][r][1]{$0.5$}
    \psfrag{ 0.6}[r][r][1]{$0.6$}
    \psfrag{ 0.7}[r][r][1]{$0.7$}
    \psfrag{ 0.8}[r][r][1]{$0.8$}
    \psfrag{ 0.9}[r][r][1]{$0.9$}
    \psfrag{1-N_l/N_t}[c][c][1]{$N_{\mathrm{buffer}}/N_{\mathrm{total}}$}
    \psfrag{pB^2/2}[c][c][1]{$p\tilde{B}^2/2$}
    \psfrag{A(x)}[r][r][0.8]{$A(x)$}
    \psfrag{sim}[r][r][0.8]{simulation}
    \psfrag{B3}[r][r][0.8]{$B=3$}
    \psfrag{B10}[r][r][0.8]{$B=10$}
    \psfrag{B30}[r][r][0.8]{$B=30$}
    \resizebox{\figwidth}{!}{\includegraphics{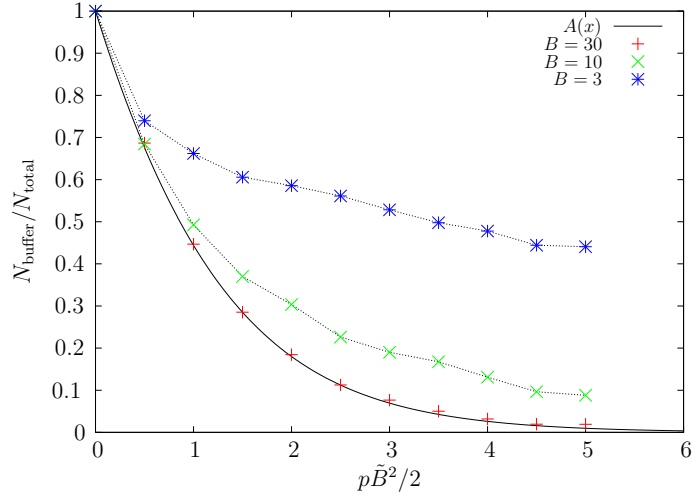}}
  \end{center}
  \caption{The ratio $N_{\mathrm{buffer}}/N_{\mathrm{total}}$, obtained from 
  \nstwo\ simulations, is plotted as the function of $p\tilde{B}^2/2$.
  The theoretical function $A(x)$ is shown for comparison. Below 
  $B\approx \alpha D$
  the buffer cannot handle packet bursts produced by the slow start algorithm,
  therefore excess packet drops appear at the buffer.  Dotted lines connecting
  data points at $B=3$ and $10$ are guides to the eye.}
  \label{fig:A-x_W}
\end{figure}

In order to demonstrate this phenomenon I carried out simulations with such a
parameter setting that $2DC/P=60$.  The buffer size was $B=3,10$, and $30$.
Simulation results are shown in Fig~\ref{fig:A-x_W}, where I compared the
theoretical formula $\A$ and the measured loss ratio
$N_{\mathrm{buffer}}/N_{\mathrm{total}}$.  Data points deviate from the
theoretical curve considerably when $B=3$.  Deviation from the theory is less
for $B=10$ than for $B=3$, but it is still significant for larger values of the
control parameter.  Finally, the measured data points fit $\A$
almost perfectly when $B=30$.
At the end of this subsection I estimate the effective loss $\ph'$ in the
$\ph\tilde{B}^{m+1}/\alpha\to\infty$ and $\ph\tilde{B}^{m+1}/\alpha\to0$
limits.  The first is the infinite buffer case, when packets get lost only on
the link.  It is evident from (\ref{eq:G}) that in the $x\to\infty$ limit the
$-e^{x}$ term dominates $G(x)$.  Therefore, $A(\ph/\alpha,\tilde{B})\approx
e^{-\frac{\ph}{\alpha}\frac{\tilde{B}^{m+1}}{m+1}}/L(c)$ if
$\ph\tilde{B}^{m+1}/\alpha\gg1$, which implies that 
\begin{equation}
  \ph'\approx
  \ph\left(1+\frac{e^{-\frac{\ph}{\alpha}\frac{\tilde{B}^{m+1}}{m+1}}}{L(c)}\right).
\end{equation}
The fraction of packets dropped at the buffer decreases at an exponential rate 
as the control parameter $\ph\tilde{B}^{m+1}/\alpha$ increases.

The second case is the ``extreme'' bottleneck buffer limit, when packets only 
get lost in the buffer.  From (\ref{eq:eff_loss}) and (\ref{eq:A}) it follows 
that
\begin{equation}
  \ph'=\ph\left(1-\frac1{L(c)\,
  G\left(\frac{\ph}{\alpha}\frac{\tilde{B}^{m+1}}{m+1}\right)}\right).
\end{equation}
With the series expansion of $L(c)\,G(x)$, derived in 
Appendix~\ref{app:LG_expansion}, we can write
\begin{equation}
  \begin{split}
    \ph'&=\ph\left(1+\frac1{1-c}\frac{\alpha}{\ph}\frac{m+1}{\tilde{B}^{m+1}}
    -\frac{1+c}2+\Ordo{\frac{\ph}{\alpha}\tilde{B}^{m+1}}\right)\\
    &=\frac{m+1}{1-c}\frac{\alpha}{\tilde{B}^{m+1}}+\frac{1-c}{2}\ph+
    \Ordo{\frac{\ph^2}{\alpha}\tilde{B}^{m+1}},
  \end{split}
\end{equation}
when $\ph^2\tilde{B}^{m+1}/\alpha\ll1$.  In particular, in an ideal \gls{lan} 
scenario 
\begin{equation}
  \label{eq:eff_loss_limit}
  p'=\frac8{3B^2}+\frac{3}{8}p+\Ordo{p^2B^2}
\end{equation}
holds for the effective packet loss probability $p'=\ph'/\alpha$ in the
$p^2B^2\ll1$ limit. The first order approximation of this formula has been
calculated in \cite{MathisSemkeMahdaviOtt97} for the same bottleneck scenario.
This is a further indication that my calculation is correct. Since I obtained
(\ref{eq:eff_loss_limit}) as a limit of my model, my work can be viewed as a
generalization of previous studies.


\subsection{Histograms and probability distributions}
\label{subsec:histo}

I continue in this section with the derivation of the congestion window 
distribution from Eqs.~(\ref{eq:sol_A})--(\ref{eq:sol_Sn}).  It is easy 
to see that the piecewise solution of (\ref{eq:sol_fn}) on the disjoint
intervals can be written in the form
\begin{equation}
  \label{eq:probe_f}
  \phi_n(w)=\frac{\ph}{\alpha}w^m 
  \sum_{k=0}^n h_{n,k}\,e^{-\frac{\ph}{\alpha}\frac{c^{-k}}{m+1}w^{m+1}}.
\end{equation}
Note that functional form of (\ref{eq:probe_f}) is the same as
(\ref{eq:W_dist}) in the infinite buffer scenario.  Substituting
(\ref{eq:probe_f}) into (\ref{eq:sol_fn}) we can derive recursive formulas for
the constants $h_{n,k}\equiv h_{n,k}(\ph/\alpha,\tilde{B})$. The constants
might depend on the parameters $\ph/\alpha$ and $\tilde{B}$ as I denoted
explicitly. After the substitution we acquire
\begin{align}
    \phi_{n+1}(w)&=\frac{\ph}{\alpha}w^m\sum_{k=0}^{n+1}
    h_{n+1,k}\,e^{-\frac{\ph}{\alpha}\frac{c^{-k}}{m+1}w^{m+1}}\notag\\
    &=\frac{\ph}{\alpha} w^me^{-\frac{\ph}{\alpha}\frac{w^{m+1}}{m+1}}
    \left(I_{n+1}+\frac{\ph}{\alpha}\sum_{k=0}^{n}h_{n,k}
    \integ_{\beta^{n+1}\tilde{B}}^{w/\beta}{w'}^m 
    e^{-\frac{\ph}{\alpha}\frac{\left(c^{-k}-c\right){w'}^{m+1}}{m+1}}\,dw'\right)\notag\\
    &=\frac{\ph}{\alpha} w^m e^{-\frac{\ph}{\alpha}\frac{w^{m+1}}{m+1}}
    \left(I_{n+1}+\sum_{k=0}^{n}\frac{h_{n,k}}{c^{-k}-c}
    e^{-c^{n+1}\left(c^{-k}-c\right)\frac{\ph}{\alpha}
    \frac{\tilde{B}^{m+1}}{m+1}}\right)\notag\\
    &-\frac{\ph}{\alpha}w^m\sum_{k=1}^{n+1}\frac{h_{n,k-1}}{c^{-k+1}-c}
    e^{\frac{\ph}{\alpha}\frac{c^{-k}}{m+1}w^{m+1}}
    \label{eq:sol_fn1}
\end{align}
It can be seen that after the recursive step in (\ref{eq:sol_fn1}) only the
required $\textit{const}\times
e^{-\frac{\ph}{\alpha}\frac{c^{-k}}{m+1}w^{m+1}}$ type terms appear. Comparing
the coefficients on both sides term by term we receive the following equations:
\begin{align}
  \label{eq:sol_C00}
  h_{0,0}
  &=A(\ph/\alpha,\tilde{B})\,e^{\frac{\ph}{\alpha}\frac{\tilde{B}^{m+1}}{m+1}}\\
  \label{eq:sol_Cn0}
  h_{n+1,0}
  &=I_{n+1}+\sum_{k=0}^n\frac{h_{n,k}}{c^{-k}-c}
  e^{-c^{n+1}\left(c^{-k}-c\right)\frac{\ph}{\alpha}\frac{\tilde{B}^{m+1}}{m+1}}\\  
  \label{eq:sol_Cnk}
  h_{n+1,k}
  &=\frac{h_{n,k-1}}{c-c^{-k+1}}
  =\frac{h_{n-k+1,0}}{c^{k}}\prod_{l=1}^{k}\frac1{1-c^{-l}}
  =L(c)\,h_k(c)\,h_{n-k+1,0},
\end{align}
where $h_k(c)$ and $L(c)$ are defined in (\ref{eq:h_l_LAN}) and (\ref{eq:L_c}).
In order to complete the system of recursive equations we have to provide
constants $I_n$.  The constant $I_0$ can be obtained from (\ref{eq:sol_A}):
\begin{align}
  \label{eq:sol_I0}
  I_0&=A(\ph/\alpha,\tilde{B})\left(e^{\frac{\ph}{\alpha}\frac{\tilde{B}^{m+1}}{m+1}}-e^{\frac{\ph}{\alpha}\frac{c\tilde{B}^{m+1}}{m+1}}\right),
\intertext{while for $n\in\mathbb{N}$ (\ref{eq:In}) can be applied:}
  I_{n+1}&=I_n-\integ_{\beta^{n+1}\tilde{B}}^{\beta^n\tilde{B}}
  \phi_n(w)\,e^{\frac{\ph}{\alpha}\frac{c w^{m+1}}{m+1}}\,dw\notag\\
  &=I_n-\sum_{k=0}^n\left(e^{-c^{n+1}\left(c^{-k}-c\right)\frac{\ph}{\alpha}\frac{\tilde{B}^{m+1}}{m+1}}
  -e^{-c^{n}\left(c^{-k}-c\right)\frac{\ph}{\alpha}\frac{\tilde{B}^{m+1}}{m+1}}\right) 
  \frac{h_{n,k}}{c^{-k}-c}
  \notag\\
  &=I_n-\sum_{k=0}^n E\left(c^n\left(c^{-k}-c\right)\frac{\ph}{\alpha}\frac{\tilde{B}^{m+1}}{m+1}\right)\frac{h_{n,k}}{c^{-k}-c}
  \label{eq:sol_In}
\end{align}

Although the number of the coefficients is infinite, we can use the
first few in practice. Since the smallest congestion window value is
$1$, no more than $\log_2B$ number of $\phi_n(w)$ functions are
relevant and the inequality $k\le n$ implies that $k$ is
also limited. Furthermore, it is obvious from (\ref{eq:sol_Cnk}) that
for every $n$ the absolute value of $h_{n,k}$ decays very quickly as $k$
increases, so $h_{n,k}\approx0$ can be supposed if $k\gtrsim3$.

So far I calculated analytically the distribution of the ``before loss''
values of the congestion window.  In practice the distribution of the
congestion window at an arbitrary moment is relevant.   I calculate this
distribution $f_W(w)$ here.  We can basically repeat the same arguments as in
Sec.~\ref{sec:OneFiniteTCP}.  In general, between losses, the congestion window
is developing according to (\ref{eq:cwnd_solv_finite}), where $\tau$ is a
uniformly distributed random variable on the \emph{random} interval $[0,\rho]$.
The conditional distribution of $\tau$---supposing that $\rho$ is given---is
$f_{\tau}(t\mid\rho=x)=\frac1x\chi_{[0,x]}(t)$.  The distribution of $\rho$ is
$f_{\rho}(x)=\frac{x}{\mathbb{E}[\delta_i]}f_{\delta_i}(x)$, similarly to the
infinite buffer scenario.  Thus, the distribution of the congestion window at
an arbitrary moment can be given by the following transformation
\begin{align}
  f_W(w)&=\integ_0^{\tilde{B}}\integ_0^{\infty}\integ_0^{\infty}
  \delta(w-T_t(w'))\,f_{\tau}(t\mid\rho=x)\,f_{\rho}(x)\,f_{\wbl}(w')
  \,dt\,dx\,dw'\notag\\
  &=\frac1{\mathbb{E}[\delta_i]}
  \integ_0^{\tilde{B}}\integ_0^{\infty}\integ_0^{x}
  \delta(w-T_t(w'))\,f_{\delta_i}(x)\,f_{\wbl}(w')\,dt\,dx\,dw',
  \label{eq:finite_cwnd_dist}
\end{align}
where $T_{\tau}(\wbl)
=\left[\wbl^{m+1}+\alpha\left(m+1\right)\tau\right]^{\frac1{m+1}}$ is the 
forward mapping of the congestion window from $\wbl$ to $W$, $\tau$ time
later.  The integration in variable $t$ can be carried out:
\begin{equation}
  \integ_0^{x}\delta(w-T_{t}(w'))\,dt
  =\frac{w^m}{\alpha}\Theta\left(x-\frac{w^{m+1}-c{w'}^{m+1}}{\alpha\left(m+1\right)}\right)
  \left[1-\Theta\left(w'-\frac{w}{\beta}\right)\right].
\end{equation}
The $w^m/\alpha$ term is from the inverse-Jacobi of $T_t(w')$, and the 
Heaviside functions correspond to the range of integration
$t=\frac{w^{m+1}-c{w'}^{m+1}}{\alpha\left(m+1\right)}\in[0,x]$.  Therefore,
(\ref{eq:finite_cwnd_dist}) can be written as follows:
\begin{equation}
  \begin{split}
    f_W(w)&=\frac1{1-A(\ph/\alpha,\tilde{B})}\frac{\ph}{\alpha}w^m\!\!\!\!
    \integ_0^{\min(\tilde{B},w/b)}\!\!\!\!\!\!
    \integ_{\frac{w^{m+1}-{w'}^{m+1}}{\alpha\left(m+1\right)}}^{\infty}\!\!\!\!\!\!
    \ph e^{-\ph x}\,f_{\wbl}(w')\,dx\,dw'\\
    &=\frac1{1-A(\ph/\alpha,\tilde{B})}\frac{\ph}{\alpha}w^m e^{-\frac{\ph}{\alpha}\frac{w^{m+1}}{m+1}}\!\!\!\!\!\!
    \integ_0^{\min(\tilde{B},w/b)}\!\!\!\!\!\!
    f_{\wbl}(w') e^{\frac{\ph}{\alpha}\frac{c{w'}^{m+1}}{m+1}}\,dw'
  \end{split}
\end{equation}
where I have used (\ref{eq:avg_tau}).  The implicit definition of
$f_{\wbl}(w)$ given in (\ref{eq:rho_transf}) and (\ref{eq:probe})
yields that
\begin{equation}
  \label{eq:cwnd_dist4}
  f_W(w)=\frac{\phi(w)}{1-A(\ph/\alpha,\tilde{B})}
\end{equation}

It can be seen that the final distribution is proportional to the regular part
of the ``before loss'' distribution, so I can apply my earlier results given
in (\ref{eq:sol_C00})\,--\,(\ref{eq:sol_In}) again.  There is no Dirac-delta
distribution in (\ref{eq:cwnd_dist4}).

In the $\tilde{B}\to\infty$ limit the derived formula (\ref{eq:cwnd_dist4})
should converge to (\ref{eq:W_dist}), the distribution derived for the infinite
buffer scenario in the previous chapter.  Let me confirm that my result is
consistent with the infinite buffer case.  I showed before that
$A(\ph/\alpha,\tilde{B}) \approx
e^{-\frac{\ph}{\alpha}\frac{\tilde{B}}{m+1}}/L(c)$ if
$\ph\tilde{B}^{m+1}/\alpha\gg1$.  Therefore, $A(\ph/\alpha,\tilde{B})\to0$ if
$\tilde{B}\to\infty$, which means that $f_W(w)=\lim_{n\to\infty}\phi_n(w)$.
Furthermore, (\ref{eq:sol_I0}) and (\ref{eq:sol_In}) imply that
$\lim_{\tilde{B}\to\infty}I_n=1/L(c)$ for all $n\in\mathbb{N}$.  Using these
results one can see from (\ref{eq:sol_C00})\,--\,(\ref{eq:sol_Cnk}) that
$\lim_{n\to\infty}\lim_{\tilde{B}\to\infty}h_{n,k}=h_k(c)$ for all
$k\in\mathbb{N}$.

\begin{figure}[p]
  \begin{center}
    \psfrag{ 0}[c][c][1]{$0$}
    \psfrag{ 10}[c][c][1]{$10$}
    \psfrag{ 20}[c][c][1]{$20$}
    \psfrag{ 30}[c][c][1]{$30$}
    \psfrag{ 40}[c][c][1]{$40$}
    \psfrag{ 50}[c][c][1]{$50$}
    \psfrag{ 60}[c][c][1]{$60$}
    \psfrag{ 70}[c][c][1]{$70$}
    \psfrag{ 0.005}[r][r][1]{$0.005$}
    \psfrag{ 0.01}[r][r][1]{$0.010$}
    \psfrag{ 0.015}[r][r][1]{$0.015$}
    \psfrag{ 0.02}[r][r][1]{$0.020$}
    \psfrag{ 0.025}[r][r][1]{$0.025$}
    \psfrag{ 0.03}[r][r][1]{$0.030$}
    \psfrag{ 0.035}[r][r][1]{$0.035$}
    \psfrag{ 0.04}[r][r][1]{$0.040$}
    \psfrag{ 0.045}[r][r][1]{$0.045$}
    \psfrag{w}[c][c][1.2]{$w$}
    \psfrag{f_W(w)}[c][c][1.2]{$f_{\tilde{W}}(w)$}
    \psfrag{sim}[l][l][0.8]{simulation}
    \psfrag{f0}[l][l][0.8]{${f_{\tilde{W}}}_0(w)$}
    \psfrag{f1}[l][l][0.8]{${f_{\tilde{W}}}_1(w)$}
    \psfrag{f2}[l][l][0.8]{${f_{\tilde{W}}}_2(w)$}
    \psfrag{f3}[l][l][0.8]{${f_{\tilde{W}}}_3(w)$}
    \psfrag{f4}[l][l][0.8]{${f_{\tilde{W}}}_4(w)$}
    \subfigure[$pB^2/2=0$]{\label{subfig:histo_LAN-0}
    \resizebox{0.7\figwidth}{!}{\includegraphics[angle=0]{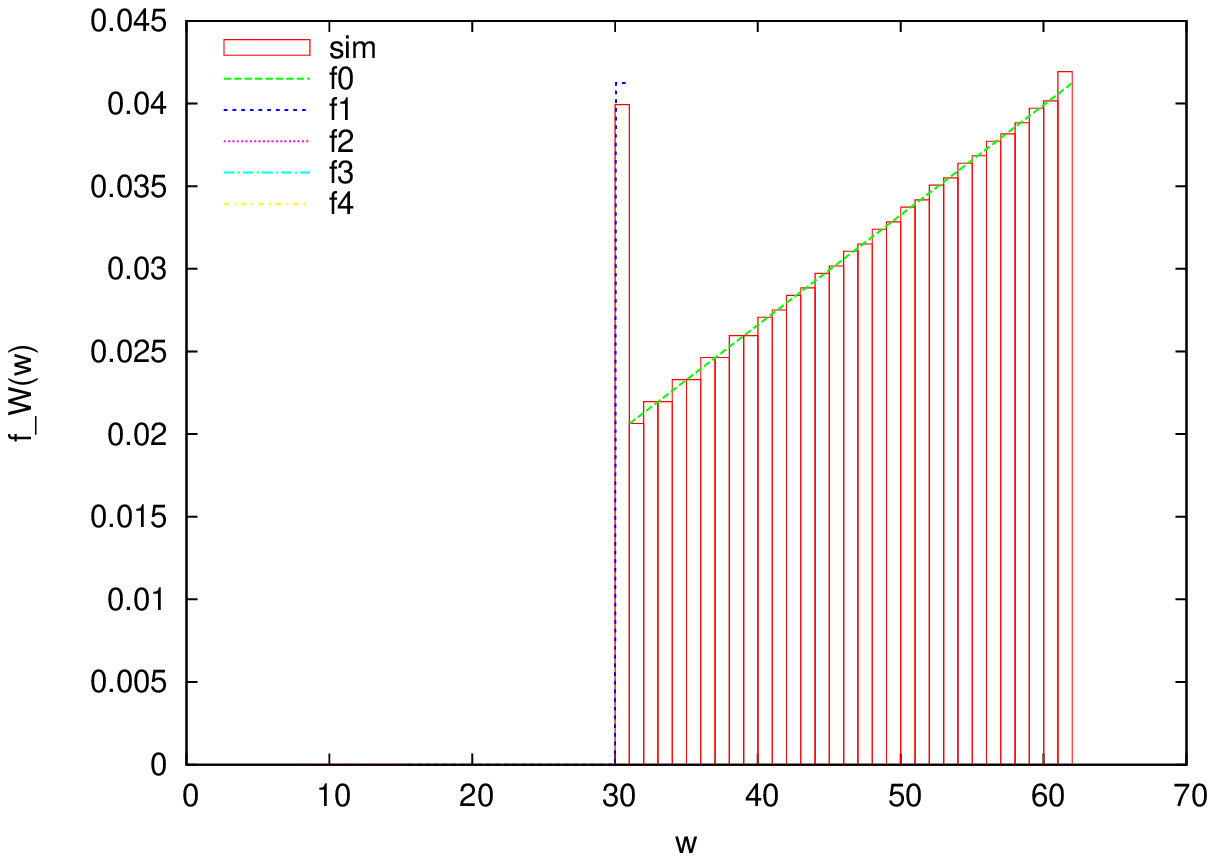}}}
    \psfrag{sim}[r][r][0.8]{simulation}
    \psfrag{f0}[r][r][0.8]{${f_{\tilde{W}}}_0(w)$}
    \psfrag{f1}[r][r][0.8]{${f_{\tilde{W}}}_1(w)$}
    \psfrag{f2}[r][r][0.8]{${f_{\tilde{W}}}_2(w)$}
    \psfrag{f3}[r][r][0.8]{${f_{\tilde{W}}}_3(w)$}
    \psfrag{f4}[r][r][0.8]{${f_{\tilde{W}}}_4(w)$}
    \subfigure[$pB^2/2=0.5$]{\label{subfig:histo_LAN-1}
    \resizebox{0.7\figwidth}{!}{\includegraphics[angle=0]{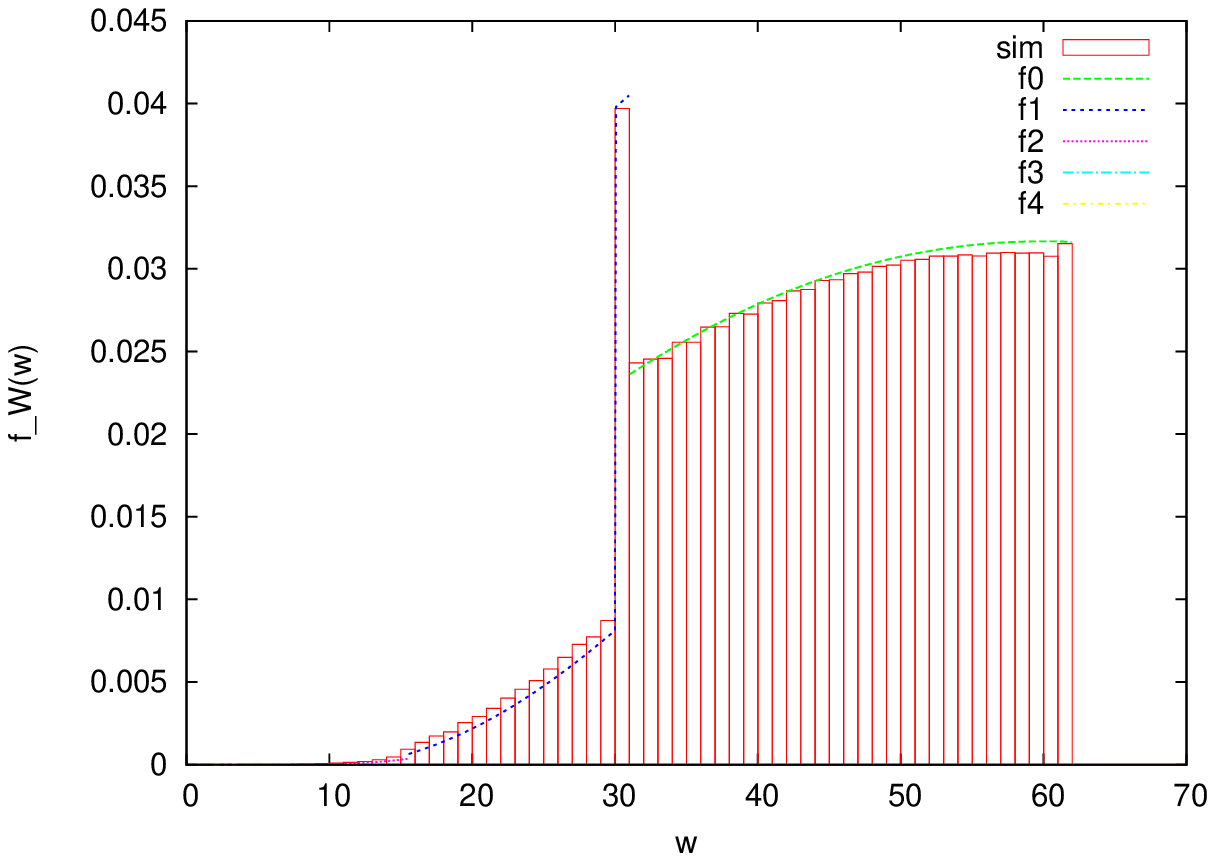}}}\\
    \subfigure[$pB^2/2=1$]{\label{subfig:histo_LAN-2}
    \resizebox{0.7\figwidth}{!}{\includegraphics[angle=0]{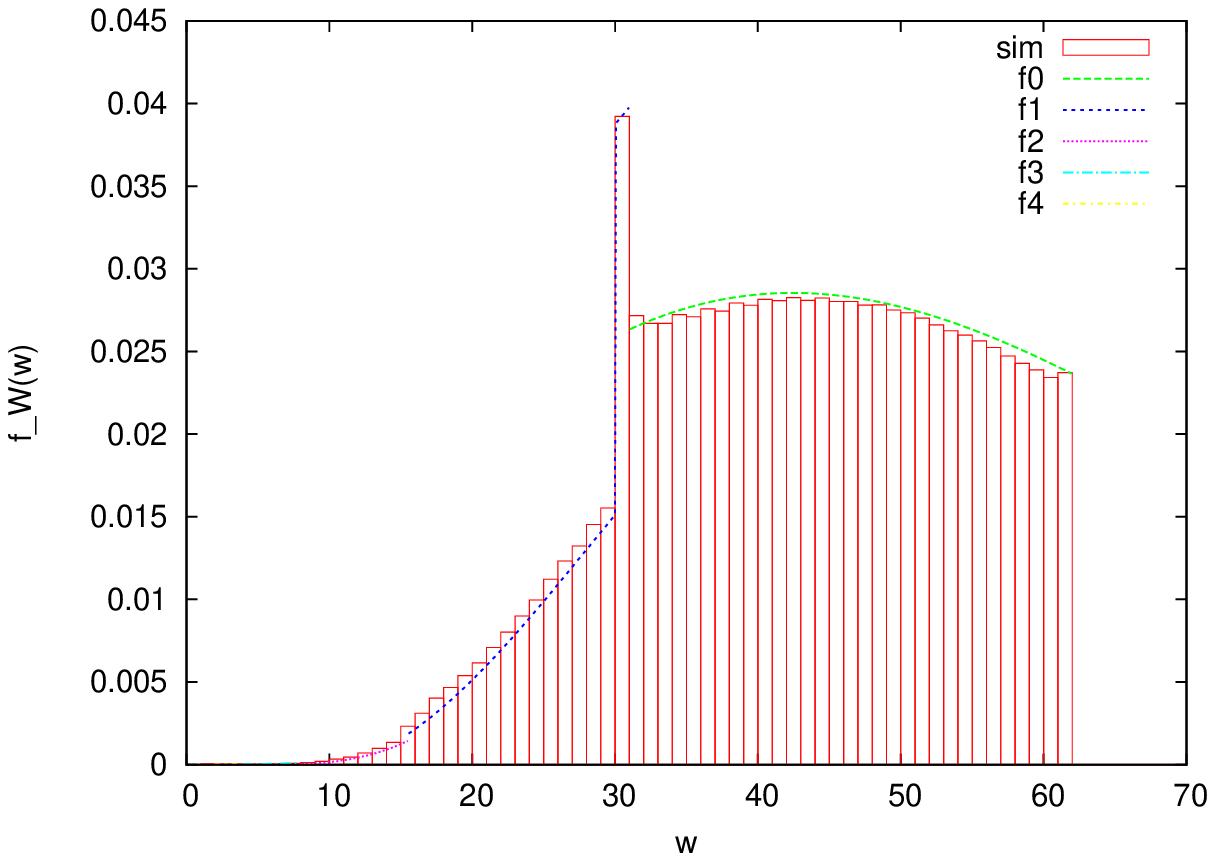}}}
    \subfigure[$pB^2/2=2$]{\label{subfig:histo_LAN-4}
    \resizebox{0.7\figwidth}{!}{\includegraphics[angle=0]{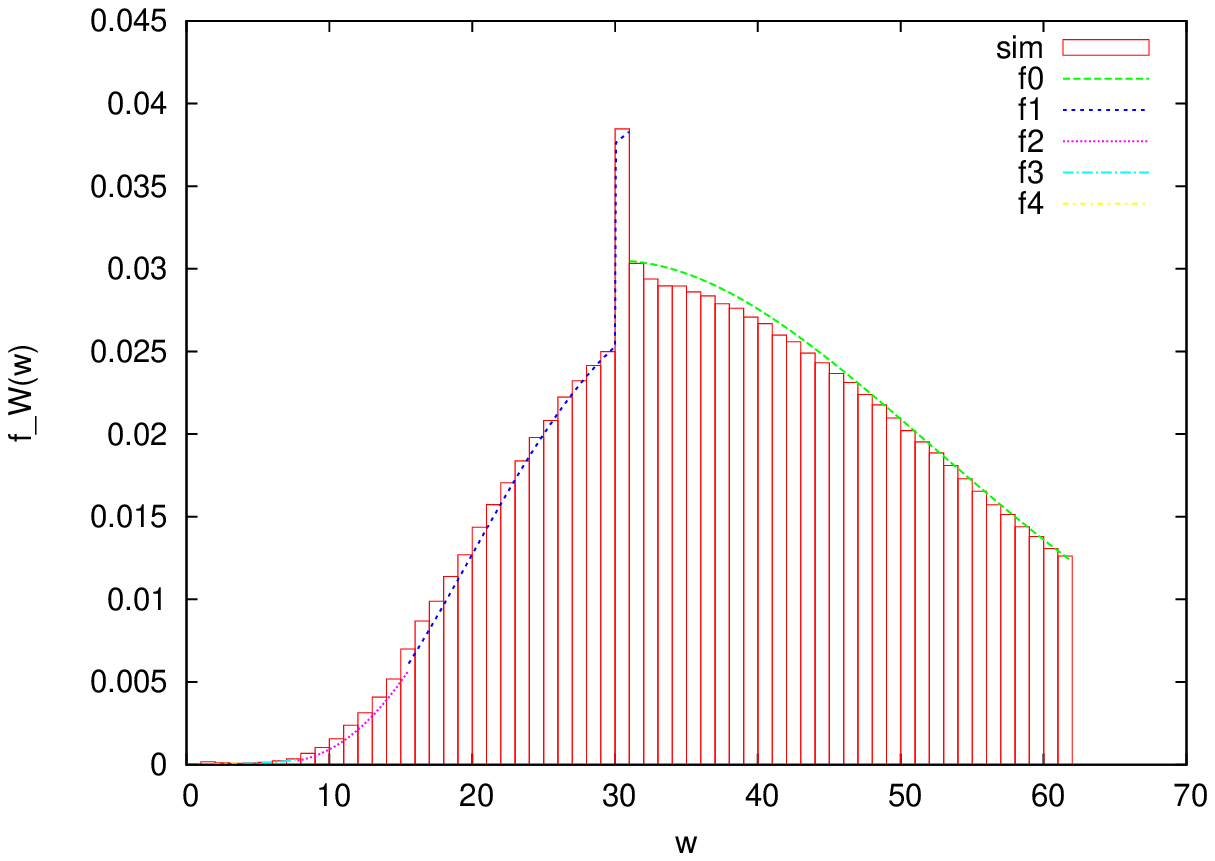}}}\\
    \subfigure[$pB^2/2=3.5$]{\label{subfig:histo_LAN-7}
    \resizebox{0.7\figwidth}{!}{\includegraphics[angle=0]{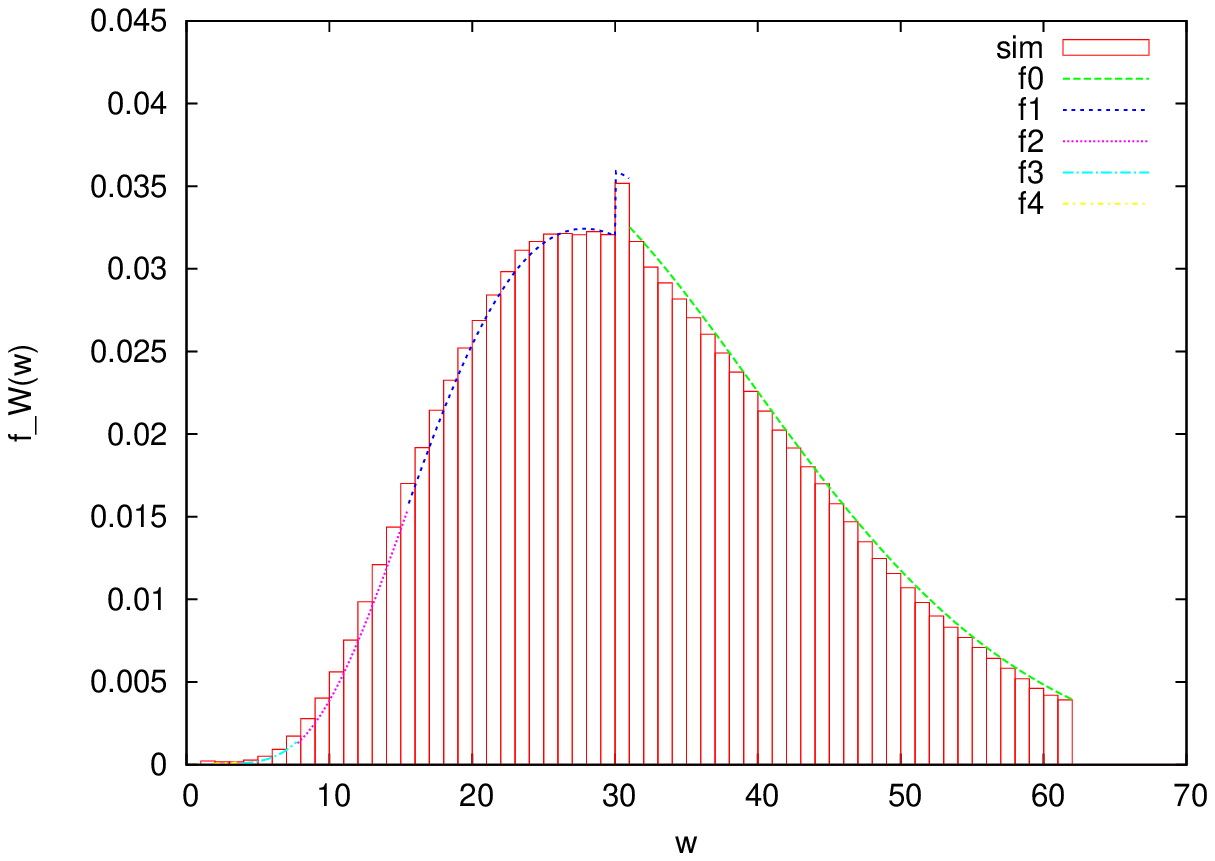}}}
    \subfigure[$pB^2/2=5$]{\label{subfig:histo_LAN-10}
    \resizebox{0.7\figwidth}{!}{\includegraphics[angle=0]{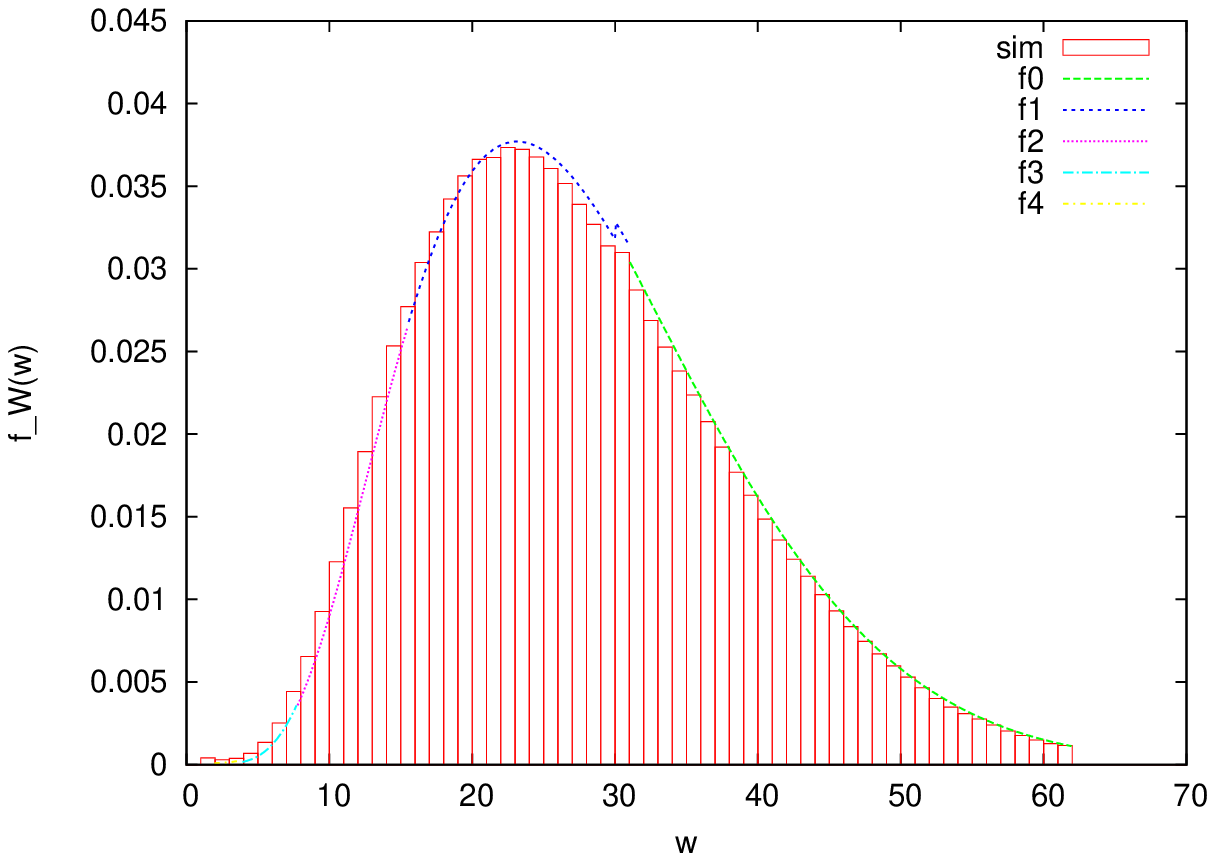}}}\\
  \end{center}
  \caption{Comparison of simulation results and theoretical model at
    buffer size $B=60$. ${f_W}_n(w),\;n\in\mathrm{N}$ denote
    piecewise solutions of the congestion window distribution 
    (\ref{eq:cwnd_dist4}).}
  \label{fig:histo_LAN}
\end{figure}

\subsubsection{Local Area Networks}

In order to verify my results I carried out simulations with \gls{ns}.  I have
applied my results for both \gls{lan} and \gls{wan} networks.  Let us consider
the \gls{lan} scenario first, where the model parameters are $m=1$,
$\beta=1/2$, $\ph/\alpha=p$ and the effective buffer size is $\tilde{B}=B$.  I
compare the simulation results and my model in Fig.~\ref{fig:histo_LAN} at
$pB^2/2=0.0, 0.5, 1.0, 2.0, 3.5$ and $5.0$ parameter values and at $B=60$ buffer
size.  The link capacity $C=256\textit{kb/s}$, link delay $D=0s$ and packet
size $P=1500\textit{byte}$ were fixed in the study of \gls{lan} and only buffer
size $B$ and loss probability $p$ were varied.  

For the interpretation of the simulation results I consider the effect of the
\gls{frfr} algorithms as well.  I showed in the previous chapter that in the
case of an infinite buffer the effect of \gls{frfr} algorithms can be taken
into consideration by the modified distribution (\ref{eq:cwnd_gendist}).  The
finite buffer case can be handled similarly, with two minor adjustments.
Firstly, the packet loss rate $\ph$ should be replaced by the total loss rate
$\ph'=\frac{\ph}{1-A}$, because plateaus of the \gls{frfr} mode appear after
packet losses happening at the buffer, too.  Secondly, the distribution of the
``after loss'' congestion window $f_{\wal}(w)$ should be calculated directly
from $\wal=\beta\wbl$ now: $f_{\wal}(w)=f_{\wbl}(\beta^{-1}w)\beta^{-1}$.
Accordingly, $f_{W}(\beta^{-1}w)$ have to be replaced by
$f_{\wbl}(\beta^{-1}w)=A(\ph/\alpha,\tilde{B})\,\delta(\beta^{-1}w-\tilde{B})
+\phi(\beta^{-1}w)$ in (\ref{eq:cwnd_gendist}).  After a variable
transformation in the delta distribution we obtain
\begin{equation}
  f_{\tilde{W}}(w)
  =\frac{f_{W}(w)+\frac{\ph}{\alpha}\tilde{B}^m
  \frac{A(\ph/\alpha,\tilde{B})}{1-A(\ph/\alpha,\tilde{B})}
  \delta(w-\beta\tilde{B})+
  \frac{\ph}{\alpha}\beta^{-\left(m+1\right)}w^m
  f_W(\beta^{-1}w)}
  {1+\frac{\ph}{\alpha}\frac{\mathbb{E}[W^m]}{1-A(\ph,\alpha,\tilde{B})}},
  \label{eq:cwnd_finite_FRFR}
\end{equation}
where (\ref{eq:cwnd_dist4}) has been used implicitly.

The most distinct consequence of \gls{frfr} algorithms is the sharp peak in the
middle of the histograms in Fig.~\ref{fig:histo_LAN}.  In analytic formula
(\ref{eq:cwnd_finite_FRFR}) the peak is represented by a Dirac-delta
distribution.  The delta-distribution has been scattered over a finite region
in Fig.~\ref{fig:histo_LAN} in order to be comparable with the peaks in the
numerical histograms.  The derived analytic expression shows very impressive
agreement with the numerical simulations.  The slight discrepancy at larger
packet loss probabilities comes from the differences between the fluid model
and the packet level simulation, discussed in Section~\ref{subsec:OneTCP_LAN}.

\subsubsection{The effect of link delay}


\begin{figure}[tb]
  \begin{center}
    \psfrag{QLAN}[c][c][1.2]{Quantiles of cwnd in LAN}
    \psfrag{QDelay}[c][c][1.2]{Quantiles of cwnd with link delay $D$}
    \psfrag{D=15}[l][l][0.8]{$B=75$, $2\alpha D=15$}
    \psfrag{D=30}[l][l][0.8]{$B=60$, $2\alpha D=30$}
    \psfrag{0}[r][r][1]{$0$}
    \psfrag{20}[r][r][1]{$20$}
    \psfrag{40}[r][r][1]{$40$}
    \psfrag{60}[r][r][1]{$60$}
    \psfrag{80}[r][r][1]{$80$}
    \psfrag{100}[r][r][1]{$100$}
    \psfrag{ 0}[c][c][1]{$0$}
    \psfrag{ 20}[c][c][1]{$20$}
    \psfrag{ 40}[c][c][1]{$40$}
    \psfrag{ 60}[c][c][1]{$60$}
    \psfrag{ 80}[c][c][1]{$80$}
    \psfrag{ 100}[c][c][1]{$100$}
    \subfigure[$p\tilde{B}^2/2=0.5$]{
      \resizebox{0.7\figwidth}{!}{\includegraphics{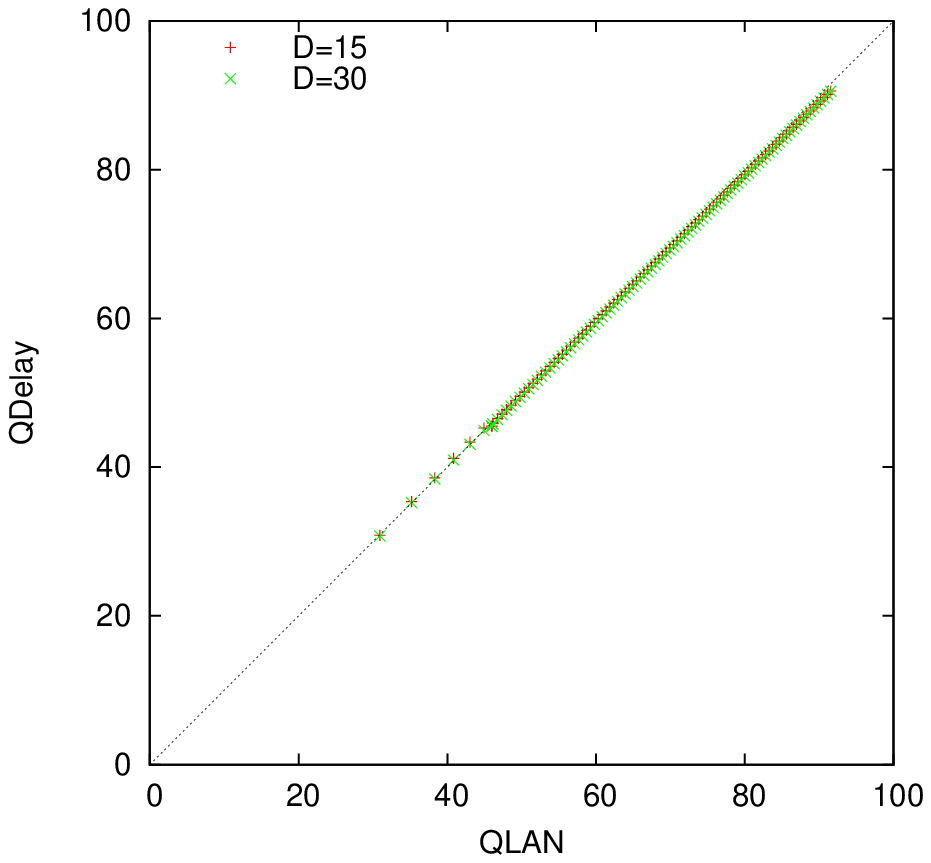}}}
    \subfigure[$p\tilde{B}^2/2=2.5$]{
      \resizebox{0.7\figwidth}{!}{\includegraphics{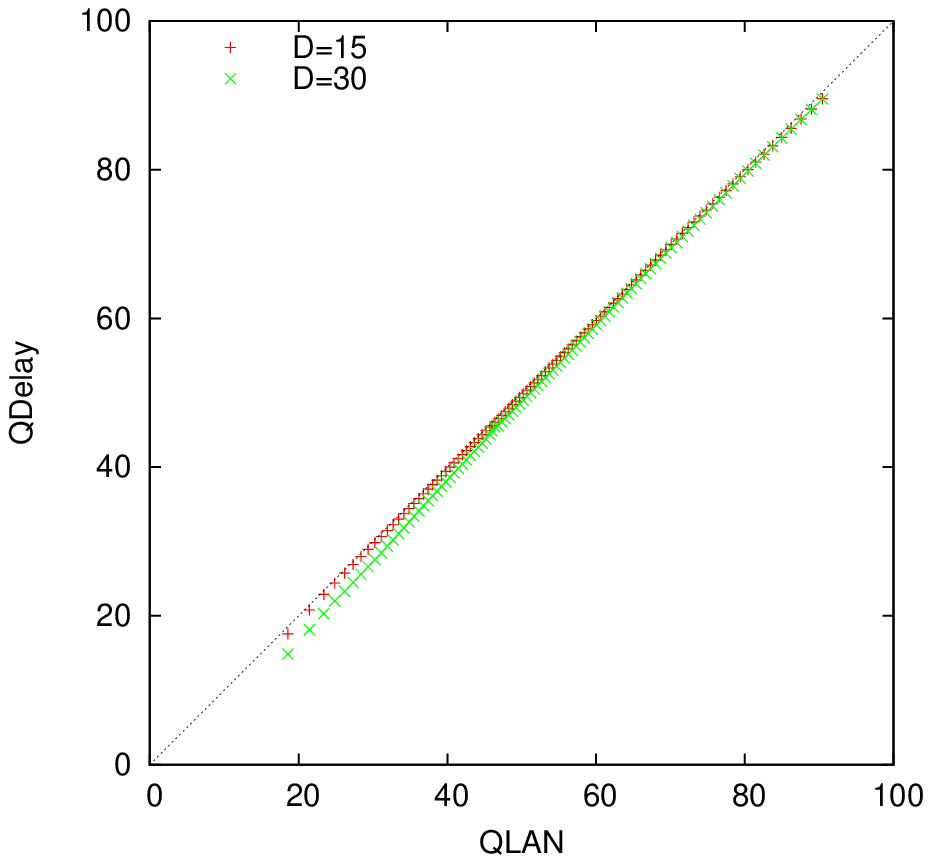}}}
  \end{center}
  \caption{Quantile-quantile plot of the congestion window in 
  network configurations with different buffer size $B$ and bandwidth-delay 
  product $2\alpha D$, but with the same effective buffer size 
  $\tilde{B}=B+2\alpha D=90$.} 
  \label{fig:delay}
\end{figure}

I assumed in the previous analysis that the link delay is zero.  My results can
be applied as approximations for situations where the link delay is non-zero,
but the probability that the buffer is empty is negligible.  In addition to the
buffer, $2\alpha D$ number of packets and acknowledgments can be found on the
link where $\alpha=C/P$.  $C$, $D$ and $P$ are the link capacity, the link
delay and the packet size respectively. The congestion window limit in this
situation must be set to the total number of packets $\tilde{B}=B+2\alpha D$ in
the system and the link can be treated as a part of the buffer. This can be
verified with simulations. In my simulation scenario $2\alpha D=15$ and
$2\alpha D=30$ number of \tcp\ and \ack\ packets could be on the link. The
buffer size was set to $B=75$ and $B=60$ respectively, so that the effective
buffer size $\tilde{B}=90$ was the same.  In Fig.~\ref{fig:delay}
quantile-quantile plots of the congestion window are shown.
Percentiles of the congestion window are plotted at the given link delay
and buffer size combinations as the function of the percentiles of cwnd in
an ideal LAN scenario.  Data was obtained from simulations at two
different control parameter values.  It can be seen that data points are close
to the diagonal, drawn by dotted lines.  This implies that the data points 
are from the same distribution when the link delay is zero and when it
is small, but not zero.  Some deviation from the diagonal can only be observed
at the lower quantiles of $p\tilde{B}^2/2=2.5$, when $B=60$, $\alpha D=30$, 
because the buffer is occasionally empty in this case.

\subsubsection{Wide Area Networks}

\begin{figure}[p]
 \begin{center} 
    \psfrag{ 0}[c][c][1]{$0$}
    \psfrag{ 10}[c][c][1]{$10$}
    \psfrag{ 20}[c][c][1]{$20$}
    \psfrag{ 30}[c][c][1]{$30$}
    \psfrag{ 40}[c][c][1]{$40$}
    \psfrag{ 50}[c][c][1]{$50$}
    \psfrag{ 60}[c][c][1]{$60$}
    \psfrag{ 70}[c][c][1]{$70$}
    \psfrag{ 80}[c][c][1]{$80$}
    \psfrag{ 0.01}[r][r][1]{$0.01$\,}
    \psfrag{ 0.02}[r][r][1]{$0.02$\,}
    \psfrag{ 0.03}[r][r][1]{$0.03$\,}
    \psfrag{ 0.04}[r][r][1]{$0.04$\,}
    \psfrag{ 0.05}[r][r][1]{$0.05$\,}
    \psfrag{ 0.06}[r][r][1]{$0.06$\,}
    \psfrag{w}[c][c][1.2]{$w$}
    \psfrag{f_W(w)}[c][c][1.2]{$f_W(w)$}
    \psfrag{sim}[r][r][0.8]{simulation}
    \psfrag{f0}[r][r][0.8]{${f_W}_0(w)$}
    \psfrag{f1}[r][r][0.8]{${f_W}_1(w)$}
    \psfrag{f2}[r][r][0.8]{${f_W}_2(w)$}
    \psfrag{f3}[r][r][0.8]{${f_W}_3(w)$}
    \psfrag{f4}[r][r][0.8]{${f_W}_4(w)$}
    \subfigure[$p\tilde{B}^2/2=0$]{\label{subfig:histo_WAN-0}
    \resizebox{0.7\figwidth}{!}{\includegraphics[angle=0]{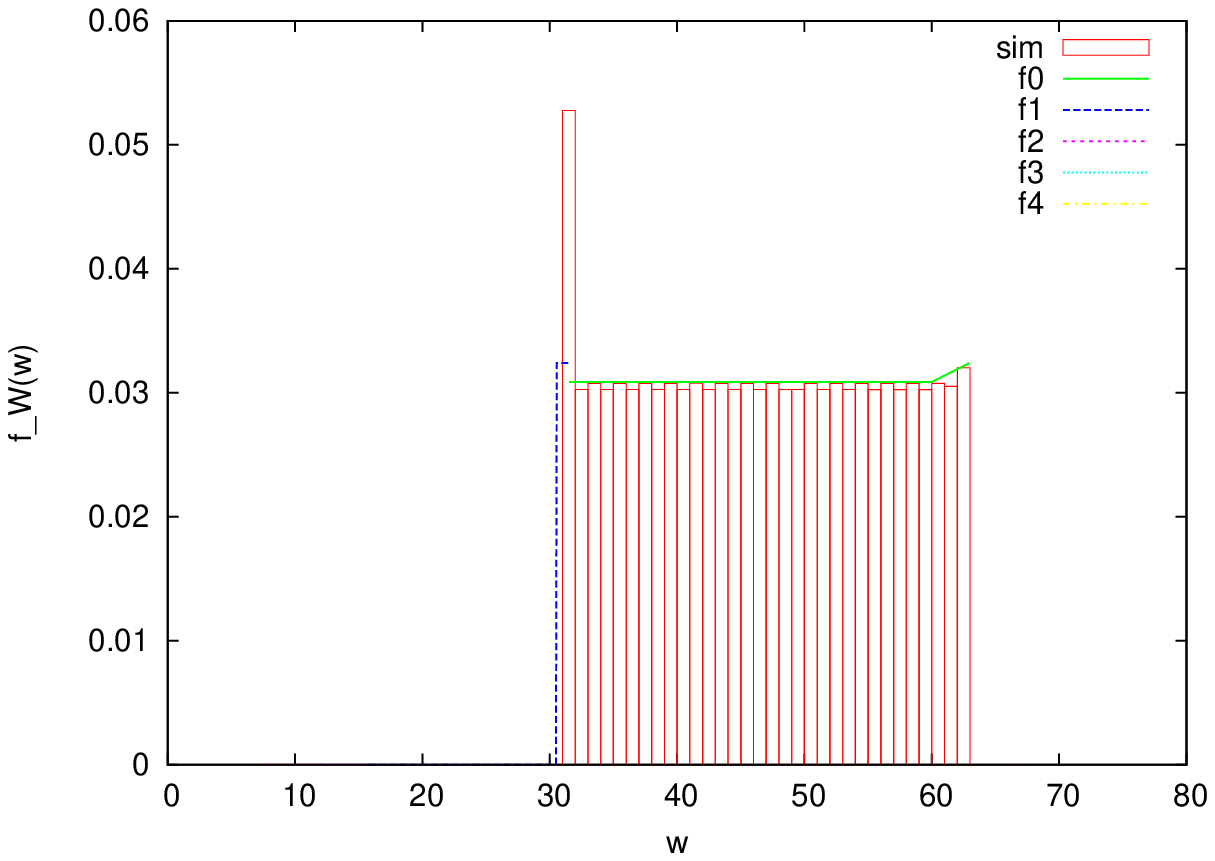}}}
    \subfigure[$p\tilde{B}^2/2=0.5$]{\label{subfig:histo_WAN-1}
    \resizebox{0.7\figwidth}{!}{\includegraphics[angle=0]{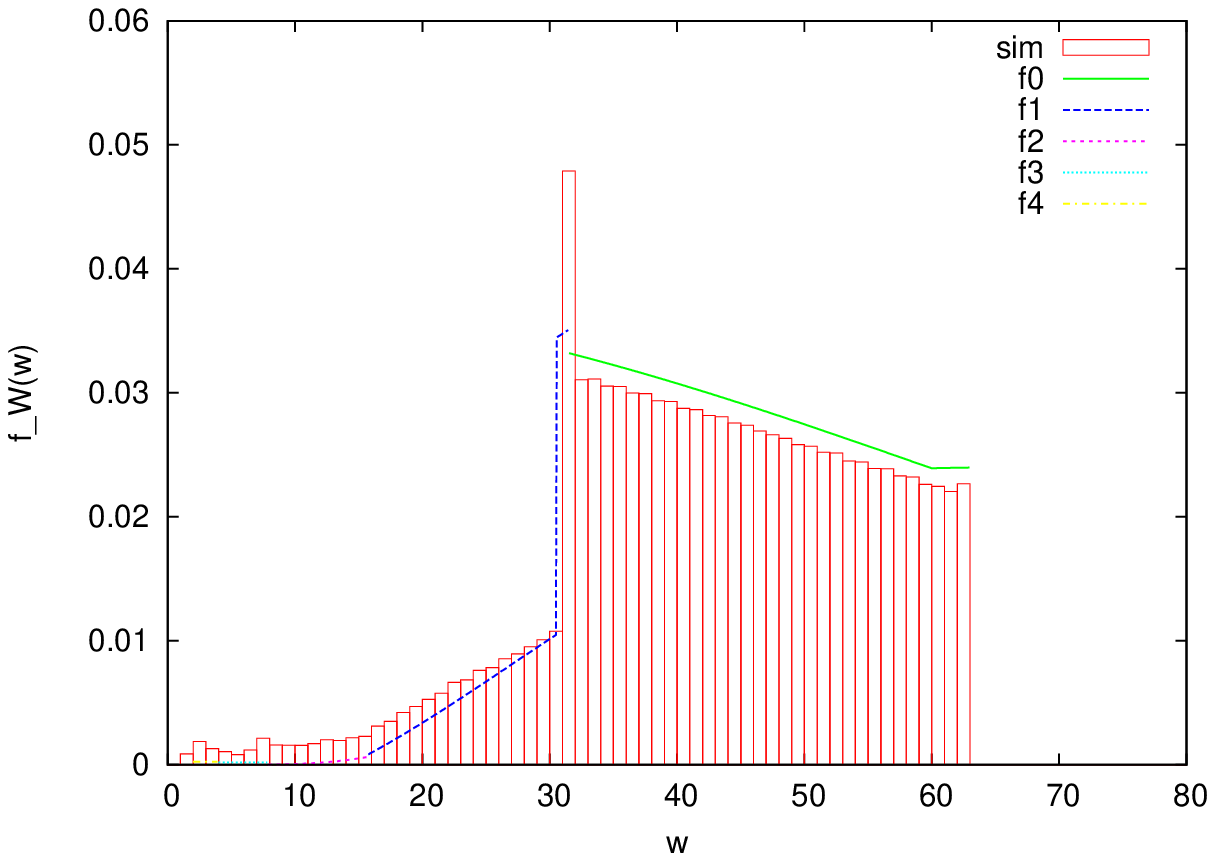}}}\\
    \subfigure[$p\tilde{B}^2/2=1$]{\label{subfig:histo_WAN-2}
    \resizebox{0.7\figwidth}{!}{\includegraphics[angle=0]{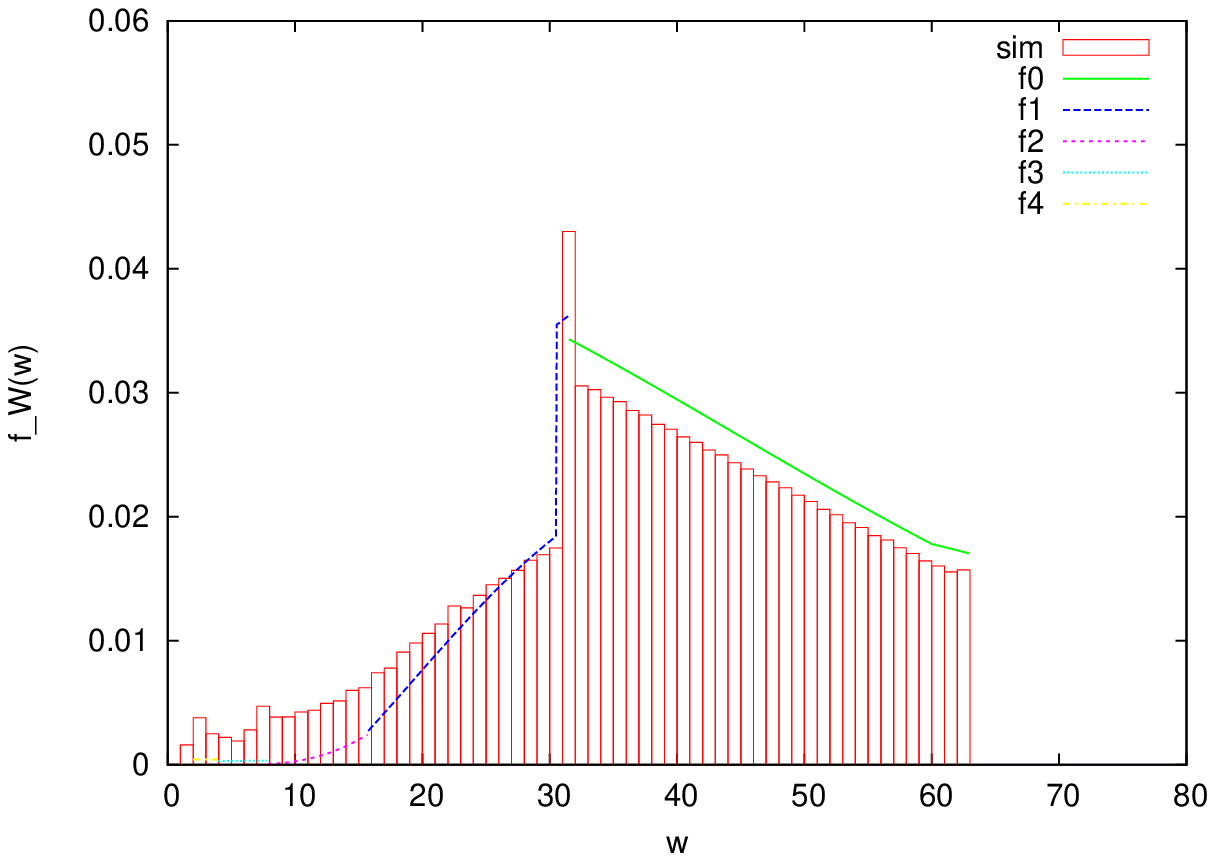}}}
    \subfigure[$p\tilde{B}^2/2=2$]{\label{subfig:histo_WAN-4}
    \resizebox{0.7\figwidth}{!}{\includegraphics[angle=0]{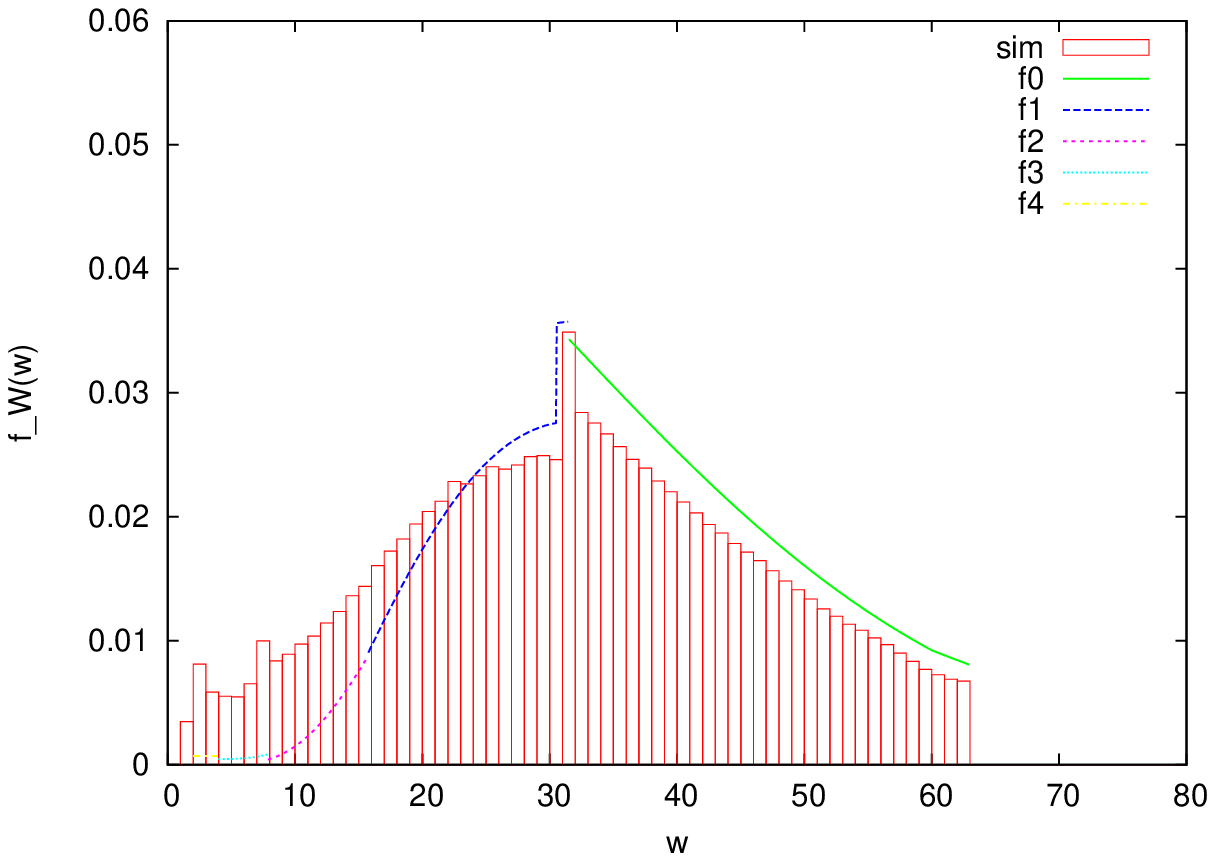}}}\\
    \subfigure[$p\tilde{B}^2/2=3.5$]{\label{subfig:histo_WAN-7}
    \resizebox{0.7\figwidth}{!}{\includegraphics[angle=0]{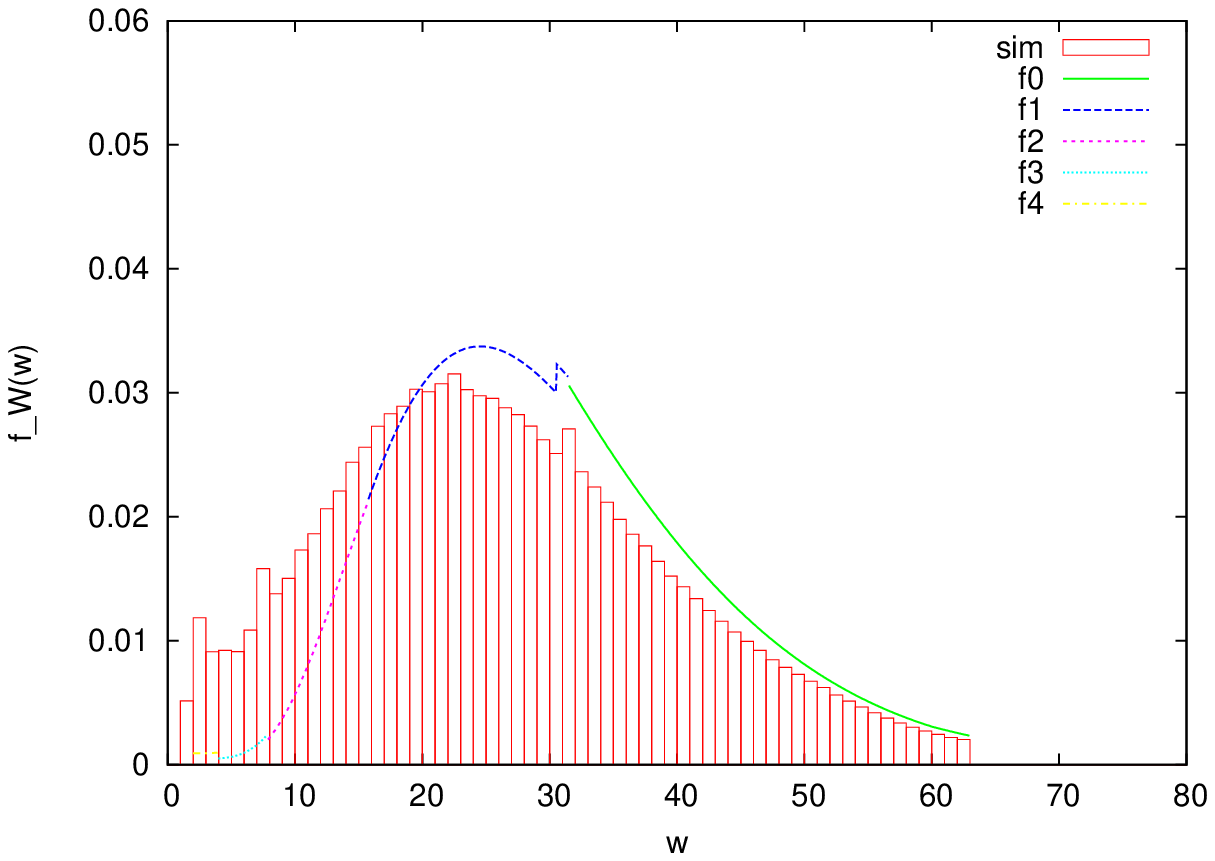}}}
    \subfigure[$p\tilde{B}^2/2=5$]{\label{subfig:histo_WAN-10}
    \resizebox{0.7\figwidth}{!}{\includegraphics[angle=0]{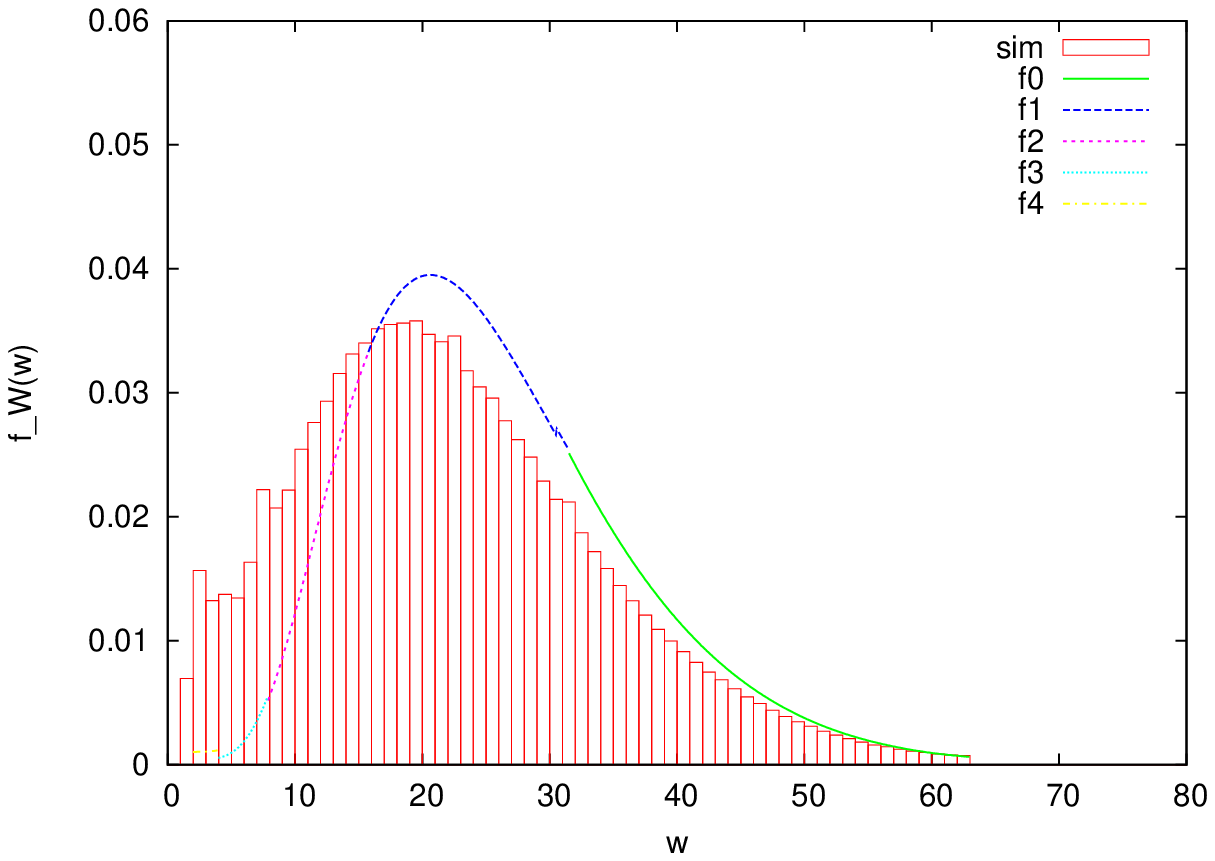}}}
    \end{center} 
    \caption{Comparison of simulation results and the theoretical
             model. The link could carry maximal $2\alpha D=60$ number of
             \tcp\ and \ack\ packets, and the buffer could store $B=3$ packets. 
             ${f_W}_i(w),\;i\in\mathbb{N}$ denote
             the piecewise solutions of the congestion window distribution.}
\label{fig:histo_WAN}
\end{figure}

\begin{figure}[p]
  \begin{center} 
    \psfrag{ 0}[c][c][1]{$0$}
    \psfrag{ 10}[c][c][1]{$10$}
    \psfrag{ 20}[c][c][1]{$20$}
    \psfrag{ 30}[c][c][1]{$30$}
    \psfrag{ 40}[c][c][1]{$40$}
    \psfrag{ 50}[c][c][1]{$50$}
    \psfrag{ 60}[c][c][1]{$60$}
    \psfrag{ 70}[c][c][1]{$70$}
    \psfrag{ 80}[c][c][1]{$80$}
    \psfrag{ 90}[c][c][1]{$90$}
    \psfrag{ 100}[c][c][1]{$100$}
    \psfrag{ 0.005}[r][r][1]{$0.005$\,}
    \psfrag{ 0.01}[r][r][1]{$0.010$\,}
    \psfrag{ 0.015}[r][r][1]{$0.015$\,}
    \psfrag{ 0.02}[r][r][1]{$0.020$\,}
    \psfrag{ 0.025}[r][r][1]{$0.025$\,}
    \psfrag{ 0.03}[r][r][1]{$0.030$\,}
    \psfrag{w}[c][c][1.2]{$w$}
    \psfrag{f_W(w)}[c][c][1.2]{$f_W(w)$}
    \psfrag{sim}[l][l][0.8]{simulation}
    \psfrag{f0}[l][l][0.8]{${f_W}_0(w)$}
    \psfrag{f1}[l][l][0.8]{${f_W}_1(w)$}
    \psfrag{f2}[l][l][0.8]{${f_W}_2(w)$}
    \psfrag{f3}[l][l][0.8]{${f_W}_3(w)$}
    \psfrag{f4}[l][l][0.8]{${f_W}_4(w)$}
    \subfigure[$p\tilde{B}^2/2=0$]{\label{subfig:histo_WAN-B30-0}
    \resizebox{0.7\figwidth}{!}{\includegraphics[angle=0]{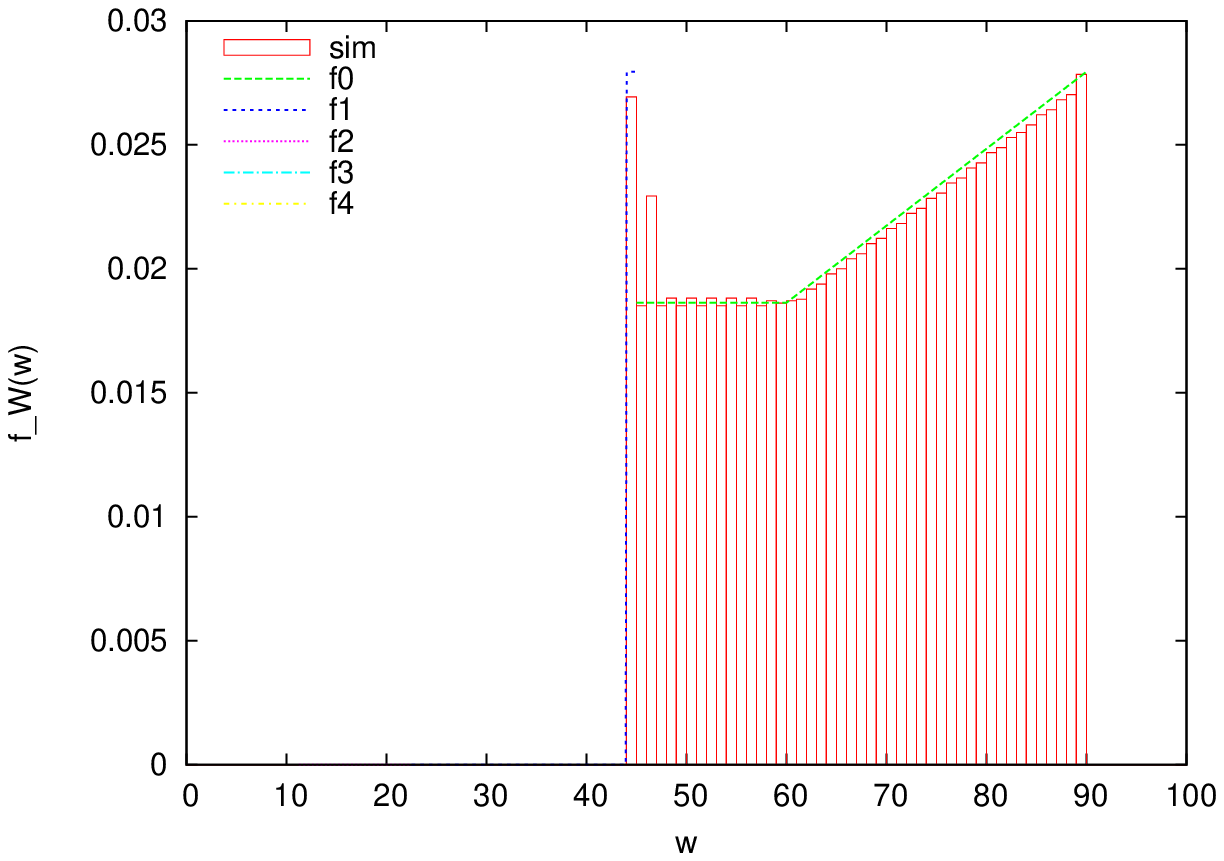}}}
    \psfrag{sim}[r][r][0.8]{simulation}
    \psfrag{f0}[r][r][0.8]{${f_W}_0(w)$}
    \psfrag{f1}[r][r][0.8]{${f_W}_1(w)$}
    \psfrag{f2}[r][r][0.8]{${f_W}_2(w)$}
    \psfrag{f3}[r][r][0.8]{${f_W}_3(w)$}
    \psfrag{f4}[r][r][0.8]{${f_W}_4(w)$}
    \subfigure[$p\tilde{B}^2/2=0.5$]{\label{subfig:histo_WAN-B30-1}
    \resizebox{0.7\figwidth}{!}{\includegraphics[angle=0]{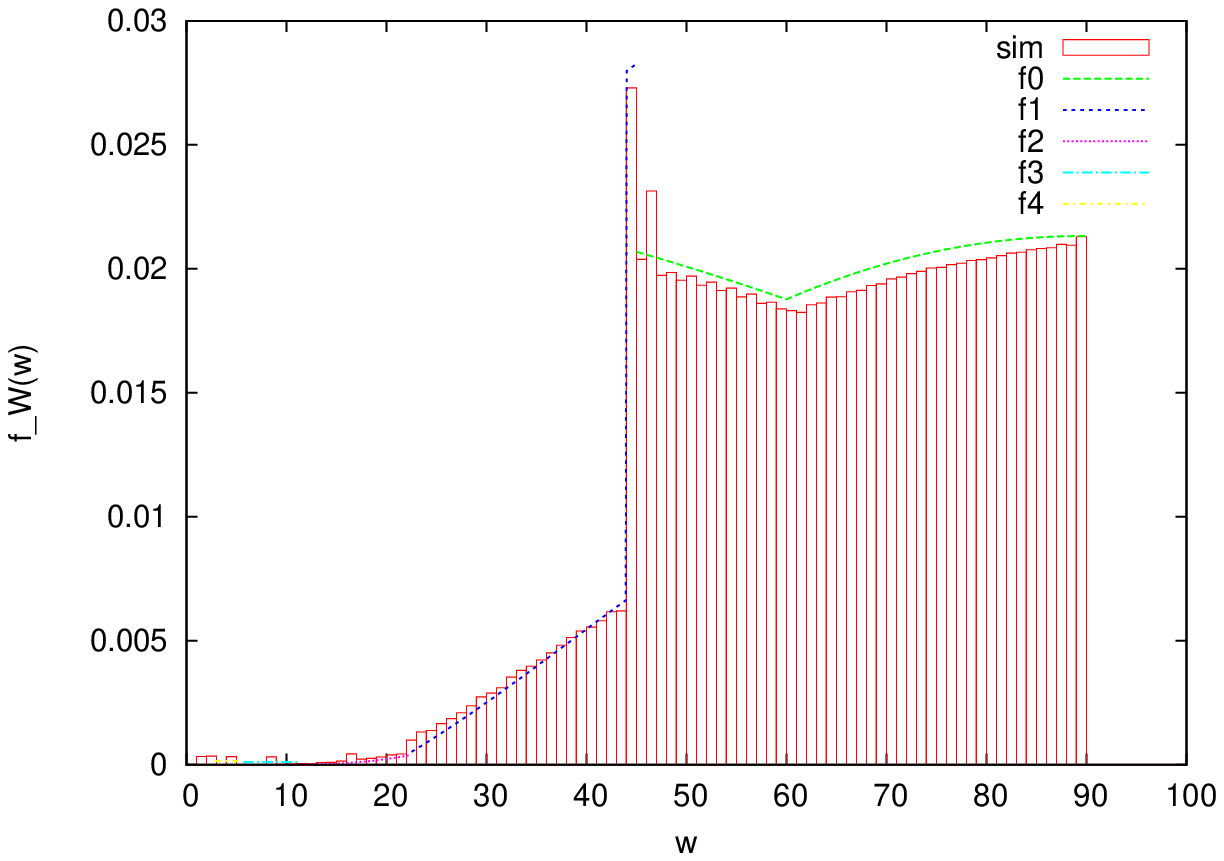}}}\\
    \subfigure[$p\tilde{B}^2/2=1$]{\label{subfig:histo_WAN-B30-2}
    \resizebox{0.7\figwidth}{!}{\includegraphics[angle=0]{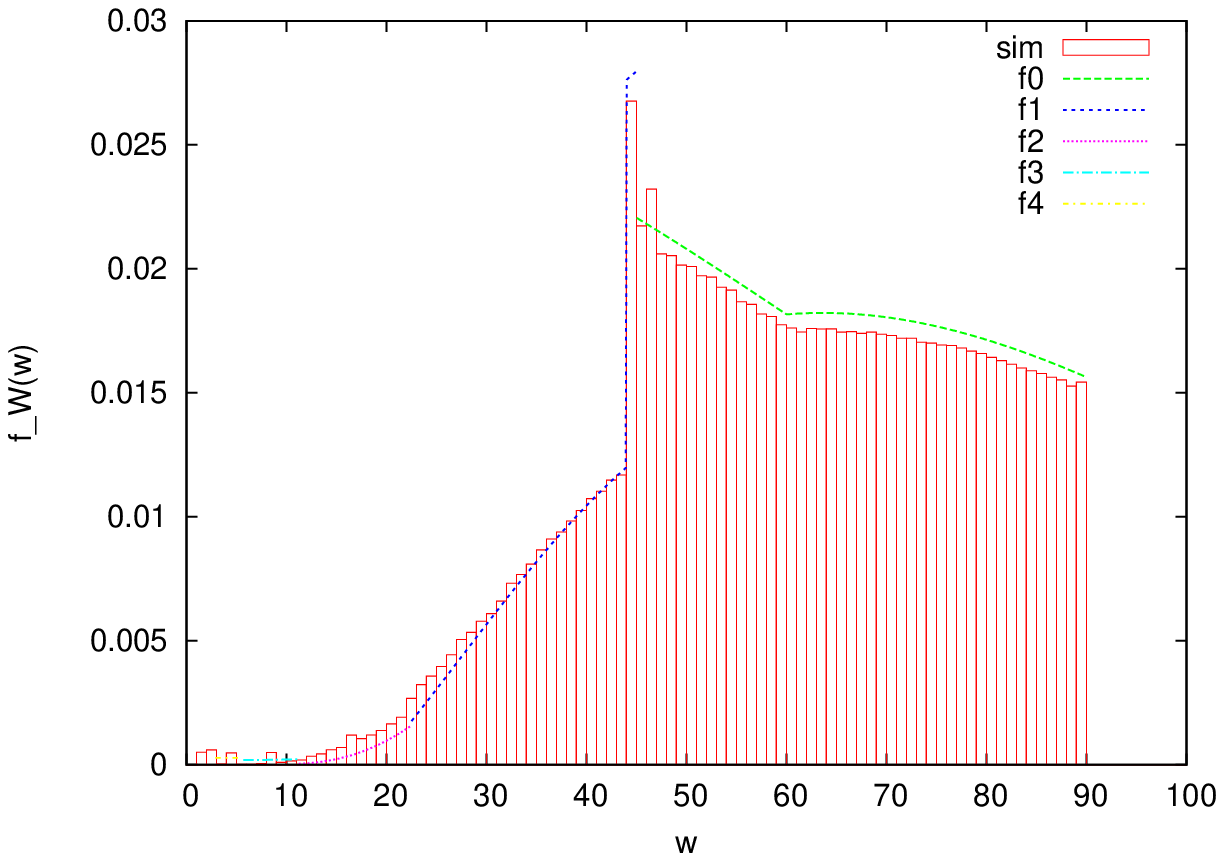}}}
    \subfigure[$p\tilde{B}^2/2=2$]{\label{subfig:histo_WAN-B30-4}
    \resizebox{0.7\figwidth}{!}{\includegraphics[angle=0]{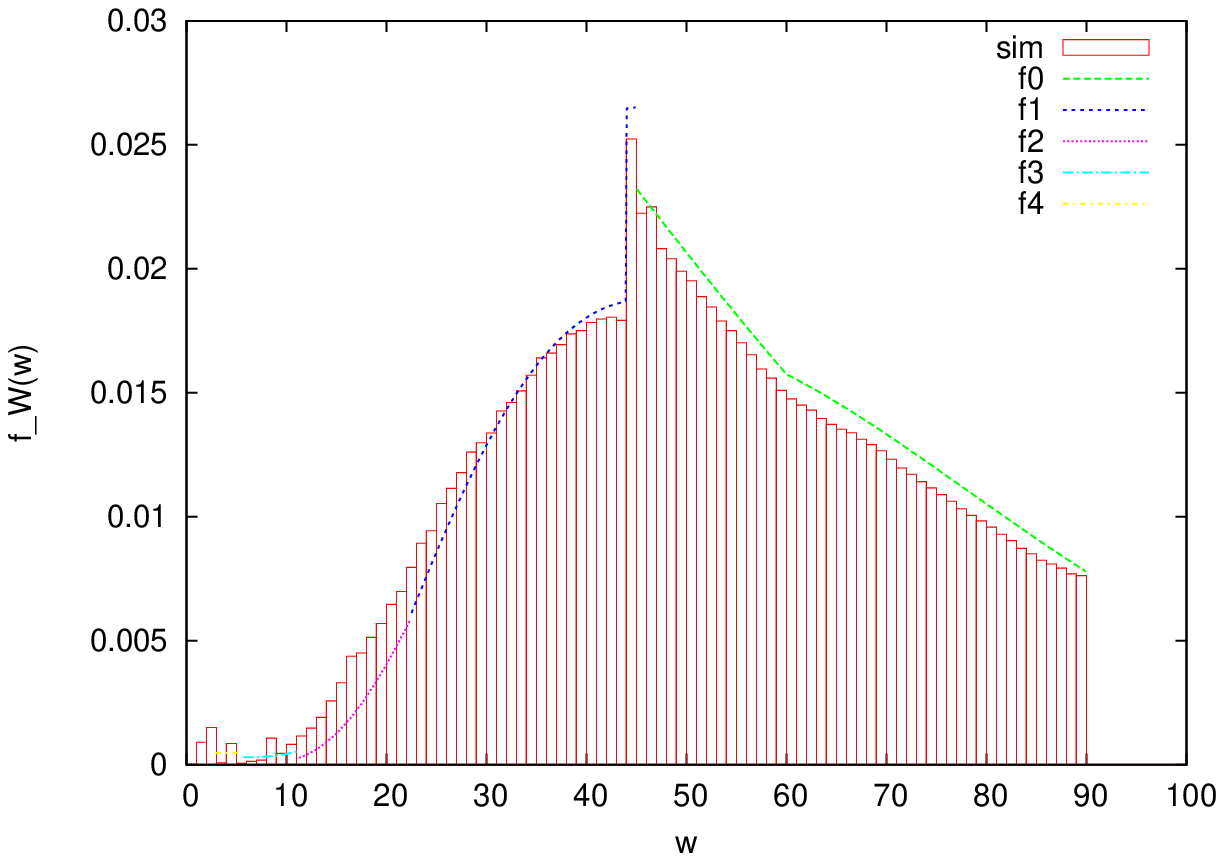}}}\\
    \subfigure[$p\tilde{B}^2/2=3.5$]{\label{subfig:histo_WAN-B30-7}
    \resizebox{0.7\figwidth}{!}{\includegraphics[angle=0]{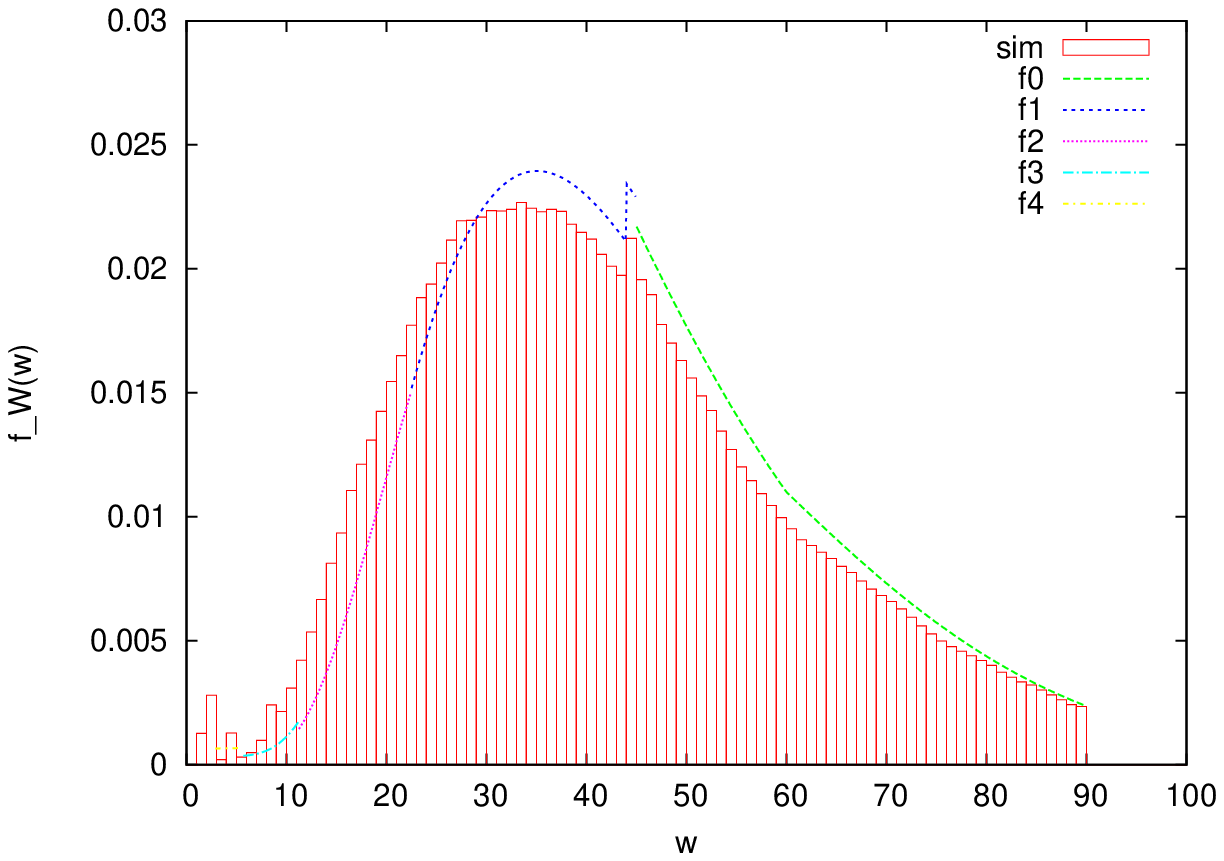}}}
    \subfigure[$p\tilde{B}^2/2=5$]{\label{subfig:histo_WAN-B30-10}
    \resizebox{0.7\figwidth}{!}{\includegraphics[angle=0]{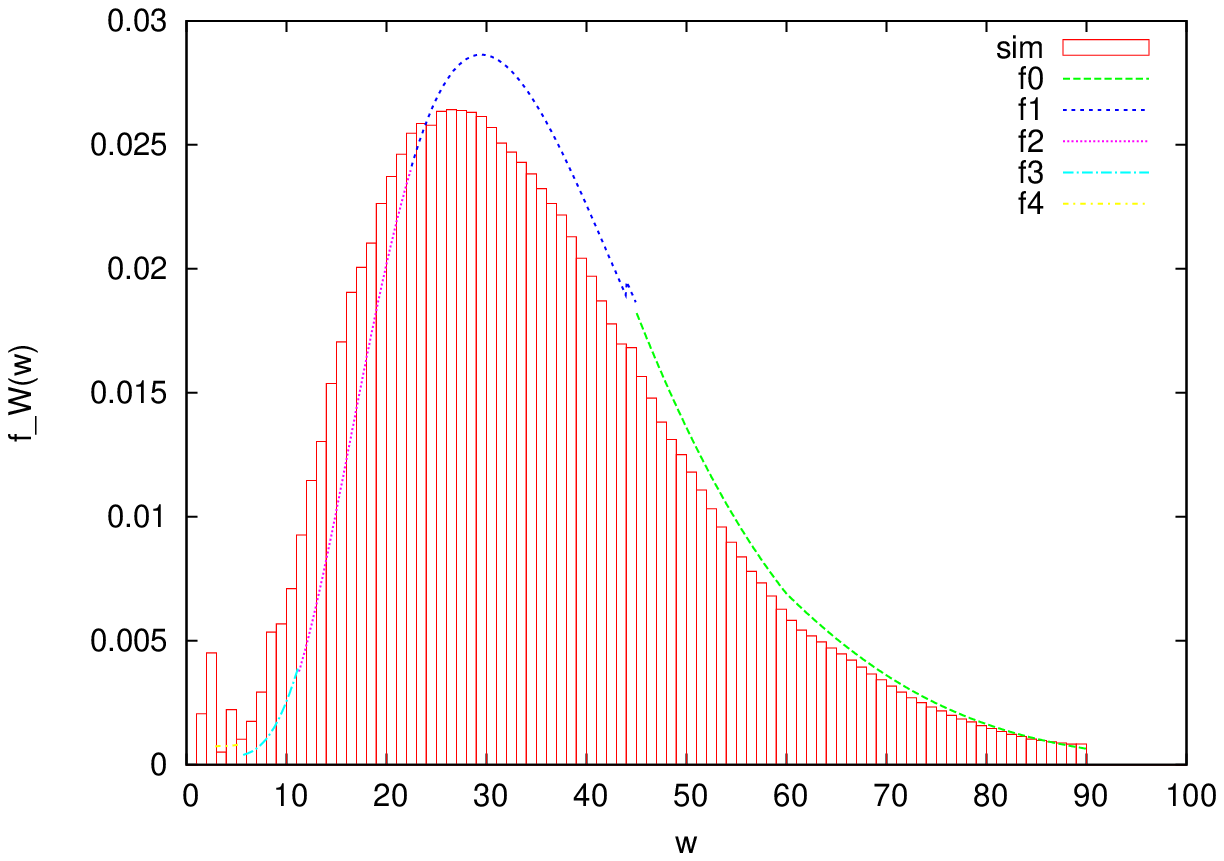}}}
    \end{center} 
    \caption{Comparison of simulation results and the theoretical
             model. The link could carry maximal $2\alpha D=60$ number of
             \tcp\ and \ack\ packets, and the buffer could store $B=30$
             packets. ${f_W}_i(w),\;i\in\mathbb{N}$ denote
             the piecewise solutions of the congestion window distribution.}
\label{fig:histo_WAN-B30}
\end{figure}

In a \gls{wan} scenario buffering delay is small compared to the link delay.
As I noted in the last section, however, it cannot be set to zero, because in a
packet level simulator packet bursts appear inevitably.  Accordingly, the link
delay was so large that it could carry $2\alpha D=60$ \tcp\ and \ack\ packets
simultaneously.  Furthermore, two buffer size values $B=3$ and $30$ were
selected for numerical simulations.  The analytical formula for the congestion
window distribution can be obtained from the modification of
(\ref{eq:cwnd_gendist_WAN}) for the \gls{frfr} algorithms, analogously to the
\gls{lan} scenario.  Other parameters of the \gls{wan} model are $m=1$,
$\ph/\alpha$, and $\beta=1/2$.  

The theoretical distributions and histograms obtained from \gls{ns} simulations
can be seen in Fig.~\ref{fig:histo_WAN} at $B=3$, which is close to the ideal
\gls{wan} scenario.  The external loss rate was varied in the $0\le
p\tilde{B}^2/2\le5$ range.  All other parameters were fixed.  One can see that
the histogram deviates from the theoretical distribution even for small values
of the control parameter.  The non-zero probability in the histogram that the
congestion window is $1$ implies that the slow start mechanism is responsible
for the discrepancy.  In Fig.~\ref{fig:histo_WAN-B30} empirical histograms are
compared with the theoretical distribution at $B=30$, which is an intermediate
configuration between \gls{lan} and \gls{wan}.  The effect of slow start mode
is much less significant than at $B=3$.

The main source of error is the macroscopic probability of slow start mode.
The other observable difference from experiments comes from the slight
discrepancy in the position of the Dirac-delta and the finite peak in the
histogram.  Despite these errors my model agrees with simulations for small
loss probabilities and gives a qualitatively correct description of the WAN
situation for larger ones.

\section{Conclusions}
\label{sec:Finite_Conclusions}

In this chapter I investigated the \tcp\ congestion avoidance algorithm in
networks where the finite buffer size limits the maximal achievable congestion
window size.  The most important development I accomplished in this study is
that the total loss felt by \tcp, including the buffer and the external packet
loss, can be predicted from the network parameters, namely the length of the
buffer and the probability of external packet loss. This formula makes it
possible to calculate the total loss along a multi-buffer, multi-link route.
The presented analytical expression, $A(x)$, can be computed numerically
without difficulty and the total loss can be calculated by a simple formula.
I also showed that $A(x)$ and the coefficients which appear in the probability
distributions depend only on a certain combination of the parameters.  This
combination is the control parameter in my model.  Networks with the same
control parameters are equivalent in the sense that the same portion of the
total packet loss occurs at the buffer, and the coefficients are the same in
the distribution function.

In addition, I derived the stationary probability distribution of the
congestion window process analytically in LAN, in WAN, and in general
situations.  New types of congestion window distributions are discovered when
the packet loss in the buffer is large compared to other sources of packet
loss.  These are different from the usual Gaussian-type single humped
distributions and my findings can help to develop a qualitative  classification
of window distributions.  I validated my calculations with computer simulations
and I showed that my analysis agrees with the simulations properly.  I also
pointed out the limits of my model.  More specifically, I demonstrated that the
effect of the slow start mechanism becomes significant if the buffer size is
small or the packet loss probability is large.




\chapter[Traffic on complex networks]{Traffic dynamics on complex networks}

The focus of the previous chapters was on \tcp\ dynamics.  The model of network
topology was very basic, consisting merely of one buffer and one link.  All
details of the network topology were concentrated into a few parameters of the
link, namely the link delay, bandwidth and packet loss probability.  These
effective parameters could be tuned freely in the model.  However, we do not
know yet how these parameters should be adjusted in a complex network 
of thousands of nodes.

Since finite buffers naturally induce packet losses a long \tcp\ session
eventually achieves an equilibrium at a certain loss probability.  For a fixed
network configuration and a system of \tcp\ connections, therefore, packet loss
probabilities are determined by the steady state of network traffic.  The
steady state of the system is heavily influenced by the allocation of the
network resources, especially the link capacity.


In this chapter I study what the optimum distribution of link capacity is in
certain types of evolving networks when the local structure of the network is
known.  The motivation behind this problem is that the Internet is basically
being developed locally.  In my model I suppose that optimum link capacity is
proportional to the mean traffic demand of the particular link.  In a
homogeneous network the average traffic demand, in turn, is proportional to the
expected number of flows that utilize a particular link.  Since routing of
packets in computer networks can be supposed to be via the shortest path
between end nodes it follows that the distribution of shortest paths is a
matter of importance.

The main subject of my investigation is the ``\emph{betweenness}'' of links,
which is to say the number of shortest paths that pass over a link.  Note that
edge betweenness is essential not only in the case of the Internet, but in
other complex networks too.  For instance, edge betweenness can measure the
``importance'' of relationships in social networks or the probability of
discovering an edge during a network survey.  Until recently, however, less
attention has been paid to edge betweenness.

The probability distribution of edge betweenness gives a rough statistical
description of links and it characterizes the network as a whole.  Therefore,
it is an important tool for an overall description of links in complex
networks.  However, if the local structure of the network is known---as I
suppose in my model---then the probability distribution of edge betweenness
under the condition of the local property provides a much finer description of
links than the total distribution.  Therefore, I will aim at the
\emph{conditional} distribution of edge betweenness.

I restrict my model to trees, that is to connected loopless graphs. The
simplicity of trees allows analytic results for edge betweenness, since the
shortest paths in trees are unique between any pair of nodes. Although trees
are special graphs, a number of real networks can be modeled by trees or by
tree-like graphs with only a negligible number of shortcuts. Important examples
of such networks are the \glspl{as} in the Internet
\cite{CaldarelliMarchettiPietronero00}. 

As a model of evolving scale-free trees I consider the \gls{ba} model extended
with initial attractiveness \cite{Szymanski87,DorogovtsevMendesSamukhin00}.
The scaling properties of the network can be finely tuned with initial
attractiveness.  Note that in the limit of initial attractiveness to infinity
the network loses its scale-free nature and becomes similar to a classical
\gls{er} network with $p_{ER}=2/N$.  Therefore, scale-free and non-scale free
networks can be compared within one model. For the sake of simplicity the
infinite limit of initial attractiveness is referred to as the ``ER
limit'' hereafter.
 
\enlargethispage{2em}
The rest of this chapter is organized as follows.  Important results of the
literature concerning network modeling are collected in
Section~\ref{sec:Internet_topology_models}.  In Section~\ref{sec:model} a short
introduction to the construction of \gls{ba} trees is given.  Simulations of
large scale complex networks are presented in
Section~\ref{sec:Large_Scale_Simulations} to illustrate the importance of
optimum capacity distribution.  My results are presented in
Section~\ref{sec:Discussion_Betweenness}.  In particular, a master equation for
the joint distribution of cluster size and in-degree of a specific edge is
derived and solved in Section~\ref{subsec:cond_joint_prob} and
Section~\ref{subsec:sol_master}, respectively. The total joint distribution of
cluster size is calculated in Section~\ref{subsec:joint_prob}. The marginal and
conditional distributions of cluster size and in-degree are derived in
Section~\ref{subsec:marginal_prob} and Section~\ref{subsec:cond_prob},
respectively.  In Section~\ref{subsec:load_prob}, the conditional distribution
of edge betweenness follows. Finally, I summarize my work in
Section~\ref{sec:conclusions}.

\section{Preliminary results of topology modeling}
\label{sec:Internet_topology_models}

\enlargethispage{2em}

In the early \glslink[format=textbf]{er}{1960's Erd\H{o}s and R\'enyi}
introduced random graphs that served as the first  mathematical model of
complex networks \cite{ErdosRenyi60}.  In their model the number of nodes is
fixed and connections are established randomly.  In one variant of the ER model
every node pair is connected independently with probability
$p_{\mathrm{ER}}(N)$.  The probability depends on the size of the network in
such a way that the average degree of nodes is fixed: $\langle
k\rangle=p_{\mathrm{ER}} N=\mathrm{const}$.  It is obvious that the
distribution of the degree of any edge is binomial, which tends to Poissonian
distribution in the $N\to\infty$ limit.  Several interesting properties of the
ER model are well understood, including the relative size of the giant
component, the threshold of connectivity, etc.  Although the ER model leads to
rich theory, it fails to predict the power law distributions observed in
scale-free networks.

\subsection{The Barab\'asi--Albert model}
\glslink[format=(textbf]{ba}{Barab\'asi and Albert} proposed a more suitable
evolving model of scale-free networks
\cite{BarabasiAlbert99,BarabasiAlbertJeong99}.  The BA model is also
based on random graph theory, but it involves two key principles in addition:
\begin{inparaenum}[\itshape a\upshape)]
\item \emph{growth}, that is, the size of the network is increasing
  during development; and 
\item \emph{preferential attachment},
  that is, new network elements are connected to higher degree nodes with
  higher probability. 
\end{inparaenum}
In the original BA model every new node connects to the core network with a
fixed number of links $m$ and the probability of attachment is proportional to
the degree of nodes.  The above rules can be translated into the following
approximating fluid equation, which describes the time  evolution of the degree
of a particular vertex: $\partial k_i(t)/\partial t=k_i/2t$.  The solution
yields $k_i(t)=m\left(t/t_i\right)^{0.5}$, where $t_i$ is the time instant when
the $i$th vertex was added to the network.  The degree distribution can be
given, supposing that new nodes are added uniformly in time, by:
$\mathbb{P}\left[k_i(t)<k\right]=\mathbb{P}\left[t_i>m^2t/k^2\right]=1-m^2t/k^2\left(t+m_0\right)$,
where $m_0$ is the number of initial vertices.  The probability density can be
obtained from $\mathbb{P}(k)=\partial \mathbb{P}\left[k_i(t)<k\right]/\partial
k$.  The stationary solution finally gives
\begin{equation}
  \mathbb{P}(k)=\frac{2m^2}{k^3}.
  \label{eq:BA_model}
\end{equation}

The BA model explained successfully the observed scale--free nature of many
networks by the ``rich-gets-richer'' phenomenon.  However, the model was too
simple to fit most measured quantities of the real Internet.  For example, the
degree scaling-exponent in (\ref{eq:BA_model}) is $\delta_{\mathrm{BA}}=3$
which is in contrast with the exponent $\delta=2.15$--$2.2$ observed in
Internet measurements
\cite{FaloutsosFaloutsosFaloutsos99,VazquezPastorSatorrasVespignani02}.  The BA
model was later refined by a number of other authors.
\citet{DorogovtsevMendes00} studied the aging of nodes.  The authors extended
the BA preferential attachment rule so that attachment probability was
proportional not only to the degree, but also to $(t-t_i)^{-\nu}$, a power law
function of age, where $\nu$ is a tunable parameter.  It has been shown 
analytically and by simulation that the scale-free structure of the network
disappears if  $\nu>1$.  Moreover, an implicit equation was derived
between the scaling exponent of the degree distribution and $\nu$ for
$-\infty<\nu<1$.  The influence of exponentially fast aging on global and local
clustering, degree--degree correlation and the diameter of the network was
analyzed by \citet{ZhuWangZhu03}.

A continuum model was developed by \citet{AlbertBarabasi00} for the study of
the effect of edge rewiring and appearance of new internal edges.  In the
extended model three operations are incorporated: 
\begin{inparaenum}[\itshape a\upshape)]
\item $m$ new edges are created with probability $p$, \label{inpar:a}
\item $m$ existing edges are rewired with probability $q$; and \label{inpar:b}
\item a new node is connected to the network with $m$ new links with probability $1-p-q$. \label{inpar:c}
\end{inparaenum}
In every step a node is chosen randomly first if \textit{\ref{inpar:a}})
applies and a random link of this node is removed if
\textit{\ref{inpar:b}}) applies.  In case of \textit{\ref{inpar:c}}) the new
node is chosen.  Then a new link is established between the selected node and 
another one which selected with the following preferential attachment rule:
\begin{equation}
  \Pi(k_i)=\frac{k_i+1}{\sum_j\left(k_i+1\right)}.
\end{equation}
The above procedure is repeated $m$ times.  

The authors have observed a transition from a scale-free regime to an 
exponential regime in the $(p,q)$ phase space.  The transition takes place on
the line $q_t=\min\left[1-p,\left(1-p+m\right)/\left(1+2m\right)\right]$.  In
the scale-free regime, where $q<q_t$, the connectivity distribution has a 
generalized power-law form:
\begin{equation}
  \mathbb{P}(k)\propto \left(k+A(p,q,m)\right)^{-\gamma(p,q,m)},
\end{equation}
where $A(p,q,m)=\left(p-q\right)\frac{2m\left(1-q\right)}{1-p-q}+1+p-q$ and
$\gamma(p,q,m)=3-2q+\frac{1-p-q}m$.  In the limit $p=q=0$ the model reduces to
the scale-free model investigated in \cite{BarabasiAlbert99}.  It can be seen
that the scaling exponent $\gamma$ changes continuously with $p$, $q$, and $m$
in the range of $2$ to $\infty$.

The classic BA model has been extended with initial attractiveness by
\citet{DorogovtsevMendesSamukhin00}.  More specifically, the probability that a
new node is connected to a given site is proportional to $A_i=A+q_i$, where
$A\ge0$ is called the \emph{initial attractiveness} and $q_i$ is the in-degree
of node $i$.  The probability distribution of the connectivity, 
\begin{equation}
  \mathbb{P}(q)=\left(1+a\right)\frac{\Gamma\left[\left(m+1\right)a+1\right]}
  {\Gamma(ma)} \frac{\Gamma(q+ma)}{\Gamma\left[q+2+\left(m+1\right)a\right]},
  \label{eq:BA_initial_attr}
\end{equation}
was derived from a Master-equation approach where $a=A/m$ and $m$ is the
number of links starting from every new node, as in the BA model.  
In the special case $a=1$ the model reproduces the original BA model with 
$A_i=k_i=q_i+m$ and the solution (\ref{eq:BA_initial_attr}) reduces to
\begin{equation}
  \mathbb{P}(k)=\frac{2m\left(m+1\right)}{k \left(k+1\right)\left(k+2\right)}.
\end{equation}
Compare this result with (\ref{eq:BA_model}), which comes from a fluid 
approach.  The two expressions converge in the $k\to\infty$ limit, but the 
constant factors are different.  For $ma+q\gg1$ the expression 
(\ref{eq:BA_initial_attr}) takes the form 
$\mathbb{P}(q)\propto\left(q+ma\right)^{-\left(2+a\right)}$, that is the
scaling exponent $\gamma=2+a$ can be tuned in the range of $2$ to $\infty$, 
similarly to the previous model.

The time evolution of the average connectivity has also been derived.  It has
been found that 
$\bar{q}(t,t_i)\propto\left(t_i/t\right)^{-1/\left(1+a\right)}$ for $t\gg t_i$.
The scaling exponent of the average connectivity of an old node is therefore
$\beta=1/\left(1+a\right)$.  It follows that scaling exponents $\gamma$ and 
$\beta$ satisfy the following scaling relation:
\begin{equation}
  \beta\left(\gamma-1\right)=1.
  \label{eq:scaling_relation}
\end{equation}
The authors have shown that (\ref{eq:scaling_relation}) is universal, since the
above scaling relation can be derived in the case of more general conditions.

Growing random networks with non-linear attractiveness have been studied in
\cite{KrapivskyRednerLeyvraz00,KrapivskyRedner01}.  It has been found that
scale-free connectivity distribution can be observed only if the attractiveness
kernel is asymptotically linear.  The authors confirmed the above findings
indicating that the scaling exponent depends on the details of the attachment
probability and can be tuned in the range of $2$ and $\infty$.  Furthermore,
the authors showed that if the attractiveness is sub-linear then the
connectivity distribution decays at an exponential rate, while if the kernel
grows more quickly than linearly then almost all nodes are connected to a
single node\glsadd[format=)]{ba}.

\subsection{Other network models}

Other mechanisms have been proposed for the formation of scale-free
networks.  \Citet{EvansSaramaki05} studied the following simple algorithm:
new vertices are connected to the end of one or more $l$-length random walk
processes.  Several variations for this general algorithm have been considered:
fixed or variable length random walks, a fixed or random number of connecting
edges, different distributions for the starting vertex of the random walk
process, edge- or vertex-wise restart of random walks, and uniform or weighted
random walks on the graph.  The authors argued that a random walk process
is a more realistic mechanism than preferential attachment, since the
random walk uses only the local properties of a network.
 
\Citet{GohKahngKim02} proposed the following stochastic model for the
evolution of Internet topology: the size of the network increased
exponentially, $N(t)=N(t_0)e^{\alpha t}$ and the connectivity of each node is
changed according to the random process
\begin{equation}
  k_i(t+1)=k_i(t)\left[1+g_{0,i}+\xi_i(t)\right],
  \label{eq:stoch_degree}
\end{equation}
where $g_{0,i}$ are constants and $\xi_i(t)$ are assumed to be independent
white noise processes representing fluctuations with mean zero and 
correlation function $\left\langle\xi_i(t)\xi_j(t')\right\rangle
=\sigma_{0,j}^2\delta(t-t')\delta_{i,j}$.  The authors 
showed that in a homogeneous case, that is when $g_0=g_{i,0}$ and 
$\sigma_0=\sigma_{i,0}$, the connectivity distribution of the network 
approximately follows a power law with exponent
\begin{equation}
  \gamma=1-\frac{g_{\mathrm{eff}}}{\sigma_{\mathrm{eff}}^2}+
  \frac{\sqrt{g_{\mathrm{eff}}^2+2\alpha\sigma_{\mathrm{eff}}^2}}
  {\sigma_{\mathrm{eff}}^2},
\end{equation}
where $g_{\mathrm{eff}}\approx g_0-\sigma_0^2/2$,
$\sigma_{\mathrm{eff}}^2\approx\sigma_0^2$.  Links are removed randomly when
the degree $k_i$ decreases and internal edges are created according to a
preferential attachment rule when the degree $k_i$ increases.  The model
includes an adaptation mechanism in which links are only rewired to nodes with
larger connectivity.  The parameters of the model have been fitted to real
\gls{as} level Internet topology.  The authors have demonstrated that their
model fits the degree-degree correlation and clustering coefficient of the real
Internet better than previous models.

In a paper by \citet{LiAldersonWillingerDoyle04} the authors argued that the
technological constraints of router design should be considered as the driving
force behind the development of the Internet.  They pointed out that the
possible bandwidth--degree combinations are restricted to a technologically
feasible region for every router.  In particular, large bandwidth links are
connected to low degree routers and as the degree increases router capacity
must be fragmented among more and more links.  A heuristic degree-preserving
rewiring algorithm has been proposed by the authors in order to take the above
technology constraint into consideration: a small number of low degree nodes
are chosen to serve as core routers first, and other high degree nodes hanging
from the core routers are selected as access routers next.  Finally, the
connections among gateway routers are adjusted in such a way that their
aggregate bandwidth to core nodes becomes almost uniform.  The resulting
\emph{Heuristically Optimal Topology (HOT)} has been compared with other
commonly used topology generators, e.g. \gls{ba} preferential attachment
network, and general random graph model.  Performance metrics and random
graph-based likelihood metrics have been defined to compare different
topologies, which are the realizations of the same degree distribution.  It has
been shown that the overall network performance of the HOT topology surpasses
the performance of other random networks.  At the same time, the ``designed''
HOT topology is very unlikely to be obtained from random graph models,
according to the defined likelihood metric.  The authors concluded that their
first-principles approach combined with engineered design should replace random
topology generators in the future.

\subsection{Earlier results regarding betweenness}

Node betweenness has been studied recently by \citet{GohKahngKim01} who argued
that it follows power law in scale-free networks, and the exponent
$\delta\approx2.2$ is independent from the exponent of the degree distribution
as long as the degree exponent is in the range $2<\gamma\le3$.  The authors
analyzed both static and evolving  networks, directed and undirected graphs as
well as a real network of collaborators in neuroscience.  Their conjecture is
based on numerical experiments.  However, \citet{Barthelemy03} presented
counter-examples to the universal behavior and demonstrated that the important
exponent is the scaling exponent of betweenness as the function of connectivity
$\eta=\left(\delta-1\right)\left(\gamma-1\right)$ instead.  In a reply
\cite{GohGhimKahngKim03} the authors argued that universality is still valid
for a restricted class of tree-like, sparse networks.  

\citet{SzaboAlavaKertesz02} used rooted deterministic trees to model scale-free
trees.  The authors have modeled \gls{ba} networks with a uniform branching process
in a mean-field approximation.  They obtained that the branching process is
$b(l)=\frac12\frac{\ln N}l$ on a layer at distance $l>0$.  The number of nodes
$n(l)$ at distance $l$ was approximated by a non-normalized Gaussian.  It has
been found that the number of shortest paths going through a node at distance
$l$ from the root node is $L(l)=\frac{\mathrm{const}}{n(l)}$, independent of
the branching process $b(l)$.  Finally, the authors showed that node
betweenness, which includes shortest paths originating to and from nodes in
excess of edge load, follows a power-law decay with a universal exponent of
$-2$.  The same scaling exponent has been found experimentally by
\citet{GohKahngKim01} for scale-free trees.  

A rigorous proof of the heuristic results of \cite{SzaboAlavaKertesz02} has
been presented by \citet{BollobasRiordan04}.  The authors showed that the 
number of shortest paths through a random vertex is
\begin{equation}
  \PP(L=l)=\frac{2N-1}{\left(2l+1\right)\left(2l+3\right)},
\end{equation}
where $N$ is the size of the network and $l\in\mathbb{N}$.  Furthermore, 
the distribution of the length of the shortest paths has been precisely 
calculated.  The asymptotic limit of the distribution was proved to be normal 
with mean and variance increasing as $\log N$. 

\section{The network model}
\label{sec:model}

The concepts of graph theory are used throughout my analysis, so I will define
briefly the terminology I use first.  A graph consists of \emph{vertices}
(nodes) and \emph{edges} (links). Edges are ordered or un-ordered pairs of
vertices, depending on whether an ordered or un-ordered graph is considered,
respectively. The \emph{order} of a graph is the number of vertices it holds,
while the \emph{degree} of a vertex counts the number of edges adjacent to it.
\emph{Path} is also defined in the most natural way: it is a vertex sequence,
in which any two consecutive elements form an edge.  A path is called a
\emph{simple path} if none of the vertices in the path are repeated. Any two
vertices in a \emph{tree} can be connected by a unique simple path.  The graph
is called connected if for any vertex pair there exists a path which starts
from one vertex and ends at the other.

The construction of the network proceeds in discrete time steps. Let us denote
time with $\tau\in\NN$, and the developed graph with
$G_{\tau}=\left(V_{\tau},E_{\tau}\right)$, where $V_{\tau}$ and $E_{\tau}$
denote the set of vertices and the set of edges at time step $\tau$,
respectively.  Initially, at $\tau=0$, the graph consists only of a single
vertex without any edges.  Then, in every time step, a new vertex is connected
to the network with a single edge. The edge is \emph{directed}, which
emphasizes that the two sides of the edge are not symmetric. The newly
connected node, which is the source of the edge, is always ``younger'' than the
target node.  The term ``younger node of a link'' is used in this sense below.
Note that the initial vertex is different from all the others, since it has
only incoming connections; I refer to it as the \emph{root vertex}. 

The target of every new edge is selected randomly from the present vertices of
the graph. The probability that a new vertex connects to an old one is
proportional to the attractiveness of the old vertex $v$, defined as
\begin{equation}
  \Att(v)=a+q,
\end{equation}
where parameter $a>0$ denotes the initial attractiveness and $q$ is the
in-degree of vertex $v$. { It has been shown in
\cite{DorogovtsevMendesSamukhin00} that the in-degree distribution is
asymptotically
$\PP(q)\simeq\left(1+a\right)\frac{\Gamma(2a+1)}{\Gamma(a)}\left(q+a\right)^{-\left(2+a\right)}$.
I will improve this result and derive the exact in-degree distribution below.}
Note that in the special case $a=0$ the attractiveness of every node is zero
except of the root vertex. It follows that every new vertex is connected to the
initial vertex in this case, which corresponds to a star topology. The special
case $a=1$ practically returns the original BA model.  Indeed, except for the
root vertex, the attractiveness of every vertex becomes equal to its degree if
$a=1$; this is exactly the definition of the attractiveness in the \gls{ba} model
\cite{BarabasiAlbert99}. Finally, if $a\to\infty$, then preferential attachment
disappears in the limit, and the model tends to a Poisson-type graph, similar
to an \gls{er} graph.

The attractiveness of sub-graph $S$ is the sum of the attractiveness
of its elements:
\begin{equation}
  \Att(S)=\sum_{v'\in S}\Att(v').
\end{equation}
I refer to a connected sub-graph as a \emph{cluster}. The attractiveness of
cluster $C$ can be given easily:
\begin{equation}
  \Att(C)=
  \left(1+a\right)|C|-1,
\end{equation}
where $|C|$ denotes the size of the cluster. It is obvious that the
overall attractiveness of the network at time step $\tau$ is
\begin{equation}
  \Att(V_{\tau})=\left(1+a\right)\left(\tau+1\right)-1.
\end{equation}

\section{Simulation of large computer networks}
\label{sec:Large_Scale_Simulations}

Before I continue with the analytic study of betweenness I would like to
illustrate the effects of different capacity allocation strategies in large
computer networks and demonstrate the importance of finding an optimum
strategy.  To this end I carried out large scale computer simulations.  
Since packet level simulations of large networks are practically impossible,
because of their huge computational requirements, I implemented a fluid
model of the network traffic based on the AIMD model, introduced below.

\subsection{The AIMD model}

\citet{BaccelliHong02} have developed the \gls[format=textbf]{aimd} model for
$N$ parallel \tcp\ flows utilizing a common bottleneck buffer. The
synchronization of the \tcp\ flows could be tuned in the range of complete
synchronization and complete randomness.  The acronym AIMD stands for
\emph{additive increase, multiplicative decrease}.  The name refers to the
basic governing principle behind \tcp\ congestion avoidance algorithm, and it
emphasizes that the details of the slow start and the \gls{frfr} algorithms are
neglected in the model.  Let $T_n$ denote the $n$th congestion time,
$\tau_{n+1}=T_{n+1}-T_n$ the elapsed time between two consecutive congestion
events, and $X_n^{(i)}$ the throughput of $i$th flow \emph{after} the $n$th
congestion event.  If instantaneous throughput is approximated by its average,
then the throughput can be related to the congestion window $W_n^{(i)}$ by the
following equation: $X_n^{(i)}=W_n^{(i)} P/R^{(i)}$, where $R^{(i)}$ is the
\rtt\ of the $i$th \tcp\ flow, and $P$ is the size of the data packets, as
above.  The evolution of the throughput can be given by \begin{equation}
\label{eq:AIMD_evolution} X_{n+1}^{(i)}
=\left[\left(1-\xi_{n+1}^{(i)}\right)+\beta^{(i)} \xi_{n+1}^{(i)}\right]
\left(X_n^{(i)}+\frac{\alpha^{(i)}P}{R^{(i)}}\tau_{n+1}\right), \end{equation}
where $\alpha^{(i)}$ and $\beta^{(i)}$ are the linear growth rate and the
multiplicative decrease factor of the congestion window, respectively, and
$\xi_n^{(i)}$ are random variables, independent in $n$, which take the value
$1$ if the $i$th \tcp\ flow experiences packet loss at the $n$th congestion
event, and $0$ otherwise.  Congestion occurs at the bottleneck, supposing
negligible or zero buffer capacity, when the total throughput reaches the
capacity of the bottleneck link $C$.  Accordingly, $\tau_{n+1}$ can be
calculated from following fluid equation: \begin{equation} \sum_{i=1}^N
\left(X_n^{(i)}+\frac{\alpha^{(i)}P}{R^{(i)}}\tau_{n+1}\right)=C.
\label{eq:AIMD_congestion} \end{equation}

The variable $\tau_{n+1}$ can be eliminated from (\ref{eq:AIMD_evolution}) and 
(\ref{eq:AIMD_congestion}), which leads to
\begin{equation}
  X_{n+1}^{(i)}=\gamma_{n+1}^{(i)}
  \left(\rho^{(i)} C + X_n^{(i)}
  -\rho^{(i)}\sum_{j=1}^N X_n^{(j)}\right),
  \label{eq:AIMD_full}
\end{equation}
where $\gamma_{n}^{(i)}=\left(1-\xi_n^{(i)}\right)+\beta^{(i)}\xi_n^{(i)}$ and 
$\rho^{(i)}=\frac{\alpha^{(i)}/R^{(i)}}{\sum_{j=1}^{N}\alpha^{(j)}/R^{(j)}}$.
The system of recursive equations (\ref{eq:AIMD_full}) can also be given in 
a simpler stochastic matrix form:
\begin{equation}
  \mathbf{X}_{n+1}=\mathbf{A}_{n+1} \cdot \mathbf{X}_n + \mathbf{B}_{n+1},
\end{equation}
where $\left(\mathbf{B}_n\right)_i=\gamma_n^{(i)}\rho^{(i)}C$ and
$\left(\mathbf{A}_n\right)_{ij}
=\gamma_n^{(i)}\left(\delta_{ij}-\rho^{(i)}\right)$, and $\delta_{ij}$ is 
the Kronecker-delta symbol.

The interaction of the competing flows is taken into account by the
synchronization rate, $r_n^{(i)}=\mathbb{E}\left[\xi_n^{(i)}\right]$.  Note
that $\xi_n^{(i)}$ are not independent at a given $n$ for $i=1\dots N$, since
at a congestion event a minimum of one \tcp\ flow must experience packet loss.  
If $\xi_n^{(i)}$ are generated independently with 
$P\left(\xi_n^{(i)}=1\right)=\pi_n^{(i)}$, but those realizations are discarded
where $\sum_{i=1}^N\xi_n^{(i)}=0$, then the synchronization rate can be 
expressed with the following conditional probability:
\begin{equation}
  r_n^{(i)}=P\left(\xi_n^{(i)}=1\Biggm|\sum_{i=1}^N\xi_n^{(i)}\neq0\right)
  =\frac{\pi_n^{(i)}}{1-\prod_{j=1}^N\left(1-\pi_n^{(i)}\right)}.
\end{equation}
For the special case $N=1$, for example, it is evident that $r_n\equiv1$.

Let us consider a simple homogeneous situation, where $\alpha^{(i)}=\alpha$,
$\beta^{(i)}=\beta$, $R^{(i)}=R$, and $r_n^{(i)}=r_n$.  It is obvious that
$\rho^{(i)}=1/N$ in this case.  Moreover, it can easily be shown that the 
expectation of the steady state throughput is
$\mathbb{E}\left[\mathbf{X_{\infty}}\right]
=\mathbb{E}\left[\mathbf{B_{\infty}}\right]$, that is
\begin{equation}
  \mathbb{E}\left[X^{(i)}\right]=\mathbb{E}\left[\gamma\right]\frac{C}{N}
  =\left[1-\left(1-\beta\right)r\right]\frac{C}{N}
\end{equation}
for all $i$.  The above formula predicts the degradation of the throughput
as the synchronization grows. This is in good agreement with simulations and
measurements.  The expected time between consecutive congestion events can
also be obtained:
\begin{equation}
  \mathbb{E}\left[\tau\right]=\left(1-\beta\right)\frac{CRr}{\alpha NP}.
\end{equation}

The ``$1/\sqrt{p}$'' formula can be derived from the extended \gls{aimd} 
model as well.  The functional form of the formula is
\begin{equation}
  X=\frac{P}{R}\frac{\sqrt{2f(r,N)}}{\sqrt{p}}.
\end{equation}
The precise form of $f(r,N)$ is rather complicated.  However, it has been shown
that $\lim_{N\to\infty}f(r,N)=1-r/4$.  This implies that for large $N$ the 
constant factor $c_0=\sqrt{2f(r,N)}$ varies in the range 
$[\sqrt{3/2},\sqrt{2}]$ with the synchronization rate.

The authors have presented a wavelet and an auto-correlation analysis for
traces of the \gls{aimd} model for a large number of \tcp\ connections.  They
concluded that the trajectory of the aggregated throughput shows multi-fractal
scaling properties on short time scales and the wavelet and auto-correlation
methods give consistent fractal dimensions.

The single-buffer \gls{aimd} model can be generalized straightforwardly for
numerical simulations of more complex networks.  One only needs to apply
(\ref{eq:AIMD_congestion}) for each link and find the minimum of possible
congestion events:
\begin{equation}
  \tau_{n+1}=\min_{e\in E}\frac{C_e-\sum_{i\in I_e}X_n^{(i)}}
  {\sum_{i\in I_e}\frac{\alpha^{(i)}P}{R^{(i)}}},
\end{equation}
where $I_e$ denotes the set of flows which utilizing link $e\in E$.  The flows
of the congested buffer are handled the same way as in the original single
buffer \gls{aimd} model.  The remaining flows in the network develop undisturbed
until the next possible congestion event.

\subsection{Performance of different bandwidth distribution strategies}
\label{subsec:Bandwidth_Strategies}

In this section different bandwidth distribution scenarios are compared using a
fluid simulator based on the above \gls{aimd} model. The underlying network
topology is the same in all scenarios: a scale-free network generated according
to the extended \gls{ba} model introduced in Section~\ref{sec:model}.  The parameter,
which controls the number of new links in the model, is set to $m=1$, that is
the resulting network is a tree.  The scaling parameter is set to $a=1$ for
numerical purposes.  In simulations link capacities are normalized in such a
way that the average capacity is the same in all scenarios. 

\begin{table}
  \caption{Link capacity and performance in case of different strategies.  The 
  assigned capacity is proportional to the quantity displayed in the second 
  column, where $q_A$, $q_B$ denote the in-degrees of the nodes which compose 
  a particular link, and $L_e$ denotes edge betweenness.}
  \label{tab:strategies}
  \begin{center}
    \begin{tabular}{|l|l|l|}
      \hline
      Strategy & $C_e$ & $\Perf [b/s]$\\
      \hline\hline
      Uniform & $\propto1$ & 740.79\\
      Maximum & $\propto\max(q_A,q_B)$ & 2391.94\\
      Minimum & $\propto\min(q_A,q_B)$ & 6574.69\\
      Product & $\propto q_A\cdot q_B$ & 5279.5\\
      Mean field & $\propto L_e$ & 11284.6\\
      \hline
    \end{tabular}
  \end{center}
\end{table}
The rules of different strategies are presented in Table~\ref{tab:strategies}.
The uniform scenario, when the capacity is the same for every link, is regarded
as a reference. It can be considered the worst case scenario, when no
information is available about the details of the network. On the contrary, the
mean field strategy---when the link capacity is proportional to the edge
betweenness---is a global optimum.  Minimum, maximum and product strategies are
a couple of naive attempts to take the local structure of the network into
account.  Note that only one global information the normalizing factor for the
average capacity is required.  In the later three cases the more connection a
link possesses, the more capacity is allocated for the particular link.  The
difference between the three strategies is whether they prefer loosely,
moderately or highly connected links, compared with the mean field allocation
strategy.

\begin{figure}[tb]
  \begin{center}
    \psfrag{load}[r][r]{mean field}
    \psfrag{min}[r][r]{minimum}
    \psfrag{max}[r][r]{maximum}
    \psfrag{uni}[r][r]{uniform}
    \psfrag{prod}[r][r]{product}
    \psfrag{Link Capacity, C [b/s]}[c][c]{Link Capacity, $C [b/s]$}
    \psfrag{Complementary CDF, Fc(C)}[c][c]{Complementary CDF, $\bar{F}(C)$}
    \psfrag{ 1e+08}[l][l]{$10^8$}
    \psfrag{ 1e+07}[l][l]{$10^7$}
    \psfrag{ 1e+06}[l][l]{$10^6$}
    \psfrag{ 100000}[c][c]{$10^5$}
    \psfrag{ 10000}[c][c]{$10^4$}
    \psfrag{ 1000}[c][c]{$10^3$}
    \psfrag{ 100}[r][r]{$10^2$}
    \psfrag{ 10}[r][r]{$10$}
    \psfrag{ 1}[r][r]{$1$}
    \psfrag{ 0.1}[r][r]{$10^{-1}$}
    \psfrag{ 0.01}[r][r]{$10^{-2}$}
    \psfrag{ 0.001}[r][r]{$10^{-3}$}
    \psfrag{ 1e-04}[r][r]{$10^{-4}$}
    \psfrag{ 1e-05}[r][r]{$10^{-5}$}
    \resizebox{0.7\textwidth}{!}{\includegraphics{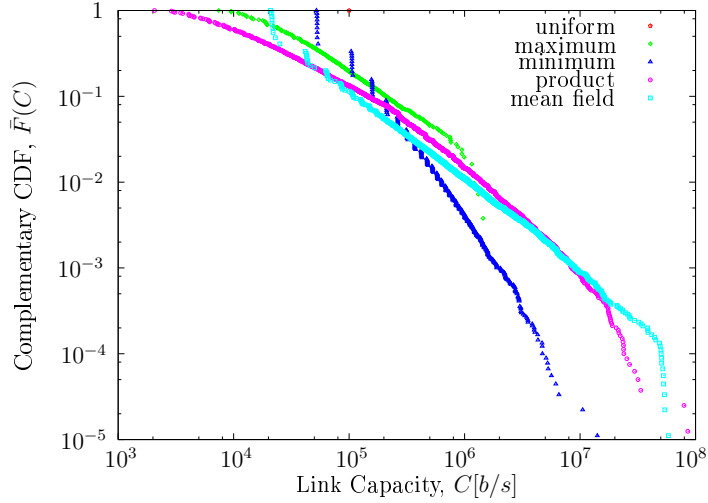}}
    \caption{Comparison of the complementary CDF of link capacity is shown for
    different bandwidth distribution strategies on log-log plot. Data is
    obtained from 10 realizations of $N=10^4$ node networks.  Average capacity 
    is set to $\bar{C}=10^5 [b/s]$ for every network. The following scenarios 
    are considered: uniform (pentagons), maximum (diamonds), minimum 
    (triangles), product (circles), and mean field (squares).}
    \label{fig:bandwidth-dist}
  \end{center}
\end{figure}

The capacity range that different strategies are more likely to prefer can be
easily determined by the complementary distribution of capacities, shown in
Fig.~\ref{fig:bandwidth-dist}.  The average capacity is set to $\left\langle
C\right\rangle=10^5$[b/s] for all cases.  The distribution of the uniform
strategy is clearly degenerated since only one capacity value is possible in
this scenario.  The maximum strategy prefers the lower bandwidths at the cost
of a cutoff at about $10^6 b/s$ capacity. The minimum strategy also prefers
lower bandwidths at the cost of high bandwidths, but no cutoff exists. The
complementary distribution of minimum strategy resembles the mean field
distribution with a different scaling exponent. The product strategy prefers
the mid-range of bandwidth, and it underestimates both the low and the high
capacity range, compared to the mean field strategy.

In order to compare different strategies one needs an ordering between them.  
Performance, the average throughput of \tcp s, provides a natural ordering 
between different strategies.  Let us define the performance of individual 
\tcp s first as the time average of their throughput $X^{(i)}(t)$:
\begin{equation}
  \Perf^{(i)}=\left\langle X^{(i)}(t)\right\rangle_t=
   \lim_{t\to\infty}\frac1{t}\int_0^t X^{(i)}(u)\,du.
\end{equation}
The global performance of a strategy is then the mean performance of the 
\tcp s operating in the network
\begin{equation}
  \Perf=\frac1{N_{\textrm{\tcp}}}\sum_{i=1}^{N_{\textrm{\tcp}}}\Perf^{(i)}.
\end{equation}

The locations of \gls{tcp} sources and destinations are distributed homogeneously
in my numerical simulations.  The length of a simulation is such that every 
\gls{tcp} connection experiences $100$ congestion epochs on average.  Network
performances obtained from simulations are shown in Table~\ref{tab:strategies}
for the different bandwidth distribution strategies.  The table shows that mean
field bandwidth allocation strategy is almost twice as effective as the
second, ``minimum strategy'', and it is more than twice as good as the
``product strategy''. The performance of a network with maximum bandwidth
distribution strategy is about one fifth the performance of the same
network when mean field strategy is used.  Moreover, the performance of uniform
scenario is even less then one third of the second worst, ``maximum strategy''.

\begin{figure}[tb] 
  \begin{center}
    \psfrag{load}[r][r]{mean field} 
    \psfrag{min}[r][r]{minimum}
    \psfrag{max}[r][r]{maximum} 
    \psfrag{uni}[r][r]{uniform}
    \psfrag{prod}[r][r]{product} 
    \psfrag{TCP Transmission Rate, T [b/s]}[c][c]{TCP performance, $\Perf^{(i)} [b/s]$} 
    \psfrag{F(T)}[c][c]{CDF, $F(\Perf^{(i)})$} 
    \psfrag{ 1e+06}[l][l]{$10^6$} 
    \psfrag{ 100000}[c][c]{$10^5$} 
    \psfrag{ 10000}[c][c]{$10^4$} 
    \psfrag{ 1000}[c][c]{$10^3$} 
    \psfrag{ 100}[r][r]{$10^2$}
    \psfrag{ 10}[r][r]{$10$} 
    \psfrag{ 1}[r][r]{$1$} 
    \psfrag{ 0}[r][r]{$0$} 
    \psfrag{ 0.2}[r][r]{$0.2$}
    \psfrag{ 0.4}[r][r]{$0.4$} 
    \psfrag{ 0.6}[r][r]{$0.6$} 
    \psfrag{ 0.8}[r][r]{$0.8$}
    \resizebox{0.7\textwidth}{!}{\includegraphics{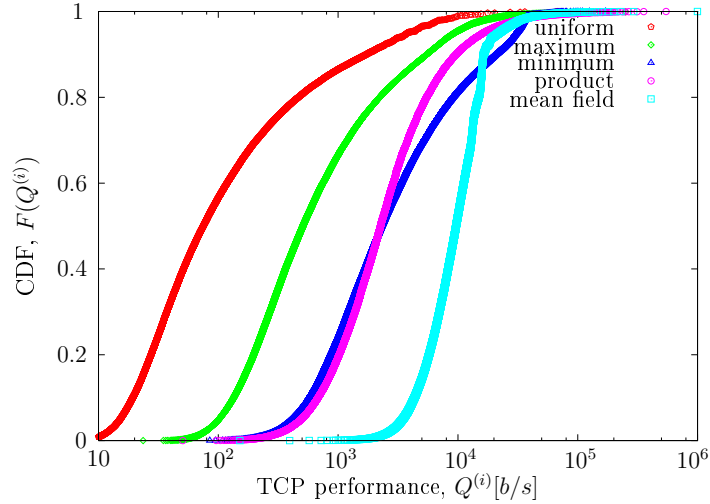}}
    \caption{Comparison of the CDF of \tcp\ performance is shown for different
    bandwidth distribution strategies on normal-log plot. Data is obtained from 
    10 realizations of $N=10^4$ node networks with scaling parameter 
    $\alpha=1/2$.  Average capacity is set to $\bar{C}=10^5 [b/s]$ for every 
    network. Simulation lasted for $100N$ congestion epochs.  The following 
    scenarios are considered: uniform (pentagons), maximum (diamonds), 
    minimum (triangles), product (circles), and mean field (squares).} 
    \label{fig:bandwidth-strategies}
  \end{center} 
\end{figure}

A more detailed picture can be gotten from the distribution of \gls{tcp}-wise
performance $F(\Perf^{(i)})$.  Simulation results of the \gls{cdf} of \tcp\
performance are shown in Figure~\ref{fig:bandwidth-strategies} for the above
mentioned bandwidth allocation strategies. The performance of mean field
strategy is clearly the best.  The bulk of the distribution is concentrated to
a relatively narrow performance interval, that is most of \glspl{tcp} can operate at
almost the same, high performance level.  The performance distribution of the
next two best performing strategies, the minimum and the product, is very
similar below their median.  Above the median the minimum strategy performs
better even though large capacity links are preferred less than the product
strategy.  It follows that the whole bandwidth range must be taken into
consideration in any bandwidth distribution strategy to reach the optimum
network performance. The performance of the maximum strategy is considerably 
worse than the previous two, mainly due to the sharp cutoff in the
capacity distribution. Finally, the uniform bandwidth distribution is the worst
of all: its performance is just a few percent of the mean field scenario's
performance. The network where this strategy is applied is heavily congested,
since the bottlenecks form in the core of the network.  

In summary, the selection of inadequate bandwidth allocation strategy can
degrade the overall performance of the network considerably.  In the following
sections I discuss analytically how additional local information could be used
to allocate capacity to links properly.  Beforehand, I introduce the
network model investigated.

\section{Discussion}
\label{sec:Discussion_Betweenness}

It is my aim to derive the probability distribution distribution of edge
betweenness in evolving scale-free trees, under the condition that the
in-degree of the ``younger'' node of any randomly selected link is known.  For
the sake of simplicity I consider the in-degree of the ``younger'' node only.
Whether a node is ``younger'' than another node or not can be defined uniquely
in evolving networks, since nodes attach to the network sequentially. Note that
the in-degree is considered instead of total degree for practical reasons only.
The construction of the network implies that the in-degree is less than the
total degree by one for every ``younger'' node.

To obtain the desired conditional distribution I calculate the exact joint
distribution of cluster size and in-degree for a \emph{specific} link first.
Then, the joint distribution of a \emph{randomly selected} link is derived,
which is comparable with the edge ensemble statistics obtained from a network
realization. The exact marginal distributions of cluster size and in-degree
follow next. After that, I give the distribution and mean of cluster size
under the condition that in-degree is known. For the sake of completeness the
conditional in-degree distribution is presented as well. Finally, the
distribution and mean of edge betweenness is derived under the condition that
the corresponding in-degree is known. Note that all of my analytic results are
\emph{exact even for finite networks}, which is valuable since the 
real networks are often much smaller than the valid range of asymptotic
formulas. Moreover, \emph{exact results for unbounded networks} are provided as
well. 

\subsection[Master equation for cluster size and in-degree]{Master equation for the joint distribution of cluster size and in-degree}
\label{subsec:cond_joint_prob}

\begin{figure}
  \begin{center}
    \psfrag{C}[c][c][2]{$C$}
    \psfrag{e}[c][c][2]{$e$}
    \psfrag{v}[c][c][2]{$v$}
    \psfrag{Root}[c][c][2]{Root}
    \resizebox{0.7\figwidth}{!}{\includegraphics{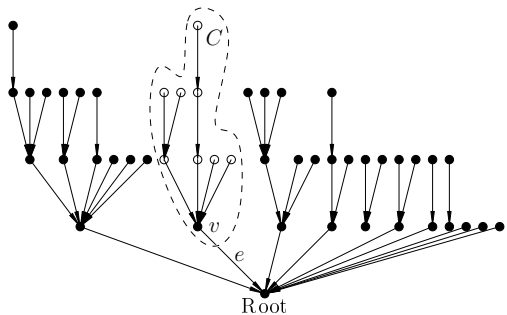}}
    \caption{Schematic illustration of the evolving network at time $\tau$. 
      Vertex $v$, connected to the network at $\tau_e$, denotes the root
      of cluster $C$. Variables $q$ and $n=|C|-1$ denote the in-degree of vertex $v$ and 
      the number of nodes in $C$ without $v$ (marked by circles), respectively.}
    \label{fig:BAmodel}
  \end{center}
\end{figure}

Let us consider the size of the network $N$, an arbitrary edge $e$, which
connected vertex $v$ to the graph at time step $\tau_e>0$, and let us denote by
$C$ the cluster that has developed on vertex $v$ until $\tau>\tau_e$
(Fig.~\ref{fig:BAmodel}).  The calculation of betweenness of the given
edge is straightforward in trees, since the number of shortest paths going
through the given edge, that is the betweenness of the edge, is obviously
$L=|C|\left(N-|C|\right)$. Therefore, it is sufficient to know the size of the
cluster on the particular edge to obtain edge betweenness.  

The development of cluster $C$ can be regarded as a Markov process. The states
of the cluster are indexed by $\left(n,q\right)$, where $n=|C|-1$ denotes the
number of vertices in cluster $C$ without $v$. The in-degree of vertex $v$ is
denoted by $q$. Transition probabilities can be obtained from the definition of
preferential attachment: 
\begin{align}
  \transprob{\tau,n,q}&=\frac{\Att\left(C_{\tau}\setminus v\right)}{\Att\left(V_{\tau}\right)}
  =\frac{n-\alpha q}{\tau+1-\alpha}\\
  \transprobq{\tau,q}&=\frac{\Att\left(v\right)}{\Att\left(V_{\tau}\right)}
  =\frac{\alpha q+1-\alpha}{\tau+1-\alpha},
\end{align}
where $\alpha=1/\left(1+a\right)\in\left]0,1\right]$ and 
$\transprob{\tau,n,q}$ denotes the transition probability 
$\left(n,q\right)\to\left(n+1,q\right)$, and
$\transprobq{\tau,q}$ denotes the transition probability 
$\left(n,q\right)\to\left(n+1,q+1\right)$, respectively. 

The Master-equation, which describes the Markov process, follows from the fact 
that cluster $C$ can develop to state $\left(n,q\right)$ obviously 
in three ways: a new vertex can be connected 
\begin{enumerate}
\item to cluster $C$ but not to vertex $v$, and the cluster was in state $\left(n-1,q\right)$,
\item to vertex $v$, and the cluster was in state $\left(n-1,q-1\right)$, or
\item to the rest of the network, and the cluster was in state $\left(n,q\right)$.
\end{enumerate}
Therefore, the conditional probability $\probi{\tau}{n,q}$ that the developed 
cluster on edge $e$ is in state $\left(n, q\right)$ satisfies the following 
Master-equation: 
\begin{align}
  \label{eq:master_eq}
  \probi{\tau}{n,q}
  &=\transprob{\tau-1,n-1,q}\,\probi{\tau-1}{n-1,q}\notag\\
  &+\transprobq{\tau-1,q-1}\probi{\tau-1}{n-1,q-1}\notag\\
  &+\left[1-\transprob{\tau-1,n,q}-\transprobq{\tau-1,q}\right]\probi{\tau-1}{n,q},
\end{align}
Since the process starts with $n=0$, $q=0$ at $\tau=\tau_e$, 
the initial condition of the above Master equation is
$\probi{\tau_e}{n,q}=\delta_{n,0}\delta_{q,0}$, where $\delta_{i,j}$ is the
Kronecker-delta symbol. 

\subsection{The solution of the master equation}
\label{subsec:sol_master}

After substituting the above transition probabilities into \eqref{eq:master_eq},
the following first order linear partial difference equation is obtained:
\begin{align}
  \notag
  \left(\tau-\alpha\right)\,\probi{\tau}{n,q}
  &=\left(n-1-\alpha q\right)\probi{\tau-1}{n-1,q}\\
  \notag
  &+\left(\alpha q+1-2\alpha\right)\probi{\tau-1}{n-1,q-1}\\
  \label{eq:master_eq_re}
  &+\left(\tau-n-1\right)\probi{\tau-1}{n,q},
\end{align} 

Let us seek a particular solution of \eqref{eq:master_eq_re}
in product form: $f(\tau)\,g(n)\,h(q)$. The following equation is obtained 
after substituting the probe function into \eqref{eq:master_eq_re}:
\begin{align}
	\left(\tau-\alpha\right)\frac{f(\tau)}{f(\tau-1)}-\tau
	&=\left(n-1-\alpha q\right)\frac{g(n-1)}{g(n)}-n-1\notag\\
	&+\left(\alpha q+1-2\alpha\right)\frac{g(n-1)}{g(n)}\frac{h(q-1)}{h(q)}.
\end{align}
The above partial difference equation can be separated into a system of three
ordinary difference equations. The solutions of the separated equations are:
\begin{align}
  f(\tau)&=\frac{\Gamma(\tau+\lambda_1)}{\Gamma(\tau-\alpha+1)},\\
  g(n)&=\frac{\Gamma(n+\lambda_2)}{\Gamma(n+\lambda_1+1)},\\
  h(q)&=\frac{\Gamma(q+1/\alpha-1)}{\Gamma(q+\lambda_2/\alpha+1)},
\end{align}
where $\lambda_1$ and $\lambda_2$ are separation parameters.

The solution of \eqref{eq:master_eq}, which fulfills the
initial conditions, is constructed from the linear combination
of the above particular solutions:
\begin{equation}
	\probi{\tau}{n,q}=\sum_{\lambda_1,\lambda_2}
	C_{\lambda_1,\lambda_2}\,f(\tau)\,g(n)\,h(q),
\end{equation}
where $C_{\lambda_1,\lambda_2}$ coefficients are independent of 
$\tau$, $n$ and $q$.

To obtain coefficients $C_{\lambda_1,\lambda_2}$, the initial condition of
\eqref{eq:master_eq} is expanded on the bases of $g(n)$ and $h(q)$. The
detailed calculation is presented in Appendix~\ref{app:kronecker_exp}. 

The solution of \eqref{eq:master_eq} is
\begin{align}
  \probi{\tau}{n,q}
  &=\frac{\Gamma(\tau-\tau_e+1)}{\Gamma(\tau_e)\,\Gamma(n+1)}
  \frac{\Gamma(\tau-n)}{\Gamma(\tau-\tau_e-n+1)}\notag\\
  &\times\frac{\Gamma(\tau_e+1-\alpha)}{\Gamma(\tau+1-\alpha)}
  \frac{\Gamma(q+1/\alpha-1)}{\Gamma(1/\alpha-1)}\,
  \sumterm{n,q}
  \label{eq:master_eq_sol_i}
\end{align}
where
$\sumterm{n,q}=\sum_{k=0}^{q}\frac{\left(-1\right)^k}{k!\left(q-k\right)!}\poch{-\alpha
k}{n}$ and $\poch{x}{n}\equiv\Gamma(n+x)/\Gamma{(x)}$ denotes Pochhammer's
symbol. Note that $\probi{\tau}{n,q}\neq0$ iff $0\le q\le n\le\tau-\tau_e$. The
conditions $0\le q$ and $n\le\tau-\tau_e$ are obvious, since $1/\Gamma(k)=0$ by
definition if $k$ is a negative integer or zero. Furthermore, the condition
$q<n$ can easily be seen if $\sumterm{n,q}$ is transformed into the following
equivalent form:
$\sumterm{n,q}=\frac{1}{q!}\frac{d^n}{dz^n}z^{n-1}\left(1-z^{-\alpha}\right)^q\bigr|_{z=1}$.
This result coincides with the fact that the size of a cluster $n$ cannot be
less than the corresponding number of in-degrees $q$.

\subsection{Joint distribution of cluster size and in-degree}
\label{subsec:joint_prob}

Equation \eqref{eq:master_eq_sol_i} provides the conditional probability that a
particular edge which was connected to the network at $\tau_e$ is in state
$\left(n,q\right)$ at $\tau>\tau_e$.  In a fully developed network, however,
the time when a particular edge is connected to the network is usually not
known. Moreover, the development of an individual link is usually not as
important as the properties of the link ensemble when it has finally developed.
Therefore, we are more interested in the total probability $\prob{\tau}{n,q}$,
that is the probability that a randomly selected edge is in state
$\left(n,q\right)$ at $\tau$, than the conditional probability
\eqref{eq:master_eq_sol_i}. The total probability can be calculated with the
help of the total probability theorem: \begin{equation}
\prob{\tau}{n,q}=\sum_{\tau_e=1}^{\tau}\probi{\tau}{n,q}\, \prob{\tau}{\tau_e},
\end{equation} where $\prob{\tau}{\tau_e}$ is the probability that a randomly
selected edge was included into the network at $\tau_e$. According to the
construction of the network one edge is added to the network at every time
step, therefore $\prob{\tau}{\tau_e}=1/\tau$. The following formula can be
obtained after the above summation has been carried
out: 
\begin{equation} 
  \prob{\tau}{n,q}=
  \frac{\tau+1-\alpha}{\tau}\frac{\poch{1/\alpha-1}{q}}{\poch{2-\alpha}{n+1}}
  \,\sumterm{n,q}, \label{eq:sol_joint} 
\end{equation} 
where $0<\alpha\le1$. In star topology, that is when $\alpha=1$, the joint
distribution $\prob{\tau}{n,q}$ evidently degenerates to
$\prob{\tau}{n,q}=\delta_{n,0}\,\delta_{q,0}$.

The \gls{er} limit of joint distribution can be obtained via the $\alpha\to0$
limit of \eqref{eq:sol_joint} (see Appendix~\ref{app:ER_limit} for details):
\begin{align}
  \lim_{\alpha\to0}\prob{\tau}{n,q}&=\frac{\tau+1}{\tau}
  \sum_{k=q-1}^{n-1}\left(-1\right)^{k+n-1}
  \frac{\binom{k}{q-1}S_{n-1}^{\left(k\right)}}{\Gamma(n+3)}
  \label{eq:sol_joint_a0}
\end{align}
where $0<q\le n<\tau$ and $S_n^{\left(m\right)}$ denote the Stirling numbers of
the first kind. Note that for the special case $n=q=0$ the \gls{er} limit is
$\lim_{\alpha\to0}\prob{\tau}{0,0}=\frac{\tau+1}{2\tau}$.

\begin{figure}[tb]
  \begin{center}
    \psfrag{n}[c][c][1.2]{Cluster size, $n$}
    \psfrag{q}[c][c][1.2]{In-degree, $q$}
    \psfrag{P(n,q)}[c][c][1.2]{$\prob{\tau}{n,q}$}
    \psfrag{q=1}[r][r][0.8]{$q=1$}
    \psfrag{q=10}[r][r][0.8]{$q=10$}
    \psfrag{q=20}[r][r][0.8]{$q=20$}
    \psfrag{n=5}[r][r][0.8]{$n=5$}
    \psfrag{n=10}[r][r][0.8]{$n=10$}
    \psfrag{n=20}[r][r][0.8]{$n=20$}
    \psfrag{n=40}[r][r][0.8]{$n=40$}
    \psfrag{1}[l][l][1]{$1$}
    \psfrag{10}[c][c][1]{$10$}
    \psfrag{100}[c][c][1]{$100$}
    \psfrag{ 1}[r][r][1]{$1$}
    \psfrag{ 0.1}[r][r][1]{$10^{-1}$}
    \psfrag{ 0.01}[r][r][1]{$10^{-2}$}
    \psfrag{ 0.001}[r][r][1]{$10^{-3}$}
    \psfrag{ 1e-04}[r][r][1]{$10^{-4}$}
    \psfrag{ 1e-05}[r][r][1]{$10^{-5}$}
    \psfrag{ 1e-06}[r][r][1]{$10^{-6}$}
    \psfrag{ 1e-07}[r][r][1]{$10^{-7}$}
    \psfrag{ 1e-08}[r][r][1]{$10^{-8}$}
    \subfigure[Joint distribution of cluster size and in-degree as the function of cluster size.]{
    	\label{subfig:deg-cl-prob-n}
    	\resizebox{0.69\figwidth}{!}{\includegraphics{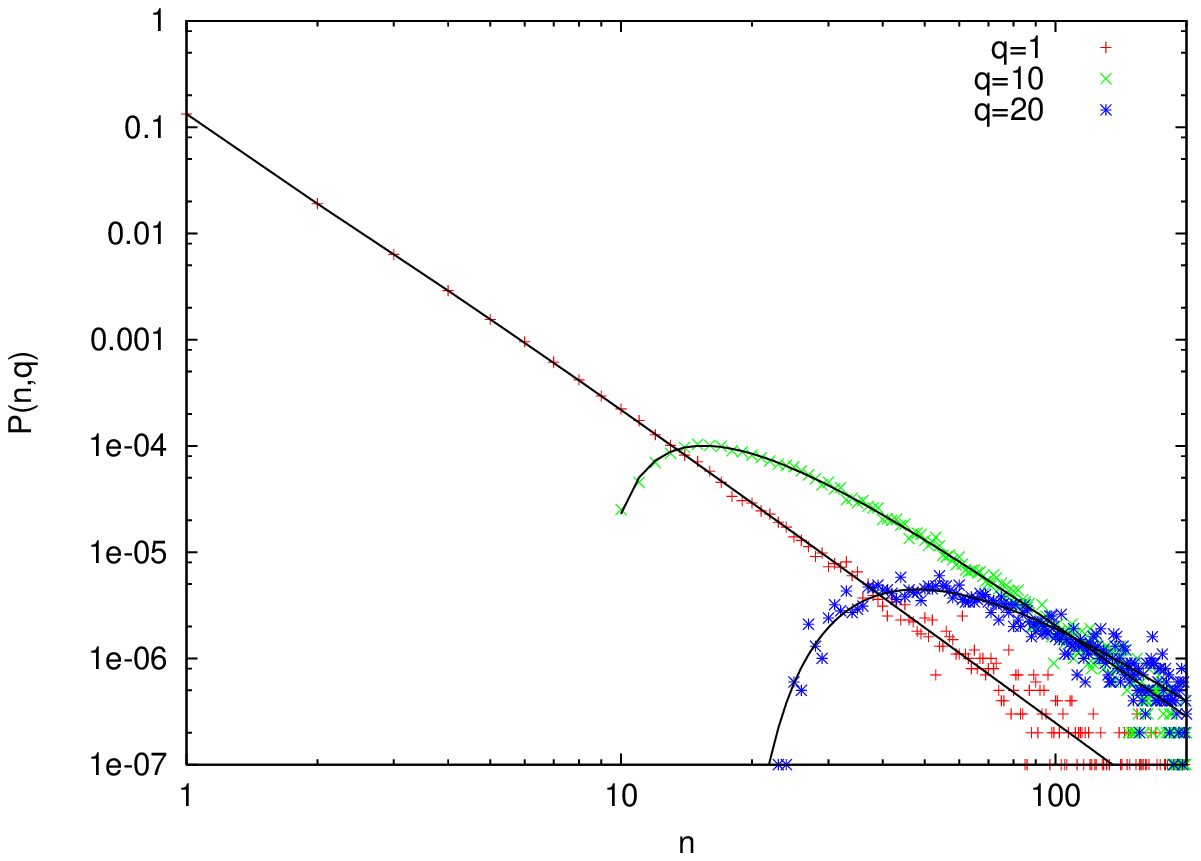}}
    }
    \subfigure[Joint distribution of cluster size and in-degree as the function of in-degree.]{
    	\label{subfig:deg-cl-prob-q}
    	\resizebox{0.69\figwidth}{!}{\includegraphics{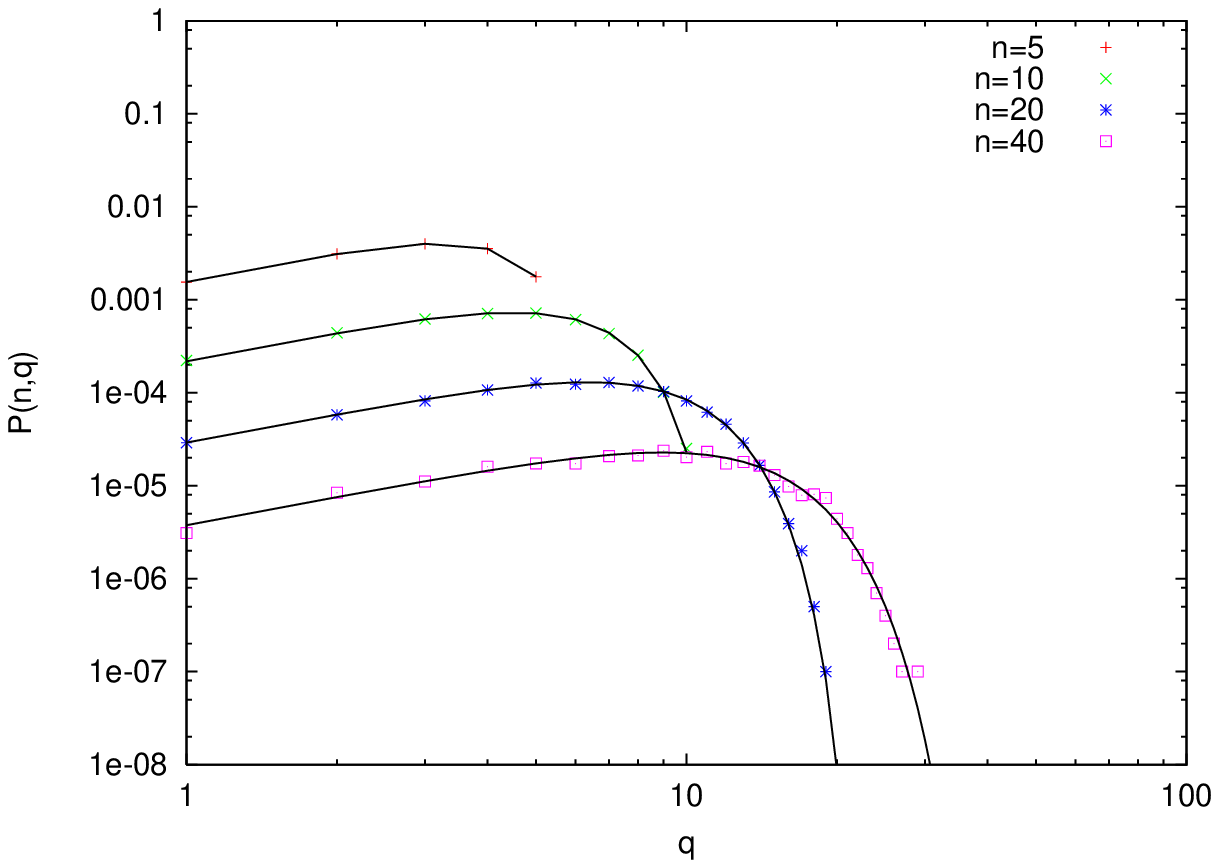}}
    }
  \end{center}
  \caption{Joint empirical distribution of cluster size and in-degree
  at $\alpha=1/2$ (symbols), and analytic formula
  (\ref{eq:sol_joint}) (solid lines) are compared 
  on double-logarithmic plot. Simulation results have been
  obtained from $100$ realizations of $10^5$ size networks. 
  }
  \label{fig:deg-cl-prob}
\end{figure}

The above formulas have been verified by extensive numerical simulations.  The
joint empirical cluster size and in-degree distribution has been compared with
the analytic formula \eqref{eq:sol_joint} for $\alpha=1/2$ in
Fig~\ref{fig:deg-cl-prob}.  Figures~\ref{subfig:deg-cl-prob-n}
and~\ref{subfig:deg-cl-prob-q} represent intersections of the joint
distribution with cutting planes of fixed in-degrees and cluster sizes,
respectively.  The figures confirm that the empirical distributions, obtained
as relative frequencies of links with cluster size $n$ and in-degree $q$ in 100
network realizations, are in complete agreement with the derived analytic
results. 

Equation~\eqref{eq:sol_joint} is the fundamental result of this section.  The
derived distribution is exact for any finite value of $\tau$, that is for any
finite \gls{ba} trees. This result is valuable for modeling a number of real
networks where the size of the network is small compared to the relevant
range of cluster size or in-degree. If the size of the network is much larger
than the relevant range of cluster size or in-degree then it is practical to
consider the network as infinitely large, that is to take the $\tau\to\infty$
limit. For the above joint distributions \eqref{eq:sol_joint} and \eqref{eq:sol_joint_a0}
the $\tau\to\infty$ limit is evident, since the $\tau$ dependent prefactors 
obviously tend to $1$ if the size of the networks grows beyond every limit.

\subsection{Distributions of cluster size and in-degree}
\label{subsec:marginal_prob}

I have derived the joint probability distribution of the cluster size and the
in-degree in the previous section.  In many cases it is sufficient to know the
probability distribution of only one random variable, since the information on
the other variable is either unavailable or not needed. It is also possible
that the one dimensional distribution is especially necessary, for example for
the calculation of a conditional distribution in 
Section~\ref{subsec:cond_prob}. 

The one dimensional (marginal) distributions $\prob{\tau}{n}$ and 
$\prob{\tau}{q}$ can be obtained from joint distribution $\prob{\tau}{n,q}$ 
as follows:
\begin{align*}
  \prob{\tau}{n}&=\sum_{q=0}^{n}\prob{\tau}{n,q}, &
  \prob{\tau}{q}&=\sum_{n=q}^{\tau-1}\prob{\tau}{n,q}.
\end{align*}
After substituting \eqref{eq:sol_joint} into the above formulas
the following expressions are obtained:
\begin{equation}
  \label{eq:sol_marginal_n}
  \prob{\tau}{n}=\frac{\tau+1-\alpha}{\tau}
  \frac{1-\alpha}{\left(n+1-\alpha\right)\left(n+2-\alpha\right)}.
\end{equation}
if $0\le n<\tau$ and $\prob{\tau}{n}=0$ if $n\ge\tau$. Furthermore,
\begin{align}
  \label{eq:sol_marginal_q}
  \prob{\tau}{q}
  &=\frac{\tau+1-\alpha}{\tau}\frac{1}{\alpha}
  \frac{\poch{1/\alpha-1}{1/\alpha}}{\poch{q+1/\alpha-1}{1/\alpha+1}}\notag\\
  &-\frac{\tau+1-\alpha}{\tau}\frac{\poch{1/\alpha-1}{q}}{\poch{2-\alpha}{\tau}}
  \sum_{k=0}^{q}\frac{\left(-1\right)^k}{k!\left(q-k\right)!}\frac{\poch{-\alpha k}{\tau}}{\alpha k+2-\alpha}.
\end{align}
if $0\le q<\tau$ and $\prob{\tau}{q}=0$ otherwise.  Rice's method
\cite{Odlyzko95} has been applied to evaluate the first term of
$\prob{\tau}{q}$ in closed form.

The \gls{er} limit of the marginal cluster size distribution can 
obviously be obtained from \eqref{eq:sol_marginal_n} at $\alpha=0$.
Furthermore, the \gls{er} limit of the marginal in-degree distribution 
can be derived analogously to the limit of the joint distribution, 
shown in Appendix~\ref{app:ER_limit}:
\begin{equation}
  \lim_{\alpha\to0}\prob{\tau}{q}
  =\frac{\tau+1}{\tau}\frac{1}{2^{q+1}}
  +\frac{\tau+1}{\tau}\frac{1}{\Gamma(\tau+2)\Gamma(q)}\frac{d^{q-1}}{d\alpha^{q-1}}
  \frac{\poch{1+\alpha}{\tau-1}}{2-\alpha}\Biggr|_{\alpha=0}\!\!\!.
\end{equation}

If the size of the network grows beyond every limit, that is if $\tau\to\infty$,
then the marginal distributions become much simpler:
\begin{align}
  \prob{\infty}{n}&=\frac{1-\alpha}{\left(n+1-\alpha\right)\left(n+2-\alpha\right)}\\
  \prob{\infty}{q}&=\frac{1}{\alpha}\frac{\poch{1/\alpha-1}{1/\alpha}}
  {\poch{q+1/\alpha-1}{1/\alpha+1}}\\
  \lim_{\alpha\to0}\prob{\infty}{q}&=\frac1{2^{q+1}}.
\end{align}

The asymptotic behavior of the cluster size and in-degree distributions differ
significantly. The tail of the cluster size distribution follows power law with
exponent $2$ either in \gls{ba} or \gls{er} network, independently of $\alpha$.
However, we learned that the tail of the in-degree distribution follows power
law with exponent $1/\alpha+1=2+a$ in \gls{ba} networks, and it falls
exponentially in \gls{er} topology, which agrees with the well known results of
previous works \cite{ErdosRenyi60}.

It is worth noting that the mean cluster size diverges logarithmically as
the size of the network tends to infinity:
\begin{equation}
  \mean{\tau}{n}=\sum_{n=0}^{\tau-1}n\,\prob{\tau}{n}=\left(1-\alpha\right)\ln\tau+\Ordo{1}.
\end{equation}
The expectation value of the in-degree, however, obviously remains finite:
$\mean{\tau}{q}=\frac{\tau}{\tau+1}<1$,
and $\mean{\infty}{q}=1$ if the size of the network is infinite. Moreover, the
variance of the in-degree can also be given exactly when the size of 
the network grows beyond every limit:
\begin{equation}
  \mean{\infty}{\left(q-1\right)^2}=\frac{2}{\left|1-2\alpha\right|}.
\end{equation}
This result implies that the fluctuations of the in-degree diverge
in a boundless network, if $\alpha=1/2$, that is in the classical BA model.

My analytic results have been verified with computer simulations.  Since
cumulative distributions are more suitable to be compared with simulations than
ordinary distributions I matched the corresponding \gls{ccdf} against
simulation data. The \gls{ccdf} of cluster size,
$\ccdf{\tau}{n}=\sum_{n'=n}^{\tau-1}\prob{\tau}{n'}$ can be calculated
straightforwardly:
\begin{equation}
  \ccdf{\tau}{n}
  =\frac{\tau+1-\alpha}{\tau}\frac{1-\alpha}{n+1-\alpha}
  -\frac{1-\alpha}{\tau},
  \label{eq:ccdf_n}
\end{equation}
where $0\le n< \tau$ and $0\le\alpha\le1$.  The \gls{ccdf} of in-degree,
$\ccdf{\tau}{q}=\sum_{q'=q}^{\tau-1}\prob{\tau}{q'}$ is more complex, however:
\begin{align}
  \notag
  \ccdf{\tau}{q}
  &=\frac{\tau+1-\alpha}{\tau}\frac{\poch{1/\alpha-1}{1/\alpha}}{\poch{q+1/\alpha-1}{1/\alpha}}
  -\frac{1-\alpha}{\tau}\\
  &+\frac{\tau+1-\alpha}{\tau}\frac{\poch{1/\alpha-1}{q}}{\poch{2-\alpha}{\tau}}
  \sum_{k=0}^{q-2}\frac{\left(-1\right)^k}{k!\left(q-2-k\right)!}
  \frac{\poch{1-\alpha-\alpha k}{\tau-1}}{\left(k+1/\alpha\right)\left(k+2/\alpha\right)}
  \label{eq:ccdf_q}
 \end{align}
where $0\le q<\tau$ and $0<\alpha\le1$. If the size of the network grows
beyond every limit, then the \glspl{ccdf} are the following:
\begin{align}
  \ccdf{\infty}{n}&=\frac{1-\alpha}{n+1-\alpha}, & 
  \ccdf{\infty}{q}&=
  \frac{\poch{1/\alpha-1}{1/\alpha}}{\poch{q+1/\alpha-1}{1/\alpha}},
\end{align}
where $0\le n$, $0\le q$ and $0<\alpha<1$.
\begin{figure}[tb]
  \begin{center}
    \psfrag{n}[c][c][1.2]{Cluster size, $n$}
    \psfrag{q}[c][c][1.2]{In-degree, $q$}
    \psfrag{Fc(n)}[c][c][1]{$\ccdf{\tau}{n}$}
    \psfrag{alpha=0}[r][r][0.8]{$\alpha=0$}
    \psfrag{alpha=1/3}[r][r][0.8]{$\alpha=1/3$}
    \psfrag{alpha=1/2}[r][r][0.8]{$\alpha=1/2$}
    \psfrag{alpha=2/3}[r][r][0.8]{$\alpha=2/3$}
    \psfrag{ 1}[c][c][1]{$1$}
    \psfrag{ 10}[c][c][1]{$10$}
    \psfrag{ 100}[c][c][1]{$10^2$}
    \psfrag{ 1000}[c][c][1]{$10^3$}
    \psfrag{ 10000}[c][c][1]{$10^4$}
    \psfrag{ 100000}[c][c][1]{$10^5$}
    \psfrag{ 1e+06}[c][c][1]{$10^6$}
    \psfrag{ 0.1}[r][r][1]{$10^{-1}$}
    \psfrag{ 0.01}[r][r][1]{$10^{-2}$}
    \psfrag{ 0.001}[r][r][1]{$10^{-3}$}
    \psfrag{ 1e-04}[r][r][1]{$10^{-4}$}
    \psfrag{ 1e-05}[r][r][1]{$10^{-5}$}
    \psfrag{ 1e-06}[r][r][1]{$10^{-6}$}
    \psfrag{ 1e-07}[r][r][1]{$10^{-7}$}
    \psfrag{ 1e-08}[r][r][1]{$10^{-8}$}
    \psfrag{ 1e-09}[r][r][1]{$10^{-9}$}
    \psfrag{ 1e-10}[r][r][1]{$10^{-10}$}
    \resizebox{\figwidth}{!}{\includegraphics{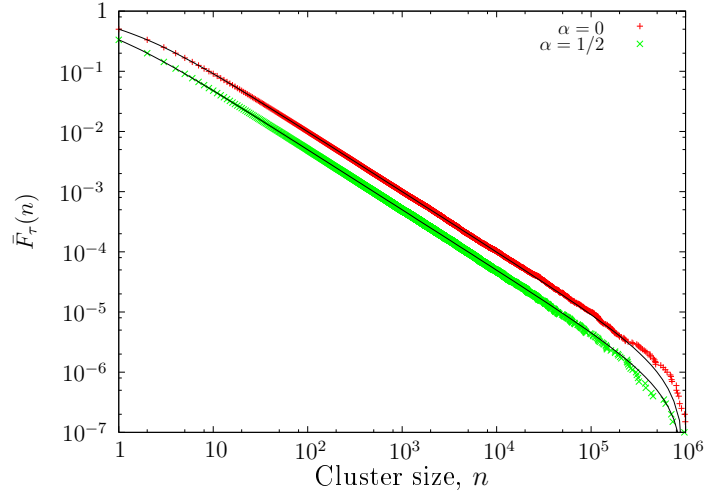}}
  \end{center}
  \caption{Figure shows comparison of empirical CCDFs of cluster size  
  distributions (points) with analytic formula (\ref{eq:ccdf_n}) 
  (lines) on logarithmic plots, at $\alpha=0$, and $1/2$. Empirical 
  distributions have been obtained from $10$ realizations of $N=10^6$ size networks.}
  \label{fig:cluster-dist}
\end{figure}

\begin{figure}[tb]
  \begin{center}
    \psfrag{n}[c][c][1.2]{Cluster size, $n$}
    \psfrag{q}[c][c][1.2]{In-degree, $q$}
    \psfrag{Fc(n)}[c][c][1.2]{$\ccdf{\tau}{n}$}
    \psfrag{Fc(q)}[c][c][1.2]{$\ccdf{\tau}{q}$}
    \psfrag{alpha=0}[r][r][0.8]{$\alpha=0$}
    \psfrag{alpha=1/3}[r][r][0.8]{$\alpha=1/3$}
    \psfrag{alpha=1/2}[r][r][0.8]{$\alpha=1/2$}
    \psfrag{alpha=2/3}[r][r][0.8]{$\alpha=2/3$}
    \psfrag{ 0}[c][c][1]{$0$}
    \psfrag{ 5}[c][c][1]{$5$}
    \psfrag{ 10}[c][c][1]{$10$}
    \psfrag{ 15}[c][c][1]{$15$}
    \psfrag{ 20}[c][c][1]{$20$}
    \psfrag{ 25}[c][c][1]{$25$}
    \psfrag{ 1}[c][c][1]{$1$}
    \psfrag{ 10}[c][c][1]{$10$}
    \psfrag{ 100}[c][c][1]{$10^2$}
    \psfrag{ 1000}[c][c][1]{$10^3$}
    \psfrag{ 10000}[c][c][1]{$10^4$}
    \psfrag{ 100000}[c][c][1]{$10^5$}
    \psfrag{ 0.1}[r][r][1]{$10^{-1}$}
    \psfrag{ 0.01}[r][r][1]{$10^{-2}$}
    \psfrag{ 0.001}[r][r][1]{$10^{-3}$}
    \psfrag{ 1e-04}[r][r][1]{$10^{-4}$}
    \psfrag{ 1e-05}[r][r][1]{$10^{-5}$}
    \psfrag{ 1e-06}[r][r][1]{$10^{-6}$}
    \psfrag{ 1e-07}[r][r][1]{$10^{-7}$}
    \resizebox{\figwidth}{!}{\includegraphics{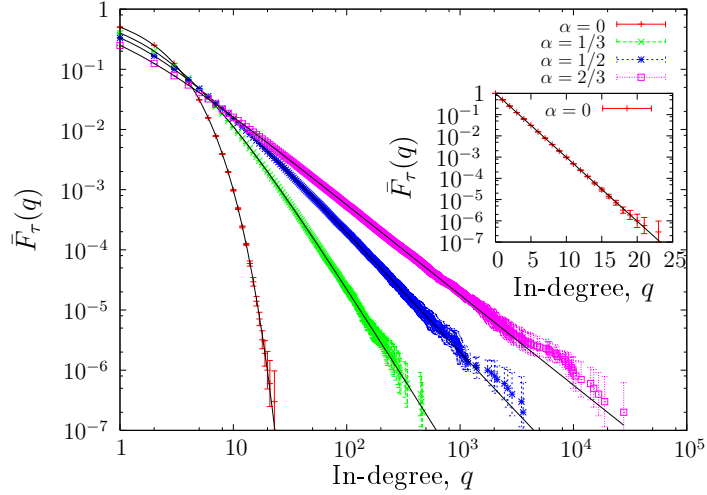}}
  \end{center}
  \caption{Figure shows comparison of empirical CCDFs of in-degree 
  distributions (points) with analytic formula (\ref{eq:ccdf_q}) 
  (lines) on logarithmic plots, at $\alpha=0$, $1/3$, $1/2$, and $2/3$. 
  Empirical distributions have been obtained from $10$ realizations of 
  $N=10^6$ size networks. 
  Inset: Comparison at $\alpha=0$ on semi-logarithmic plot.}
  \label{fig:deg-dist}
\end{figure}

Comparisons of analytic \gls{ccdf} of cluster size \eqref{eq:ccdf_n} and
empirical distributions are shown in Figure~\ref{fig:cluster-dist} for
$\alpha=0$, $1/3$, $1/2$, and $2/3$. Experimental data has been collected from
$10$ realizations of $10^6$ node networks.  Figure~\ref{fig:cluster-dist} shows
that simulations fully confirm my analytic result.

On Figure~\ref{fig:deg-dist} analytic formula \eqref{eq:ccdf_q} and the
empirical \glspl{ccdf} of in-degree, obtained from the same $10^6$ node realizations,
are compared.  Note the precise match of the simulation and the theoretical
distribution on almost the whole range of data. Some small discrepancy can be
observed around the low probability events.  This deviation is caused by the
aggregation of errors on the cumulative distribution when some rare event occurs
in a finite network.

\subsection{Conditional probabilities and expectation values}
\label{subsec:cond_prob}

In the previous sections exact joint and marginal distributions of cluster size
and in-degree were analyzed for both finite and infinite networks. All
these distributions provide general statistics of the network.  In this section
I proceed further, and I investigate the scenario when the ``younger''
in-degree of a randomly selected link is known. I ask the cluster size
distribution under this condition, that is the conditional distribution
$\prob{\tau}{n\mid q}$. The results of the previous sections are referred to
below to obtain the conditional probability distribution, and eventually the
conditional expectation of cluster size. For the sake of completeness, the
conditional distribution and expectation of in-degree are also given at the
end of this section.

The conditional cluster size distribution can be given by the quotient of the
joint and the marginal in-degree distributions by definition:
\begin{equation}
  \prob{\tau}{n\mid q}=\frac{\prob{\tau}{n,q}}{\prob{\tau}{q}}.
\end{equation}
The exact conditional distribution for any finite network can be obtained after
substituting (\ref{eq:sol_joint}) and (\ref{eq:sol_marginal_q}) into the above
expression. For a boundless network the conditional distribution takes the
simpler form:
\begin{equation}
  \prob{\infty}{n\mid q}=\alpha\frac{\poch{2/\alpha-1}{q+1}}{\poch{2-\alpha}{n+1}}\sumterm{n,q},
  \label{eq:mean_cond_cl_inf}
\end{equation}
where $0\le q\le n$. If $n\gg1$, then $\prob{\infty}{n\mid q}\sim
\alpha\poch{2/\alpha-1}{q+1}/n^{3}+\Ordo{1/n^4}$, that is the conditional
cluster size distribution falls faster than the ordinary cluster size
distribution. It follows that the mean
of the conditional cluster size distribution will not diverge like the
mean of the ordinary distribution.

What is the expected size of a cluster under the condition that the
in-degree of its root is known? For practical reasons, I do not 
calculate $\meanq{\tau}{n}$ directly, but I calculate
$\meanq{\tau}{n+2-\alpha}=\meanq{\tau}{n}+2-\alpha$ instead:
\begin{equation}
  \meanq{\tau}{n+2-\alpha}
  =\frac{1}{\prob{\tau}{q}}\sum_{n=q}^{\tau-1}\left(n+2-\alpha\right)\prob{\tau}{n,q}.
\end{equation}
Since $\left(n+2-\alpha\right)\prob{\tau}{n,q}=\frac{\tau+1-\alpha}{\tau}
\frac{\poch{1/\alpha-1}{q}}{\poch{2-\alpha}{n}}\sumterm{n,q}$, the above 
summation can be given similarly to the marginal distribution $\prob{\tau}{q}$ in 
(\ref{eq:sol_marginal_q}):
\begin{align*}
  \notag
  \sum_{n=q}^{\tau-1}\left(n+2-\alpha\right)\prob{\tau}{n,q}
  &=\frac{\tau+1-\alpha}{\tau}\frac{1/\alpha-1}{q+1/\alpha-1}\\
  &-\frac{\tau+1-\alpha}{\tau}\frac{\poch{1/\alpha-1}{q}}{\poch{2-\alpha}{\tau-1}}
  \sum_{k=0}^{q}\frac{\left(-1\right)^k}{k!\left(q-k\right)!}\frac{\poch{-\alpha k}{\tau}}{\alpha k+1-\alpha}
\end{align*}
After replacing the above sum in $\meanq{\tau}{n}$, the following equation can be obtained:
\begin{equation}
  \meanq{\tau}{n+2-\alpha}
  =\left(1-\alpha\right)\frac{\poch{q+1/\alpha}{1/\alpha}}{\poch{1/\alpha-1}{1/\alpha}}G_{\tau}(q),
  \label{eq:mean_q}
\end{equation}
where
\begin{equation}
  G_{\tau}(q)=\frac{\displaystyle 1-\frac{\poch{1/\alpha-1}{q+1}}{\poch{1-\alpha}{\tau}}
    \sum_{k=0}^{q}\frac{\left(-1\right)^k}{k!\left(q-k\right)!}\frac{\poch{-\alpha k}{\tau}}{k+1/\alpha-1}}
    {\displaystyle 1-\frac{\poch{2/\alpha-1}{q+1}}{\poch{2-\alpha}{\tau}}
    \sum_{k=0}^{q}\frac{\left(-1\right)^k}{k!\left(q-k\right)!}\frac{\poch{-\alpha k}{\tau}}{k+2/\alpha-1}}.
\end{equation}
The identity $\lim_{\tau\to\infty}G_{\tau}(q)\equiv1$ implies that
$G_{\tau}(q)$ involves the finite scale effects, and the factors preceding
$G_{\tau}(q)$ give the asymptotic form of $\meanq{\tau}{n+2-\alpha}$: 
\begin{equation}
  \meanq{\infty}{n+2-\alpha}
  =\left(1-\alpha\right)\frac{\poch{q+1/\alpha}{1/\alpha}}{\poch{1/\alpha-1}{1/\alpha}}.
  \label{eq:mean_q_inf}
\end{equation}
It can be seen that the expectation of cluster size, under the condition that
the in-degree is known, is finite in an unbounded network. This stands in
contrast to the unconditional cluster size, discussed in the previous section,
which diverges logarithmically as the size of the network grows beyond every
limit.
\begin{figure}[tb]
  \begin{center}
    \psfrag{q}[c][c][1.2]{In-degree, $q$}
    \psfrag{E(n|q)}[c][c][1.2]{$\meanq{\tau}{n}$}
    \psfrag{alpha=0}[r][r][0.8]{$\alpha=0$}
    \psfrag{alpha=1/3}[r][r][0.8]{$\alpha=1/3$}
    \psfrag{alpha=1/2}[r][r][0.8]{$\alpha=1/2$}
    \psfrag{alpha=2/3}[r][r][0.8]{$\alpha=2/3$}
    \psfrag{ 1}[c][c][1]{$1$}
    \psfrag{ 10}[c][c][1]{$10$}
    \psfrag{ 100}[c][c][1]{$10^2$}
    \psfrag{ 1000}[c][c][1]{$10^3$}
    \psfrag{ 10000}[c][c][1]{$10^4$}
    \psfrag{ 100000}[c][c][1]{$10^5$}
    \psfrag{1}[r][r][1]{$1$}
    \psfrag{10}[r][r][1]{$10$}
    \psfrag{100}[r][r][1]{$10^2$}
    \psfrag{1000}[r][r][1]{$10^3$}
    \psfrag{10000}[r][r][1]{$10^4$}
    \psfrag{100000}[r][r][1]{$10^5$}
    \resizebox{\figwidth}{!}{\includegraphics{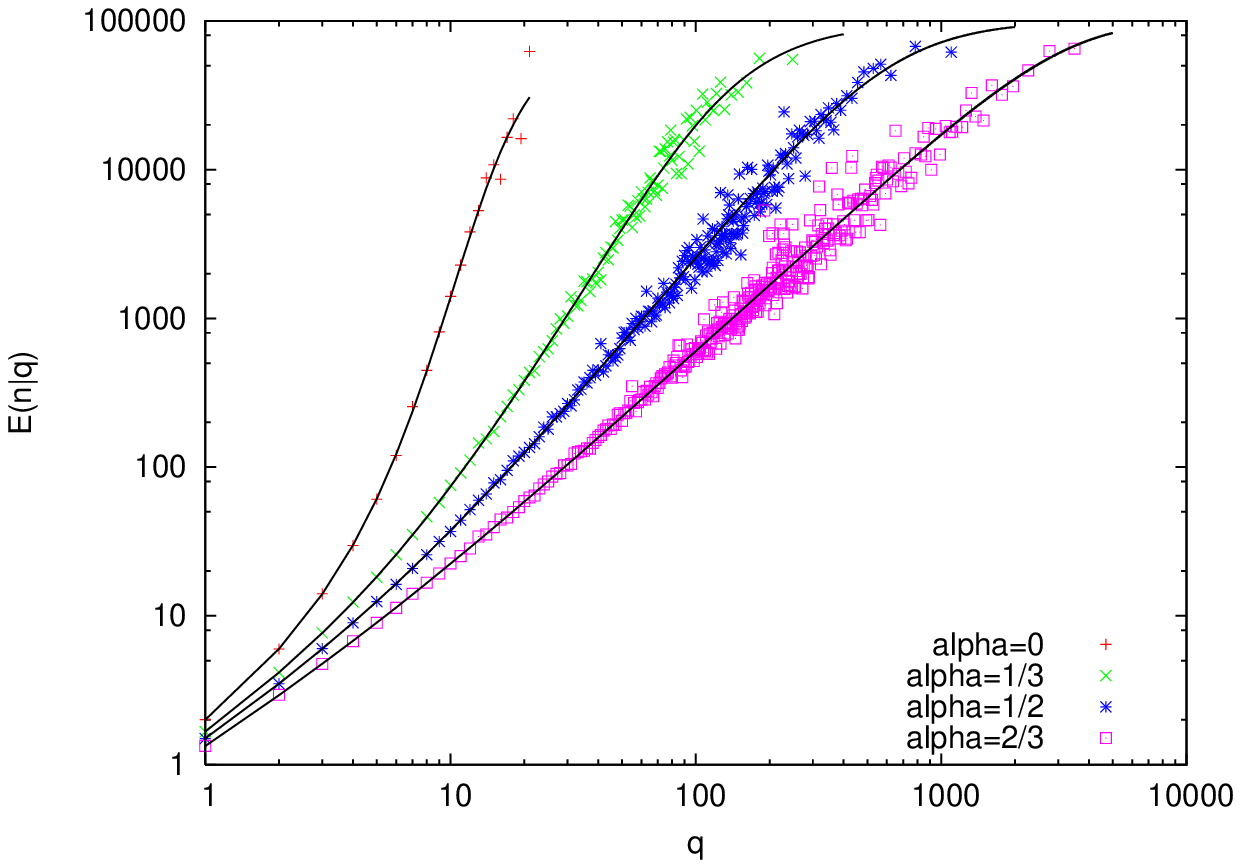}}
  \end{center}
  \caption{Figure shows the average cluster size as the function of the 
  in-degree $q$, obtained from 100 realizations of $10^5$ size networks.
  Simulation data has been collected at $\alpha=0$, $1/3$, $1/2$, and $2/3$ 
  parameter values. Analytical result (\ref{eq:mean_q}) of conditional 
  expectation $\meanq{\tau}{n}$ is shown with continuous lines.}
  \label{fig:deg-cluster-avg}
\end{figure}

In the \gls{er} limit the expected conditional cluster size becomes
\begin{equation}
  \lim_{\alpha\to0}\meanq{\infty}{n+2}=2^{q+1}.
  \label{eq:mean_q_inf_a0}
\end{equation}
The fundamental difference between the scale-free and non-scale-free networks
can be observed again. In the scale-free case the expected conditional cluster
size asymptotically grows with the in-degree to the power of $1/\alpha$, while
in the latter case it grows exponentially. On Figure~\ref{fig:deg-cluster-avg}
the exact analytic formula (\ref{eq:mean_q}) is compared with simulation
results at $\alpha=0$, $1/3$, $1/2$, and $2/3$. The simulations clearly justify
my analytic solution.

\begin{figure}[tb]
  \begin{center}
    \psfrag{n}[c][c][1.2]{Cluster size, $n$}
    \psfrag{E(q|n)}[c][c][1.2]{$\meann{\tau}{q}$}
    \psfrag{alpha=0}[r][r][0.8]{$\alpha=0$}
    \psfrag{alpha=1/3}[r][r][0.8]{$\alpha=1/3$}
    \psfrag{alpha=1/2}[r][r][0.8]{$\alpha=1/2$}
    \psfrag{alpha=2/3}[r][r][0.8]{$\alpha=2/3$}
    \psfrag{1}[r][r][1]{$1$}
    \psfrag{10}[r][r][1]{$10$}
    \psfrag{100}[r][r][1]{$10^2$}
    \psfrag{1000}[r][r][1]{$10^3$}
    \psfrag{ 1}[c][c][1]{$1$}
    \psfrag{ 10}[c][c][1]{$10$}
    \psfrag{ 100}[c][c][1]{$10^2$}
    \psfrag{ 1000}[c][c][1]{$10^3$}
    \psfrag{ 10000}[c][c][1]{$10^4$}
    \psfrag{ 100000}[c][c][1]{$10^5$}
    \psfrag{ 1e+06}[c][c][1]{$10^6$}
    \resizebox{\figwidth}{!}{\includegraphics{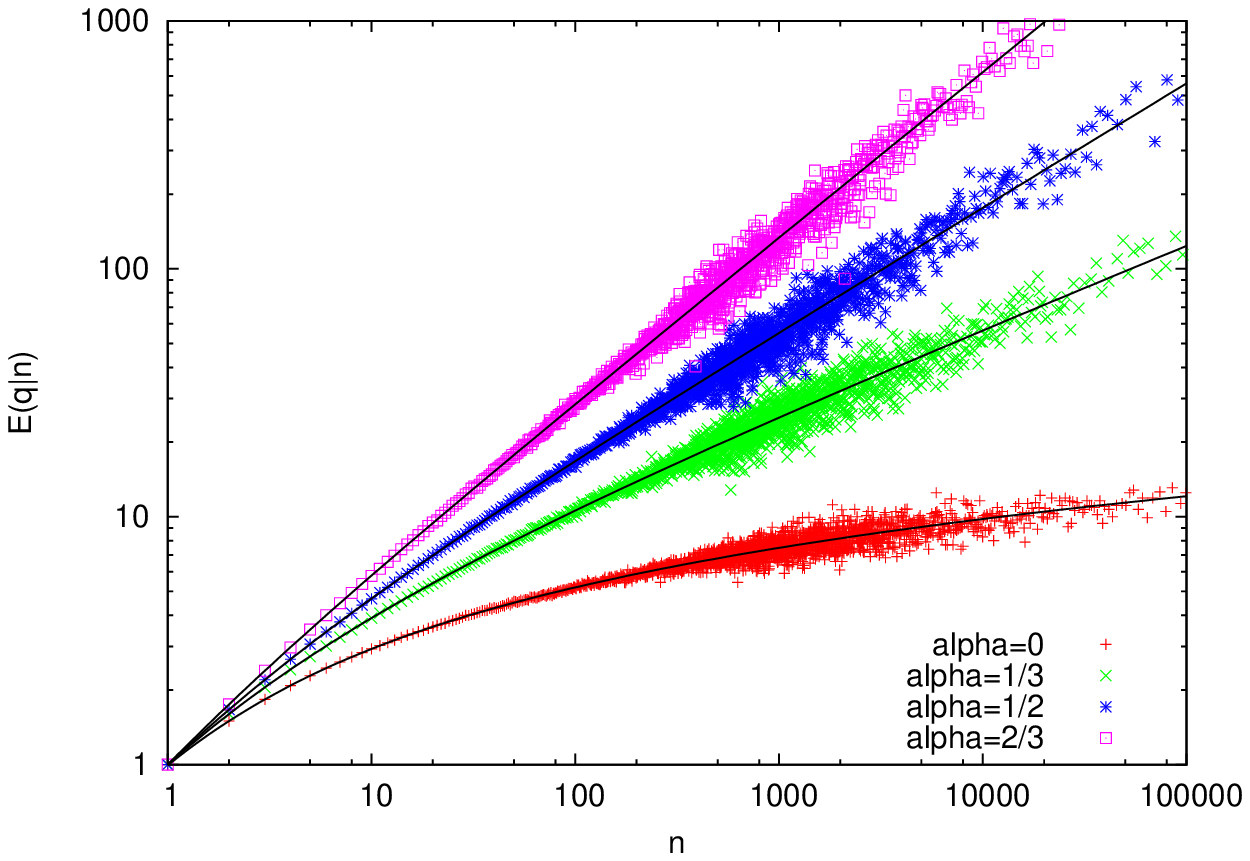}}
  \end{center}
  \caption{Figure shows the average in-degree as the function of the cluster 
  size $n$, obtained from 100 realizations of $10^5$ size networks.
  Simulation data has been collected at $\alpha=0$, $1/3$, $1/2$, and $2/3$ 
  parameter values. Analytical result (\ref{eq:mean_n}) of conditional 
  expectation $\meann{\tau}{q}$ is shown with continuous lines.}
  \label{fig:cluster-deg-avg}
\end{figure}

Let us briefly investigate the opposite scenario, that is when the cluster size
is known and the statistics of the in-degree are sought under this condition.
The conditional distribution can be obtained from the combination of
Eqs.~(\ref{eq:sol_joint}), (\ref{eq:sol_marginal_n}) and the definition
\begin{equation}
  \prob{\tau}{q\mid n}=\frac{\prob{\tau}{n,q}}{\prob{\tau}{n}}.
\end{equation}
The conditional expectation of in-degree can be acquired by the same technique 
as the conditional expectation of cluster size. Let us calculate 
\begin{equation}
  \meann{\tau}{q+1/\alpha-1}=\meann{\tau}{q}+1/\alpha-1
\end{equation}
instead of $\meann{\tau}{q}$ directly:
\begin{align}
  \notag
  \meann{\tau}{q+1/\alpha-1}
  &=\frac{1}{\prob{\tau}{n}}\sum_{q=0}^{n}\left(q+1/\alpha-1\right)\prob{\tau}{n,q}\\
  &=\frac{\Gamma(2-\alpha)}{\alpha}\poch{n+1-\alpha}{\alpha},
  \label{eq:mean_n}
\end{align}
where $0\le n<\tau$. Note that the conditional expectation of in-degree is
independent of $\tau$, that is of the size of the network. 
In the \gls{er} limit the expectation of the in-degree becomes
\begin{equation}
  \lim_{\alpha\to0}\meann{\tau}{q}=\Psi(n+1)+\gamma,
\end{equation}
where $\Psi(x)=\frac{d}{dx}\ln\Gamma(x)$ denotes the digamma function, and
$\gamma=-\Psi(1)\approx 0.5772$ is the Euler--Mascheroni constant.
Asymptotically the expectation of the in-degree in a scale-free tree grows with
the cluster size to the power of $\alpha$, while in a \gls{er} tree it grows only
logarithmically, since $\Psi(n+1)=\log n+\Ordo{1/n}$.  Therefore, conditional
in-degree and conditional cluster size are mutually inverses
\emph{asymptotically}. Figure~\ref{fig:cluster-deg-avg} shows the analytic
solution (\ref{eq:mean_n}) and simulation data at $\alpha=0$, $1/3$, $1/2$, and
$2/3$ parameter values. Simulation data has been collected from 100
realizations of $10^5$ size networks.

\subsection{Conditional distribution of edge betweenness}
\label{subsec:load_prob}

Using the results of the previous sections, I am finally ready to answer the
problem which motivated my work, that is the distribution of the edge
betweenness under the condition that the in-degree of the ``younger'' node of
the link is known. As I noted at the beginning of
Section~\ref{subsec:cond_joint_prob}, the edge betweenness can be expressed
with cluster size: 
\begin{equation}
  L=\left(n+1\right)\left(\tau-n\right).
  \label{eq:load_cluster}
\end{equation}
Therefore, conditional edge betweenness can be given formally by the following transformation
of random variable $n$:
\begin{equation}
  \prob{\tau}{L\mid q}=\sum_{n=0}^{\tau-1}\delta_{L,\left(n+1\right)\left(\tau-n\right)}\prob{\tau}{n\mid q}.
\end{equation}
Obviously, $\prob{\tau}{L\mid q}$ is non-zero only at those values of $L$, 
where \eqref{eq:load_cluster} has an integer solution for $n$. If 
\begin{equation}
  n_L=\frac{\tau-1}{2}-\sqrt{\frac{\left(\tau+1\right)^2}{4}-L}
  \label{eq:cluster_load}
\end{equation}
is such an integer solution of the quadratic equation \eqref{eq:load_cluster}, 
and $L\neq \left(\tau+1\right)^2/4$, then 
\begin{equation}
  \prob{\tau}{L\mid q}=\prob{\tau}{n_L\mid q}+\prob{\tau}{\tau-1-n_L\mid q}.
\end{equation} 
If $L=\left(\tau+1\right)^2/4$ is integer, then $\prob{\tau}{L\mid q}=\prob{\tau}{n_L\mid q}$.

The conditional expectation of edge betweenness can be obtained from \eqref{eq:load_cluster}:
\begin{equation}
  \label{eq:meanq_load_expand}
  \meanq{\tau}{L}=\tau\meanq{\tau}{n+1}-\meanq{\tau}{\left(n+1\right)n}.
\end{equation}
Therefore, for the exact calculation of $\meanq{\tau}{L}$ the first and the
second moment of the conditional cluster size distribution are required. The
first moment, that is the mean, has been derived in the previous section. 
In order to calculate the second moment let us use the technique I have 
developed in the previous sections. Let us consider:
\begin{equation}
  \meanq{\tau}{\left(n+2-\alpha\right)\left(n+1-\alpha\right)}
  =\frac{\tau+1-\alpha}{\tau}\frac{\poch{1/\alpha-1}{q}}{\prob{\tau}{q}}
  \sum_{n=q}^{\tau-1}\frac{\sumterm{n,q}}{\poch{2-\alpha}{n-1}}.
\end{equation}
We shall be cautious when the summation for $n$ is evaluated.
The $k=1$ term in 
$\sumterm{n,q}=\sum_{k=0}^{q}\frac{\left(-1\right)^k}{k!\left(q-k\right)!}\poch{-\alpha k}{n}$
must be treated separately to avoid a divergent term:
  \begin{align*}
    \sum_{n=q}^{\tau-1}\frac{\sumterm{n,q}}{\poch{2-\alpha}{n-1}}
    &=\frac{1-\alpha}{\left(q-1\right)!}\left[\alpha\Psi(\tau-\alpha)-\alpha\Psi(1-\alpha)-\Psi(q)-\gamma\right]\\
    &-\frac{1}{\alpha}\frac{1}{\poch{2-\alpha}{\tau-2}}\sum_{k=2}^{q}\frac{\left(-1\right)^k}{k!\left(q-k\right)!}
    \frac{\poch{-\alpha k}{\tau}}{k-1}
\end{align*}
The exact formula for $\meanq{\tau}{L}$ can be obtained straightforwardly, after
\eqref{eq:mean_q} and the above expressions have been substituted into
\eqref{eq:meanq_load_expand}. 

Let us consider the scenario when the size of the network tends to infinity.
Equation~\eqref{eq:load_cluster} implies that edge betweenness diverges as
$\tau\to\infty$, therefore $L$ should be rescaled for an infinite network.
From the asymptotics of the digamma function
$\Psi(\tau-\alpha)=\ln\tau+\Ordo{1/\tau}$ it follows that
$\meanq{\tau}{\left(n+2-\alpha\right)\left(n+1-\alpha\right)}$ grows only
logarithmically, slower than the linear growth of
$\tau\meanq{\tau}{n+2-\alpha}$. Therefore, edge betweenness asymptotically
grows linearly as the size of the network grows beyond every limit.
Let us rescale edge betweenness
\begin{equation}
  \Lambda_{\tau}=\frac{L(\tau)}{\tau+1}
\end{equation}
and let us consider the limit 
$\Lambda=\lim_{\tau\to\infty}\Lambda_{\tau}=n_{\Lambda}+1$.
The \gls{ccdf} of the rescaled edge betweenness
can be given by 
\begin{equation}
  \ccdf{\infty}{\Lambda\mid q}
  =\lim_{\tau\to\infty}\sum_{n=n_{\Lambda_{\tau}}}^{\tau-1-n_{\Lambda_{\tau}}}\prob{\tau}{n\mid q}
  =\frac{1}{\prob{\infty}{q}}\sum_{n=\Lambda-1}^{\infty}\prob{\infty}{n,q}.
\end{equation} 
When the summation has been carried out, the following equation is obtained:
\begin{equation}
  \ccdf{\infty}{\Lambda\mid q}
  =\frac{\poch{2/\alpha-1}{q+1}}{\poch{2-\alpha}{\Lambda-1}}
  \sum_{k=0}^{q}\frac{\left(-1\right)^k}{k!\left(q-k\right)!}
  \frac{\poch{-\alpha k}{\Lambda-1}}{k+2/\alpha-1},
  \label{eq:ccdf_load}
\end{equation}
where $q+1\le\Lambda$. 
{ If $1<q\ll\Lambda$, then only the first term of the sum should be taken into
account, and it is easy to see that
\begin{equation}
  \ccdf{\infty}{\Lambda\mid q}=\frac{\alpha^2\left(1-\alpha\right)}{2\Gamma(2/\alpha-1)}
  \frac{q^{2/\alpha}}{\Lambda^2}+\Ordo{1/\Lambda^{2+\alpha}}.
  \label{eq:ccdf_load_approx}
\end{equation}
It can be seen that the scaling exponent $-2$ is independent of $\alpha$.
The above asymptotic formula has been obtained for infinite networks.
The same power law scaling can be observed in finite size networks as \eqref{eq:ccdf_load_approx}
if $\Lambda_{\tau}\ll\tau$. However, $\ccdf{\tau}{\Lambda_{\tau}\mid q}\equiv0$ 
if $\Lambda_{\tau}>\tau$ in finite networks, therefore asymptotic formula 
\eqref{eq:ccdf_load_approx} evidently  becomes invalid if 
$\Lambda_{\tau}\approx\tau$.

It is obvious that as the size of the network grows larger and larger,
asymptotic formula \eqref{eq:ccdf_load} becomes more and more accurate. One can
ask how fast this convergence is. From elementary estimations of
$\ccdf{\tau}{\Lambda_{\tau}\mid q}$ one can show that for fixed
$\Lambda_{\tau}$:
\begin{equation}
  \ccdf{\tau}{\Lambda_{\tau}\mid q}=\ccdf{\infty}{\Lambda_{\tau}\mid q}
  -\left(1-\ccdf{\infty}{\Lambda_{\tau}\mid q}\right) 
  \frac{\alpha^2\left(1-\alpha\right)}{2}\frac{1}{\tau^2}
  +\Ordo{1/\tau^{2+\alpha}},
   \label{eq:ccdf_size_approx}
\end{equation}
that is corrections to the asymptotic formula decrease with $\tau^{-2}$
for large $\tau$.
}

\begin{figure}[tb]
  \begin{center}
    \psfrag{Lambda}[c][c][1.2]{Rescaled edge betweenness, $\Lambda$}
    \psfrag{Fc(L|q)}[c][c][1.2]{$\ccdf{\infty}{\Lambda\mid q}$}
    \psfrag{N=1e4}[r][r][0.8]{$N=10^4$}
    \psfrag{N=1e5}[r][r][0.8]{$N=10^5$}
    \psfrag{N=1e6}[r][r][0.8]{$N=10^6$}
    \psfrag{q=1}[r][r][0.8]{$q=1$}
    \psfrag{q=2}[r][r][0.8]{$q=2$}
    \psfrag{ 1}[r][r][1]{$1$}
    \psfrag{ 0.1}[r][r][1]{$10^{-1}$}
    \psfrag{ 0.01}[r][r][1]{$10^{-2}$}
    \psfrag{ 0.001}[r][r][1]{$10^{-3}$}
    \psfrag{ 1e-04}[r][r][1]{$10^{-4}$}
    \psfrag{ 1e-05}[r][r][1]{$10^{-5}$}
    \psfrag{ 1e-06}[r][r][1]{$10^{-6}$}
    \psfrag{ 10}[c][c][1]{$10$}
    \psfrag{ 100}[c][c][1]{$10^2$}
    \psfrag{ 1000}[c][c][1]{$10^3$}
    \resizebox{\figwidth}{!}{\includegraphics{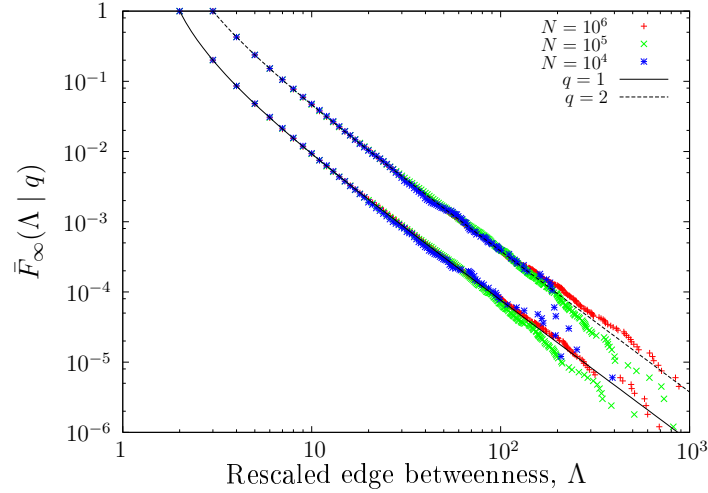}}
  \end{center}
  \caption{Figure shows CCDF of edge betweenness under the condition that
  the in-degree $q$ is known. Empirical CCDF has been obtained from $100$ 
  realizations of $N=10^4$ and $N=10^5$, and $10$ realizations of $N=10^6$ size
  networks at $\alpha=1/2$ parameter value. Continuous lines show analytic 
  result of infinite network limit (\ref{eq:ccdf_load}).}
  \label{fig:load-dist-1}
\end{figure}

\begin{figure}[tb]
  \begin{center}
    \psfrag{q}[c][c][1.2]{In-degree, $q$}
    \psfrag{E(L|q)}[c][c][1.2]{$\meanq{\infty}{\Lambda}$}
    \psfrag{N=1e4}[r][r][0.8]{$N=10^4$}
    \psfrag{N=1e5}[r][r][0.8]{$N=10^5$}
    \psfrag{N=1e6}[r][r][0.8]{$N=10^6$}
    \psfrag{alpha=0}[r][r][0.8]{$\alpha=0$}
    \psfrag{alpha=1/2}[r][r][0.8]{$\alpha=1/2$}
    \psfrag{ 0}[c][c][1]{$0$}
    \psfrag{ 1}[c][c][1]{$1$}
    \psfrag{ 5}[c][c][1]{$5$}
    \psfrag{ 10}[c][c][1]{$10$}
    \psfrag{ 15}[c][c][1]{$15$}
    \psfrag{ 20}[c][c][1]{$20$}
    \psfrag{ 100}[c][c][1]{$10^2$}
    \psfrag{ 1000}[c][c][1]{$10^3$}
    \psfrag{1}[r][r][1]{$1$}
    \psfrag{10}[r][r][1]{$10$}
    \psfrag{100}[r][r][1]{$10^2$}
    \psfrag{1000}[r][r][1]{$10^3$}
    \psfrag{10000}[r][r][1]{$10^4$}
    \psfrag{100000}[r][r][1]{$10^5$}
    \psfrag{1e+06}[r][r][1]{$10^6$}
    \resizebox{\figwidth}{!}{\includegraphics{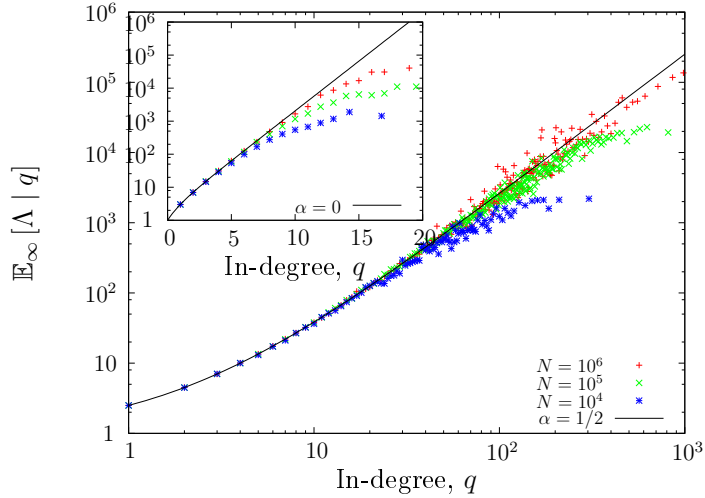}}
  \end{center}
  \caption{Figure shows average edge betweenness under the condition 
  that the in-degree $q$ is known as the function of $q$ on log-log plot. 
  Numerical data has been collected from $100$ 
  realizations of $N=10^4$ and $N=10^5$, and $10$ realizations of $N=10^6$ size
  networks at $\alpha=1/2$ parameter value. Inset shows the same scenario at 
  $\alpha=0$ parameter value on semi-logarithmic plot. Continuous lines show 
  analytic results of the infinite network limit \eqref{eq:deg-load} and 
  \eqref{eq:deg-load_a0}.}
  \label{fig:deg-load-avg}
\end{figure}

On Figure~\ref{fig:load-dist-1} comparison of analytic formula
\eqref{eq:ccdf_load} with simulation results is presented for $q=1$ and $q=2$.
The empirical \gls{ccdf} of rescaled edge betweenness, under the condition that
in-degree $q$ is known, is shown for $10^4$, $10^5$, and $10^6$ size networks,
at $\alpha=1/2$ parameter value. The empirical \glspl{ccdf} of rescaled edge
betweenness evidently collapse to the same curve for different size networks,
and they coincide precisely with my analytic result.

The expectation of the rescaled edge betweenness under the condition that in-degree
$q$ is known can be given by $\meanq{\infty}{\Lambda}=\meanq{\infty}{n_{\Lambda}+1}$.
Using \eqref{eq:mean_q_inf} and \eqref{eq:mean_q_inf_a0} I receive 
\begin{align}
  \label{eq:deg-load}
  \meanq{\infty}{\Lambda}
  &=\left(1-\alpha\right)\frac{\poch{q+1/\alpha}{1/\alpha}}{\poch{1/\alpha-1}{1/\alpha}}-1+\alpha,\\
  \lim_{\alpha\to0}\meanq{\infty}{\Lambda}&=2^{q+1}-1.
  \label{eq:deg-load_a0}
\end{align}
One can see that $\meanq{\infty}{\Lambda}\sim q^{1/\alpha}$ for $q\gg1$ if $\alpha>0$ and 
$\meanq{\infty}{\Lambda}\sim e^{q}$ for $q\gg1$ if $\alpha\to0$.

Analytic results \eqref{eq:deg-load} and \eqref{eq:deg-load_a0}, and simulation
data are shown in Figure~\ref{fig:deg-load-avg} at $\alpha=1/2$ and $\alpha=0$
parameter values. Numerical data has been collected from the same $10^4$,
$10^5$, and $10^6$ size networks as above. As the size of the network grows
a larger and larger range of the rescaled empirical data collapses to the same
analytic curve. On the high degree region some discrepancy can be observed due
to the finite scale effects.

Finally, let us note that the precise unconditional distribution of edge betweenness
$\prob{\tau}{L}=\sum_{n=0}^{\tau-1}\delta_{L,\left(n+1\right)\left(\tau-n\right)}\prob{\tau}{n}$
can be obtained from \eqref{eq:sol_marginal_n} as well. Furthermore, \gls{ccdf}
of the unconditional betweenness
$\ccdf{\tau}{L}=\sum_{n=n_L}^{\tau-n_L-1}\prob{\tau}{n}$ can be derived in
closed form:
\begin{equation}
  F_{\tau}^{c}(L)=\frac{\tau+1-\alpha}{\tau}
  \frac{\left(1-\alpha\right)\left(\tau-2n_L\right)}{\left(n_L+1-\alpha\right)\left(\tau-n_L+1-\alpha\right)}.
\end{equation}

For the sake of simplicity I have assumed during my calculations that
in-degrees of the ``younger'' nodes are provided. However, it is possible that
even though both in-degrees of every link are known, we cannot distinguish
them from each other, that is we cannot tell which is the ``younger'' node. How
could I extend my results to this scenario?  Let us consider a new edge when
it is connected to the network. The in-degree of the new node is obviously $0$.
The in-degree of the other node, to which the new node is connected, is equal
to or larger than one.  Due to preferential attachment the larger the in-degree
is the faster it grows. Even if preferential attachment is absent, the growth
rate of every in-degree is the same. Therefore, it is expected that the initial
deficit in the in-degree of the ``younger'' node grows or remains at the same
level during the evolution of the network. It follows that it is a reasonable
approximation to substitute the in-degree of the ``younger'' node $q$ with
$q_{\text{min}}=\min(q_1,q_2)$ in my formulas.

\section{Conclusions}
\label{sec:conclusions}

A typical network construction problem is to design network
infrastructure without wasting precious resources at places where not needed.
An appropriate design strategy is to allocate network resources
proportionally to the expected traffic. In a mean field approximation
the expected traffic is proportional to the number of shortest paths
going through a certain network element, that is the betweenness.  

The precise calculation of all the betweenness requires complete information on
the network structure.  In real life, however, the number of shortest paths is
often impossible to tell because the structure of the network is not fully
known. One of the practical results of this chapter is that the expectation of
edge betweenness can be estimated precisely when only limited local information
on network structure---the in-degree of the ``younger'' node---is available.

Another difficulty of network design is that the size of real networks is
finite. Moreover, the size of real networks is often so small that asymptotic
formulas can be applied only with unacceptable error. The other important
novelty of my results is that the derived formulas are exact even for finite
networks, which allows better design of realistic finite size networks.

Various statistical properties of evolving random trees have been investigated
in this chapter. I have focused on the cluster size, the in-degree and the edge
betweenness. I have considered the $m=1$ case of the BA model extended with
initial attractiveness for modeling random trees. Initial attractiveness allows
fine tuning of the scaling parameter. Moreover, in the limit of the tuning
parameter $\alpha\to0$ the applied model tends to a non-scale-free
structure, which is in many aspects similar to the classical ER model.
I was therefore able to investigate both the scale-free and the
non-scale-free scenario within the same framework.

I also presented conditional expectations of cluster size and in-degree for
both finite and unbounded networks.  I have found that asymptotically the
conditional cluster size grows with in-degree to the power of $1/\alpha$ and
the conditional in-degree grows with cluster size to the power of $\alpha$,
respectively. The ER limit has been discussed as well.  I have shown that the
conditional cluster size grows exponentially and the conditional in-degree
grows logarithmically when $\alpha\to0$.

I have derived the distribution of edge betweenness under the condition that
the corresponding in-degree is known. I have found that the conditional
expectation of edge betweenness grows linearly with the size of the network.
For the analysis of unbounded networks I have defined the rescaled edge
betweenness $\Lambda$, and derived its distribution and expectation under the
condition that in-degree $q$ is provided. My analytic results have been
verified at different network sizes and parameter values by extensive numerical
simulations.  I have demonstrated that numerical simulations fully confirm my
analytic results.

\chapter{Concluding remarks}
\label{cha:Conclusions}

The study of complex networks has evolved considerably in recent years.  An
interesting example of complex networks is the Internet, which has become part
of everyday life.  Two important aspects of the Internet, namely the properties
of its topology and the characteristics of its data traffic, have attracted
growing attention of the physics community.  My thesis has considered problems
of both aspects.

In the introduction I briefly presented an overview of the basic components of
Internet structure and traffic.  The workings of the Transmission Control
Protocol (\tcp), the primary algorithm governing traffic in the current
Internet, were discussed in more detail, since they are the main focus of
the first part of my analysis.  Most of the terminologies I use in this thesis
were also defined here.

In the next chapter I studied the stochastic behavior of \tcp\ in an elementary
network scenario consisting of a standalone infinite-sized buffer and an access
link.  This simple model might constitute the building blocks of more complex
Internet traffic.  I calculated the stationary distribution of the stochastic
congestion window process, which regulates the traffic of \tcp.  My analysis
not only considered the ideal congestion window dynamics, but also included the
effect of the fast recovery and fast retransmission (FR/FR) algorithms of \tcp.
Furthermore, I showed that my model can be extended further analytically to
involve the effect of link propagation delay, characteristic of Wide Area
Networks.  Various moments of the congestion window process were calculated.
An important achievement is that all the parameters are at hand in the entire
model, and no parameter fitting is necessary.  I also applied the mean field
approximation to describe many parallel \tcp\ flows.  My analytic results were
validated against packet level numerical simulations and the simulations agreed
to a high degree with the analytic formulas I derived.

I continued my thesis with the investigation of finite-sized semi-bottleneck
buffers, where packets can be dropped not only at the link, but also at the
buffer.  I demonstrated that the behavior of the system depends only on a
certain combination of the parameters.  Moreover, an analytic formula was
derived that gives the ratio of packet loss rate at the buffer to the total
packet loss rate.  This formula makes it possible to treat buffer-losses as if
they were link-losses.  I considered the effect of the FR/FR algorithms and I
calculated the probability distribution of the congestion window for both Local
and Wide Area Network scenarios.  I showed that a sharp peak might appear in
the window distribution due to the FR/FR mode of \tcp.  My analytical results
matched numerical simulations properly in the case of large buffer sizes and
small packet loss probabilities.  In the opposite range of parameters, however,
the slow start mechanism of \tcp\ plays a more important role and it cannot be
neglected completely from the precise description of the congestion window
dynamics.  Nevertheless, my calculations gave qualitatively correct results 
in these cases as well.  Hopefully, my methods, developed in this chapter, can 
be applied later for modeling the slow start mechanism.

In the last part of my thesis I studied computer networks from a structural
perspective.  The scaling exponent of the node connectivity could be tuned in
the network model that I  investigated.  In addition, the non-scale-free limit
of the node connectivity could also be investigated.  I demonstrated through
fluid simulations that the distribution of resources, specifically the link
bandwidth, has a serious impact on the global performance of a computer
network.  Then I analyzed the distribution of edge betweenness in a growing
scale-free tree under the condition that a local property, the in-degree of the
``younger'' node of an arbitrary edge, is known in order to find an optimum
distribution of link capacity.  The derived formula is exact even for
finite-sized networks.  I also calculated the conditional expectation of edge
betweenness, rescaled for infinite networks.  My analytic results were 
compared to numerical simulations that confirmed my calculations appropriately.

\begin{appendix}

\chapter[Mathematical proofs]{Mathematical proofs of the applied identities}

\section{Series expansion of $L(c)\,G(x)$}
\label{app:LG_expansion}

In this section the series expansion of $L(c)G(x)$ in $x$ is derived,
where $L(c)$ is defined in (\ref{eq:L_c}) and $G(x)$ in (\ref{eq:G}).  We 
prove two lemmas first:

\textbf{Lemma 1}
\textit{For $c\in\mathbb{R}$, $c\neq1$
  \begin{equation}
    \label{eq:G_expansion_1}
    \sum_{k=0}^{N}c^k\prod_{l=1}^{k}\frac1{1-c^l}=\prod_{k=1}^N\frac1{1-c^k}.
  \end{equation}
}
\begin{proof}
We prove the lemma by induction for $N$.  Indeed, for $N=1$ the formula is 
evidently true: $1+\frac{c}{1-c}=\frac1{1-c}$.  As the induction hypothesis 
suppose that the formula is true for $N$.  Then
\begin{align}
    \sum_{k=0}^{N+1}c^k\prod_{l=1}^{k}\frac1{1-c^l}
    &=\sum_{k=0}^{N}c^k\prod_{l=1}^{k}\frac1{1-c^l}
    +c^{N+1}\prod_{l=1}^{N+1}\frac1{1-c^l}\notag\\
    &=\prod_{l=1}^N\frac1{1-c^l}
    +c^{N+1}\prod_{l=1}^{N+1}\frac1{1-c^l}\\
    &=\left(1-c^{N+1}\right)\prod_{l=1}^{N+1}\frac1{1-c^l}
    +c^{N+1}\prod_{l=1}^{N+1}\frac1{1-c^l}
    =\prod_{l=1}^{N+1}\frac1{1-c^l}.
    \notag
\end{align}
\end{proof}

\textbf{Lemma 2}
\textit{For $c\in[0,1[$ 
  \begin{equation}
    \prod_{k=1}^{n-1}\left(1-c^k\right)=\sum_{k=0}^{\infty}c^{kn}
    \prod_{l=k+1}^{\infty}\left(1-c^l\right).
  \end{equation}
}
\begin{proof}
  This lemma is proven by induction for $n$.  For $n=1$ Lemma 1 proves the
  formula.  Indeed, multiply both sides of (\ref{eq:G_expansion_1}) by
  $\prod_{k=1}^N\left(1-c^k\right)$.  If $c\in[0,1[$ then we can take the 
  $N\to\infty$ limit, which provides exactly the formula to be proven.

  As the induction hypothesis let us suppose that the formula is true for
  $n$.  Then
  \begin{align}
      \prod_{k=1}^n\left(1-c^k\right)
      &=\left(1-c^n\right)\prod_{k=1}^{n-1}\left(1-c^k\right)
      =\prod_{k=1}^{n-1}\left(1-c^k\right)-c^n\sum_{k=0}^{\infty}c^{kn}
      \prod_{l=k+1}^{\infty}\left(1-c^l\right)\notag\\
      &=\prod_{k=1}^{n-1}\left(1-c^k\right)
      -\sum_{k=0}^{\infty}c^{\left(k+1\right)n}
      \prod_{l=k+1}^{\infty}\left(1-c^l\right)\notag\\
      &=\prod_{k=1}^{n-1}\left(1-c^k\right)
      -\sum_{k=0}^{\infty}c^{kn}\left(1-c^k\right)
      \prod_{l=k+1}^{\infty}\left(1-c^l\right)\notag\\
      &=\sum_{k=0}^{\infty}c^{k\left(n+1\right)}
      \prod_{l=k+1}^{\infty}\left(1-c^l\right),
  \end{align}
  which proves the formula for $n+1$.
\end{proof}

\textbf{Theorem}
\textit{For $x\in\mathbb{R}$ and $c\in[0,1[$
  \begin{equation}
    \label{eq:LG_expansion}
    L(c)\,G(x)=-\sum_{n=1}^{\infty}\frac1{n!}\prod_{l=1}^n
    \left(1-c^l\right)x^n.
  \end{equation}
}

\begin{proof}
  From the series expansion of the exponential function it follows that
  $E(x)=e^{-cx}-e^{-x}=-\sum_{n=1}^{\infty}\left(-1\right)^n
  \frac{1-c^n}{n!}x^n$.  Let us substitute this expression into the definition
  of $G(x)=\sum_{k=0}^{\infty}F(-c^kx)\prod_{l=1}^{k}\frac1{1-c^l}$\,:
  \begin{equation}
    \begin{split}
      L(c)\,G(x)&=-\prod_{l=1}^{\infty}\left(1-c^l\right)
      \sum_{k=0}^{\infty}\sum_{n=1}^{\infty}\left(-1\right)^n
      \frac{1-c^n}{n!}\left(-c^k x\right)^n
      \prod_{l=1}^{k}\frac1{1-c^l}\\
      &=-\sum_{n=1}^{\infty}\frac{1-c^n}{n!}\sum_{k=0}^{\infty}c^{kn}
      \prod_{l=k+1}^{\infty}\left(1-c^l\right)x^n\\
      &=-\sum_{n=1}^{\infty}\frac1{n!}\prod_{l=1}^n\left(1-c^l\right)x^n,
    \end{split}
  \end{equation}
  where we applied the Lemma~2 in the last equation in order to prove the 
  theorem.
\end{proof}

\section{Expansion of the Kronecker-delta function}
\label{app:kronecker_exp}
We have seen that the general solution of Eq.~(\ref{eq:master_eq}) is
\begin{equation}
  \probi{\tau}{n,q}=\sum_{\lambda_1,\lambda_2}C_{\lambda_1,\lambda_2}\,
  f(\tau)\,g(n)\,h(q),
\end{equation}
and the initial condition is $\probi{\tau_e}{n,q}=\delta_{n,0}\,\delta_{q,0}$, where
\begin{equation}
  \delta_{n,m}=
  \begin{cases}
    1, &\textrm{if $n=m$},\\
    0, &\textrm{if $n\neq m$}
  \end{cases}
\end{equation}
is the Kronecker-delta function, and $n$ and $m$ are integers.
Coefficients $C_{\lambda_1,\lambda_2}$ are calculated in this 
section.  First we show that
\begin{equation}
  \label{eq:delta_exp}
  \delta_{n,0}=\sum_{k=0}^n\frac{\left(-1\right)^k}{k!}\frac1{\Gamma(n-k+1)}.
\end{equation}
Note that we can consider $m=0$ without any loss of generality, since
$\delta_{n,m}\equiv\delta_{n-m,0}$. 

If $n<0$, then the summand in (\ref{eq:delta_exp}) is indeed zero by
definition. If $n>0$, then
\begin{equation}
  \sum_{k=0}^n\frac{\left(-1\right)^k}{k!}\frac1{\Gamma(n-k+1)}
  =\frac1{n!}\sum_{k=0}^n\binom{n}{k}\left(-1\right)^k=0
\end{equation}
follows from the binomial theorem. Finally, for $n=0$,
\begin{equation}
  \sum_{k=0}^0\frac{\left(-1\right)^k}{k!}\frac1{\Gamma(-k+1)}=
  \frac{\left(-1\right)^0}{0!}\frac1{\Gamma(1)}=1.
\end{equation}

Coefficients $C_{\lambda_1,\lambda_2}$ can be obtained from the
term by term comparison of $\probi{\tau_e}{n,q}
=\sum_{\lambda_1,\lambda_2}C_{\lambda_1,\lambda_2}f(\tau_e)\,g(n)\,h(q)$
with the expansion of the initial condition $\delta_{n,0}\,\delta_{q,0}$,
shown above. One can easily confirm with the help of identity 
$f(n)\delta_{n,0}\equiv f(0)\delta_{n,0}$ that the same terms appear on 
both sides, if $\lambda_1=-k_1$, and 
$\lambda_2=-\alpha k_2$, and coefficients $C_{k_1,k_2}$ are the following:
\begin{equation}
  C_{k_1,k_2}=\frac{\left(-1\right)^{k_1+k_2}}{k_1!\,k_2!}
  \frac{\Gamma(\tau_e+1-\alpha)}{\Gamma(\tau_e-k_1)}\frac1{\Gamma(-\alpha k_2)}
  \frac{1}{\Gamma(1/\alpha-1)}.
\end{equation}

Finally, to obtain (\ref{eq:sol_joint}) the summation for $k_1$ can be 
carried out explicitly:
\begin{align*}
  \sum_{k_1=0}^{n}\frac{\left(-1\right)^{k_1}}{k_1!\,\Gamma(n-k_1+1)}\frac{\Gamma(\tau-k_1)}{\Gamma(\tau_e-k_1)}
  =\frac{\Gamma(\tau-\tau_e+1)}{\Gamma(n+1)\Gamma(\tau_e)}\frac{\Gamma(\tau-n)}{\Gamma(\tau-\tau_e-n+1)}
\end{align*}

\section{The $\alpha\to0$ limit of joint distribution $\prob{\tau}{n,q}$}
\label{app:ER_limit}

In this section we prove that the ER limit of the joint probability
$\prob{\tau}{n,q}$ is (\ref{eq:sol_joint_a0}).

\textbf{Theorem}
\textit{
  Let us consider $\prob{\tau}{n,q}$ as defined in (\ref{eq:sol_joint}), where
  $0<q<n<\tau$ are integers. Then the following limit holds:
  \begin{equation}
    \lim_{\alpha\to0}\prob{\tau}{n,q}=\frac{\tau+1}{\tau\,\Gamma(n+3)}
    \sum_{k=q-1}^{n-1}\left(-1\right)^{n-1-k}
    S_{n-1}^{\left(k\right)}\binom{k}{q-1}.
  \end{equation}
  \label{thm:ER_limit}
  }
  
\begin{proof}
First, let us note that $\sumterm{n,q}$ in (\ref{eq:sol_joint}) can be
rewritten in the following equivalent form: $\sumterm{n,q}=\alpha\sum_{k=0}^{q-1}
\frac{\left(-1\right)^k\poch{1-\alpha-\alpha k}{n-1}}{k!\left(q-1-k\right)!}$.
Next, Pochhammer's symbol $\poch{1/\alpha-1}{q}$ is replaced with its asymptotic form:
$\poch{1/\alpha-1}{q}=1/\alpha^q\left(1+\Ordo{\alpha}\right)$.
After the obvious limits have been evaluated the following equation is obtained:
\begin{equation}
  \lim_{\alpha\to0}\prob{\tau}{n,q}=\frac{\tau+1}{\tau\,\Gamma(n+3)}
  \lim_{\alpha\to0}\frac{
    \sum_{k=0}^{q-1}\frac{\left(-1\right)^k\poch{1-\alpha-\alpha k}{n-1}}{k!\left(q-1-k\right)!}}
  {\alpha^{q-1}}.
\end{equation}

The above limit, by definition, can be substituted with $q-1$ order differential at $\alpha=0$, 
if all the lower order derivates of the sum are zero at $\alpha=0$. Indeed,
\begin{align*}
  \lim_{\alpha\to0}\frac{
    \sum_{k=0}^{q-1}\frac{\left(-1\right)^k\poch{1-\alpha-\alpha k}{n-1}}{k!\left(q-1-k\right)!}}
  {\alpha^{q-1}}
  &=\frac{1}{m!}\frac{d^m}{d\alpha^m}
  \sum_{k=0}^{q-1}\frac{\left(-1\right)^k\poch{1-\alpha-\alpha k}{n-1}}{k!\left(q-1-k\right)!}
  \Biggr|_{\alpha=0}\\
  &=\frac{1}{m!}\frac{d^m\poch{1+\alpha}{n-1}}{d\alpha^m}
  \Biggr|_{\alpha=0}
  \sum_{k=0}^{q-1}\frac{\left(-1\right)^k\left(-k-1\right)^m}{k!\left(q-1-k\right)!},
\end{align*}
where the sum is $0$ if $m<q-1$ and $1$ if $m=q-1$. Therefore, the limit can be
transformed to
\begin{equation}
  \lim_{\alpha\to0}\prob{\tau}{n,q}=\frac{\tau+1}{\tau\,\Gamma(n+3)}\frac{1}{\left(q-1\right)!}
  \frac{d^{q-1}\poch{1+\alpha}{n-1}}{d\alpha^{q-1}}\biggr|_{\alpha=0}.
\end{equation}
Finally, let us consider the power expansion of Pochhammer's symbol:
$\poch{x}{m}=\sum_{k=0}^{m}\left(-1\right)^{n-k} S_m^{\left(k\right)}x^k$,
where $S_m^{\left(k\right)}$ are the Stirling numbers of the first kind.
The expansion formula has been applied at $x=1+\alpha$ and $m=n-1$, which implies
\begin{align*}
  \lim_{\alpha\to0}\prob{\tau}{n,q}
  &=\frac{\tau+1}{\tau\,\Gamma(n+3)}\sum_{k=q-1}^{n-1}\frac{\left(-1\right)^{n-1-k}S_m^{\left(k\right)}}
  {\left(q-1\right)!}\frac{d^{q-1}\left(1+\alpha\right)^k}{d\alpha^{q-1}}\biggr|_{\alpha=0}\\
  &=\frac{\tau+1}{\tau\,\Gamma(n+3)}\sum_{k=q-1}^{n-1}\left(-1\right)^{n-1-k}S_{n-1}^{\left(k\right)}\binom{k}{q-1}.
\end{align*}
\end{proof}

\end{appendix}
\backmatter
\clearpage

\printglossaries

\addcontentsline{toc}{chapter}{Bibliography}
\bibliographystyle{unsrtnat}
\bibliography{thesis}

\cleardoublepage
\addcontentsline{toc}{chapter}{Summary}
\pagestyle{empty}
\enlargethispage{3em}
\begin{center}
\LARGE \textbf{Summary}
\end{center}
\smallskip

The study of complex networks has evolved considerably in recent years.  An
interesting example of complex networks is the Internet, which has become part
of everyday life.  Two important aspects of the Internet, namely the properties
of its topology and the characteristics of its data traffic, have attracted
growing attention of the physics community.  My thesis has considered problems
of both aspects.

First I studied the stochastic behavior of \tcp, the primary algorithm
governing traffic in the current Internet, in an elementary network scenario
consisting of a standalone infinite-sized buffer and an access link.  I
calculated the stationary distribution of the stochastic congestion window
process, which regulates the traffic of \tcp.  My analysis not only considered
the ideal congestion window dynamics, but also included the effect of the fast
recovery and fast retransmission (FR/FR) algorithms.  Furthermore, I showed
that my model can be extended further to involve the effect of link propagation
delay, characteristic of WAN.  An important achievement is that no parameter
fitting is necessary in my model.  I also applied the mean field approximation
to describe many parallel \tcp\ flows.  After having been validated against
packet level numerical simulations, my analytic results agreed almost
perfectly.

I continued my thesis with the investigation of finite-sized semi-bottleneck
buffers, where packets can be dropped not only at the link, but also at the
buffer.  I demonstrated that the behavior of the system depends only on a
certain combination of the parameters.  Moreover, an analytic formula was
derived that gives the ratio of packet loss rate at the buffer to the total
packet loss rate.  This formula makes it possible to treat buffer-losses as if
they were link-losses.  I calculated the probability distribution of the
congestion window for both LAN and WAN scenarios.  I showed that a sharp peak
might appear in the window distribution due to the FR/FR mode of \tcp.  My
analytical results matched numerical simulations properly.

In the last part of my thesis I studied computer networks from a structural
perspective.  I demonstrated through fluid simulations that the distribution of
resources, specifically the link bandwidth, has a serious impact on the global
performance of the network.  Then I analyzed the distribution of edge
betweenness in a growing scale-free tree under the condition that a local
property, the in-degree of the ``younger'' node of an arbitrary edge, is known
in order to find an optimum distribution of link capacity.  The derived formula
is exact even for finite-sized networks.  I also calculated the conditional
expectation of edge betweenness, rescaled for infinite networks.

\cleardoublepage
\selectlanguage{magyar}
\enlargethispage{3em}
\begin{center}
\LARGE \textbf{Összefoglalás}
\end{center}
\smallskip

Az elmúlt években a komplex hálózatok kutatása rendkívül sokat fejlődött.  A
komplex hálózatok egyik legérdekesebb példája a mára a mindennapi élet részévé
vált \emph{internet}.  Az internet két fontos területe -- a topológiájának
tulajdonságai és a rajta folyó adatforgalom jellemzői -- iránt az utóbbi időben
a fizikusok körében is egyre nagyobb az érdeklődés.  Dolgozatomban az említett
két terület néhány kérdését vizsgáltam.

Elsőként a jelenlegi internet legfontosabb forgalomszabályozó algoritmusának, a
TCP-nek a sztochasztikus viselkedését tanulmányoztam egy elemi hálózati
konfigurációban, mely egy egyedülálló bufferből, és egy hozzá kapcsolódó
vezetékből állt. Meghatároztam a TCP-forgalmat szabályozó torlódási ablaknak a
stacionárius eloszlását. Vizsgálatomban nem csupán az ideális torlódási ablak
dinamikát tekintettem, hanem figyelembe vettem a FR/FR (fast recovery/fast
retransmission) algoritmusok hatását is. Megmutattam továbbá, hogy a modellem
hogyan általánosítható úgy, hogy a vezetékeken fellépő csomagkésleltetést is
figyelembe vegye.  Fontos eredmény, hogy nem szükséges ismeretlen
paramétert illenszteni a modellben.  A modellt párhuzamosan működő TCP-k
leírására átlagtér-közelítésben alkalmaztam. Az analítikus eredményeket
csomag-szintű numerikus szimulációkkal összevetve rendkívül jó egyezést kaptam.

A disszertációt véges méretű bufferek vizsgálatával folytattam, ahol a csomagok
nem csak a vezetéken, hanem a bufferben is elveszhetnek.  Megmutattam, hogy a
rendszer viselkedése csupán a paraméterek egy bizonyos kombinációjától függ.
Levezettem továbbá egy analítikus formulát, mely megadja a bufferben eldobott,
és a vezetéken elveszett csomagok arányát. Ez a formula lehetővé teszi, hogy a
bufferben történő csomagvesztéseket úgy tekintsük, mintha az a vezetéken
történt volna. Kiszámoltam a torlódási ablak eloszlását mind lokális (LAN),
mind tág (WAN) hálózati környezetben. Megmutattam, hogy az FR/FR algoritmusok
miatt egy éles csúcs jelenik meg az eloszlásban. Az analítikus eredmények jól
egyeztek a szimulációkkal.

Dolgozatom utolsó részében szerkezeti szempontból vizsgáltam komplex
számítógépes hálózatokat. Folyadékközelítésű szimulációkkal bemutattam, hogy az
erőforrások, különösképpen a vezetékek sávszélességének elosztása, jelentősen
befolyásolja a hálózat összteljesítményét.  Ezután annak érdekében, hogy az
él-kapacitások optimális elrendezését meghatározzam, az él-köztesség
eloszlását vizsgáltam növekvő skála-független fában azzal a feltétellel, hogy
tetszőleges él ,,fiatalabb'' csúcsának bejövő fokszáma ismert. A levezetett
formula még véges méretű hálózatokra is egzakt.  Végül megadtam a végtelen
hálózatokra átskálázott él-köztesség feltételes várható értékét.

\end{document}